\documentclass[twocolumn]{aastex62}

\usepackage{color}
\usepackage{amsmath}

\newcommand{\Mjup}{\mbox{$M_\mathrm{Jup}$}}
\newcommand{\Msun}{\mbox{$M_{\odot}$}}

\shorttitle{Eccentricity Distributions of Imaged Exoplanets and Brown Dwarf Companions}
\shortauthors{Bowler, Blunt, \& Nielsen}

\begin{document}

\title{Population-Level Eccentricity Distributions of Imaged Exoplanets and Brown Dwarf Companions: \\ Dynamical Evidence for Distinct Formation Channels\footnote{Some 
of the data presented herein were obtained at the W.M. Keck Observatory, which is operated as a scientific partnership 
among the California Institute of Technology, the University of California and the National Aeronautics and Space Administration. 
The Observatory was made possible by the generous financial support of the W.M. Keck Foundation.  
Based in part on data collected at Subaru Telescope, which is operated by the National Astronomical Observatory of Japan.
}}

\correspondingauthor{Brendan P. Bowler}
\email{bpbowler@astro.as.utexas.edu}

\author[0000-0003-2649-2288]{Brendan P. Bowler}
\affiliation{Department of Astronomy, The University of Texas at Austin, Austin, TX 78712, USA}

\author{Sarah C. Blunt}
\affiliation{Department of Astronomy, California Institute of Technology, Pasadena, CA, USA}
\affiliation{Center for Astrophysics $|$ Harvard \& Smithsonian, Cambridge, MA, USA}
\affiliation{NSF Graduate Research Fellow}

\author{Eric L. Nielsen}
\affiliation{Kavli Institute for Particle Astrophysics and Cosmology, Stanford University, Stanford, CA 94305, USA}

\begin{abstract}

The orbital eccentricities of directly imaged exoplanets and brown dwarf companions provide clues about their
formation and dynamical histories.    
We combine new high-contrast imaging observations of substellar companions
obtained primarily with Keck/NIRC2 
together with astrometry from the literature to test for differences in the
population-level eccentricity distributions of 27 long-period giant planets and brown dwarf
companions between 5--100 AU using hierarchical Bayesian modeling.  
Orbit fits are performed in a uniform manner for companions with short orbital arcs;
this typically results in broad constraints for individual eccentricity distributions, but together as an ensemble these 
systems provide valuable insight into their collective underlying orbital patterns.  
The shape of the eccentricity distribution function for our full sample 
of substellar companions is approximately flat from $e$=0--1.  When subdivided by companion mass and mass ratio, the 
underlying distributions for giant planets and brown dwarfs show significant differences.
Low mass ratio companions preferentially have low eccentricities, similar to the orbital properties of warm Jupiters
found with radial velocities and transits.  
We interpret this as evidence for \emph{in situ} formation on largely undisturbed orbits within massive, extended disks.
Brown dwarf companions exhibit a broad peak 
at $e$ $\approx$ 0.6--0.9 with evidence for a dependence on orbital period.
This closely resembles the orbital properties and period-eccentricity trends of wide (1--200 AU) stellar binaries,
suggesting that brown dwarfs in this separation range predominantly form in a similar fashion.
We also report evidence that the ``eccentricity dichotomy'' observed at small separations extends to planets on wide orbits:
the mean eccentricity for the multi-planet system HR 8799 is lower than for systems with single planets.
In the future, larger samples and continued astrometric orbit monitoring will help establish whether these eccentricity distributions
correlate with other parameters such as stellar host mass, multiplicity, and age.

\end{abstract}

\keywords{brown dwarfs --- planets and satellites: gaseous planets --- planets and satellites: formation}

\section{Introduction}{\label{sec:intro}}

The orbital eccentricities of exoplanets directly trace their formation and dynamical histories.
Planets are expected to form on circular, coplanar orbits within protoplanetary disks, but can
develop non-zero eccentricities through dynamical interactions 
with other planets (e.g., \citealt{Rasio:1996ie}; \citealt{Weidenschilling:1996ts}; \citealt{Juric:2008db}; \citealt{Ford:2008jo}; \citealt{Dawson:2013wf}; \citealt{Petrovich:2016ee}),
secular Kozai-Lidov perturbations with a massive outer companion (e.g., \citealt{Naoz:2016cr}; \citealt{Mustill:2017jra}),
or planet-disk interactions (e.g., \citealt{Goldreich:2003cc}).
Over time, the eccentricities of close-in planets can be damped due to tidal dissipation with the host star (e.g., \citealt{Ogilvie:2014ia}) and 
at young ages as a result of torques and dynamical friction within a gas or planetesimal disk (\citealt{Duffell:2015cj}; \citealt{Morbidelli:2018hra}).

The observed eccentricities of giant planets measured from radial velocity surveys 
span the entire range of bound orbits ($0 \le e < 1$), 
in stark contrast to the nearly circular orbits of gas and ice giants in the Solar System ($e$$<$0.05).
The eccentricity-period distribution of exoplanets within $\approx$3 AU is consistent with having been 
shaped by tidal circularization at 
short orbital periods and planet-planet scattering at longer orbital periods,
with more massive planets having higher eccentricities, on average (e.g., \citealt{Chatterjee:2008gd}; \citealt{Winn:2015jt}).
\citet{Ma:2014cc} find that this trend continues into the brown dwarf regime:
companions above $\approx$40~\Mjup \ and beyond the tidal circularization radius 
exhibit an approximately flat eccentricity distribution resembling the orbits of binary stars 
(e.g., \citealt{Raghavan:2010gd}; \citealt{Duchene:2013il}).
This similarity suggests that high-mass brown dwarfs within a few AU predominantly form like stellar binaries.

At wider separations, high-contrast imaging has uncovered over one hundred substellar companions spanning
a large range of separations ($\approx$5--8000 AU) and masses ($\approx$2--75~\Mjup; see 
compilations by, e.g., \citealt{Zuckerman:2009gc}, \citealt{Faherty:2010gt}, 
\citealt{Bowler:2016jk}, \citealt{Chauvin:2018fi}, and \citealt{Deacon:2014ey}).  
Many formation routes have been proposed for these brown dwarf and planetary-mass companions:
core-nucleated or pebble-assisted accretion 
(e.g., \citealt{Pollack:1996jp}; \citealt{Alibert:2005ee}; \citealt{Lambrechts:2012gr}; \citealt{Lambrechts:2014iq});
gravitational instabilities in protoplanetary disks (e.g., \citealt{Boss:1997di}; \citealt{Durisen:2007wg}; \citealt{Vorobyov:2013dl}); 
fragmentation of collapsing molecular cloud cores (e.g., \citealt{Boss:2001vw}; \citealt{Bate:2002iq}; \citealt{Bate:2010wq});
and gravitational outward scattering by closer-in, higher-mass companions  
(e.g., \citealt{Boss:2006ge}; \citealt{Scharf:2009eq}; \citealt{Veras:2009br}).
Each formation mechanism operates over large overlapping windows of 
companion mass, orbital separation, and time, which has made it
difficult to observationally distinguish the dominant origin of this population.
For example, the discovery of extremely low-mass (but high mass ratio) binaries implies that cloud fragmentation can produce
few-$\Mjup$ objects at the opacity limit for fragmentation (e.g., 2M1207--3932 b, \citealt{Chauvin:2004cy}; 
2M0441+2301 Bb, \citealt{Todorov:2010cn}, \citealt{Bowler:2015en}; 2M1119--1137 AB, \citealt{Best:2017bra})\footnote{The discovery 
of isolated planetary-mass objects bolsters this conclusion, although these objects could also represent
ejected planets (e.g., \citealt{Forgan:2014epa}).}
whereas the orbital architectures of some directly imaged planetary systems indicate they formed within a disk 
(e.g., HR 8799 bcde, \citealt{Marois:2008ei}, \citealt{Marois:2010gpa}; $\beta$ Pic b, \citealt{Lagrange:2010fsa}; 
HD 95086 b, \citealt{Rameau:2013ds}, \citealt{Rameau:2016dx}, \citealt{Chauvin:2018ib}).

These formation channels predict two broad patterns for the orbital eccentricities of companions:
objects that assembled in protoplanetary disks  
(and without subsequent orbital evolution) should have low eccentricities, whereas those
that formed from cloud fragmentation or migrated via outward scattering should exhibit a
broad range of eccentricities (\citealt{ambartsumian:1937wd}; \citealt{Veras:2009br}; \citealt{Bate:2012hy}).  
The orbital periods of most widely-bound substellar companions are prohibitively long---1000 years 
at 100 AU for a Sun-like host star and over 10$^4$ years at 500 AU---to 
detect orbital motion given the limited time baselines since their discoveries.
Indeed, orbital motion has only been measured for a few substellar companions beyond 100 AU  
(e.g., GQ Lup B, \citealt{Ginski:2014ef}, \citealt{Wu:2017kd}; 
ROXs12 B, \citealt{Bryan:2016eo}; GSC 6214-210 B, \citealt{Pearce:2019iv}).
On the other hand, the majority of imaged planets and brown dwarfs within 100 AU 
have shown slight but significant orbital motion after only a few years of monitoring.

In this study we combine new adaptive optics imaging observations of substellar companions 
with astrometry from the literature
to uniformly constrain the orbits and 
underlying population-level eccentricity distributions of directly imaged giant planets ($\lesssim$15~\Mjup)
and brown dwarf companions ($\approx$15--75~\Mjup).
Each system typically traces out a short orbit arc, resulting in a broad eccentricity posterior distribution, but 
assembling them into a large sample allows us to 
infer population-level properties of these objects 
using hierarchical Bayesian inference.

\begin{deluxetable*}{lccccccccc}[!htb]
\renewcommand\arraystretch{0.9}
\tabletypesize{\footnotesize}
\setlength{ \tabcolsep } {.1cm} 
\tablewidth{0pt}
\tablecolumns{10}
\tablecaption{Observations and Astrometry of Substellar Companions\label{tab:newastrometry}}
\tablehead{
       \colhead{Name} & \colhead{Telescope/}  & \colhead{UT Date} & \colhead{Epoch} & \colhead{Filter/} &  \colhead{$N$ $\times$ Coadds $\times$ $t_\mathrm{exp}$}  &  \colhead{$\theta_\mathrm{rot}$} &  \colhead{Separation}  & \colhead{P.A.}  & \colhead{Comp.} \\
       \colhead{}        & \colhead{Instrument}  & \colhead{(Y-M-D)} &  \colhead{(UT)} & \colhead{Coronagraph} & \colhead{(s)}  &  \colhead{($\degr$)} & \colhead{(mas)}  & \colhead{($\degr$)} & \colhead{SNR}
        }   
\startdata
HD 49197 B &  Subaru/HiCIAO & 2011-12-28      &     2011.990  & $K_S$/cor300  &  60 $\times$ 1 $\times$ 30 &  19 &    913 $\pm$ 15  &          77.1 $\pm$ 0.7  &  26 \\
HD 49197 B &  Keck/NIRC2 & 2014-12-04          &    2014.924   & $K_S$/cor600  &  40 $\times$ 1 $\times$ 60  &  17  &   875 $\pm$ 5  &          75.9 $\pm$ 0.3  &  816 \\
HD 49197 B &  Keck/NIRC2 & 2016-03-22          &     2016.223  &  $K_S$/cor600  & 40 $\times$ 2 $\times$ 15  &  11  & 874 $\pm$ 5  &         76.5 $\pm$ 0.3  & 488 \\
HD 49197 B &  Keck/NIRC2 & 2018-01-30          &     2018.080  &  $K_S$/cor600  &  27 $\times$ 1 $\times$ 30  &15  &   845 $\pm$ 5  &         76.1 $\pm$ 0.3  & 84  \\
GJ 504 B      &  Keck/NIRC2 & 2016-03-22          &     2016.223  &  $H$/cor600  &  100 $\times$ 10 $\times$ 3  &  91  &  2504 $\pm$ 5   &         322.7 $\pm$ 0.4  &  13  \\
GJ 504 B      &  Keck/NIRC2 & 2018-01-30          &     2018.081  &  $H$/cor600  &  160 $\times$ 10 $\times$ 3  &  119  &   2503 $\pm$ 5   &         320.8 $\pm$ 0.3  & 16  \\
HD 19467 B  &  Keck/NIRC2 & 2018-01-30          &     2018.080  &  $H$/cor600 &  81 $\times$ 6 $\times$ 5  & 23  &  1628 $\pm$ 5 &     239.5 $\pm$ 0.3  &  26  \\
$\kappa$ And B &  Keck/NIRC2 & 2016-06-27      &     2016.489  &  $H$/cor600  &  10 $\times$ 10 $\times$ 2  &  3  &   965 $\pm$ 5   &     51.3 $\pm$ 0.3  &  9 \\
HD 1160 B    &  Keck/NIRC2 & 2018-01-30          &     2018.080  &  $K_S$/cor600  &  17 $\times$ 1 $\times$ 5 & $\cdots$ &  790 $\pm$ 5  &      245.1 $\pm$ 0.3  & 31  \\
1RXS0342+1216 B &  Keck/NIRC2 & 2018-01-30 &    2018.080  &   $K_S$/cor600  &  12 $\times$ 1 $\times$ 20 & $\cdots$  &  772.3 $\pm$ 1.8 &        19.6 $\pm$ 0.10  &  360 \\
CD-35 2722 B &  Keck/NIRC2 &  2018-01-30        &    2018.080  &  $K_S$/cor600  & 10 $\times$ 1 $\times$ 30  &  $\cdots$ &   2925 $\pm$  2     &      241.07 $\pm$ 0.10  &  116\\
DH Tau B       &  Keck/NIRC2 &   2018-01-30        &    2018.080   &   $K_S$/cor600  &  3 $\times$ 1 $\times$ 60  & $\cdots$ &  2354 $\pm$  2      &      138.46 $\pm$  0.10  &  90\\
HD 23514 B   &  Keck/NIRC2 &   2018-01-30        &    2018.080    &  $K_S$/cor600  &  16 $\times$ 1 $\times$ 30  &  $\cdots$  &  2648 $\pm$  2    &        227.13 $\pm$ 0.10  & 34\\
Ross 458 B    &  Keck/NIRC2 &   2016-03-22        &    2016.223    & $K_S$  &  10 $\times$ 100 $\times$ 0.014  &  $\cdots$ & 505.4 $\pm$ 1.7  &       51.63 $\pm$ 0.10  &  33 \\
Ross 458 B    &  Keck/NIRC2 &   2018-01-30        &    2018.081    &  $K_S$  &  39 $\times$ 100 $\times$ 0.01  &  $\cdots$ &  361.6 $\pm$ 1.7  &       21.71 $\pm$ 0.10  &  10  \\
TWA 5 B        &  Keck/NIRC2 &    2018-01-30        &    2018.081    &   $K_S$/cor600  &  10 $\times$ 1 $\times$ 10  &  $\cdots$ &  1852.1 $\pm$ 1.9  &  353.09 $\pm$ 0.10  & 60 \\
2M1559+4403 B  &  Keck/NIRC2 & 2018-01-30      &   2018.081    &  $K_S$/cor600  &  10 $\times$ 10 $\times$ 3  &  $\cdots$ & 5609 $\pm$ 3         &      284.27 $\pm$ 0.10  & 646  \\
1RXS2351+3127 B  &  Keck/NIRC2 & 2019-07-07     &   2019.513    &  $H$/cor600  &  5 $\times$ 1 $\times$ 30  &  $\cdots$ & 2395 $\pm$ 2         &      90.71 $\pm$ 0.10  & 195  \\
\enddata
\end{deluxetable*}

In Section~\ref{sec:obs} we describe our adaptive optics imaging from Keck Observatory and Subaru Telescope.
Our updated orbit fits for systems with new data are summarized in Section~\ref{sec:orbits}.
We provide uniform orbit fits for our sample of substellar companions and 
present results using hierarchical Bayesian modeling in Section~\ref{sec:eccentricities}.
Implications are discussed  in the broader context of formation scenarios in Section~\ref{sec:discussion}.
We summarize our findings in Section~\ref{sec:conclusions}.

\section{Observations}{\label{sec:obs}}

\subsection{Subaru/HiCIAO Adaptive Optics Imaging}{\label{sec:hiciao}}

We targeted HD 49197 using the High Contrast Instrument for the
Subaru Next Generation Adaptive Optics (HiCIAO; \citealt{Hodapp:2008cd}; \citealt{Suzuki:2010by}) near-infrared imager
coupled with the AO188 adaptive optics system (\citealt{Hayano:2010kj})
at Subaru Telescope on UT 2011 December 28 (see Table~\ref{tab:newastrometry} for details).
Conditions were photometric but the seeing was poor and variable; UKIRT reported
$K$-band natural seeing measurements between 1.5--2$''$ during our observations.
We acquired a total of 60 frames in $K_S$ band, each with an integration time of 30 s.
Observations were taken using natural guide star adaptive optics in pupil-tracking 
(angular differential imaging; \citealt{Liu:2004kk}; \citealt{Marois:2006df})
mode, which uses field rotation to distinguish speckles from real point sources.
HD 49197 was placed behind the 300 mas diameter opaque Lyot coronagraph during the ADI sequence,
which spanned 19$\degr$ of sky rotation.

\begin{figure*}
  \vskip -.9 in
  \hskip -0.7 in
  \resizebox{9.5in}{!}{\includegraphics{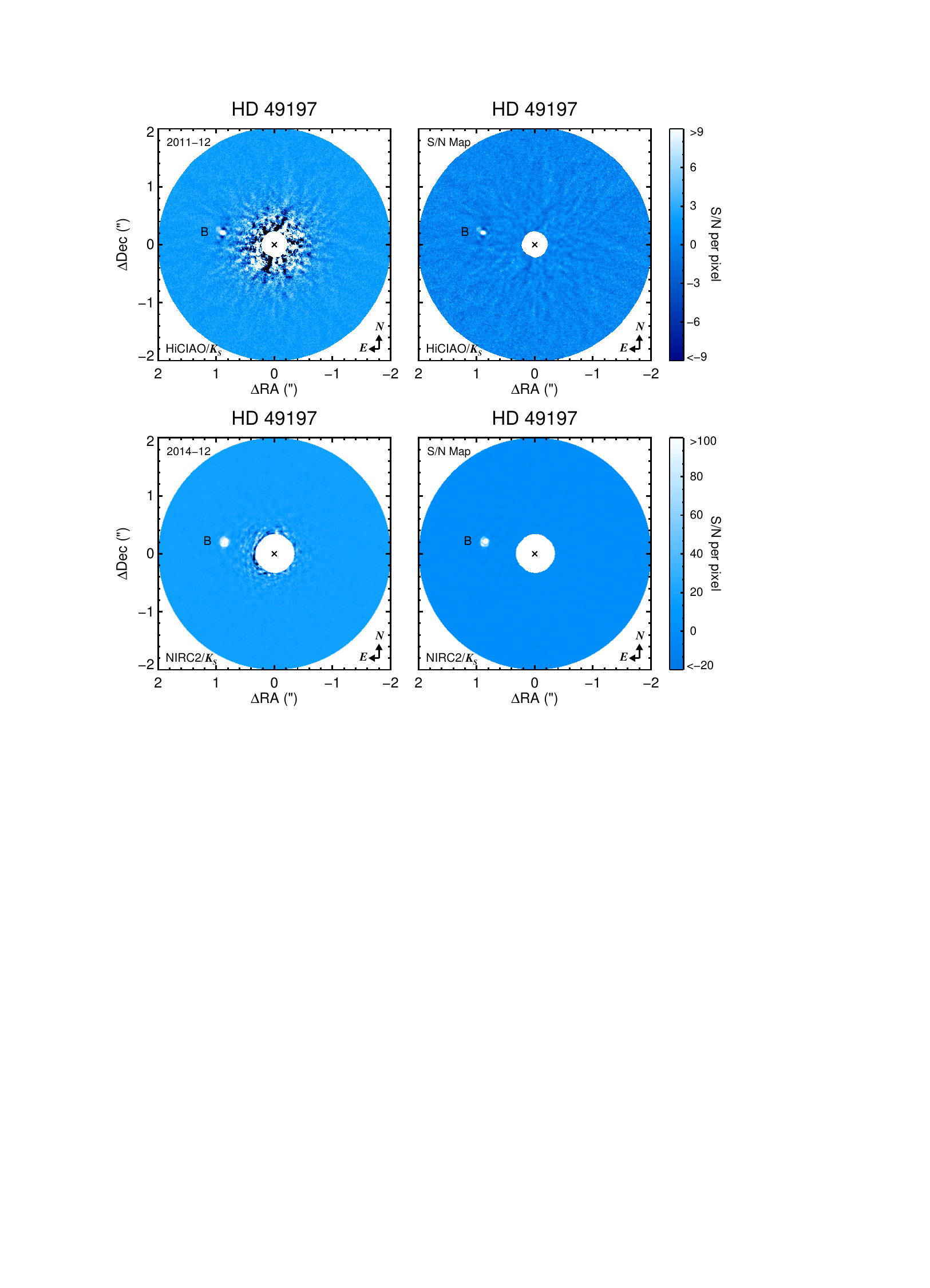}}
  \vskip -5.9 in
  \caption{$K_S$-band observations of HD 49197 with Subaru/HiCIAO in 2011 (top) and Keck/NIRC2 in 2014 (bottom).
  The brown dwarf companion HD 49197 B is clearly recovered in the PSF-subtracted, median-combined frames (left) and 
  signal-to-noise maps (right) with a signal to noise of 26 in the 2011 HiCIAO epoch and 816 in the 2014 NIRC2 epoch. 
  North is up and East is to the left. The color bars reflect the signal-to-noise pixel values of the S/N maps. \label{fig:adi1} } 
\end{figure*}

Data reduction and PSF subtraction follow the steps detailed in \citet{Bowler:2015ja}.
Systematic bias stripes from the detector readout electronics are measured and subtracted,
cosmic rays and bad pixels are removed, then images are divided by a normalized flat field.
The $K_S$-band distortion solution from \citet{Bowler:2015ja} is applied to each image to correct for
optical aberrations.  
The corresponding $K_S$-band plate scale of 9.67 $\pm$ 0.03 mas pixel$^{-1}$ is adopted for this data set.
The typical residual rms on the distortion correction is 1.2 pix, or 11.6 mas. 
Celestial north was found to be aligned with the detector columns to within the measurement
errors, so no rotation was applied and a value of 0.0 $\pm$ 0.1$\degr$ is adopted.
Images are registered using a 2D elliptical Gaussian fit to the PSF wings surrounding
the coronagraph and then assembled into a data cube.
Finally, PSF subtraction is carried out using ``aggressive'' and ``conservative'' implementations of PSF subtraction with 
the Locally-Optimized Combination of Images (LOCI) algorithm (\citealt{Lafreniere:2007bg}).
The aggressive subtraction using LOCI parameters $W$=8, $N_A$=300, $g$=1, $N_{\delta}$=0.5, $dr$=2
produced a higher signal-to-noise ratio for the modest-contrast ($\Delta$$K_S$$\approx$8~mag) companion
HD 49197 B so we adopt this version of the reduction here.
The final processed image is shown in Figure~\ref{fig:adi1}.


\begin{figure*}
  \vskip -.9 in
  \hskip -0.7 in
  \resizebox{9.5in}{!}{\includegraphics{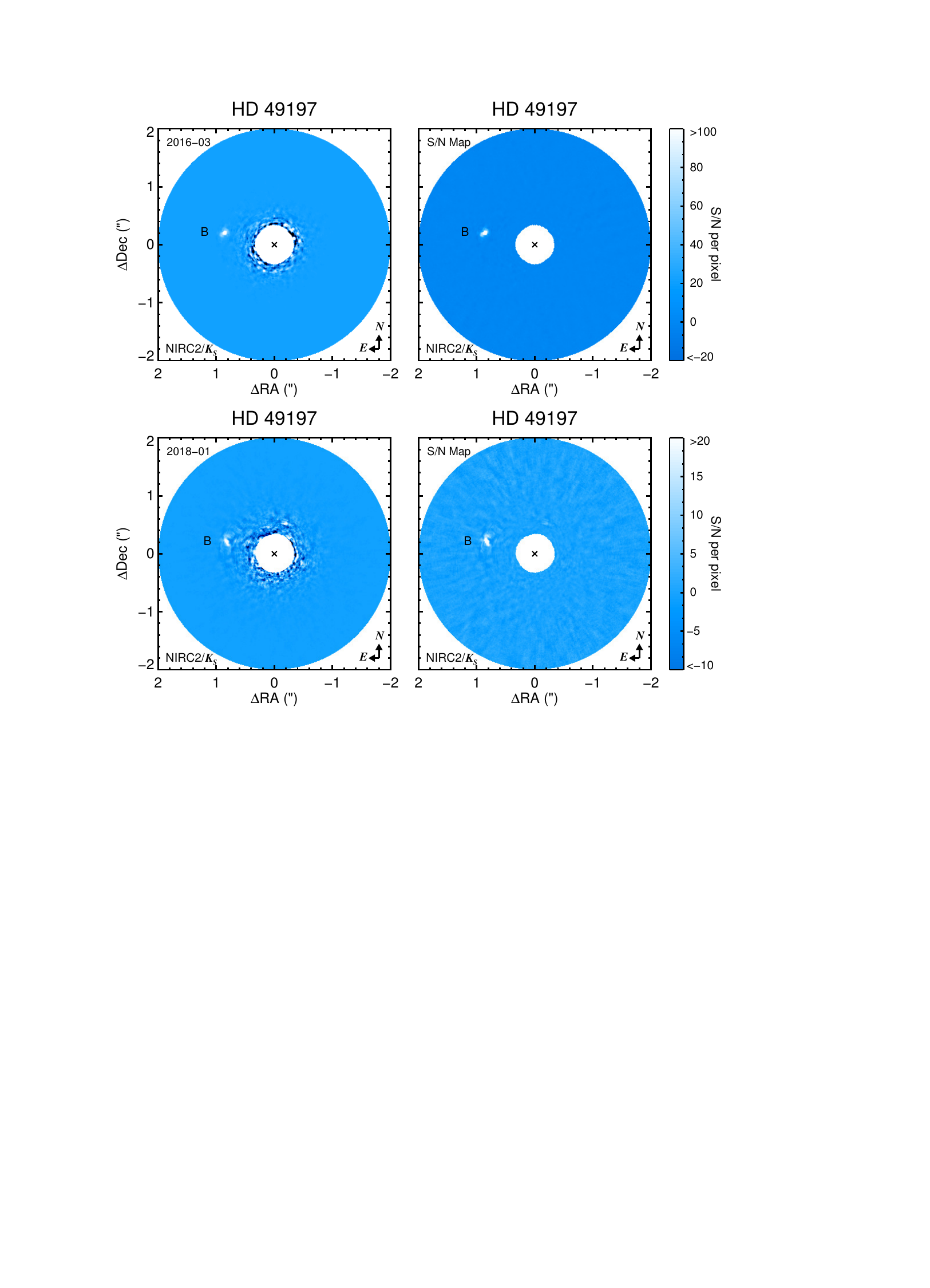}}
  \vskip -5.9 in
  \caption{Keck/NIRC2 $K_S$-band observations of HD 49197 in 2016 (top) and 2018 (bottom).  The brown dwarf 
  HD 49197 B is clearly recovered in the PSF-subtracted, median-combined frames (left) and 
  signal-to-noise maps (right) with a signal to noise of 488 in the 2016 epoch and 84 in the 2018 epoch. 
  North is up and East is to the left. The color bars reflect the signal-to-noise pixel values of the S/N maps. \label{fig:adi2} } 
\end{figure*}


\begin{figure*}
  \vskip -.9 in
  \hskip -0.7 in
  \resizebox{9.5in}{!}{\includegraphics{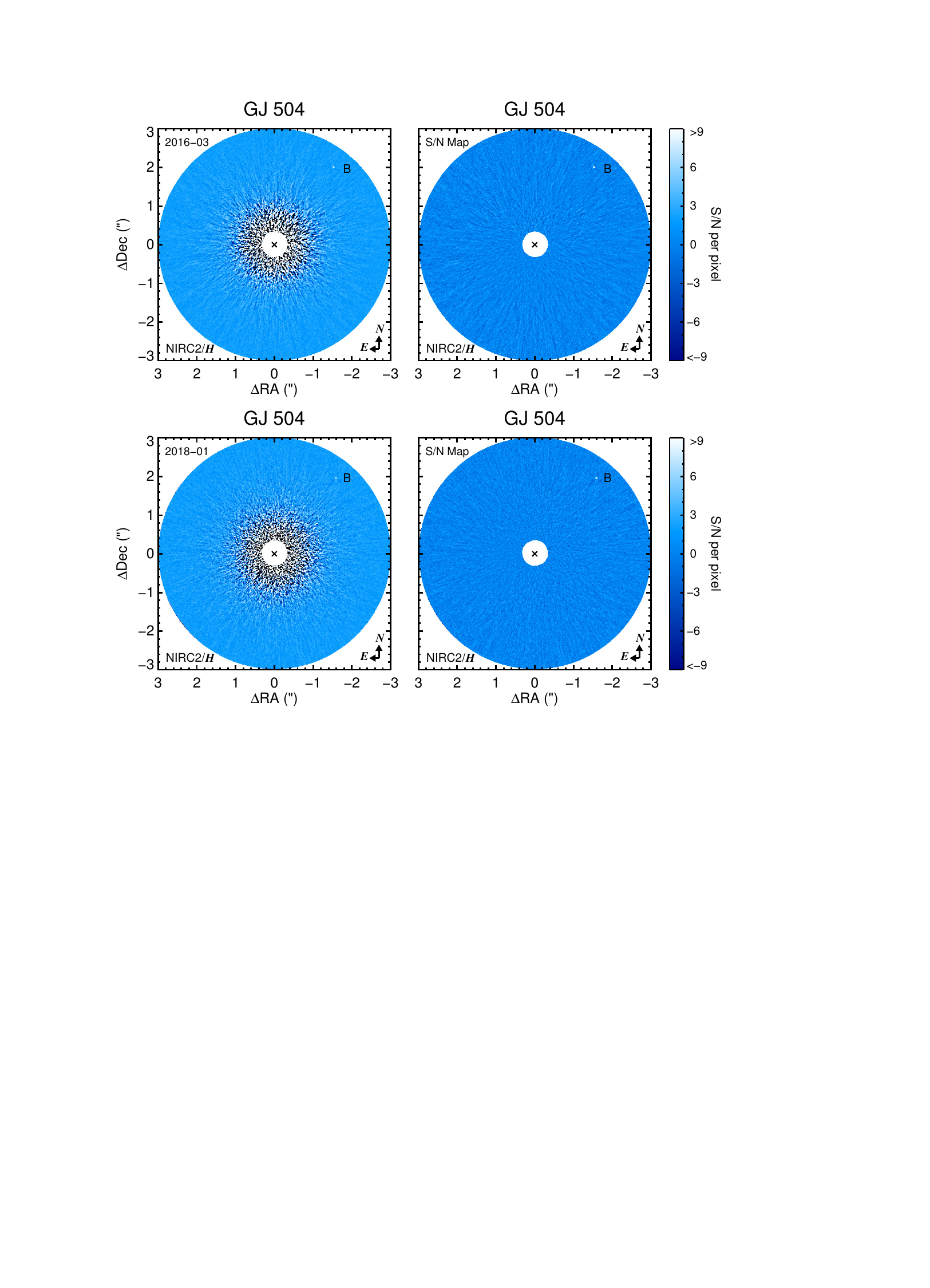}}
  \vskip -5.9 in
  \caption{Keck/NIRC2 $H$-band observations of GJ 504 in 2016 (top) and 2018 (bottom).  The substellar companion 
  GJ 504 B is clearly
  recovered in the PSF-subtracted, median-combined frames (left) and in the signal-to-noise maps (right)
  with a signal to noise of 13 in the 2016 epoch and 16 in the 2018 epoch. 
  North is up and East is to the left.  The color bars reflect the signal-to-noise pixel values of the S/N maps. \label{fig:adi3} } 
\end{figure*}


\begin{figure*}
  \vskip -0.9 in
  \hskip -0.7 in
  \resizebox{9.5in}{!}{\includegraphics{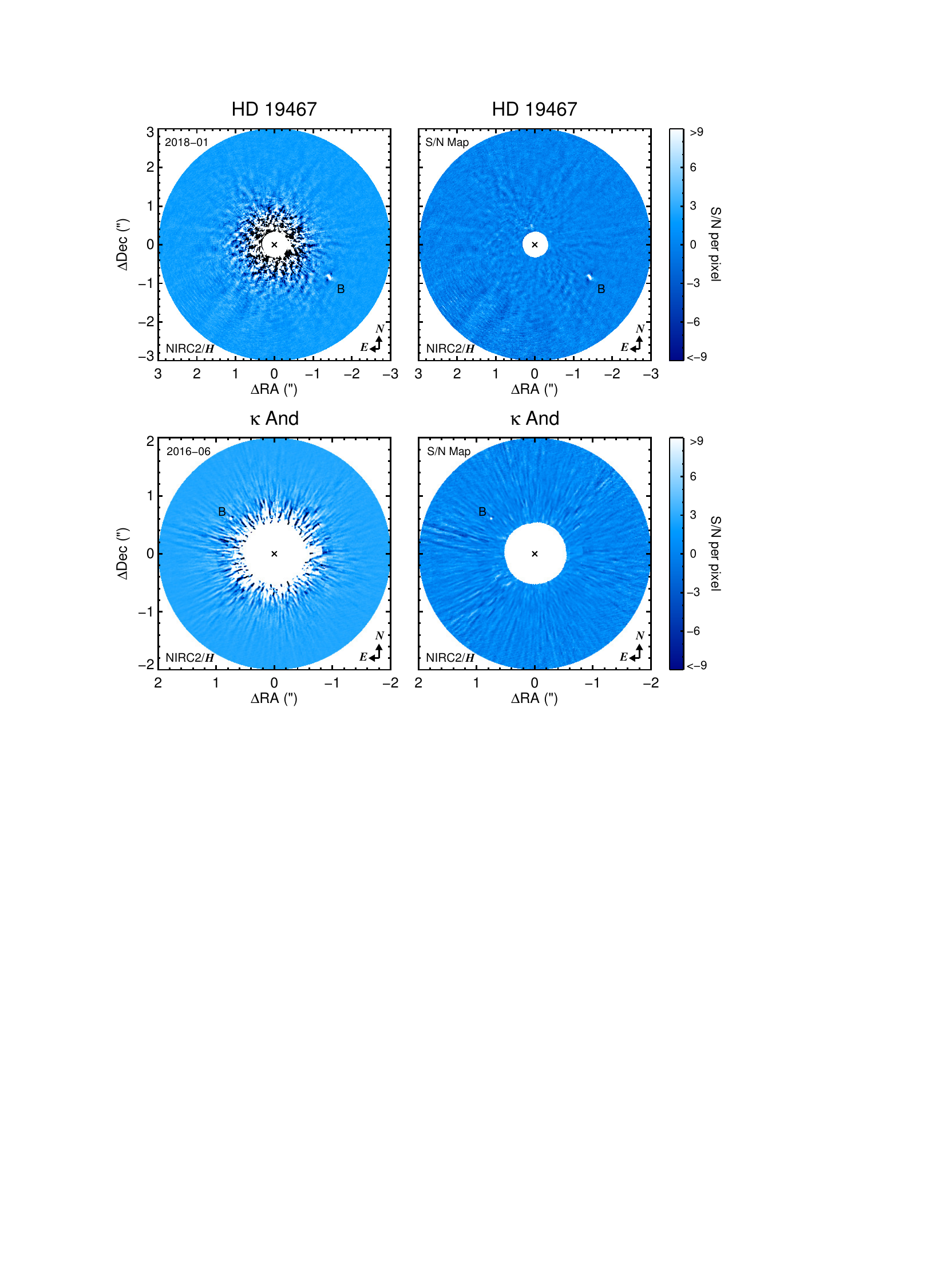}}
  \vskip -5.9 in
  \caption{Keck/NIRC2 $H$-band observations of HD 19467 in 2018 (top) and $\kappa$ And in 2016 (bottom).  
  Both substellar companions are 
  recovered in the PSF-subtracted, median-combined frames (left) and in the signal-to-noise maps (right)
  with a signal to noise of 26 for HD 19467 B and 9 for $\kappa$ And B. 
  North is up and East is to the left.  The color bars reflect the signal-to-noise pixel values of the S/N maps. \label{fig:adi4} } 
\end{figure*}


\begin{figure*}
  \vskip -1. in
  \hskip -0.7 in
  \resizebox{9.5in}{!}{\includegraphics{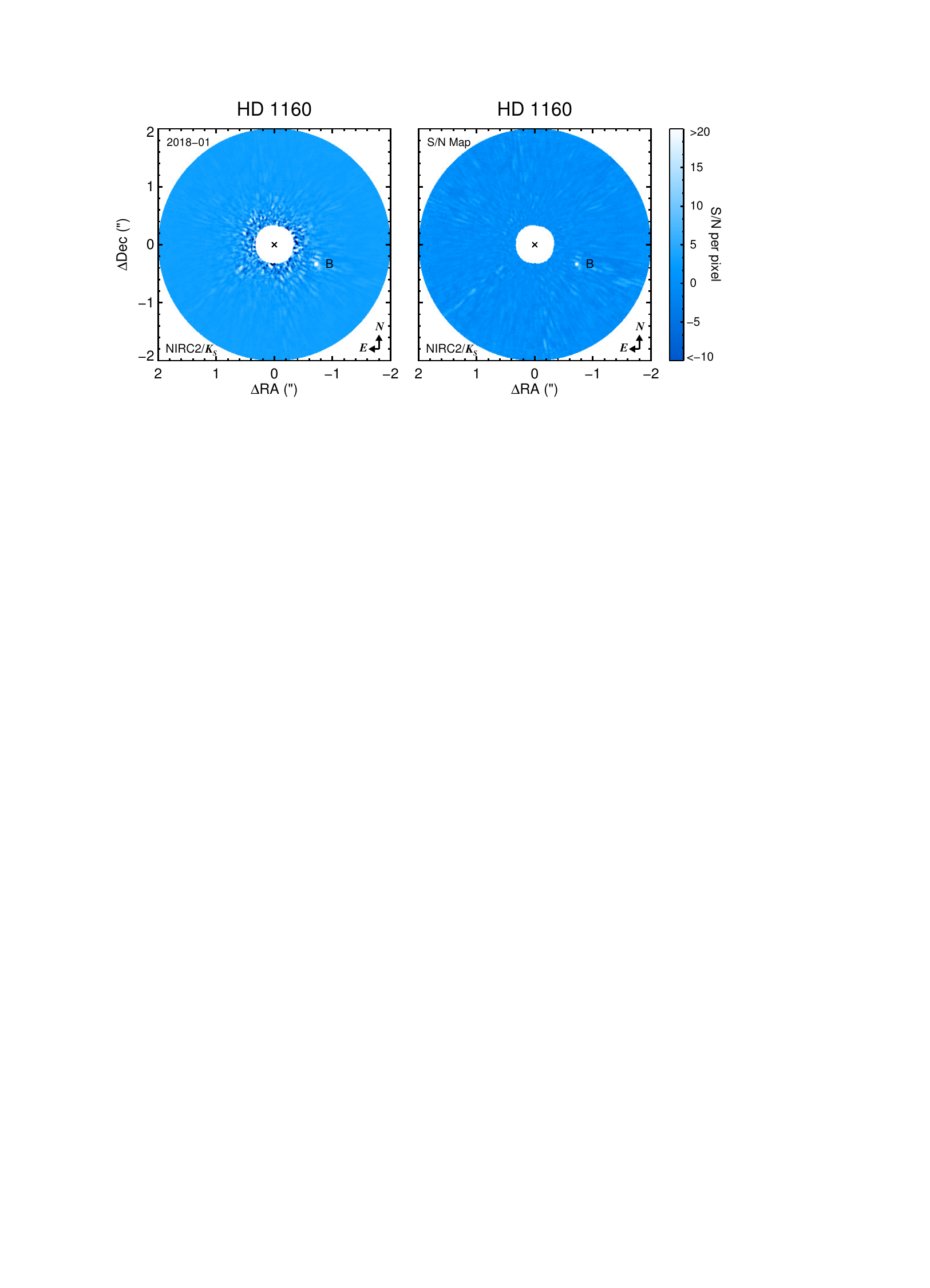}}
  \vskip -9.2 in
  \caption{Keck/NIRC2 $K_S$-band observations of HD 1160 in 2018.  
  The substellar companion HD 1160 B is 
  recovered in the PSF-subtracted, median-combined frame (left) and in the signal-to-noise map (right)
  with a signal to noise of 31. 
  North is up and East is to the left. The color bar reflects the signal-to-noise pixel values of the S/N map. \label{fig:adi5} } 
\end{figure*}

\subsection{Keck/NIRC2 Adaptive Optics Imaging}{\label{sec:nirc2}}

We observed 13 targets with substellar companions between 2014 and 2019 (Table~\ref{tab:newastrometry})
with the NIRC2 camera behind natural guide star adaptive optics at Keck Observatory (\citealt{Wizinowich:2000hl}).
All observations were acquired with the narrow camera mode, which provides a plate scale of $\approx$10 mas pix$^{-1}$ 
and a field of view of 10$\farcs$2 $\times$ 10$\farcs$2.
Observations of HD 19467, $\kappa$ And, GJ 504, and HD 49197 were acquired in pupil tracking 
mode to facilitate standard post processing PSF subtraction. 
Total on-source integration times ranged from 3 min to 81 min 
and the field of view rotation angle ranged from 3$\degr$ to 119$\degr$.
The brown dwarf companions to HD 1160, 1RXS0342+1216, CD 35--2722, DH Tau, HD 23514, 
Ross 458, TWA 5, 2M1559+4403, and 1RXS2351+3127 have lower contrasts and
were observed with shorter integration times.  The partly transparent 600-mas 
diameter focal plane coronagraph was used for all systems except Ross 458.
Details of the observations can be found in Table~\ref{tab:newastrometry}.

After removing bad pixels and cosmic rays, images are bias subtracted, flat fielded, and corrected for
optical distortions using the distortion solution from \citet{Yelda:2010ig} for observations taken before April 2015
and from \citet{Service:2016gk} for observations after that date.
The corresponding plate scales are 9.952 $\pm$ 0.002 mas pixel$^{-1}$ and 9.971 $\pm$ 0.004 mas pixel$^{-1}$ 
for these respective dates, and the P.A. offsets are 
0.252 $\pm$ 0.009$\degr$ and 0.262 $\pm$ 0.020$\degr$
with respect to the detector columns.
The typical rms residual after distortion correction is $\approx$1 mas.
PSF subtraction for the ADI sequences is carried out following the description in \citet{Bowler:2015ja}: 
images are first registered using the position of the star visible behind the focal plane mask,
then PSF subtraction is performed using LOCI.
The median-combined PSF-subtracted image is then north aligned and a noise map is created by measuring the rms 
in annuli centered on the host star with a width of 3 pixels. 
The final PSF-subtracted images and signal-to-noise maps are shown in Figures \ref{fig:adi1}--\ref{fig:adi5}.
Note that HD 1160 was observed in field-tracking mode; PSF subtraction for this target entailed de-rotating
to a common pupil angle, implementing PSF subtraction with LOCI, then re-rotating to a common sky position angle
before coadding the individual frames.

\begin{figure*}
  \vskip -0.5 in
  \hskip -.6 in
  \resizebox{8.5in}{!}{\includegraphics{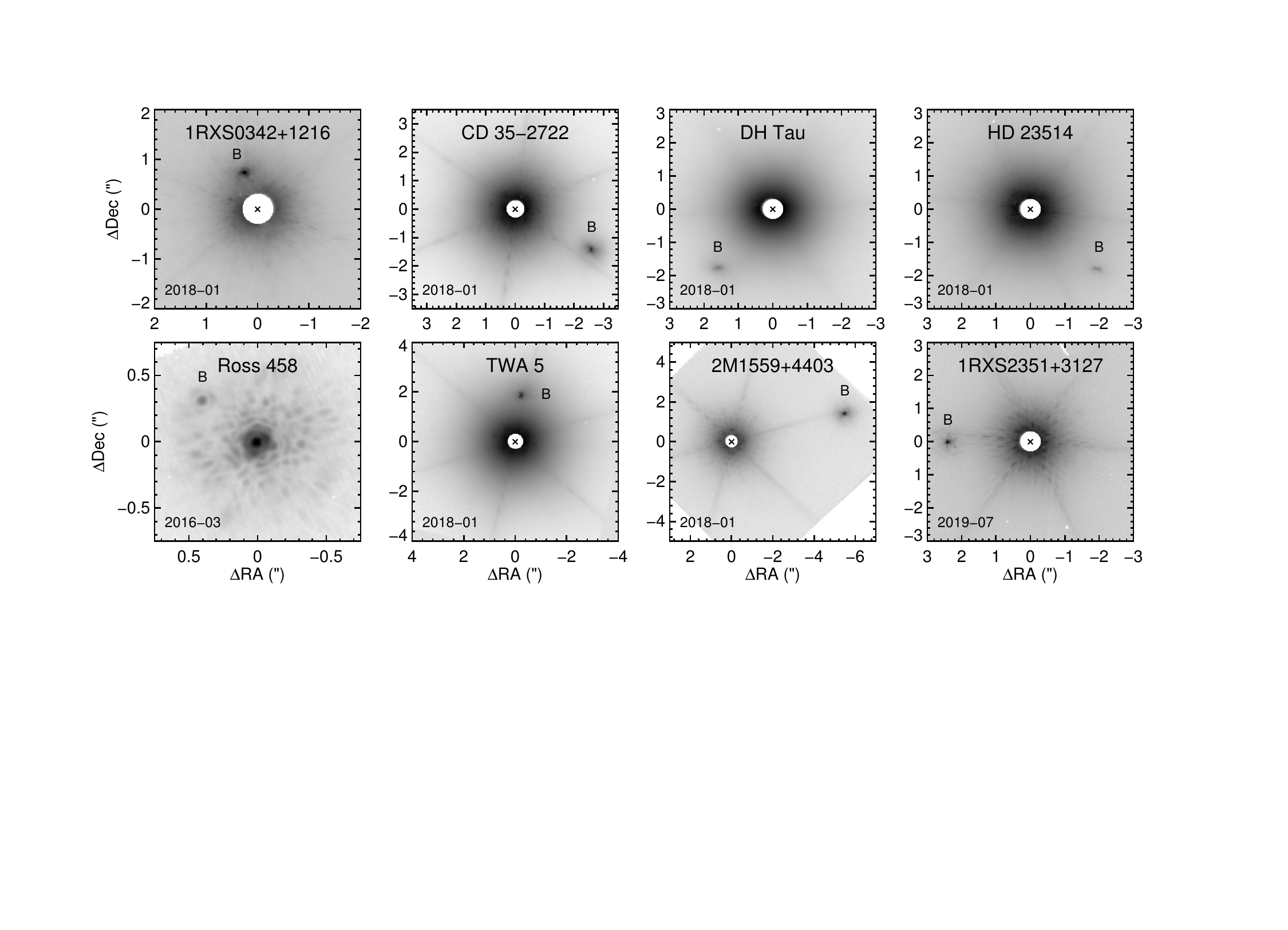}}
  \vskip -2.3 in
  \caption{NIRC2 $K_S$-band observations of 1RXS0342+1216, CD-35 2722, DH Tau, HD 23514, Ross 458, TWA 5, 2M1559+4403, and 1RXS2351+3127.
  No PSF subtraction is needed to recover the substellar companions for these modest-contrast systems. 
  North is up and East is to the left. \label{fig:mosaic} } 
\end{figure*}

No PSF subtraction is necessary to recover the other substellar companions with modest flux ratios, 
which are readily visible in the raw frames.
After basic image reduction these frames are registered, north aligned, 
then coadded to produce the final images shown in Figure~\ref{fig:mosaic}.


\subsection{Astrometry and Orbital Motion}{\label{sec:astrometry}}

Relative astrometry is measured in each final processed image.  Separations in pixels are converted to 
angular distances using the detector plate scale.  Following \citet{Bowler:2018gy}, 
uncertainties 
take into account random positional measurement errors, estimated to be 0.5 pix for HiCIAO and 0.1 pix for NIRC2
based on end-to-end injection-recovery tests of companions and host star 
positions in \citet{Bowler:2015ja} and \citet{Bowler:2018gy};
uncertainties in the plate scale; and rms errors from the distortion correction.
Similarly, the P.A. uncertainty incorporates random measurement errors, uncertainty in the 
absolute orientation of celestial north on the detector, rms errors from the distortion correction, 
and angular uncertainty associated with average azimuthal shearing of the PSF within each image 
for observations taken in pupil-tracking mode.
\citet{Bowler:2018gy} found that systematic errors can dominate the astrometric uncertainty budget
for ADI observations with NIRC2 using the 600 mas coronagraph.  We therefore adopt conservative 
errors of 5 mas and 0.3 deg for our ADI observations in this work.\footnote{Note that we do not expect systematic 
errors to be as severe for relative astrometry with short exposures when the 
companion is bright and the host star is behind the coronagraph mask.  See Appendix B of \citet{Bowler:2015ja} 
for calibration tests with a binary system.}
Our final measurements are reported in Table~\ref{tab:newastrometry}.

The signal-to-noise ratio (SNR) of each companion detection is calculated using aperture photometry.
Aperture radii of $\approx$$\lambda$/$D$ are chosen to encapsulate the central Airy disk and minimize potential 
effects of oversubtraction at larger radii. 
Here $D$ is the telescope diameter and $\lambda$ is the central wavelength of the filter.
We adopt a 6-pix aperture for observations using HiCIAO in $K_S$ band, a 5-pix aperture for NIRC2 in $K_S$ band,
and a 4-pix aperture for NIRC2 in $H$ band.
Noise levels are derived using counts in 100 circular apertures at the same separation but different
position angles in each image.  These SNR measurements range from 9 (for $\kappa$ And B)
to 816 for our 2014 epoch of HD 49197 (see Table~\ref{tab:newastrometry}).

\begin{figure*}
  \vskip -0.0 in
  \hskip 1.2 in
  \resizebox{4.5in}{!}{\includegraphics{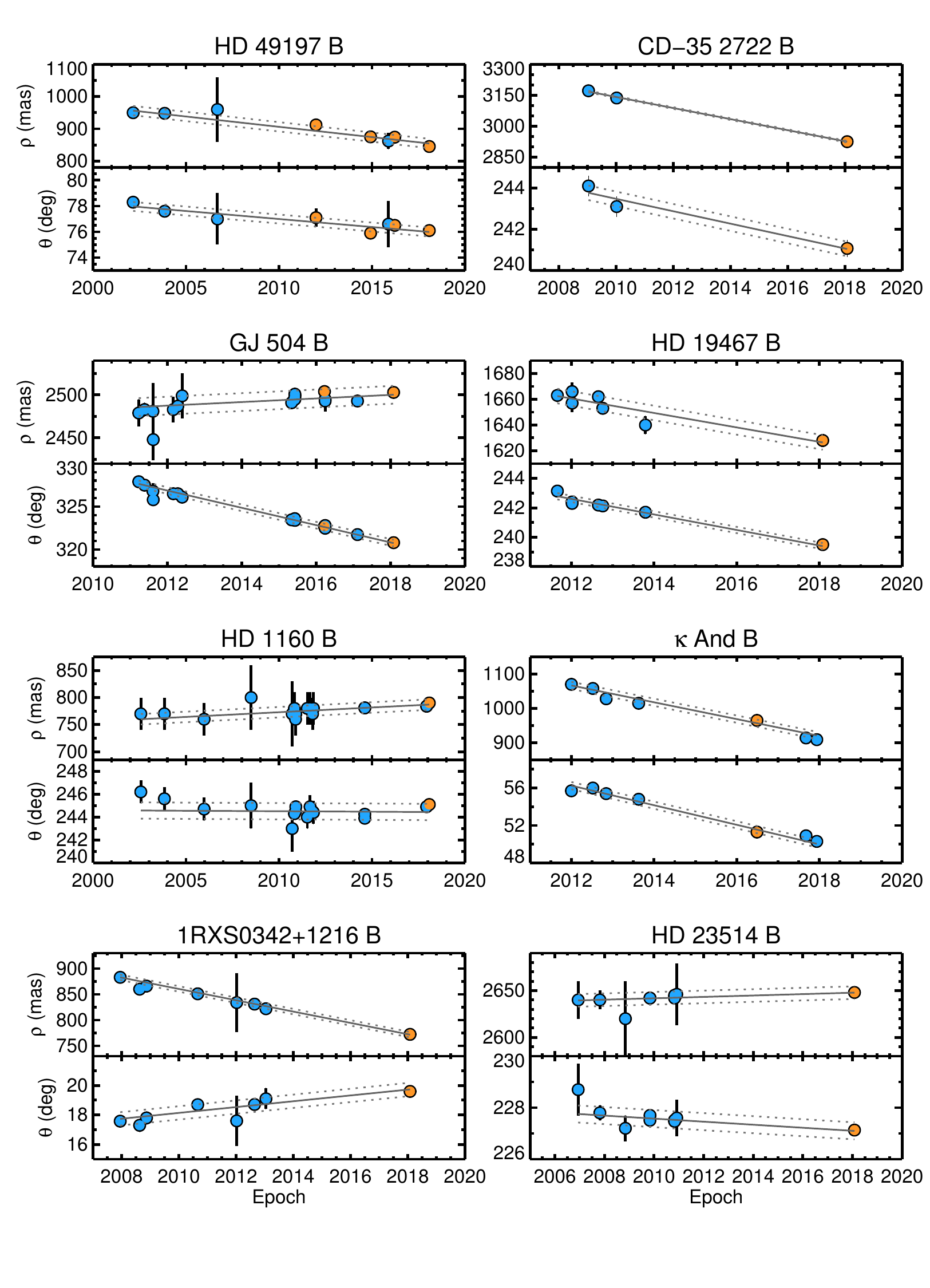}}
  \vskip -.4 in
  \caption{Orbital motion for substellar companions with new astrometry.  Our observations are shown in orange; blue 
  circles are from the literature (see Appendix \ref{sec:litastrometry}).  Thick error bars are raw (uncorrected) quoted uncertainties, while thin
  error bars show the uncertainty needed to bring the reduced $\chi^2$ value of a linear fit to unity.  In most cases
  these are smaller than the symbol size.  Linear fits are shown in gray and are listed in Table~\ref{tab:linearfit}.
  The rms error about each fit is indicated with a dotted gray line.  \label{fig:orbitalmotion1} }
\end{figure*}

\begin{figure*}[htb!]
  \vskip -0 in
  \hskip 1.2 in
  \resizebox{4.5in}{!}{\includegraphics{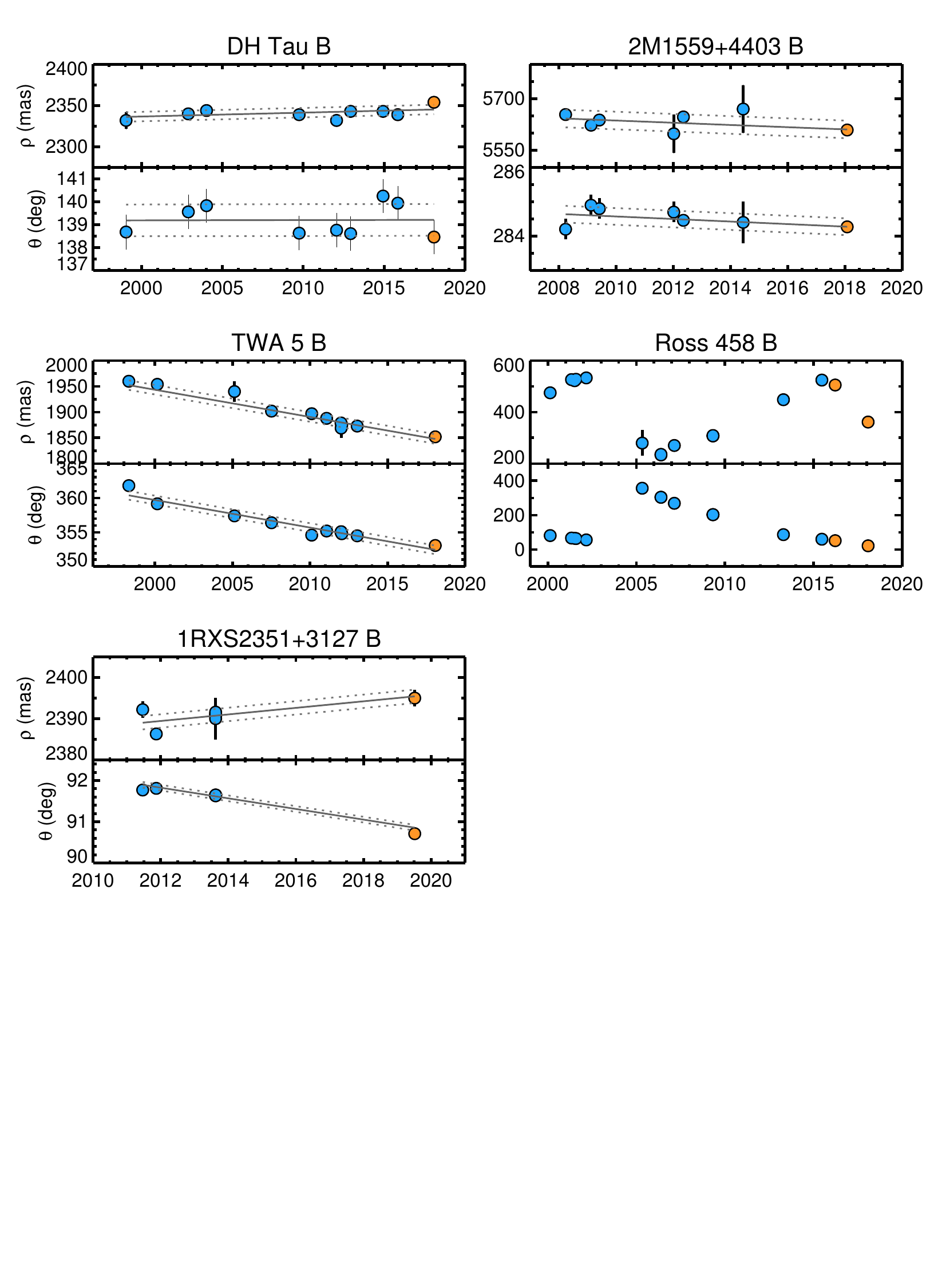}}
  \vskip -1.9 in
  \caption{Orbital motion for substellar companions with new astrometry.  Our observations are shown in orange; blue 
  circles are from the literature (see Appendix \ref{sec:litastrometry}).  Thick error bars are raw (uncorrected) quoted uncertainties, while thin
  error bars show the uncertainty needed to bring the reduced $\chi^2$ value of a linear fit to unity.  In most cases
  these are smaller than the symbol size with the notable exception of the P.A. measurements for DH Tau B.  
  Linear fits are shown in gray and are listed in Table~\ref{tab:linearfit}. 
  The rms error about each fit is indicated with a dotted gray line. \label{fig:orbitalmotion2} } 
\end{figure*}

Figures~\ref{fig:orbitalmotion1} and \ref{fig:orbitalmotion2} compare our new astrometry (Table~\ref{tab:newastrometry}) to published
values in the literature (Appendix \ref{sec:litastrometry}).  Orbital motion is clearly detected in many systems and generally extends and refines  
linear evolution in separation and P.A. with time, with the exception of Ross 458 which has a well-determined orbit.   
The addition of recent observations is especially important for 
systems like CD--35 2722 B, for which no astrometry has been published since its discovery by \citet{Wahhaj:2011by}.

There are several instances in which our observations were taken close in time with other published epochs,
which provides an opportunity to compare both measurements for mutual consistency.
Our astrometry of HD 49197 from UT 2016 March 22 was taken four months after the observations
by \citet{Bottom:2017bv} on 22 November 2015 UT with the Stellar Double Coronagraph at Palomar Observatory.
Our measured separation of 874 $\pm$ 5 mas and P.A. of 76.5 $\pm$ 0.3$\degr$ are in good agreement
with the values of 862 $\pm$ 25 mas and 76.6 $\pm$ 1.8$\degr$ from Bottom et al.
We targeted GJ 504 about one week before an observation taken with SPHERE, which was reported in \citet{Bonnefoy:2018ch}.
They find \{$\rho$=2495 $\pm$ 2 mas, $\theta$=322.48 $\pm$ 0.05$\degr$\} 
on UT 2016 March 29;
our measurements of \{$\rho$=2504 $\pm$ 5 mas, $\theta$=322.7 $\pm$ 0.4$\degr$\} were obtained on UT 2016 March 22.
These P.A.s

\begin{rotatetable*}
\begin{deluxetable*}{lcccccccccc}
\renewcommand\arraystretch{0.9}
\tabletypesize{\footnotesize}
\setlength{ \tabcolsep }{.1cm}
\tablewidth{0pt}
\tablecolumns{11}
\tablecaption{Linear evolution of separation and P.A.\label{tab:linearfit}}
\tablehead{
\colhead{Target}  &  \colhead{Sep. $a_0$}   & \colhead{Sep. $a_1$} &  \colhead{Sep. rms} &  \colhead{$\sigma_\mathrm{jit, \rho}$} & \colhead{Sep.} &   \colhead{P.A. $b_0$} & \colhead{P.A. $b_1$} & \colhead{P.A. rms} & \colhead{$\sigma_\mathrm{jit, \theta}$}  &  \colhead{P.A.} \\
\colhead{Name}    &  \colhead{(mas)}   & \colhead{(mas yr$^{-1}$)}  &  \colhead{(mas)}  & \colhead{(mas)}  &  \colhead{$\chi^2_{\nu}$}  & \colhead{($\degr$)}  & \colhead{($\degr$ yr$^{-1}$)}  & \colhead{($\degr$)}  & \colhead{($\degr$)} & \colhead{$\chi^2_{\nu}$}  
  }
\startdata
        HD 984 B    &  --24198.3242  $\pm$    6652.9297    &    12.1124  $\pm$    3.3012     &   13.1     &     3.2     &   0.99     &   17555.8809  $\pm$     867.8156     &  --8.6683  $\pm$   0.4306     &   0.85     &   0.00     &   0.49    \\
       HD 1160 B    &   --2723.1067  $\pm$    1785.6223    &     1.7393  $\pm$    0.8863     &    9.7     &     0.0     &   0.10     &     261.4736  $\pm$      72.9082     &  --0.0084  $\pm$   0.0362     &   0.71     &   0.24     &   0.99    \\
      HD 19467 B    &    12941.7510  $\pm$    2004.5902    &   --5.6069  $\pm$    0.9957     &    5.6     &     1.8     &   0.99     &    1304.3900  $\pm$     107.5698     &  --0.5277  $\pm$   0.0534     &   0.21     &   0.06     &   0.99    \\
 1RXS0342+1216 B    &    22939.0391  $\pm$     327.6723    &  --10.9844  $\pm$    0.1632     &    5.3     &     0.0     &   0.94     &   --378.7789  $\pm$      63.8557     &    0.1975  $\pm$   0.0317     &   0.45     &   0.23     &   0.96    \\
        51 Eri b    &     6720.5132  $\pm$    2516.3901    &   --3.1086  $\pm$    1.2477     &    3.8     &     0.0     &   0.31     &   11322.8174  $\pm$     343.0927     &  --5.5345  $\pm$   0.1701     &   0.39     &   0.00     &   0.26    \\
    CD-35 2722 B    &    57221.8008  $\pm$    1387.3359    &  --26.9053  $\pm$    0.6888     &    4.3     &     3.5     &   0.98     &     849.6935  $\pm$     127.4335     &  --0.3016  $\pm$   0.0633     &   0.35     &   0.40     &   1.00    \\
      HD 49197 B    &    13782.7002  $\pm$     843.4277    &   --6.4062  $\pm$    0.4196     &   14.4     &     4.6     &   0.98     &     327.0609  $\pm$      44.3513     &  --0.1244  $\pm$   0.0220     &   0.35     &   0.00     &   0.81    \\
       HR 2562 B    &  --50161.6797  $\pm$    2145.8904    &    25.1873  $\pm$    1.0638     &    1.2     &     0.0     &   0.28     &   --202.7896  $\pm$     257.1518     &    0.2482  $\pm$   0.1275     &   0.17     &   0.00     &   0.44    \\
       HR 3549 B    &    17045.1328  $\pm$    9505.5195    &   --8.0338  $\pm$    4.7161     &    0.9     &     0.0     &   0.02     &    1324.7207  $\pm$     498.3914     &  --0.5798  $\pm$   0.2473     &   0.29     &   0.00     &   0.20    \\
      HD 95086 b    &   --1867.0281  $\pm$    1941.0756    &     1.2346  $\pm$    0.9629     &    5.2     &     0.0     &   0.31     &    2351.5354  $\pm$     193.0089     &  --1.0930  $\pm$   0.0957     &   0.32     &   0.00     &   0.32    \\
        GJ 504 B    &   --1710.9081  $\pm$    1831.3406    &     2.0867  $\pm$    0.9085     &   10.4     &     0.6     &   0.99     &    2354.0681  $\pm$      51.8511     &  --1.0076  $\pm$   0.0257     &   0.39     &   0.03     &   1.00    \\
     HIP 65426 b    &     6849.5601  $\pm$    2470.6326    &   --2.9850  $\pm$    1.2246     &    8.0     &     0.0     &   0.82     &     511.8435  $\pm$     213.0702     &  --0.1794  $\pm$   0.1056     &   0.55     &   0.12     &   0.98    \\
        PDS 70 b    &     2706.3250  $\pm$    3142.1650    &   --1.2465  $\pm$    1.5588     &    6.2     &     0.0     &   0.45     &    5107.3291  $\pm$     817.6153     &  --2.4576  $\pm$   0.4055     &   1.35     &   0.00     &   0.28    \\
        PZ Tel B\tablenotemark{a}    &  --58956.1328  $\pm$     835.7415    &    29.5029  $\pm$    0.4153     &    6.0     &     3.8     &   0.98     &     326.9745  $\pm$      62.2090     &  --0.1328  $\pm$   0.0309     &   0.47     &   0.16     &   0.98    \\
     HD 206893 B    &    16907.2207  $\pm$    2168.6890    &   --8.2524  $\pm$    1.0749     &    2.4     &     0.0     &   0.81     &   18744.5137  $\pm$     537.0878     &  --9.2639  $\pm$   0.2663     &   0.64     &   0.29     &   0.98    \\
  $\kappa$ And B    &    50347.6562  $\pm$    3829.7366    &  --24.4935  $\pm$    1.9010     &   10.6     &     5.2     &   0.99     &    2160.1741  $\pm$     167.6283     &  --1.0457  $\pm$   0.0832     &   0.38     &   0.00     &   0.63    \\
      HD 23514 B    &     1142.6167  $\pm$     531.7578    &     0.7459  $\pm$    0.2644     &    6.7     &     0.0     &   0.39     &     346.3043  $\pm$      28.8196     &  --0.0591  $\pm$   0.0143     &   0.33     &   0.05     &   0.88    \\
        DH Tau B    &     1396.2018  $\pm$     757.6589    &     0.4703  $\pm$    0.3766     &    5.7     &     4.9     &   0.98     &     136.6982  $\pm$      81.7527     &    0.0012  $\pm$   0.0407     &   0.69     &   0.74     &   0.99    \\
   2M1559+4403 B    &    12130.2617  $\pm$    2763.0696    &   --3.2307  $\pm$    1.3737     &   25.6     &    10.5     &   1.00     &     359.0865  $\pm$      37.8433     &  --0.0371  $\pm$   0.0188     &   0.24     &   0.00     &   0.82    \\
         TWA 5 B    &    12507.5928  $\pm$     623.8889    &   --5.2820  $\pm$    0.3101     &    9.2     &     2.3     &   0.99     &    1161.5425  $\pm$      68.8053     &  --0.4009  $\pm$   0.0342     &   0.63     &   0.58     &   1.00    \\
 1RXS2351+3127 B    &      788.8538  $\pm$     609.4363    &     0.7955  $\pm$    0.3026     &    1.6     &     0.3     &   0.99     &     352.0048  $\pm$      30.3868     &  --0.1293  $\pm$   0.0151     &   0.07     &   0.05     &   0.90    \\
\enddata
\tablecomments{This table provides relations that describe the linear evolution of separation and P.A. as a function of calendar year ($t$): $\rho$($t$) = $a_0$ + $a_1$$t$
and $\theta$($t$) = $b_0$ + $b_1$$t$.  These incorporate astrometric jitter terms ($\sigma_\mathrm{jit, \rho}$ and $\sigma_\mathrm{jit, \theta}$) which are added in quadrature with the quoted astrometric errors  
to ensure reduced $\chi^2$ values are $\le$1.0.}
\tablenotetext{a}{Note that PZ Tel B shows signs of slight curvature, which is not reflected in this linear fit.}
\end{deluxetable*}
\end{rotatetable*}


\noindent are consistent at the $<$1$\sigma$ level and our separation measurement is 
consistent within 2$\sigma$ (9.0 $\pm$ 5.4 mas).
Our UT 2018 January 30 astrometry of HD 1160 B (\{$\rho$=790.0 $\pm$ 5 mas, $\theta$=245.1 $\pm$ 0.3$\degr$\})
are within 1$\sigma$ of the values also taken with NIRC2 by \citet{Currie:2018hj} on UT 2017 
December 9 (\{$\rho$=784 $\pm$ 6 mas, $\theta$=244.9 $\pm$ 0.3$\degr$\}), 
about two months before our observations.
Despite this good agreement with previous observations, 
below we 
attempt to address any potential systematic errors that could result from combining 
measurements taken with different instruments.
This is especially important because systematic errors or underestimated uncertainties
can mimic a strong acceleration, which is most readily accounted for in orbits fits through a 
high eccentricity and time of periastron close to the epoch of observations (see, for example,
results for HR 8799~e from \citealt{Konopacky:2016dk}).

Linear evolution in the astrometry is generally expected for most companions because of their 
long orbital periods relative to the time baseline of the observations ($\approx$5--20 yr).
We adopt reduced $\chi^2$ values as a metric to assess
the fidelity of the astrometric uncertainties, both for our own measurements and those from the literature.
These sample a wide range of instruments, PSF subtraction algorithms, and approaches to measuring astrometry,
all of which have the potential to introduce systematic errors in the final astrometry.
Linear fits generally provide good matches to the data ($\chi^2_{\nu}$ $\approx$ 1), but in a few instances the reduced $\chi^2$ value
is unreasonably large, indicating biased astrometry or underestimated errors.
DH Tau is an especially acute example: $\chi^2_{\nu}$ = 6.8 for $\rho ( t )$ and $\chi^2_{\nu}$ = 96 for $\theta ( t )$.
The source of this scatter is most likely from instrument-to-instrument calibration errors in distortion correction, plate scale, and 
north alignment.

To mediate these biases for our orbit fits we introduce a ``jitter'' term added in quadrature with the astrometric uncertainties
($\sigma_\mathrm{jit, \rho}$ for separation and $\sigma_\mathrm{jit, \theta}$ for P.A.).
This effectively increases the astrometric errors in a systematic fashion until they are consistent with the expected level 
due to random scatter about linear evolution in separation and P.A. with time (Figures~\ref{fig:orbitalmotion1} and \ref{fig:orbitalmotion2}).  
For fits with $\chi^2_{\nu}$ $>$ 1, jitter is derived by iteratively increasing $\sigma_\mathrm{jit, \rho}$ and $\sigma_\mathrm{jit, \theta}$ 
until the linear fits result in $\chi^2_{\nu}$ = 1.
Most systems do not require any additional additive error term.  When needed, the typical jitter level for separation
is $\approx$2--6 mas and for P.A. is $\approx$0.1--0.4$\degr$.
For DH Tau, values of $\sigma_\mathrm{jit, \rho}$ (4.9 mas) and $\sigma_\mathrm{jit, \theta}$ (0.74$\degr$) 
imply that especially strong systematic errors are present in this dataset. 
Linear fits of the separation and P.A. over time incorporating this added uncertainty are provided in Table~\ref{tab:linearfit}.  
Residuals of these fits do not show any trends, suggesting the primary source of the large $\chi^2_\nu$ 
values is from systematic astrometric errors or underestimated uncertainties, 
rather than our assumption of linear evolution for curved orbital motion (acceleration).

\section{Updated Orbit Fits}{\label{sec:orbits}}

We use our new observations (Table~\ref{tab:newastrometry}) and published astrometry in the literature 
(Table~\ref{tab:astrometry} in Appendix~\ref{sec:litastrometry}) to update 
the Keplerian orbit fits for the 13 systems we observed.  In many cases no astrometric epochs have been
reported for these systems over the past several years.  Our new data reveal slight but significant orbital motion
for the majority of targets.  In a few instances these represent the first clear indications of orbital motion for these systems 
(HD 49197 B, HD 23514 B, 2M1559+4403 B, and 1RXS2351+3127).

Orbit fits are carried out using the \texttt{orbitize!} package for fitting orbits of directly imaged planets\footnote{https://github.com/sblunt/orbitize; https://zenodo.org/record/3242703\#.XRfV-5NKiu4} 
(\citealt{Blunt:2019vq}).  \texttt{orbitize!} implements the
``Orbits for the Impatient'' (\texttt{OFTI}) Bayesian rejection sampling algorithm detailed in \citet{Blunt:2017eta}
and the \texttt{ptemcee} parallel-tempered Markov Chain Monte Carlo (MCMC) approach to sampling the posterior from \citet{ForemanMackey:2013io} and \citet{Vousden:2015jb}.
\texttt{OFTI} is computationally more efficient than MCMC for
mapping complex multimodal posterior shapes, 
for constraining the orbital elements of systems with relative astrometry that spans only a small fraction of
their orbital periods,
and when the posteriors are similar to the priors, 
whereas MCMC is faster for well-constrained orbits.
However, both approaches produce similar results when using the same parameter priors.
We utilize \texttt{OFTI} and MCMC in approximately equal proportions for orbit constraints in this study.

Astrometric errors include jitter values listed in Table~\ref{tab:linearfit}
for separation and P.A. as described in Section~\ref{sec:astrometry}.
The following uninformative priors are adopted for the orbital elements: 
Jeffreys prior (1/$a$) for semi-major axis ($a$) from 0.001 to 10$^7$~AU;
uniform for eccentricity ($e$) from $e$=0 to 1;
uniform in cos($i$) for inclination ($i)$ from 0 to $\pi$; 
uniform for argument of periastron ($\omega$);
uniform for longitude of ascending node ($\Omega$);
and uniform for time of periastron passage ($\tau$) from $\tau$=0 to 1,
here expressed in units of period fraction past MJD=0.
Stellar masses and parallaxes are allowed to vary but are dominated by the following priors:
normal distribution for stellar mass 
and normal distribution for parallax measurements taken from $Gaia$ DR2 (\citealt{GaiaCollaboration:2018io}).\footnote{For
orbit fits in this work
we assume the total system mass is equal to stellar mass plus the mass of the imaged companion.
}

Results for the updated orbit fits using our new astrometry are summarized in Table~\ref{tab:orbitize}
and described in detail for each system below.

\subsection{HD 49197 B}

\citet{Metchev:2004kl} discovered this companion as part of their Palomar AO search for substellar companions to young
stars in the solar neighborhood.  They measure a spectral type of  L4 $\pm$ 1 from a 
$K$-band spectrum and infer a mass of $\approx$63~\Mjup \ for HD 49197 B 
based on its absolute magnitude.
We assume a stellar mass of 1.11 $\pm$ 0.06 \Msun \ from \citet{Mints:2017di}, 
a total system mass of 1.17 $\pm$ 0.06 \Msun, and a parallax of 
23.99 $\pm$ 0.05 mas from $Gaia$ DR2 in our orbit analysis.

Two epochs of astrometry were obtained by \citet{Metchev:2004kl} in 2002 and 2003, 
at which point the separation was 0$\farcs$95 and P.A. was $\approx$78$\degr$.
Additional epochs were obtained in 2006 by \citet{Serabyn:2009ca} and in 2015 by \citet{Bottom:2017bv}, who
first noted orbital motion toward the star.  We have been monitoring this companion at a low cadence since 2011 and 
confirm this motion to smaller separations at a rate of --6.4 mas yr$^{-1}$ based on four epochs 
with HiCIAO and NIRC2.  We also detect significant evolution in P.A. for the first time at a rate of --0.12$\degr$ yr$^{-1}$ 
in the clockwise direction.
Significant astrometric jitter is required to lower $\chi^2_{\nu}$ to $\approx$1 for the separation measurements 
($\sigma_\mathrm{jit, \rho}$ = 4.6 mas).
Overall HD 49197 B has moved inward by about 0$\farcs$1 and moved by about 2$\degr$ over the past 15 years.

Our best-fitting orbital solution has a semi-major axis of 29$^{+7}_{-10}$~AU
and an orbital period of 150$^{+50}_{-70}$ yr (2$\sigma$ credible interval: 80--450 yr).
This implies that $\approx$10\% of its orbit has currently been mapped. 
The eccentricity of HD 49197 B  is poorly constrained.
Continued monitoring will be needed to refine the orbital elements for this companion.
Results from the orbit fit are shown in Figures~\ref{fig:orbitfits1} and \ref{fig:hd49197corner}.

\subsection{CD--35 2722 B}

\citet{Wahhaj:2011by} identified this young L4 $\pm$ 1 brown dwarf companion 
as part of the Gemini NICI Planet-Finding Campaign (\citealt{Liu:2010hn}).
The host is an M1 member of the $\approx$120~Myr AB Dor association,
implying a mass of 31 $\pm$ 8 \Mjup \ for CD--35 2722 B based on its luminosity and young age.
We adopt a host mass of 0.40 $\pm$ 0.05 \Msun \ from \citet{Wahhaj:2011by},
a total system mass of 0.43 $\pm$ 0.05 \Msun,
and a parallax of 44.635 $\pm$ 0.026 mas from $Gaia$ DR2 in our orbit analysis.

Wahhaj et al. presented two epochs from 2009 and 2010, but no additional astrometry
has been reported since then.  Our epoch from 2018 reveals significant orbital
motion at a rate of --27 mas yr$^{-1}$ toward the host star in separation and --0.3$\degr$ yr$^{-1}$ 
in P.A. in the clockwise direction (Figure~\ref{fig:orbitalmotion1}).    A modest level of additional jitter is 
needed---$\sigma_\mathrm{jit, \rho}$ = 3.5 mas and $\sigma_\mathrm{jit, \theta}$ = 0.40$\degr$---although
we note that with only three epochs of data it is difficult to precisely 
determine the magnitude of any systematic error. 

\begin{figure*}
  \gridline{\fig{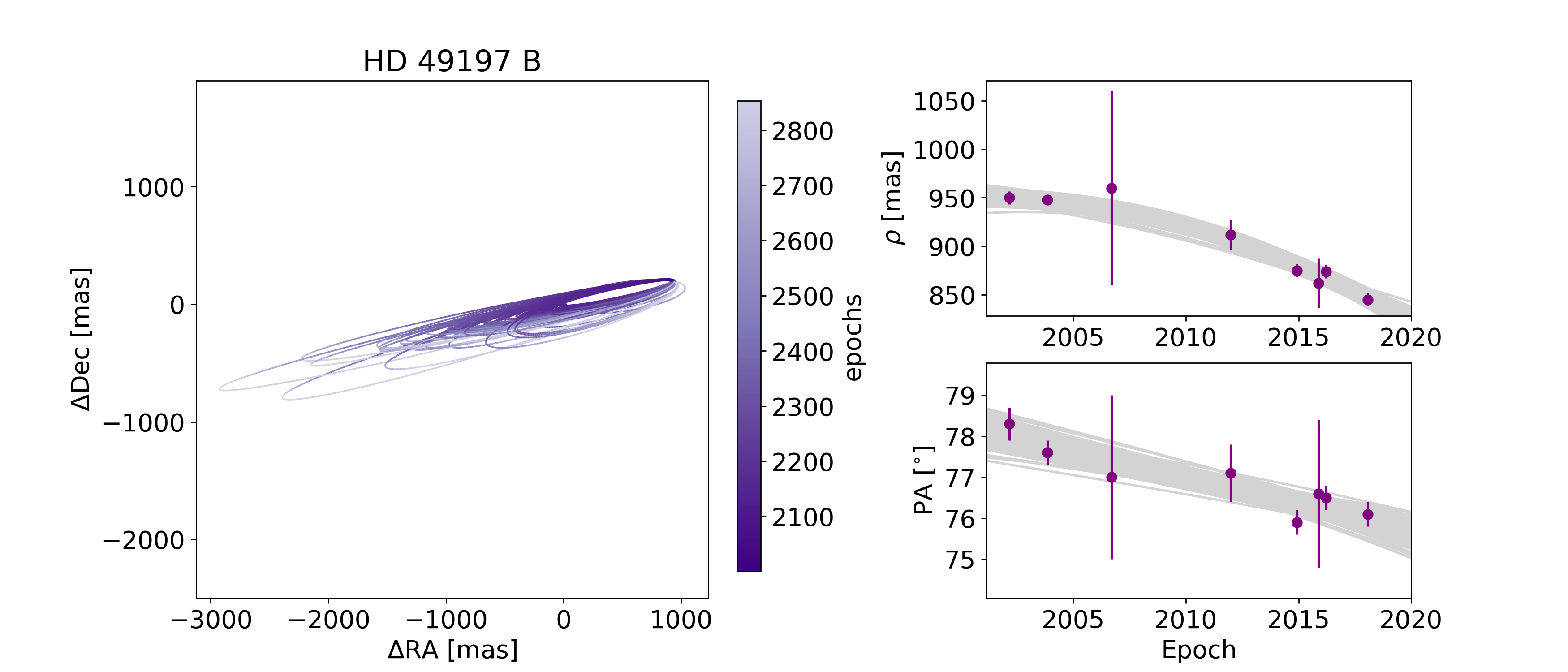}{0.5\textwidth}{(a) HD 49197 B}
                \fig{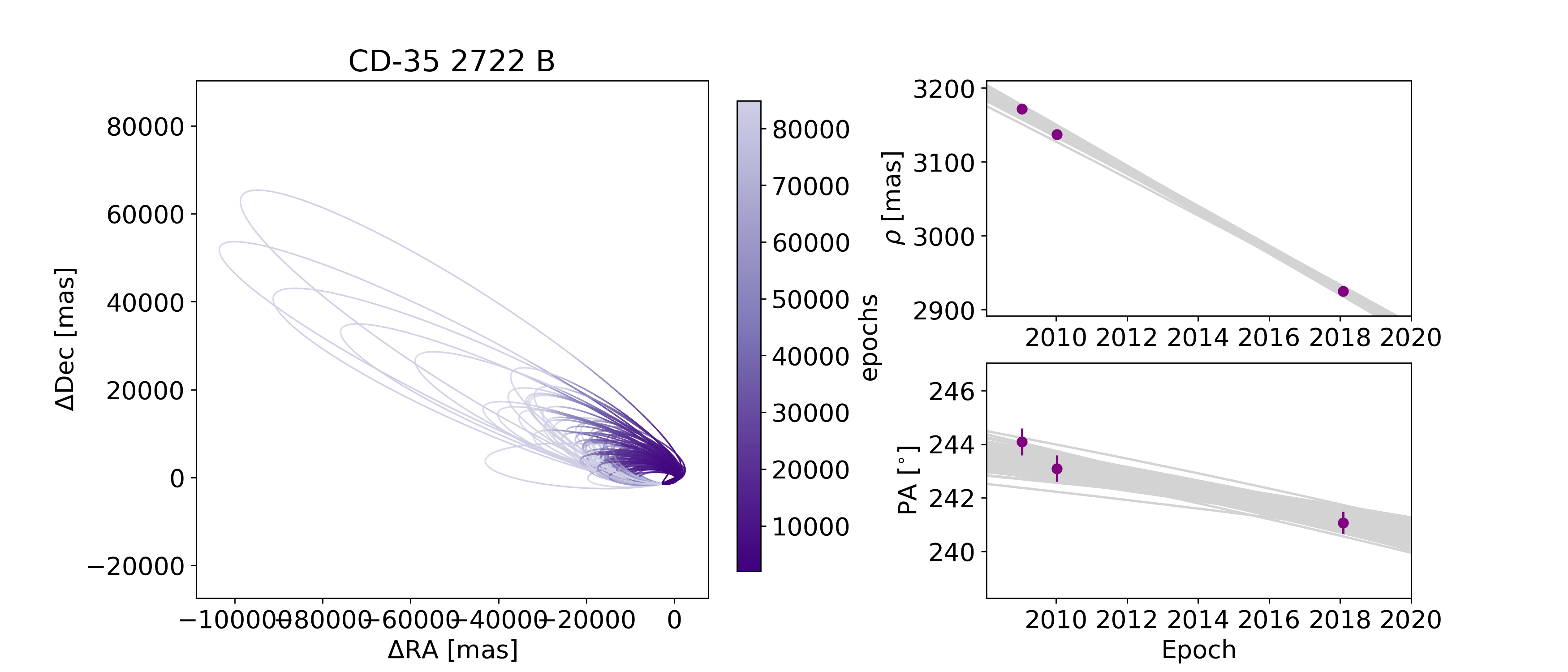}{0.5\textwidth}{(b) CD-35 2722 B}}                
  \gridline{\fig{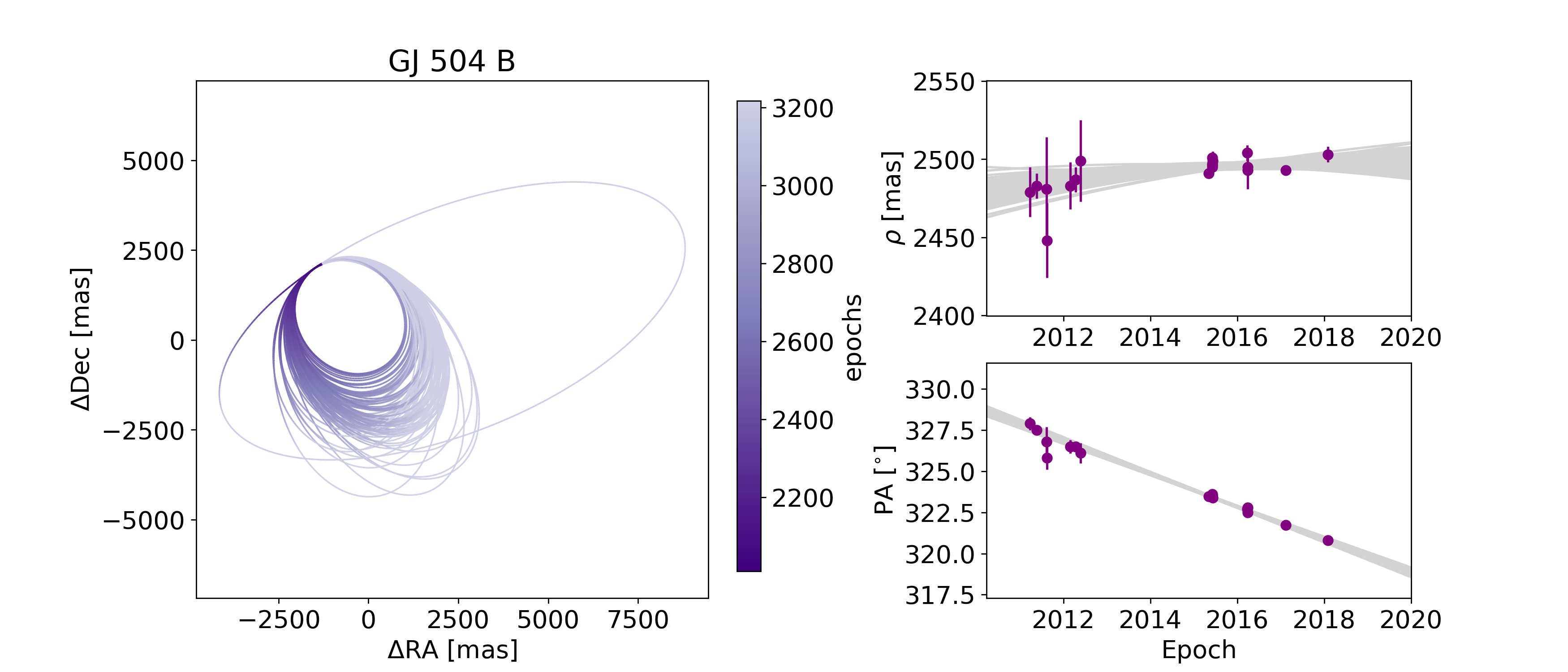}{0.5\textwidth}{(c) GJ 504 B}
                \fig{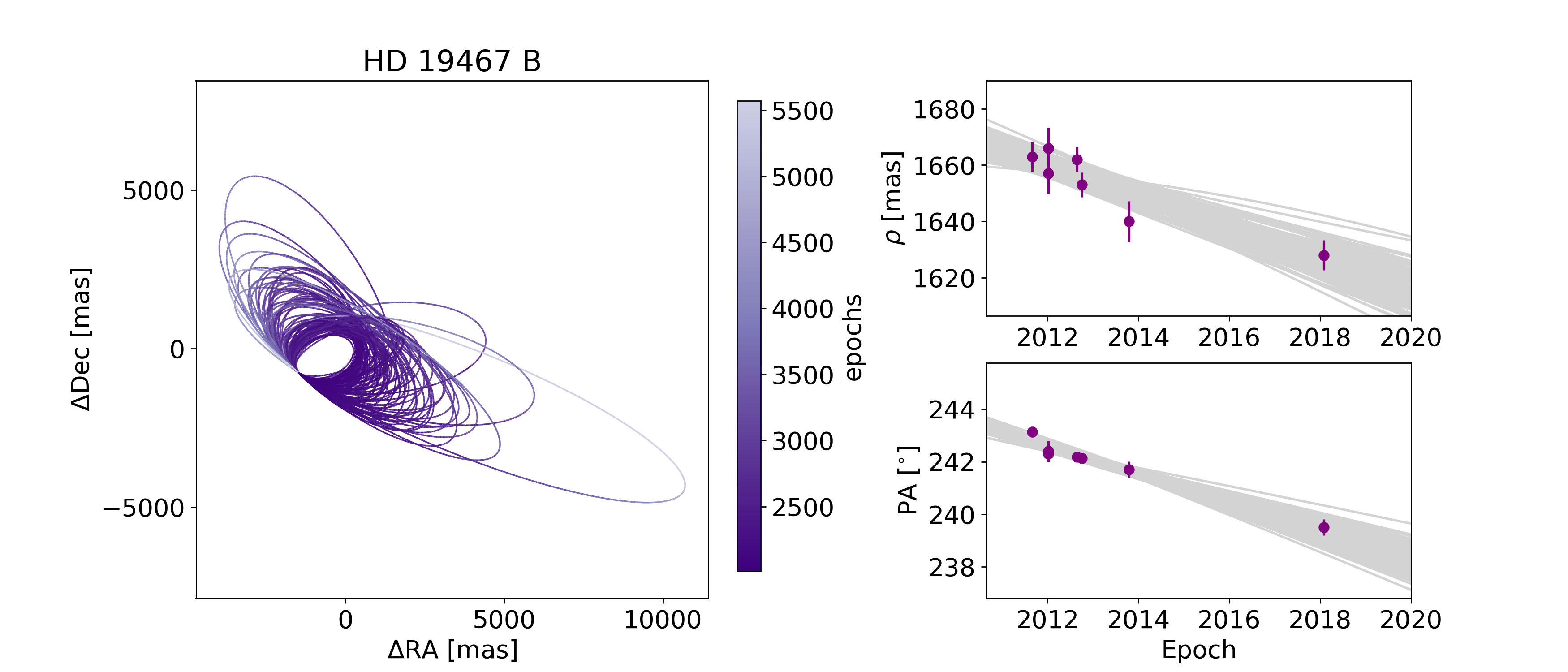}{0.5\textwidth}{(d) HD 19467 B}}
  \gridline{\fig{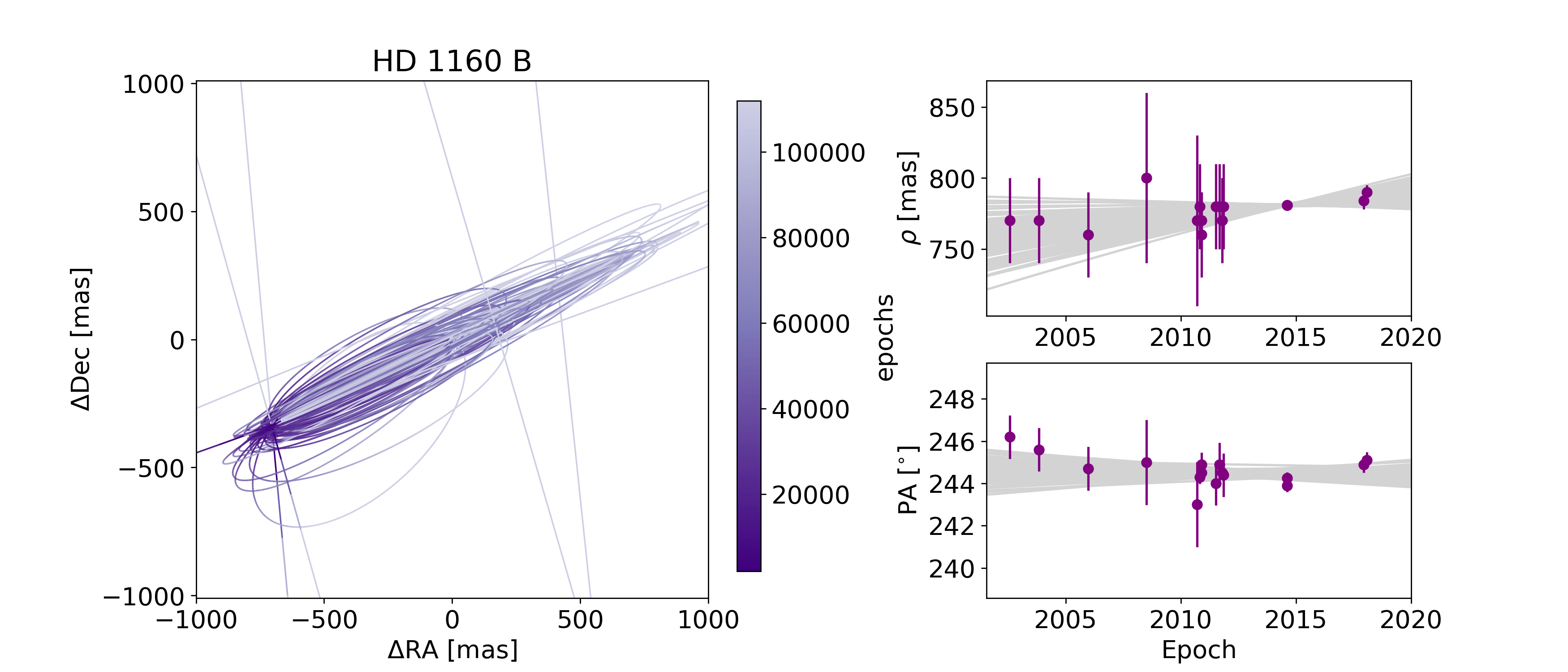}{0.5\textwidth}{(e) HD 1160 B}
                \fig{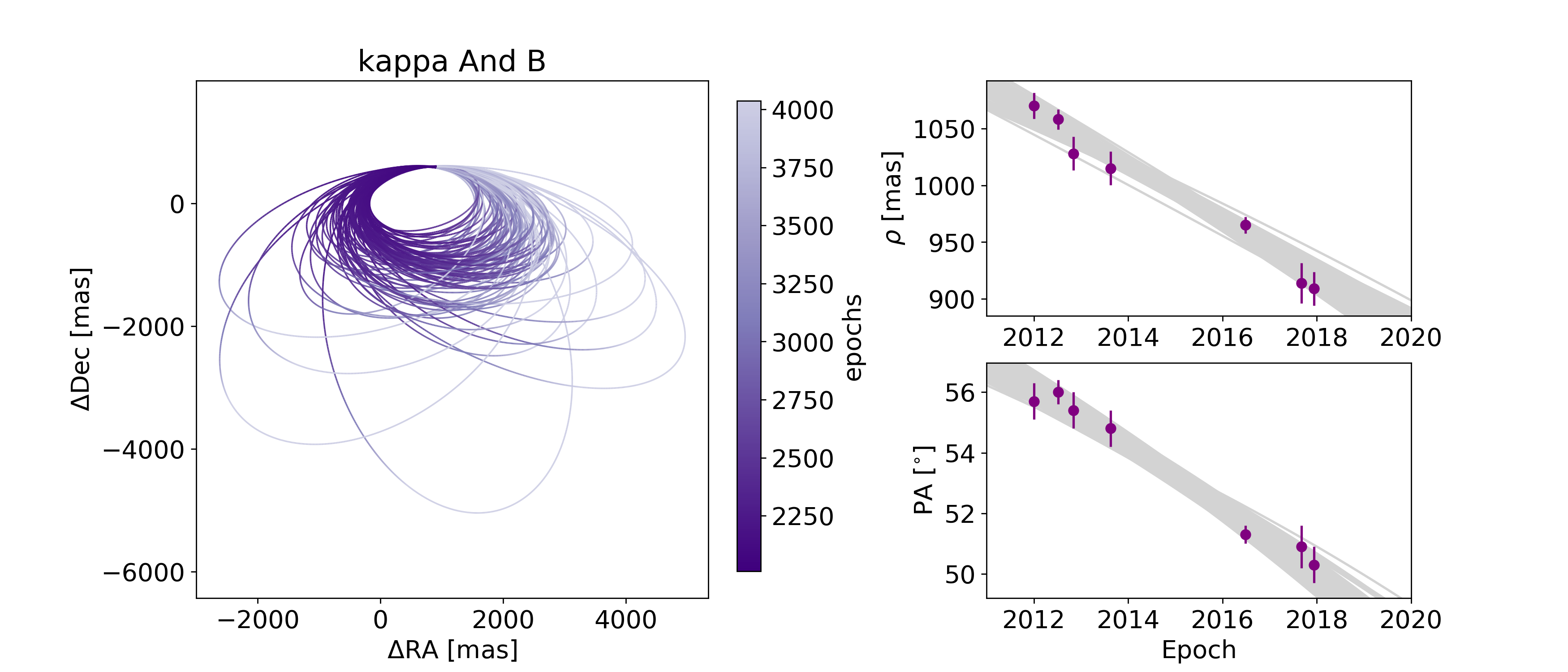}{0.5\textwidth}{(f) $\kappa$ And B}}
  \vskip 0 in
  \caption{Orbit fits for HD 49197 B (a), CD-35 2722 B (b), GJ 504 B (c), HD 19467 B (d), HD 1160 B (e), 
  and $\kappa$ And B (f). 
  For each object the left panel shows 100 
  randomly drawn orbits from the posterior distributions of orbital elements, 
  color coded to show the expected orbital location over time.  The right panels show measured separation  
  and P.A. of the companion compared to randomly drawn orbits (gray).  Note that jitter values
  have been added in quadrature to the plotted uncertainties.
    \label{fig:orbitfits1} } 
\end{figure*}

Our orbit fit is shown in Figures~\ref{fig:orbitfits1} and \ref{fig:cd352722corner}, and summarized in Table~\ref{tab:orbitize}.
We find a semimajor axis of 240$^{+90}_{-130}$ AU, a high eccentricity of $\approx$0.94, 
and an orbital period of 5400$^{+2900}_{-3900}$ yr (2$\sigma$ credible interval: 1000--32000 yr).
These results are largely consistent with the orbit constraints from \citet{Blunt:2017eta}, 
which only made use of the 2009 and 2010 epochs, but our new epoch now excludes low-eccentricity 
solutions.  


\begin{figure*}
  \gridline{\fig{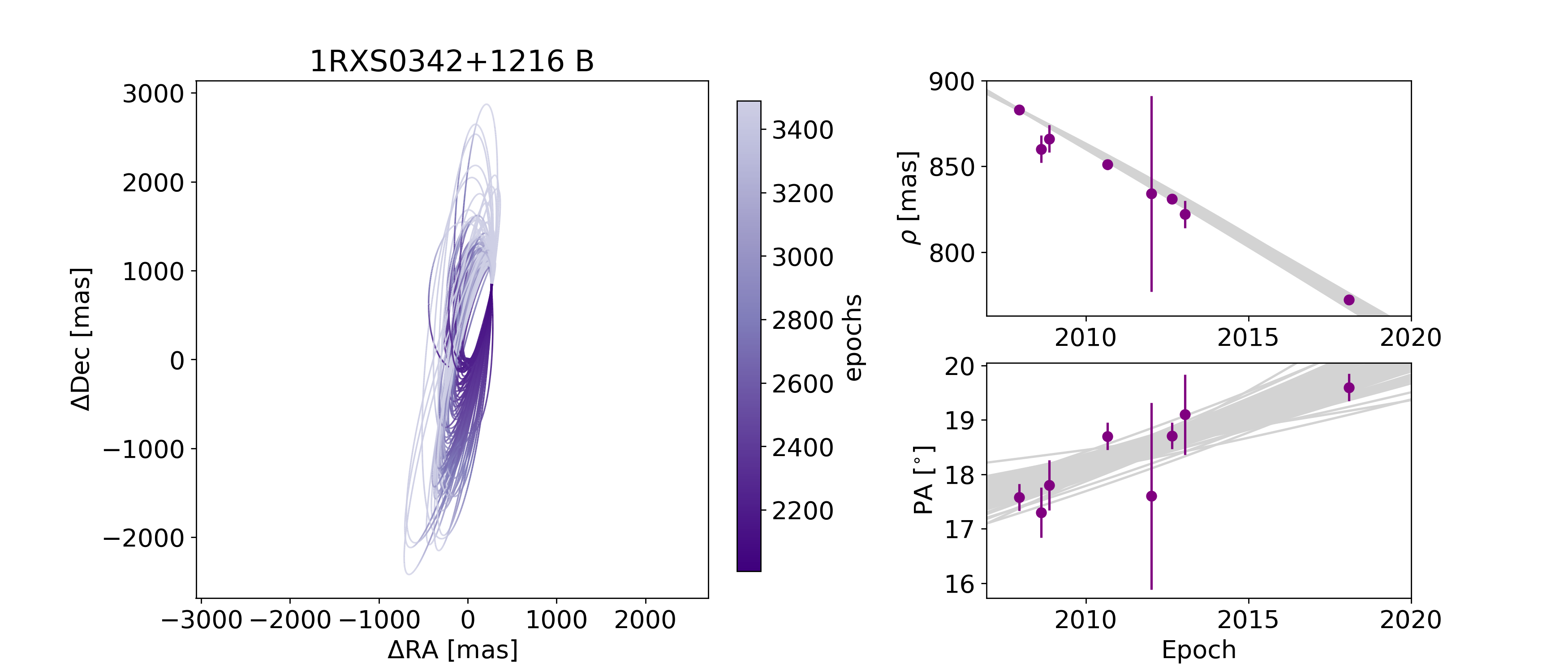}{0.5\textwidth}{(a) 1RXS0342+1216 B}
                \fig{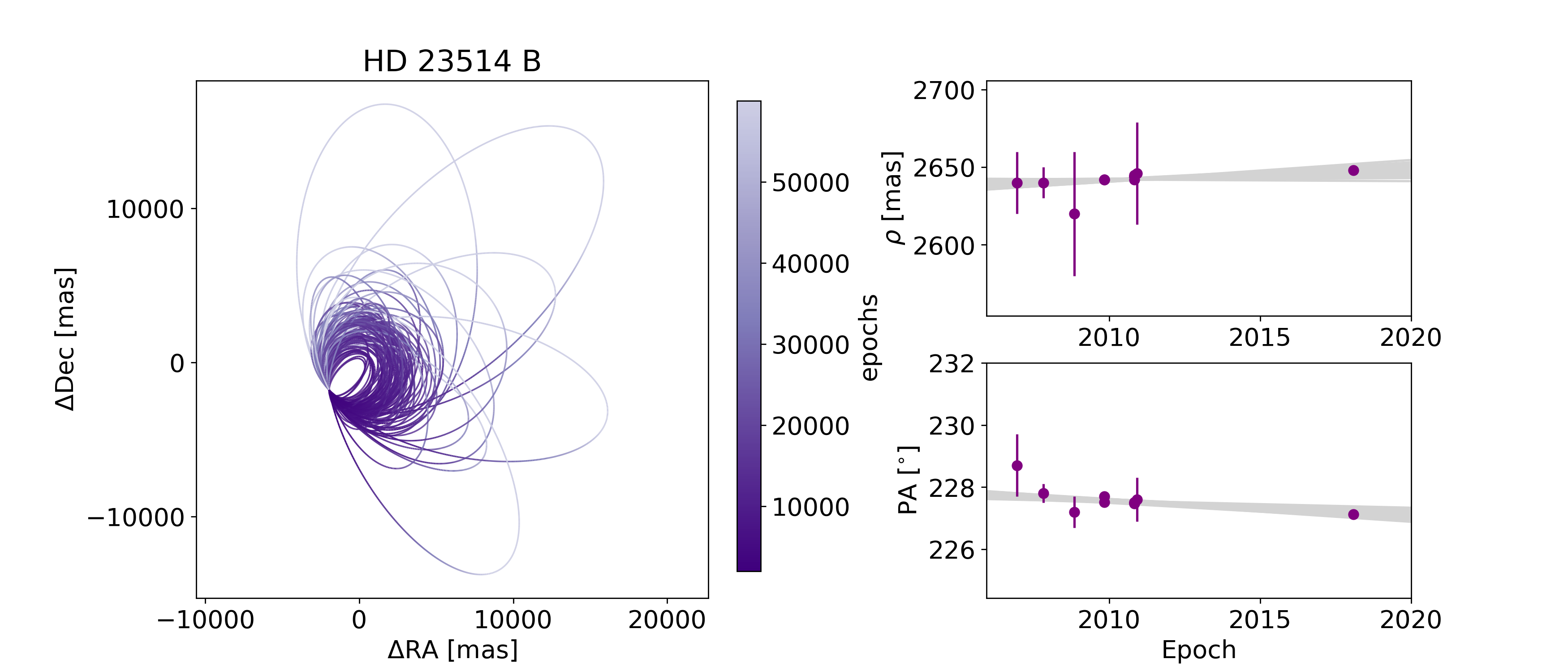}{0.5\textwidth}{(b) HD 23514 B}}
  \gridline{\fig{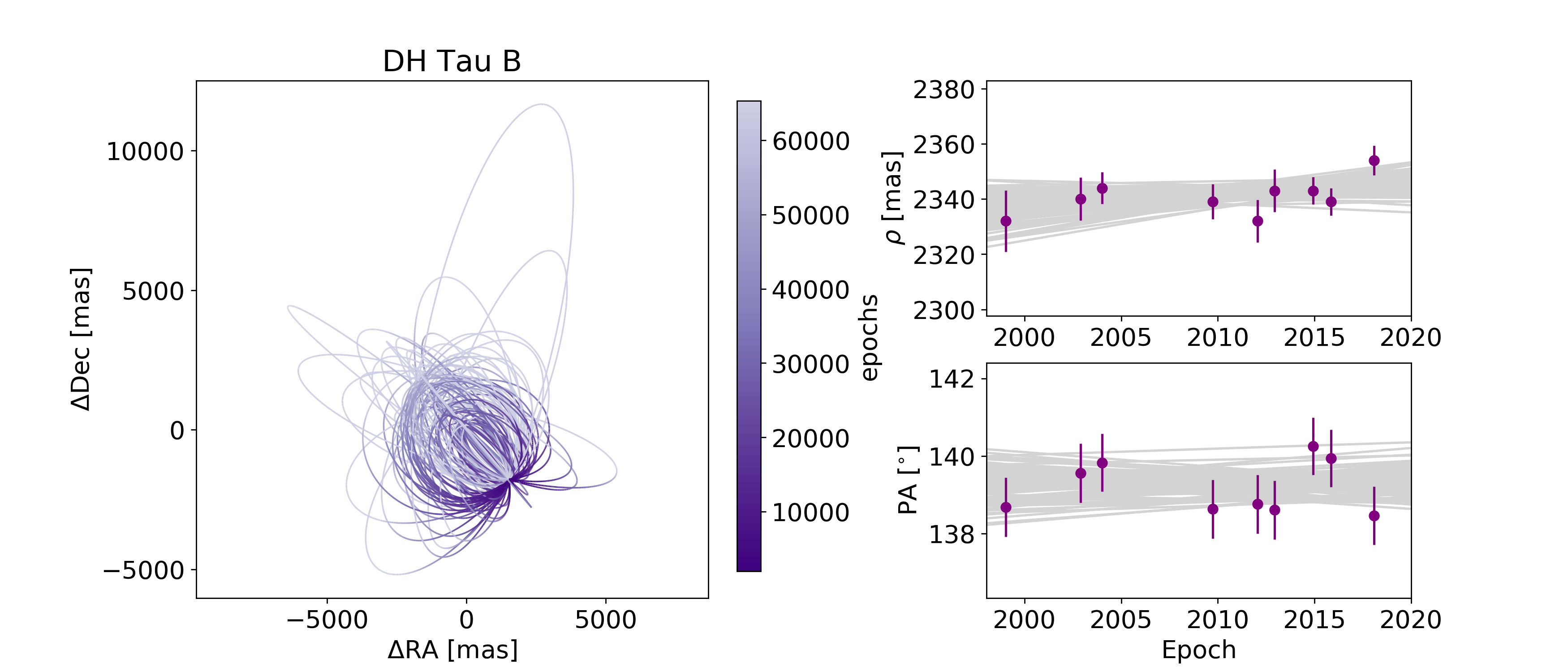}{0.5\textwidth}{(c) DH Tau B}
                \fig{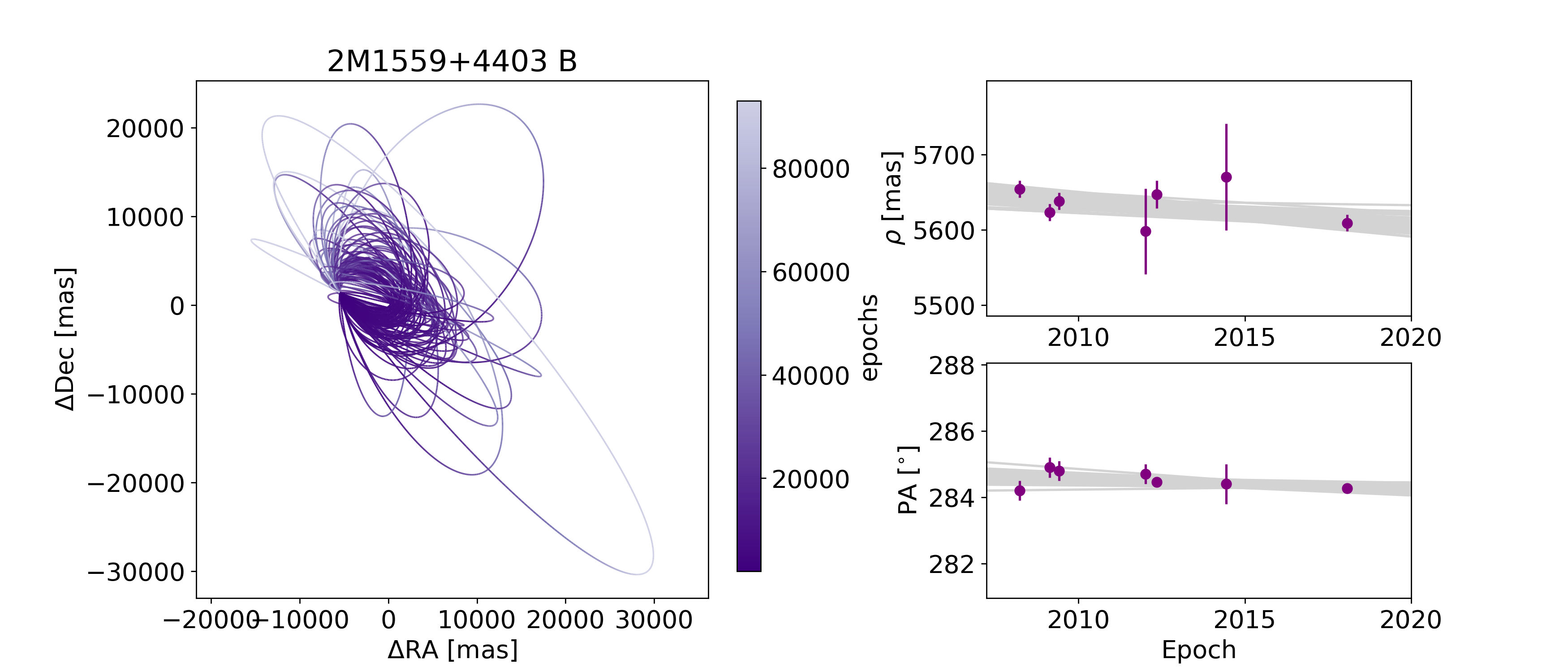}{0.5\textwidth}{(d) 2M1559+4403 B}}                
  \gridline{\fig{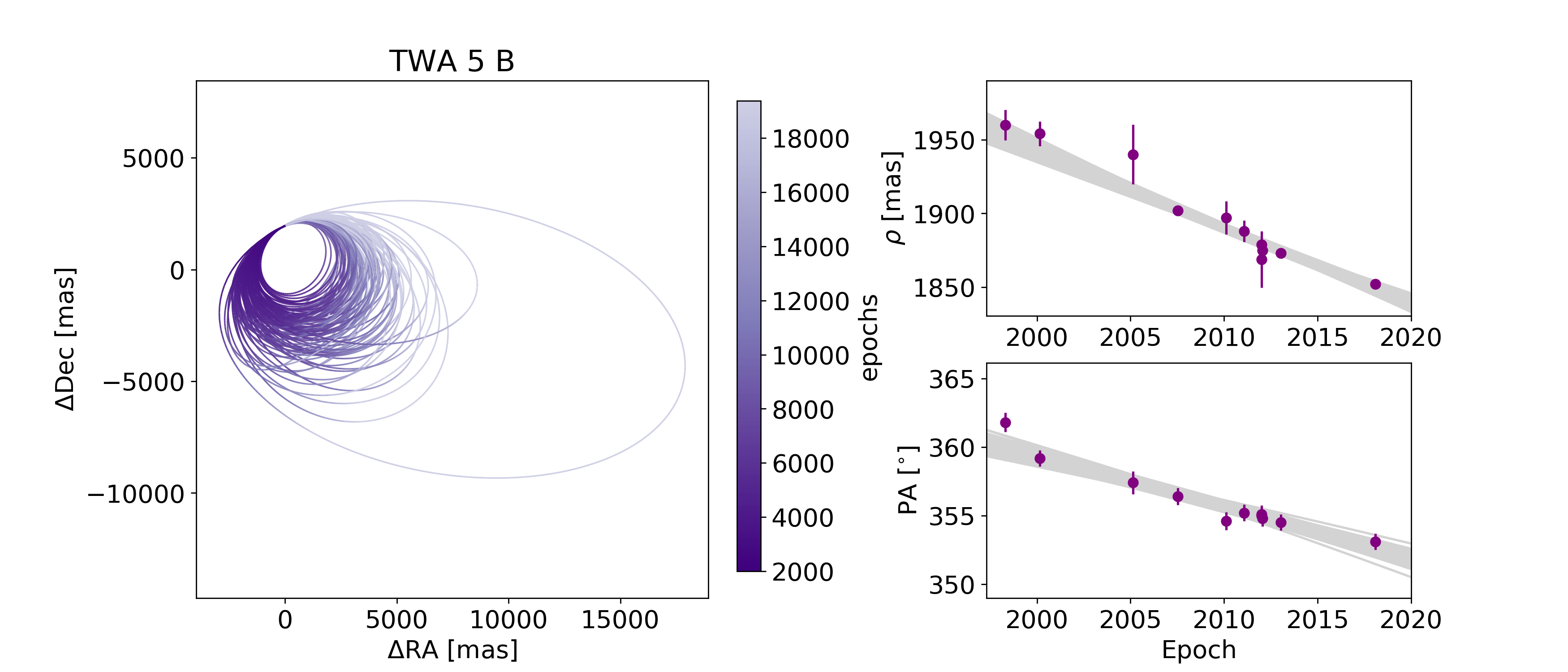}{0.5\textwidth}{(e) TWA 5 B}
                \fig{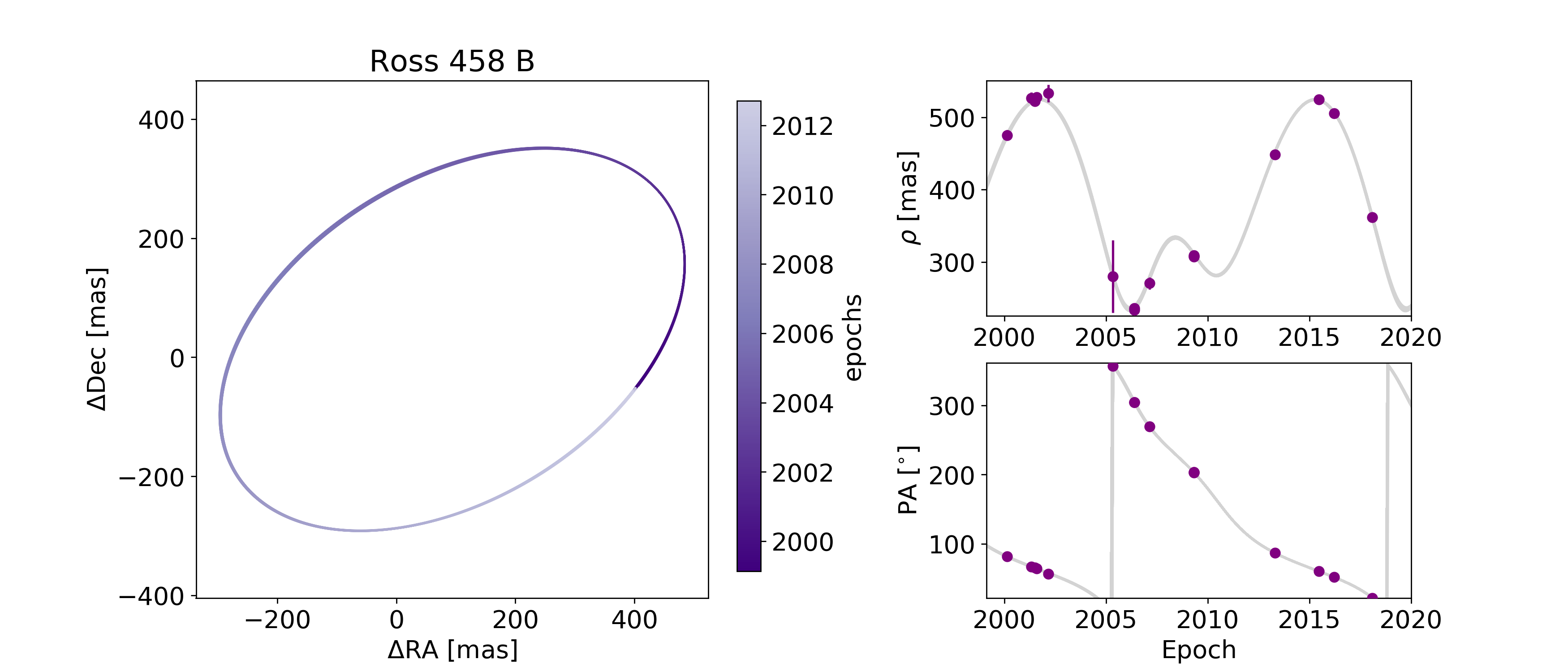}{0.5\textwidth}{(f) Ross 458 B}}
  \gridline{\fig{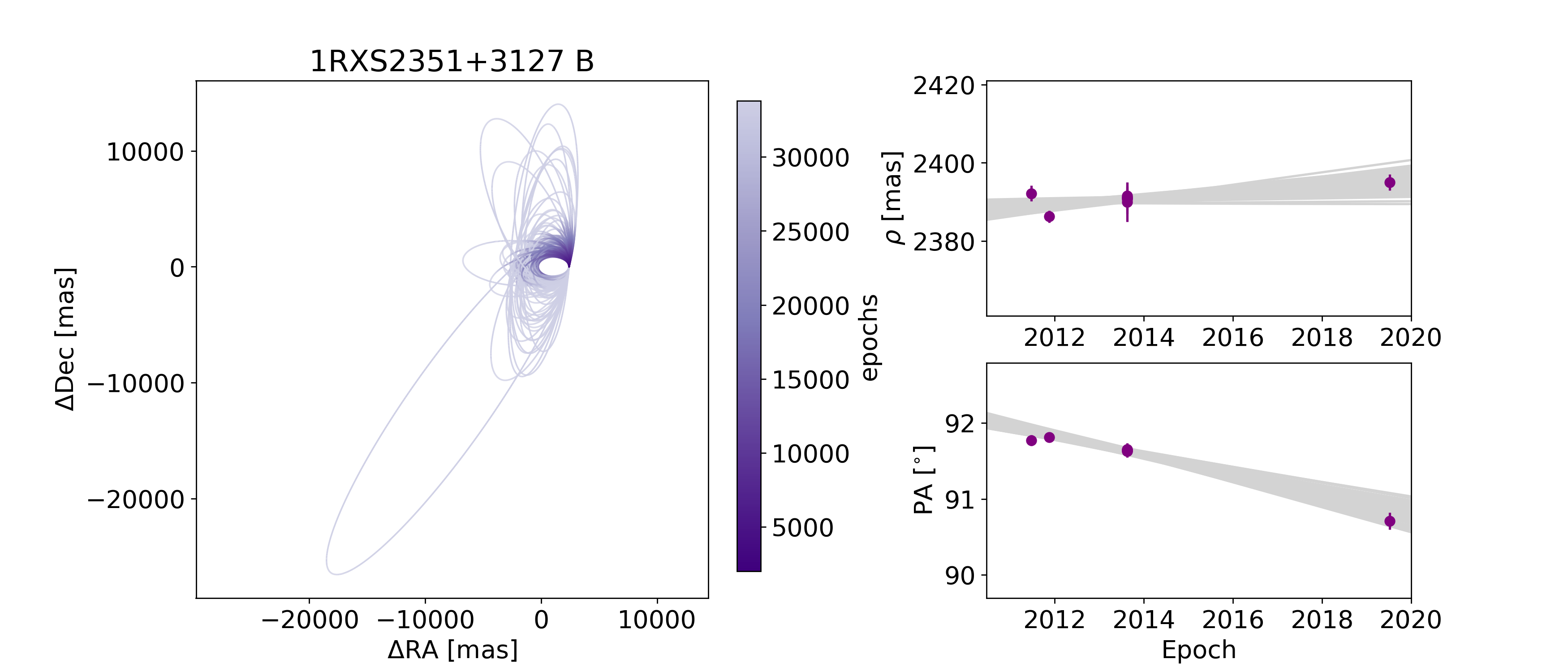}{0.5\textwidth}{(g) 1RXS2351+3127 B}}
  \vskip 0 in
  \caption{Orbit fits for 1RXS0342+1216 B (a), HD 23514 B (b), DH Tau B (c), 2M1559+4403 B (d), TWA 5 B (e), Ross 458 B (f), and 1RXS2351+3127 B (g).  See Figure~\ref{fig:orbitfits1} for details.
    \label{fig:orbitfits2} } 
\end{figure*}

\subsection{GJ 504 B}

GJ 504 B was discovered by \citet{Kuzuhara:2013jz} as part of the Strategic Exploration of Exoplanets
and Disks with the Subaru Telescope direct imaging survey (\citealt{Tamura:2016jg}).
Depending on the system age (21 $\pm$ 2 Myr or 4.0 $\pm$ 1.8 Gyr), 
and assuming hot-start evolutionary models, this T8--T9.5 companion has a mass that may be as 
low as $\approx$1~\Mjup \ or as high as $\approx$33~\Mjup \ 
(\citealt{Bonnefoy:2018ch}).
For this study we assume a stellar host mass of 1.10--1.25 \Msun \ from \citet{Bonnefoy:2018ch}, 
a companion mass of $\approx$23~\Mjup \ from \citet{Bonnefoy:2018ch},
a total system mass of 1.20 $\pm$ 0.04 \Msun,
and a parallax of 57.02 $\pm$ 0.25 mas from $Gaia$ DR2.

Our observations continue the astrometric trends noted by \citet{Kuzuhara:2013jz} and \citet{Bonnefoy:2018ch}
from their data spanning 2011 to 2017:
GJ 504 B is slowing moving away from the host star at a rate of 2.1 mas yr$^{-1}$, while 
most of the motion is in the P.A. with a rate of --1.0$\degr$ yr$^{-1}$ in the clockwise direction.
Modest jitter levels of $\sigma_\mathrm{jit, \rho}$ = 0.6 mas and $\sigma_\mathrm{jit, \theta}$ = 0.03$\degr$ are added in quadrature
with the available astrometry for this system.
Our orbit fit for GJ 504 B indicates a semi-major axis of 41$^{+4}_{-10}$ AU, a modest eccentricity of 0.22$^{+0.11}_{-0.17}$, and 
a period of 240$^{+40}_{-80}$ yr (2$\sigma$ credible interval: 140--470 yr).
Our posteriors generally resemble those from \citet{Blunt:2017eta} and 
\citealt{Bonnefoy:2018ch}), with slightly improved constraints due to the extended astrometric coverage.
Results from the orbit fit are shown in Figures~\ref{fig:orbitfits1} and \ref{fig:gj504corner}.

\subsection{HD 19467 B}

HD 19467 B was first imaged by \citet{Crepp:2014ce} based on
a long-term radial acceleration of its Sun-like host star.  
Photometry and spectroscopy from \citet{Crepp:2015gt} imply a mid-T spectral 
type and a mass of at least 52~\Mjup.  
We assume a stellar mass of 0.95 $\pm$ 0.02 \Msun \ from \citet{Crepp:2014ce},
a total system mass of 1.00 $\pm$ 0.02 \Msun,
and a parallax of 31.226 $\pm$ 0.041 mas from $Gaia$ DR2 in our orbit analysis.

Published astrometry from \citet{Crepp:2014ce} and 
\citet{Crepp:2015gt} between 2011 and 2014 showed clockwise orbital motion and
 motion toward the host star.
Our new astrometry in 2018 confirm this trend, revealing evolution in P.A. at a rate of
--0.5$\degr$ yr$^{-1}$ and in separation at a rate of --5.6 mas yr$^{-1}$.
A modest amount of additional jitter ($\sigma_\mathrm{jit, \rho}$ = 1.8 mas, $\sigma_\mathrm{jit, \theta}$ = 0.06$\degr$) 
is needed to bring the $\chi^2_{\nu}$ to unity for linear evolution of separation and P.A.
HD 19467 B has an orbital period of 420$^{+170}_{-250}$ yr (2$\sigma$ credible interval: 160--1530 yr)
with a semimajor axis of 56$^{+15}_{-25}$ AU.  It does not appear to have a high eccentricity ($e$$<$0.8).
Results from the orbit fit to the astrometry for this system are shown in Figures~\ref{fig:orbitfits1} and \ref{fig:hd19467corner}.

\subsection{HD 1160 B}

\citet{Nielsen:2012jk} identified this low-mass companion to the A0 star HD 1160 
as part of the Gemini-NICI Planet-Finding Campaign, along with the low-mass stellar 
companion HD 1160 C at a wider separation (5$\farcs$2).
\citet{Garcia:2017dk} found a mid-M spectral type for HD 1160 B and mass range between 35 and 90 \Mjup \ depending
on the system age. 
Recently \citet{Curtis:2019jx} found that this system may belong to the proposed $\approx$120 Myr Psc-Eri stream, 
which stretches 120$\degr$ across the sky and about 400 pc in space.  At this age, the implied mass for the companion HD 1160 B
would be about 0.12~\Msun.  However, details about the origin, metallicity, membership probabilities, 
and potential for age gradients in this new proposed stream have not yet been established.  For this study
we adopt the more uncertain age and mass constraints for HD 1160 B from \citet{Nielsen:2012jk} and \citet{Garcia:2017dk}
but caution that this companion may reside above the hydrogen burning limit if the age is indeed older.
The mass we adopt for HD 1160 A is 1.95 $\pm$ 0.05 \Msun \ (\citealt{Blunt:2017eta}), 
the total system mass we use is 2.01 $\pm$ 0.06 \Msun,
and the $Gaia$ DR2 parallax is 7.942 $\pm$ 0.076 mas.

Astrometry from 2002 to 2018 is presented by \citet{Nielsen:2012jk}, \citet{Maire:2016go}, and \citet{Currie:2018hj}.
Our new epoch in 2018 is consistent with that of \citet{Currie:2018hj}, which was taken about two months earlier.
We find slow motion away from HD 1160 A at a rate of 1.7 mas yr$^{-1}$ but no significant 
change in P.A. over time.
There is no evidence that the separation measurements errors are underestimated, but we find 
a jitter level of 0.24$\degr$ is needed to inflate the P.A. measurements.
Our orbit fit (Figures~\ref{fig:orbitfits1} and \ref{fig:hd1160corner}) implies a semimajor axis of 
80$^{+20}_{-30}$ AU and an orbital period of 520$^{+200}_{-270}$ yr 
(2$\sigma$ credible interval: 230--3700 yr).  
The eccentricity is poorly constrained.
These results are consistent with those from \citet{Blunt:2017eta}.

\subsection{$\kappa$ And B} 

$\kappa$ And B (\citealt{Carson:2013fw}) is a substellar companion orbiting a  young B9 star.
The companion mass falls near the brown dwarf/planetary mass boundary at 22 $\pm$ 9 \Mjup.
Follow up photometry and spectroscopy from \citet{Hinkley:2013ko}, \citet{Bonnefoy:2014dx}, and \citet{Currie:2018hj}
indicate an early-L dwarf with red colors similar to low-gravity planets and brown dwarfs.
We adopt a host star mass of 2.8$\pm$ 0.1 \Msun \ (\citealt{Jones:2016hg}),
a total system mass of 2.82 $\pm$ 0.10 \Msun,
and a parallax of 19.975 $\pm$ 0.342 mas from $Gaia$ DR2.

Our astrometry from 2016 combined with measurements from the literature obtained between 2012 and 2018 
reveal rapid motion toward the star at a rate of --24 mas yr$^{-1}$ and P.A. evolution at a rate of --1.0$\degr$ yr$^{-1}$
in the clockwise direction.
Significant jitter is required for the separation measurements ($\sigma_\mathrm{jit, \rho}$ = 5.2 mas) but
no jitter is needed for the P.A. uncertainties.
Our orbit fit is shown in Figure~\ref{fig:orbitfits1} and the corner plot is displayed in Figure~\ref{fig:kappaandcorner}; 
we find a semimajor axis of 80$^{+20}_{-30}$ AU, 
an orbital period of 420$^{+150}_{-210}$ yr (2$\sigma$ credible interval: 170--1300 yr), and a high  
eccentricity of 0.74$^{+0.10}_{-0.08}$.  These results are consistent with the orbit fit from \citet{Currie:2018hj}.

\subsection{1RXS0342+1216 B} 

This 35 $\pm$ 8 \Mjup \ companion to the M4 dwarf 1RXS J034231.8+121622 (2MASS J03423180+1216225) 
was discovered by \citet{Bowler:2015ja}
as part of the Planets Around Low-Mass Stars (PALMS) survey.
\citet{Bowler:2015ch} determine a spectral type of L0 from near-infrared spectroscopy and identified significant orbital motion.
For our orbit analysis we adopt a host mass of 0.20 $\pm$ 0.05 \Msun \ (\citealt{Bowler:2015ch}),
a total system mass of 0.23 $\pm$ 0.05 \Msun, and parallax of 
30.308 $\pm$ 0.067 mas from $Gaia$ DR2.

\begin{rotatetable*}
\begin{deluxetable*}{lcccccccccc}
\renewcommand\arraystretch{0.9}
\tabletypesize{\scriptsize}
\setlength{ \tabcolsep } {.1cm} 
\tablewidth{0pt}
\tablecolumns{11}
\tablecaption{Results from Orbit Fits\label{tab:orbitize}}
\tablehead{
       \colhead{Name}         & \colhead{$P$}   & \colhead{$a$}  &  \colhead{$e$}   & \colhead{$i$}     & \colhead{$\omega$\tablenotemark{a}} & \colhead{$\Omega$\tablenotemark{a}} & \colhead{$\tau$\tablenotemark{b}} \\ 
       \colhead{}             & \colhead{(yr)}  & \colhead{(AU)} &  \colhead{}      & \colhead{($\degr$)} & \colhead{($\degr$)}  & \colhead{($\degr$)}  & \colhead{}    
        }   
\startdata
                  HD 984 B    &                 67.5$^{+23}_{-43}$ (14.6--260)    &         17.6$^{+4.3}_{-8.1}$ (6.98--43.9)    &                  0.23$^{+0.11}_{-0.23}$ (0.0--0.63)    &            120$^{+6.1}_{-7.3}$ (108--137)     &                         111$^{+69}_{-41}$     &                        26.5$^{+11}_{-16}$     &                   0.504$^{+0.32}_{-0.36}$    \\
                 HD 1160 B    &                523$^{+200}_{-270}$ (228--3700)    &            81.9$^{+20}_{-31}$ (48.7--302)    &                 0.78$^{+0.22}_{-0.23}$ (0.076--1.0)    &          92.0$^{+8.7}_{-9.3}$ (61.6--137)     &                        48.1$^{+20}_{-48}$     &                      64.6$^{+4.8}_{-3.6}$     &                  0.114$^{+0.070}_{-0.11}$    \\
                HD 19467 B    &                417$^{+170}_{-250}$ (162--1530)    &            55.8$^{+15}_{-25}$ (29.9--133)    &                0.39$^{+0.26}_{-0.18}$ (0.034--0.74)    &             125$^{+9.4}_{-14}$ (105--159)     &                        66.4$^{+32}_{-44}$     &                         113$^{+16}_{-41}$     &                   0.373$^{+0.36}_{-0.37}$    \\
           1RXS0342+1216 B    &                470$^{+150}_{-240}$ (157--1040)    &          36.5$^{+8.1}_{-14}$ (18.8--61.4)    &                0.34$^{+0.25}_{-0.34}$ (0.026--0.98)    &         80.8$^{+4.7}_{-4.0}$ (38.8--87.0)     &                         116$^{+64}_{-42}$     &                      12.9$^{+3.0}_{-5.0}$     &                   0.469$^{+0.22}_{-0.26}$    \\
                  51 Eri b    &                34.5$^{+9.6}_{-15}$ (17.1--117)    &         12.8$^{+2.4}_{-3.8}$ (8.14--28.9)    &                0.50$^{+0.11}_{-0.083}$ (0.15--0.68)    &             132$^{+9.1}_{-10}$ (115--156)     &                        85.7$^{+25}_{-27}$     &                        72.6$^{+34}_{-73}$     &                   0.503$^{+0.44}_{-0.24}$    \\
              CD-35 2722 B    &            5380$^{+2900}_{-3900}$ (988--32200)    &            241$^{+85}_{-130}$ (87.8--785)    &               0.94$^{+0.033}_{-0.026}$ (0.86--0.99)    &              151$^{+20}_{-15}$ (119--177)     &                         136$^{+33}_{-21}$     &                         260$^{+21}_{-22}$     &                0.0451$^{+0.021}_{-0.039}$    \\
                HD 49197 B    &                  146$^{+53}_{-67}$ (75.0--449)    &         29.1$^{+6.7}_{-9.7}$ (19.1--61.8)    &                 0.73$^{+0.27}_{-0.22}$ (0.072--1.0)    &          96.9$^{+4.1}_{-4.1}$ (91.9--133)     &                         138$^{+42}_{-18}$     &                      75.6$^{+2.7}_{-2.5}$     &                   0.503$^{+0.19}_{-0.50}$    \\
                 HR 2562 B    &                  114$^{+52}_{-78}$ (35.6--630)    &          26.3$^{+7.5}_{-14}$ (12.2--82.3)    &                 0.45$^{+0.28}_{-0.45}$ (0.037--1.0)    &         86.3$^{+3.2}_{-2.3}$ (59.3--91.7)     &                         125$^{+55}_{-63}$     &                       301$^{+3.7}_{-3.3}$     &                   0.439$^{+0.20}_{-0.44}$    \\
                 HR 3549 B    &                599$^{+260}_{-380}$ (168--3030)    &            94.3$^{+28}_{-43}$ (42.1--278)    &                  0.43$^{+0.17}_{-0.41}$ (0.0--0.88)    &             132$^{+21}_{-21}$ (97.5--172)     &                        87.3$^{+34}_{-87}$     &                         264$^{+96}_{-46}$     &                   0.436$^{+0.20}_{-0.43}$    \\
                HD 95086 b    &                 350$^{+90}_{-120}$ (175--1030)    &            59.2$^{+10}_{-13}$ (38.6--122)    &                 0.14$^{+0.067}_{-0.14}$ (0.0--0.48)    &              150$^{+12}_{-13}$ (127--174)     &                         105$^{+74}_{-38}$     &                        88.1$^{+55}_{-39}$     &                   0.271$^{+0.11}_{-0.21}$    \\
                  GJ 504 B    &                   242$^{+37}_{-84}$ (138--468)    &          41.3$^{+4.0}_{-10}$ (29.0--64.1)    &                  0.22$^{+0.11}_{-0.17}$ (0.0--0.44)    &            141$^{+8.0}_{-8.7}$ (128--169)     &                        61.0$^{+37}_{-61}$     &                        153$^{+9.5}_{-12}$     &                 0.357$^{+0.097}_{-0.078}$    \\
               HIP 65426 b    &                649$^{+300}_{-409}$ (219--5280)    &            93.9$^{+28}_{-45}$ (45.7--380)    &                0.55$^{+0.42}_{-0.22}$ (0.027--0.97)    &             112$^{+14}_{-18}$ (84.8--158)     &                        75.2$^{+39}_{-72}$     &                         329$^{+31}_{-56}$     &                   0.661$^{+0.34}_{-0.18}$    \\
                  PDS 70 b    &                  131$^{+41}_{-65}$ (50.8--476)    &         23.6$^{+5.3}_{-7.9}$ (13.1--56.2)    &                  0.23$^{+0.10}_{-0.23}$ (0.0--0.59)    &              141$^{+14}_{-15}$ (115--171)     &                        82.4$^{+33}_{-82}$     &                        76.1$^{+50}_{-76}$     &                   0.495$^{+0.19}_{-0.49}$    \\
                  PZ Tel B    &                  108$^{+39}_{-35}$ (61.3--210)    &         24.8$^{+5.3}_{-5.5}$ (17.6--38.6)    &                 0.89$^{+0.10}_{-0.048}$ (0.74--1.0)    &          93.4$^{+1.2}_{-1.7}$ (91.0--100)     &                        168$^{+12}_{-4.7}$     &                     57.1$^{+1.2}_{-0.88}$     &                   0.595$^{+0.40}_{-0.18}$    \\
               HD 206893 B    &              28.1$^{+4.4}_{-9.2}$ (17.4--55.6)    &         10.2$^{+1.1}_{-2.3}$ (7.43--16.1)    &                  0.25$^{+0.17}_{-0.14}$ (0.0--0.44)    &             143$^{+9.7}_{-10}$ (128--170)     &                        74.3$^{+33}_{-74}$     &                         255$^{+31}_{-37}$     &                   0.522$^{+0.47}_{-0.20}$    \\
            $\kappa$ And B    &                417$^{+150}_{-210}$ (174--1300)    &            78.9$^{+18}_{-28}$ (44.8--169)    &               0.74$^{+0.098}_{-0.075}$ (0.53--0.88)    &              139$^{+13}_{-18}$ (115--170)     &                         134$^{+35}_{-24}$     &                        72.0$^{+21}_{-16}$     &                   0.449$^{+0.20}_{-0.22}$    \\
                HD 23514 B    &           7010$^{+2500}_{-3900}$ (2120--27400)    &           417$^{+100}_{-160}$ (201--1040)    &                  0.29$^{+0.11}_{-0.28}$ (0.0--0.68)    &              140$^{+17}_{-16}$ (111--172)     &                         106$^{+74}_{-37}$     &                        259$^{+100}_{-43}$     &                   0.740$^{+0.26}_{-0.10}$    \\
                  DH Tau B    &           7460$^{+3200}_{-4800}$ (2360--59800)    &            330$^{+89}_{-160}$ (158--1320)    &               0.49$^{+0.19}_{-0.49}$ (0.0023--0.96)    &            79.9$^{+38}_{-40}$ (15.5--154)     &                        99.0$^{+65}_{-46}$     &                         129$^{+51}_{-38}$     &                   0.623$^{+0.31}_{-0.18}$    \\
             2M1559+4403 B    &           6510$^{+2800}_{-4300}$ (1830--33500)    &             289$^{+80}_{-140}$ (131--859)    &                  0.47$^{+0.28}_{-0.34}$ (0.0--0.90)    &             122$^{+17}_{-21}$ (91.4--165)     &                        79.2$^{+40}_{-79}$     &                         122$^{+33}_{-21}$     &                   0.297$^{+0.13}_{-0.25}$    \\
                   TWA 5 B    &               1810$^{+540}_{-840}$ (737--4420)    &             146$^{+30}_{-45}$ (85.5--265)    &                0.43$^{+0.099}_{-0.10}$ (0.23--0.67)    &              151$^{+12}_{-13}$ (129--175)     &                         123$^{+41}_{-31}$     &                        48.0$^{+24}_{-23}$     &                 0.159$^{+0.066}_{-0.095}$    \\
                Ross 458 B    &       13.528$^{+0.02}_{-0.014}$ (13.49--13.56)    &   4.95$^{+0.0096}_{-0.0095}$ (4.93--4.97)    &           0.2419$^{+0.001}_{-0.001}$ (0.240--0.244)    &    130.0$^{+0.14}_{-0.13}$ (129.7--130.3)     &                   156.5$^{+0.40}_{-0.39}$     &                    56.8$^{+0.13}_{-0.12}$     &                 0.985$^{+0.015}_{-0.012}$    \\
           1RXS2351+3127 B    &              1490$^{+630}_{-869}$ (568--10100)    &             102$^{+27}_{-43}$ (56.8--367)    &               0.46$^{+0.25}_{-0.17}$ (0.0030--0.73)    &              127$^{+13}_{-14}$ (103--161)     &                        96.2$^{+47}_{-51}$     &                        80.4$^{+18}_{-76}$     &                   0.686$^{+0.24}_{-0.20}$    \\
\enddata
\tablecomments{Values represent the median and 68.3\% credible interval of each marginalized posterior.  95.4\% credible intervals are listed in parentheses.}
\tablenotetext{a}{Here $\omega$ amd $\Omega$ are defined on the interval [0,2$\pi$).}
\tablenotetext{b}{Time of periastron passage expressed as fraction of orbital period after MJD=0.}
\end{deluxetable*}
\end{rotatetable*}


Our new epoch was obtained in 2018, significantly extending the orbital coverage since the last published epoch 
acquired in 2013.
1RXS0342+1216 B continues to approach its host star at a rate of --11.0 mas yr$^{-1}$ 
in a counter-clockwise direction at 0.2$\degr$ yr$^{-1}$.  
We find a jitter term of $\sigma_\mathrm{jit, \theta}$ = 0.23$\degr$ is needed to inflate the P.A. uncertainties.
Our orbit fit is shown in Figures~\ref{fig:orbitfits2} and \ref{fig:rxs0342corner}; we find a 
semimajor axis of 37$^{+8}_{-14}$ AU and an orbital period of
470$^{+150}_{-240}$ yr (2$\sigma$ credible interval: 
160--1040 yr).
At present the orbital eccentricity is effectively unconstrained.

\subsection{HD 23514 B} 

This substellar companion to the unusually dusty Pleiad HD 23514 was first identified by \citet{Rodriguez:2012ef}.
\citet{Bowler:2015ch} determined a spectral type of M8 for HD 23514 B from $H$- and $K$-band spectroscopy. 
We adopt a host star mass of 1.42 $\pm$ 0.15 from \citet{Huber:2016fs},
a total system mass of 1.48 $\pm$ 0.15 \Msun,
and parallax of 7.210 $\pm$ 0.054 mas from $Gaia$ DR2.

When combined with published astrometry by \citet{Rodriguez:2012ef} and \citet{Yamamoto:2013gu},
our new astrometry from 2018 establishes the first signs of orbital motion for HD 23514 B: the separation
is increasing by 0.7 mas yr$^{-1}$ and the P.A. is evolving by --0.06$\degr$ yr$^{-1}$
in the clockwise direction.  Over the $\approx$11 year baseline since
the first detection of this system in 2006, this amounts to a change of about 8 mas and 0.7$\degr$.
No jitter is necessary for the separation measurements and only a slight amount (0.05$\degr$) is needed for the P.A.s.
Results from the orbit fit are shown in Figures~\ref{fig:orbitfits2} and \ref{fig:hd23514corner}.
Despite the limited orbit coverage, eccentricities above $\approx$0.8 can be ruled out and current astrometry favors 
modest values near 0.3.  The semi-major axis is 420$^{+100}_{-160}$ AU with a long and highly uncertain orbital
period of 7000$^{+2500}_{-3900}$ yr (2$\sigma$ credible interval: 2100--27400 yr).

\subsection{DH Tau B} 

\citet{Itoh:2005ii} discovered this brown dwarf companion to DH Tau, a young ($\approx$2~Myr) M1 
star in Taurus.  \citet{Bonnefoy:2014dh} determine a spectral type of M9.25 and infer a mass of 8--21~\Mjup.
\citet{Ginski:2014ef} examined astrometry for this system but did not find signs of orbital motion at that time.
For our orbit fit we adopt a host mass of 0.64 $\pm$ 0.04 \Msun \ from \citet{Kraus:2009hc},
a total system mass of 0.65 $\pm$ 0.04 \Msun,
and a parallax of 7.388 $\pm$ 0.069 mas from $Gaia$ DR2.

Our new observations from 2018 add to extensive published astrometry for this system dating back to 1999.
However these published values are highly discrepant---especially among the P.A. measurements---most likely indicating
significant instrument-to-instrument calibration errors or underestimated uncertainties.
This is evidenced by the excess astrometric jitter needed to bring the $\chi^2_{\nu}$ value for a linear fit to unity: 4.9 mas for the 
separations and 0.74$\degr$ for the P.A.s.  We find modest evidence for slight motion away from the host star at a rate of 0.5 mas yr$^{-1}$
but no signs of any change in P.A. over time.  
Nevertheless, these data offer some broad constraints 
on the orbital properties of DH Tau B (Figures~\ref{fig:orbitfits2} and \ref{fig:dhtaucorner}).  
The semi-major axis is 330$^{+90}_{-160}$ AU and the orbital period is
7500$^{+3200}_{-4800}$ yr (2$\sigma$ credible interval: 2400--60000 yr).  
The eccentricity is unconstrained.

\subsection{2M1559+4403 B} 

2MASS J15594729+4403595 B is a widely-separated (5$\farcs$6; 250 AU) brown dwarf companion 
to a young M1.5 star and was first identified by \citet{Janson:2012dc}.  
\citet{Bowler:2015ch} measured strong lithium absorption in the optical 
spectrum of the companion and found a spectral type of M7.5.  
The system age ($\lesssim$200 Myr) and luminosity of 2M1559+4403 B
imply a mass of 43 $\pm$ 9 \Mjup.  The host is likely a single-lined spectroscopic binary (\citealt{Bowler:2015ch}),
which is bolstered by the strong astrometric excess noise (1.3 mas) reported from the $Gaia$ DR2 astrometric fit.
For this work we adopt a host mass of 0.54 $\pm$ 0.10 \Msun \ from \citet{Muirhead:2018io},
a total system mass of 0.58 $\pm$ 0.10 \Msun,
and the $Gaia$ DR2 parallax of the host (22.340 $\pm$ 0.218 mas) for our orbit analysis.

\citet{Janson:2012dc}, \citet{Bowler:2015ch}, \citet{Janson:2014gz}, and \citet{Bowler:2015ch} 
presented astrometry for this companion spanning 2008 to 2014, which did not reveal signs of orbital motion.
We find evidence for orbital motion toward the host star in separation with our new epoch from 2018
at a rate of --3.2 mas yr$^{-1}$. 2M1559+4403 B also
shows changes in P.A. at the 2$\sigma$ level with a rate of --0.04$\degr$ yr$^{-1}$.
Astrometric jitter is required for the separation measurement uncertainties  
($\sigma_\mathrm{jit, \rho}$ = 10.5 mas) but not for the P.A. errors.
Our orbit constraints for this system are presented in Figures~\ref{fig:orbitfits2} and \ref{fig:2m1559corner}.
2M1559+4403 B has a semimajor axis of 290$^{+80}_{-140}$ AU and an orbital period of
6500$^{+2800}_{-4300}$ yr (2$\sigma$ credible interval: 1800--33500 yr).  
Its eccentricity is poorly constrained but values above $\sim$0.9 are disfavored.

\subsection{TWA 5 B}

TWA 5 is a young ($\approx$10~Myr) triple system comprising a close ($\approx$3~AU) stellar binary, 
TWA 5 Aab (\citealt{Macintosh:2001tr}), and 
a wider brown dwarf companion located at $\approx$80 AU (\citealt{Lowrance:1999ck}; \citealt{Webb:1999kf}).
The 6-year orbit of TWA 5 Aab has been well mapped over the past two decades 
(\citealt{Konopacky:2007hy}; \citealt{Kohler:2013im}) but no orbit determination has been made for TWA 5 B.
TWA 5 Aab is unresolved in our observations in 2018, but these data combined with 
published astrometry dating back to 1998 provide a twenty-year 
baseline to assess the orbit of TWA 5 B\footnote{Note that 
we have assumed the published relative astrometry for TWA 5 B are  
with respect to the photocenter of TWA Aab.  This is explicitly stated 
for astrometry from \citet{Kohler:2013im} and implied when Aab is unresolved.}.  
For our orbit analysis we adopt the dynamical  mass of 0.9 $\pm$ 0.1 \Msun \ from \citet{Kohler:2013im} for the inner binary TWA 5 Aab,
a total system mass of 0.92 $\pm$ 0.10 \Msun,
and a parallax of 20.252 $\pm$ 0.059 mas from $Gaia$ DR2.

TWA 5 B is approaching the host pair at a rate of --5.3 mas yr$^{-1}$ and we measure
evolution in P.A. at a rate of --0.4$\degr$ yr$^{-1}$ in the clockwise direction, 
which takes into account astrometric jitter levels of 
$\sigma_\mathrm{jit, \rho}$ = 2.3 mas and $\sigma_\mathrm{jit, \theta}$ = 0.58$\degr$.
The orbit of TWA 5 B about Aab is shown in Figures~\ref{fig:orbitfits2} and \ref{fig:twa5corner}, 
and the orbital elements are summarized in Table~\ref{tab:orbitize}.
We find a semi-major axis of 150$^{+30}_{-50}$ AU, a moderate eccentricity of 0.43$^{+0.10}_{-0.10}$, and 
an orbital period of 1810$^{+540}_{-840}$ yr (2$\sigma$ credible interval: 740--4400 yr).
The orbital inclination of TWA 5 B about Aab is 151$^{+12}_{-13}$$\degr$, 
which is misaligned with the measured orbital inclination of TWA 5 Aab (97.5 $\pm$ 0.1$\degr$)
from \citet{Kohler:2013im} at the 4$\sigma$ level.  
It therefore appears that the orbits of TWA 5 Aab and TWA 5 B about Aab
are not coplanar.

\subsection{Ross 458 B}

Ross 458 is a triple system comprising a close binary, Ross 458 AB (\citealt{Heintz:1990wr}), and a comoving 
late-T dwarf planetary-mass companion at a projected separation of 
about 1200 AU (\citealt{Goldman:2010ct}; \citealt{Scholz:2010cy}).
The system age is estimated to be between 150 and 800 Myr (\citealt{Burgasser:2010bu}).
From its absolute magnitude of $M_K$$\approx$9.7 (\citealt{Beuzit:2004fq}), Ross 458 B resides below the 
hydrogen burning limit based on the \citet{Baraffe:2015fwa} evolutionary models 
if its age is younger than about 300 Myr.
If the age is closer to 800 Myr then its implied mass is about 0.1~\Msun.

\citet{Heintz:1990wr} first detected unresolved astrometric perturbations of Ross 458 with an orbital period of 13.5 years.
This system has been since monitored with adaptive optics for more than a complete orbital cycle.  
We adopt a host star mass of 0.61 $\pm$ 0.03 \Msun \ from \citet{Neves:2013cm},
a total system mass of 0.68 $\pm$ 0.03 \Msun,
 and a parallax of 86.857 $\pm$ 0.152 mas from $Gaia$ DR2.
Our observations in 2016 and 2018 supplement astrometry from \citet{Mann:2019ey}  and 
\citet{WardDuong:2015ej} which together extend astrometric coverage back to 2000.
The orbit for Ross 458 B is therefore very well determined (see Figures~\ref{fig:orbitfits2} and \ref{fig:ross458corner}); we 
measure an orbital period of 13.528$^{+0.02}_{-0.014}$ yr (2$\sigma$ credible interval: 13.49--13.56 yr), 
a semi-major axis of 4.95 $\pm$ 0.01 AU, and an orbital eccentricity of 0.242 $\pm$ 0.001.

\subsection{1RXS2351+3127 B}

1RXS J235133.3+312720 B is an early-L type substellar companion 
discovered as part of the PALMS survey (\citealt{Bowler:2012cs}).  The M2 host is a member
of the $\approx$120 Myr AB Dor moving group (\citealt{Shkolnik:2012cs}; \citealt{Schlieder:2012gu}; \citealt{Malo:2013gn});
at this age the inferred mass of 1RXS2351+3127 B is 32 $\pm$ 6 \Mjup.
We adopt a host mass of 0.45 $\pm$ 0.05 \Msun \ from \citet{Bowler:2012cs},
a total system mass of 0.48 $\pm$ 0.05 \Msun,
and parallax of 23.218 $\pm$ 0.052 mas from $Gaia$ DR2.

\citealt{Bowler:2015ch} detected hints of orbital motion based on observations spanning 2011 to 2013.  
Our epoch from 2019 clearly establishes evolution in the astrometry.  
1RXS2351+3127 B is moving away from its host star at a rate of 0.8 mas yr$^{-1}$ 
in a clockwise direction at --0.13$\degr$ yr$^{-1}$.  
We find jitter terms of $\sigma_\mathrm{jit, \rho}$ = 0.3 mas and $\sigma_\mathrm{jit, \theta}$ = 0.05$\degr$. 
1RXS2351+3127 B has a semimajor axis of 100$^{+30}_{-40}$ AU and an orbital period of
1490$^{+630}_{-870}$ yr (2$\sigma$ credible interval: 570--10100 yr).  
Eccentricities above $\sim$0.8 are disfavored.


\begin{figure}
  \vskip -0.5 in
  \hskip -.6 in
  \resizebox{4.9in}{!}{\includegraphics{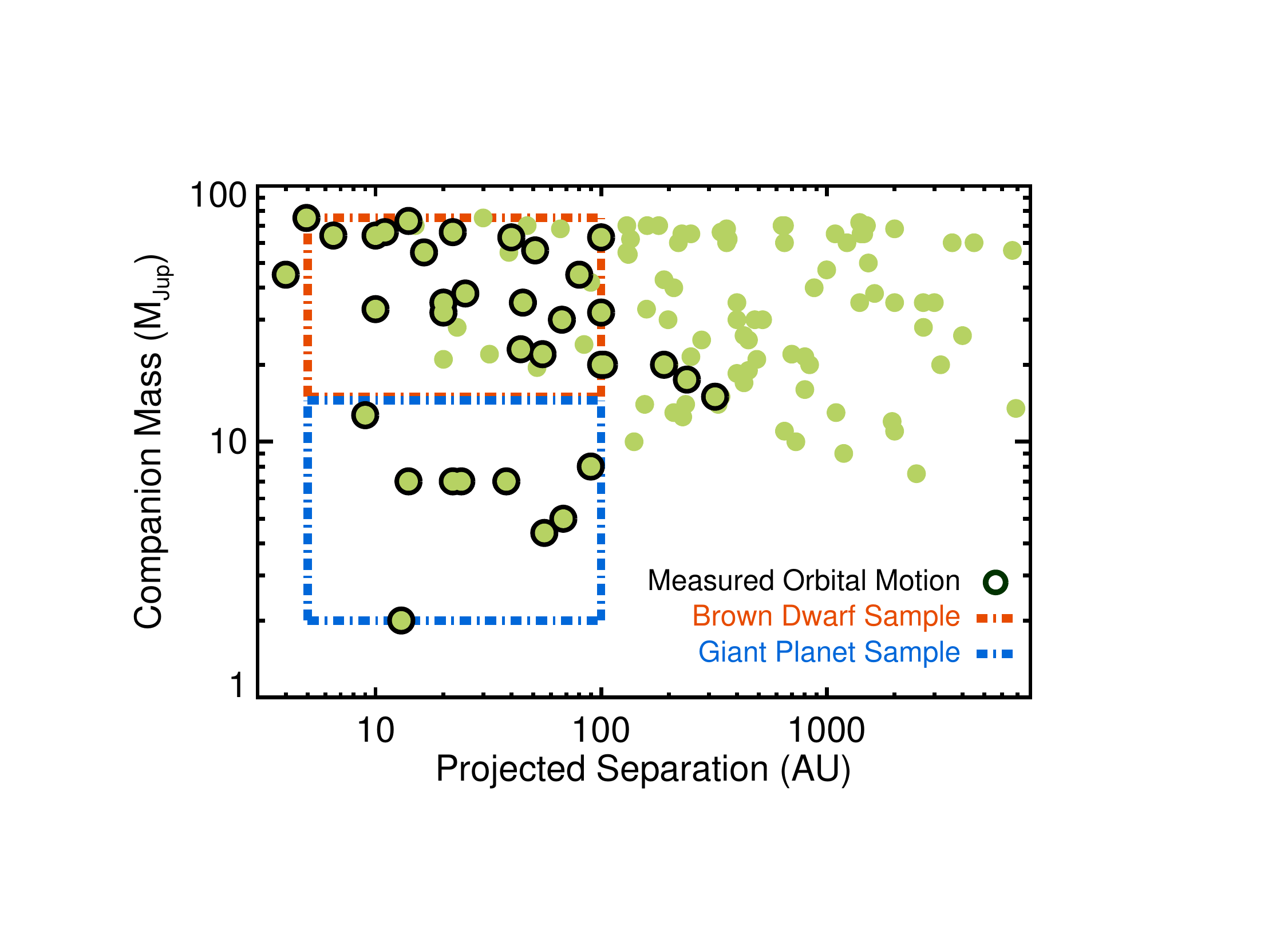}}
  \vskip -.4 in
  \caption{Sample selection for this study based on a literature compilation of 125 known companions with mass estimates
  below the hydrogen burning boundary and orbiting stars, as of May 2019.  
  Most systems with projected separations beyond $\sim$200 AU were found in seeing-limited 
  red-optical and infrared surveys like 2MASS, \emph{WISE}, and Pan-STARRS,
  whereas those closer in have typically been found with the aid of adaptive optics.  Companions with measured
  orbital motion are highlighted in black.  Among these, we isolate a sample of 27 systems with masses between
  2--75~\Mjup, separations between 5--100 AU, and whose host stars are expected to be single to study 
  the underlying population-level eccentricity distribution of substellar companions.  
  This is further divided into subsamples of 18 brown dwarfs (15--75~\Mjup) 
  and 9 giant planets (2--15~\Mjup) shown in the red and blue boxes, respectively.  
     \label{fig:sample} } 
\end{figure}

\section{Population-Level Eccentricities of Brown Dwarfs and Giant Planets}{\label{sec:eccentricities}}

The goal of this study is to assess whether imaged planets and brown dwarf companions on wide
orbits form or dynamically evolve in similar ways by comparing their population-level eccentricity distributions.
Over one hundred substellar companions to stars have been discovered over the past quarter century spanning 2--75~\Mjup.
The majority of these objects are located at wide separations beyond 100 AU and were largely found in seeing-limited
infrared surveys like 2MASS, Pan-STARRS, and \emph{WISE}.
The subset of this population located at smaller angular separations and generally within $\sim$200 AU were predominantly 
discovered with the aid of adaptive optics.  
Most of these brown dwarfs and giant planets 
have been continuously observed since their discoveries and
typically exhibit orbital motion after a few years 
of astrometric monitoring (Figure~\ref{fig:sample}).

In this Section we make use of companions undergoing orbital motion to assess the underlying 
eccentricity distributions of brown dwarf companions and giant planets at the population levels.  
We first describe our sample selection and experimental design for this analysis. 
For systems with small fractional orbital coverage we perform new
orbit fits in a uniform manner with \texttt{orbitize!} using published astrometry.
Finally, the underlying eccentricity distributions for giant planets and brown dwarfs are 
inferred using hierarchical Bayesian inference and compared with the eccentricity distributions of warm Jupiters
and binary stars.

\subsection{Defining the Sample}{\label{sec:sample}}

We have constructed a large (and to our knowledge complete) sample of 125 imaged
substellar companions to stars.  This list draws from 
previous catalogs of low-mass companions from \citet{Zuckerman:2009gc}, \citet{Faherty:2010gt}, \citet{Deacon:2014ey}, 
and \citet{Bowler:2016jk}, together with additional discoveries compiled from the literature.
Orbital motion is detected for 36 companions based on multi-epoch astrometry, both from the literature
and from our new observations in this work (Figure~\ref{fig:sample}).

The following criteria are used to define our samples of brown dwarfs and giant planets, with the goal
of identifying targets that are most representative of the initial dynamical conditions of this population
and least likely to have been influenced by
significant orbital migration as a result of dynamical encounters with a third body.
Companions must have projected
separations between 5 and 100 AU at the time of discovery and
have measured orbital motion from multi-epoch astrometry.
The hosts must be stars ($>$75~\Mjup), rather than brown dwarfs, to prevent
biasing the samples towards binary star-like mass ratio distributions.
Since the eccentricities of substellar companions can also be influenced by post-formation interactions  
in hierarchical triple systems 
(e.g., \citealt{Fabrycky:2007jh}; \citealt{Allen:2012dn}; \citealt{Reipurth:2015dz}),
we require the host stars not be known binaries.
We similarly limit the sample to companions that are themselves single.
Companions whose existence or characteristics are a matter of ongoing debate---in particular LkCa15 bcd (\citealt{Kraus:2012gk}; \citealt{Ireland:2014jm}; \citealt{Sallum:2015ej}; \citealt{Thalmann:2016kd}; \citealt{Currie:2019ko})
and HD 100546 bc (e.g., \citealt{Quanz:2013ii}; \citealt{Currie:2015jk}; \citealt{Rameau:2017js})---are excluded from this analysis.
This amounts of 27 systems which represent our full sample of directly imaged substellar companions undergoing orbital motion.
Finally, we isolate two main subsamples within 5--100~AU based on their masses as inferred from 
hot-start
evolutionary models: 18 brown dwarfs spanning 15--75~\Mjup \ and 9 
giant planets between 2 and 15~\Mjup.
Characteristics of the host stars and companions are summarized Table~\ref{tab:sample}.

\subsection{Uniform Orbits with Literature Astrometry}{\label{sec:neworbits}}

We compiled all available astrometry of targets in our full sample with small fractional orbit coverage (see Appendix \ref{sec:litastrometry})
to reassess their orbits in a uniform manner,
rather than rely on orbit determinations in the literature which can be influenced by different priors and 
approaches to fitting short orbital arcs.
For the remaining targets with well- (or moderately well-) constrained orbits we adopt eccentricity posteriors from the literature.
This includes the HR 8799 planets (bcde), for which we use the ``unconstrained'' eccentricity posteriors from \citet{Wang:2018bb},
Gl 229 B from \citet{Brandt:2019kp}, 
and approximately Gaussian-shaped eccentricity distributions for 
the following five systems:
$\beta$ Pic b ($e$ = 0.24 $\pm$ 0.06; \citealt{Dupuy:2019cy}), 
HD 4113 C ($e$ = 0.38 $\pm$ 0.06; \citealt{Cheetham:2018ha}),
HD 4747 B ($e$ = 0.735 $\pm$ 0.003; \citealt{Brandt:2019ey}),
Gl 758 B ($e$ = 0.40 $\pm$ 0.09; \citealt{Brandt:2019ey}),
and HR 7672 ($e$ = 0.542 $\pm$ 0.018; \citealt{Brandt:2019ey}).
Note that because the orbits of these systems are well constrained, the choice
of priors does not meaningfully influence the

\begin{longrotatetable}
\begin{deluxetable*}{lcccccccccccccc}
\renewcommand\arraystretch{0.9}
\tabletypesize{\scriptsize}
\setlength{ \tabcolsep }{.1cm}
\tablewidth{0pt}
\tablecolumns{15}
\tablecaption{Sample of Substellar Companions Between 2--75~\Mjup \ and 5--100 AU Exhibiting Orbital Motion \label{tab:sample}}
\tablehead{
\colhead{Target}  &  \colhead{$\alpha$}  & \colhead{$\delta$} & \colhead{Host}  & \colhead{Age} & \colhead{$\pi$\tablenotemark{a}}  &  \colhead{$M_*$}  & \colhead{$M_\mathrm{comp}$\tablenotemark{b}}  & \colhead{$M_2$/$M_1$}   &    \colhead{Comp.}  &  \colhead{Sep.}   &  \colhead{Sep.}  &   \colhead{IWA/$a$\tablenotemark{c}} & \colhead{$\Delta t$/$P$\tablenotemark{d}} & \colhead{Ref.}  \\
\colhead{Name}  &  \colhead{(J2000.0)}  & \colhead{(J2000.0)} & \colhead{SpT}  & \colhead{(Myr)}  & \colhead{(mas)}  &  \colhead{($M_{\odot}$)} &  \colhead{($M_\mathrm{Jup}$)}  &  \colhead{($\times$10$^{-2}$)} &\colhead{SpT}  &  \colhead{($''$)}  &  \colhead{(AU)}  & \colhead{} &  \colhead{} &  \colhead{}  
  }
\startdata
HD 984 B        &  00 14 10.25 & --07 11 56.8   &  F7 & 30--200  &   21.7806 $\pm$  0.0564  &  1.15 $\pm$ 0.06  & 34--94          & 5.3$^{+2.0}_{-1.4}$  & M6.5 $\pm$ 1.5   & 0.2      & 10   & 0.26 & 0.05 &  1, 2, 3    \\
HD 1160 B       &  00 15 57.30 &  +04 15 04.0   &  A0 & 80--125 &   7.9417 $\pm$  0.0764  &  1.95 $\pm$ 0.10  & 35--90          & 3.1$^{+1.0}_{-0.8}$  & M5.5$^{+1.0}_{-0.5}$   & 0.8      & 100  & 1.18  & 0.03 &  4, 5, 6, 7   \\
HD 4113 C       &  00 43 12.60 & --37 58 57.47  &  G5 & 5000$^{+1300}_{-1700}$ & 23.8531 $\pm$  0.0536  &  1.05 $\pm$ 0.03  &  66 $\pm$ 5 & 6.0 $\pm$ 0.5    &  T9 $\pm$ 1 &  0.5    &  22    & 0.23 & 0.008 &  8  \\
HD 4747 B       &  00 49 26.76 & --23 12 44.9   &  G9 & 3300$^{+2300}_{-1900}$ &  53.1836 $\pm$  0.1264  &  0.82 $\pm$ 0.08 & 66 $\pm$ 3 & 7.7 $\pm$ 0.9   & L9--T1   & 0.6      & 11     & 0.56  & 0.27 &  9, 10, 11    \\
HD 19467 B      &  03 07 18.57 & --13 45 42.4   &  G3 & 4600--10000 &  31.2255 $\pm$  0.0410  &  0.95 $\pm$ 0.02  &  57$^{+5}_{-7}$  & 5.7 $\pm$ 0.6 & T5.5 $\pm$ 1   & 1.6      & 51   &  0.47 & 0.02 & 12, 13    \\
1RXS0342+1216 B  &  03 42 31.80 &  +12 16 22.5   &  M4 & 60--300 &  30.3075 $\pm$  0.0668  &  0.20 $\pm$ 0.05  & 35 $\pm$ 8 & 17 $\pm$ 5 &  L0 $\pm$ 1  &  0.8  &  19  & 0.12 & 0.02 &  14, 15 \\
51 Eri b        &  04 37 36.13 & --02 28 24.7   &  F0 &  23 $\pm$ 3 &  33.5770 $\pm$  0.1354  &  1.75 $\pm$ 0.05 &  2 $\pm$ 1  &  0.11 $\pm$ 0.05  &  T4--T8  &  0.45 &  13  & 0.71 & 0.11 & 16, 17, 18, 19 \\
$\beta$ Pic b   &  05 47 17.09 & --51 03 59.4   &  A6 & 23 $\pm$ 3 &  50.6231 $\pm$  0.3339  &  1.84 $\pm$ 0.05  & 13 $\pm$ 3 &  0.7 $\pm$ 0.16      &  L1 $\pm$ 1 &  0.4  &  9  & 0.33  & 0.53 & 19, 20, 21, 22  \\
CD--35 2722 B   &  06 09 19.21 & --35 49 31.1   &  M1 & 133$^{+15}_{-20}$ &  44.6346 $\pm$  0.0262  &  0.40 $\pm$ 0.05  &  31 $\pm$ 8 & 7 $\pm$ 2 & L4 $\pm$ 1 & 3.1      & 67    & 0.08   & 0.002  &  23, 24  \\
Gl 229 B        &  06 10 34.62 & --21 51 52.7   &  M1 & 2000--6000  & 173.6955 $\pm$  0.0457  &  0.54 $\pm$ 0.04  &  72 $\pm$ 5 & 12.7 $\pm$ 1.3  &  T7p $\pm$ 0.5 & 6.8  &  39 &  0.79 & 0.05 &  25, 26, 27, 28  \\
HD 49197 B      &  06 49 21.33 &  +43 45 32.8   &  F5 & 260--790 &  23.9903 $\pm$  0.0486  &  1.11 $\pm$ 0.06  &  63$^{+13}_{-21}$ &  5.4 $\pm$ 1.4 & L4 $\pm$ 1   & 1        & 40   & 1.6  & 0.11 &  29, 30, 3    \\
HR 2562 B       &  06 50 01.01 & --60 14 56.9   &  F5 & 200--750 &  29.3767 $\pm$  0.0411  &  1.37 $\pm$ 0.02  & 32 $\pm$ 14  &  2.2 $\pm$ 1.0 & T2--T3   & 0.6      & 20   &  0.37  & 0.01 &  31, 32   \\
HR 3549 B       &  08 53 03.78 & --56 38 58.1   &  A0 & 100--150 &  10.4864 $\pm$  0.0898  &  2.3$^{+0.2}_{-0.1}$ &  40--50  & 1.9 $\pm$ 0.2  & M9--L0   & 0.9     & 80   & 0.46  & 0.005 &  33, 34   \\
HD 95086 b      &  10 57 03.01 & --68 40 02.4   &  A8 & 17 $\pm$ 4 &  11.5684 $\pm$  0.0325  &  1.7 $\pm$ 0.1\tablenotemark{e}   &  4.4 $\pm$ 0.8  & 0.25 $\pm$ 0.05  & L1--T3  &  0.6  &  52 & 0.79  & 0.02 & 35, 36, 37 \\
GJ 504 B\tablenotemark{f}  &  13 16 46.52 &  +09 25 26.9   &  G0 & 4000 $\pm$ 1800 &  57.0186 $\pm$  0.2524  &  1.10--1.25 &  23$^{+10}_{-9}$  & 1.8 $\pm$ 0.8 & T8--T9.5    & 2.5  & 44  &  0.79 & 0.03 &  38, 39  \\
HIP 65426 b     &  13 24 36.09 & --51 30 16.0   &  A2 &  14 $\pm$ 4 &  9.1566 $\pm$  0.0626  &  1.96 $\pm$ 0.04  & 8 $\pm$ 1  & 0.39 $\pm$ 0.05  & mid-L  & 0.8  &  90  & 0.36 & 0.003 & 40, 41  \\
PDS 70 b        &  14 08 10.15 & --41 23 52.57  &  K7 &  5 $\pm$ 1 &  8.8159 $\pm$  0.0405  &  0.76 $\pm$ 0.02  & 4--10  & 0.9 $\pm$ 0.3  &  L: &  0.2 & 22 & 0.76  & 0.05 & 42, 43  \\
PZ Tel B        &  18 53 05.88 & --50 10 49.9   &  K0 &  23 $\pm$ 3 & 21.2186 $\pm$  0.0602  &  1.25$^{+0.05}_{-0.20}$ &  38--72  & 4.2$^{+1.0}_{-0.8}$ & M7 $\pm$ 1   & 0.3      & 17   &  0.52  & 0.07 & 44, 45, 19, 5   \\
Gl 758 B        &  19 23 34.01 &  +33 13 19.1   &  G8 & 6000--10000 &  64.0623 $\pm$  0.0218  &  0.83 $\pm$ 0.11   & 38.1$^{+1.7}_{-1.5}$  &  4.4 $\pm$ 0.6  & T7--T8   & 1.8      & 25  &  0.26   &  0.04  &  46, 47, 48, 49, 11     \\
HR 7672 B       &  20 04 06.22 &  +17 04 12.6   &  G0 & 3000--5000 &  56.4256 $\pm$  0.0690  &  0.96 $\pm$ 0.05  &  72.7 $\pm$ 0.8  &  7.2 $\pm$ 0.4  & L4.5 $\pm$ 1.5  & 0.8     & 14   &  0.63  & 0.11 &  50, 51, 11   \\
HR 8799  b      &  23 07 28.72 &  +21 08 03.3   &  A5 & 40 $\pm$ 5 &  24.2175 $\pm$  0.0881  &  1.48 $\pm$ 0.05 & 5 $\pm$ 1 & 0.32 $\pm$ 0.07  &  L/Tpec  &  1.7  &  68  & 0.48 & 0.03 &  52, 53, 54, 55  \\
HR 8799  c      &  23 07 28.72 &  +21 08 03.3   &  A5 & 40 $\pm$ 5 &  24.2175 $\pm$  0.0881  &  1.48 $\pm$ 0.05 & 7 $\pm$ 1 & 0.45 $\pm$ 0.07  &  L/Tpec  &  0.95  &  38  & 0.90 & 0.08 &  52, 53, 54, 55  \\
HR 8799  d      &  23 07 28.72 &  +21 08 03.3   &  A5 & 40 $\pm$ 5 &  24.2175 $\pm$  0.0881  &  1.48 $\pm$ 0.05 & 7 $\pm$ 1 & 0.45 $\pm$ 0.07  &  L6--L8pec  &  0.6  &  24  & 0.75 & 0.16 & 52, 53, 54, 55 \\
HR 8799  e      &  23 07 28.72 &  +21 08 03.3   &  A5 & 40 $\pm$ 5 & 24.2175 $\pm$  0.0881  &  1.48 $\pm$ 0.05 & 7 $\pm$ 1 & 0.45 $\pm$ 0.07  &  L6--L8pec  &  0.4  &  14  & 0.54 & 0.13  &  52, 53, 54, 55, 56  \\
HD 206893 B     &  21 45 21.90 & --12 47 00.1   &  F5 & 250$^{+450}_{-200}$ &  24.5062 $\pm$  0.0639  &  1.32 $\pm$ 0.02  &  15--50 &  2.3 $\pm$ 0.9  &  L3--L5pec & 0.3  &  10 & 0.72 & 0.10  &  57, 58 \\
$\kappa$ And B  &  23 40 24.51 &  +44 20 02.2   &  B9 &  50$^{+30}_{-40}$ & 19.9751 $\pm$  0.3418  &  2.8 $\pm$ 0.1  &  22 $\pm$ 9  &  0.8 $\pm$ 0.3 &  L0--L1 &  1.1  &  55  & 0.41 & 0.01 & 59, 60, 61 \\
1RXS2351+3127 B &  23 51 33.66 & +31 27 22.9  & M2 & 133$^{+15}_{-20}$ &  23.2183 $\pm$ 0.0524  &  0.45 $\pm$ 0.05  &  32 $\pm$ 6  & 6.8 $\pm$ 1.5  &  L0$\pm$1  &  2.4  &  100 & 0.26 & 0.005 & 62, 24 \\
\enddata
\tablenotetext{a}{Parallaxes from $Gaia$ DR2.}
\tablenotetext{b}{Assumes hot-start cooling history.}
\tablenotetext{c}{IWA/$a$ refers to the inner working angle at the contrast of the companion and at the time of discovery, divided by the maximum \emph{a posteriori} of the 
semi-major axis distribution.  The IWA was visually estimated based on the discovery papers.  This value---IWA/$a$---represents a useful metric
to assess the potential impact of discovery bias on the inferred population-level eccentricity distribution (see \citealt{Dupuy:2011ip} and Section~\ref{sec:discoverybias}).
Values above 1.0, between 0.5--1.0, and below 0.5 indicate strong, moderate, and minimal discovery bias, respectively. }
\tablenotetext{d}{Fractional orbital coverage, computed from the time baseline over which this companion has been imaged ($\Delta t$) and the best-fit orbital period ($P$).}
\tablenotetext{e}{Uncertainty in the host mass is approximated from multiple assessments in the literature (\citealt{Chen:2014dt}; \citealt{Meshkat:2013fz}).}
\tablenotetext{f}{\citet{Bonnefoy:2018ch} find two degenerate solutions for the host age (21 $\pm$ 2 Myr or 4.0 $\pm$ 1.8 Gyr), which results in two 
solutions for the companion mass (1.3$^{+0.6}_{-0.3}$ or 23$^{+10}_{-9}$ \Mjup) and mass ratio (0.11 $\pm$ 0.04 $\times$10$^{-2}$ or 1.8 $\pm$ 0.8 $\times$10$^{-2}$).  
For this study we adopt the older age and correspondingly higher companion mass. }
\tablerefs{
(1) \citet{Meshkat:2015hd}; 
(2) \citet{JohnsonGroh:2017kh}; 
(3) \citet{Mints:2017di}; 
(4) \citet{Nielsen:2012jk}; 
(5) \citet{Maire:2016go}; 
(6) \citet{Garcia:2017dk};
(7) \citet{Blunt:2017eta}; 
(8) \citet{Cheetham:2018ha}; 
(9) \citet{Crepp:2016fg}; 
(10) \citet{Crepp:2018gf}; 
(11) \citet{Brandt:2019ey}; 
(12) \citet{Crepp:2014ce};
(13) \citet{Crepp:2015gt}; 
(14) \citet{Bowler:2015ch}; 
(15) \citet{Bowler:2015ja}; 
(16) \citet{Macintosh:2015fw}; 
(17) \citet{Simon:2011ix}; 
(18) \citet{Rajan:2017hq}; 
(19) \citet{Mamajek:2014bf}; 
(20) \citet{Lagrange:2010fsa}; 
(21) \citet{Dupuy:2019cy}; 
(22) \citet{Chilcote:2017fv}; 
(23) \citet{Wahhaj:2011by}; 
(24) \citet{Gagne:2018ks};
(25) \citet{Nakajima:1995bb}; 
(26) \citet{Oppenheimer:1995wl}; 
(27) \citet{Burgasser:2006cf}; 
(28) \citet{Brandt:2019kp}; 
(29) \citet{Metchev:2004kl};
(30) \citet{Metchev:2009ky}; 
(31) \citet{Konopacky:2016dk}; 
(32) \citet{Mesa:2018er}; 
(33) \citet{Mawet:2015kk}; 
(34) \citet{Mesa:2016kn}; 
(35) \citet{Rameau:2013vh};
(36) \citet{Meshkat:2013fz}; 
(37) \citet{DeRosa:2014bu}; 
(38) \citet{Kuzuhara:2013jz};
(39) \citet{Bonnefoy:2018ch}; 
(40) \citet{Chauvin:2017hl}; 
(41) \citet{Cheetham:2019ju}; 
(42) \citet{Keppler:2018dd}; 
(43) \citet{Muller:2018jr}; 
(44) \citet{Biller:2010ku};
(45) \citet{Mugrauer:2010cp}; 
(46) \citet{Thalmann:2009ca};
(47) \citet{Brewer:2016gf}; 
(48) \citet{Nilsson:2017hm}; 
(49) \citet{Bowler:2018gy}; 
(50) \citet{Liu:2002fx}; 
(51) \citet{Crepp:2012eg}; 
(52) \citet{Marois:2008ei}; 
(53) \citet{Wang:2018bb}; 
(54) \citet{Barman:2015dy}; 
(55) \citet{Bonnefoy:2016gx}; 
(56) \citet{Marois:2010gpa}; 
(57) \citet{Milli:2017fs}; 
(58) \citet{Delorme:2017hl}
(59) \citet{Carson:2013fw}; 
(60) \citet{Jones:2016hg}; 
(61) \citet{Currie:2018hj};
(62) \citet{Bowler:2012cs}.
}
\end{deluxetable*}
\end{longrotatetable}


\begin{figure}[th!]
  \vskip -0.5 in
  \hskip -.6 in
  \resizebox{4.9in}{!}{\includegraphics{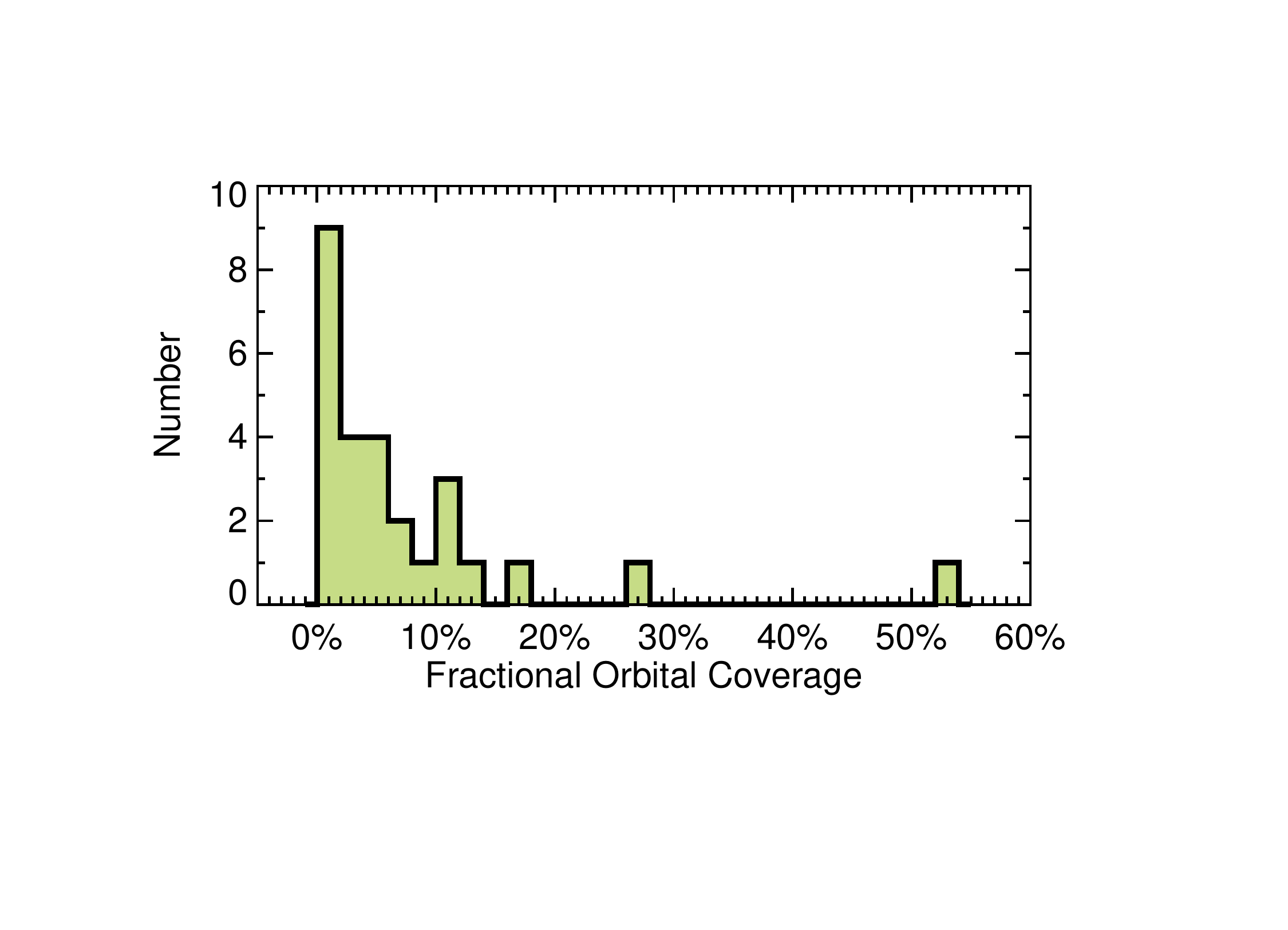}}
  \vskip -0.9 in
  \caption{Fractional orbital coverage for the 27 systems in our final sample.
  These are determined using the first and latest published epochs of astrometry along with periods
  from this paper, when available, or the most recent orbit fit in literature.  Most systems have only
  completed a few percent of their orbits from the time they were discovered to the latest observation.
  $\beta$ Pic b has the highest fractional coverage at 53\%, followed by HD 4747 B at 27\%.
     \label{fig:coverage} } 
\end{figure}

\noindent posteriors.
Figure \ref{fig:coverage} shows the distribution of fractional orbital coverage for our sample.  Most
systems have traced out $<$10\% of their orbits.  Only 11 have been imaged for 
over 5\% of their orbital periods.


\begin{figure*}[htb!]
  \gridline{\fig{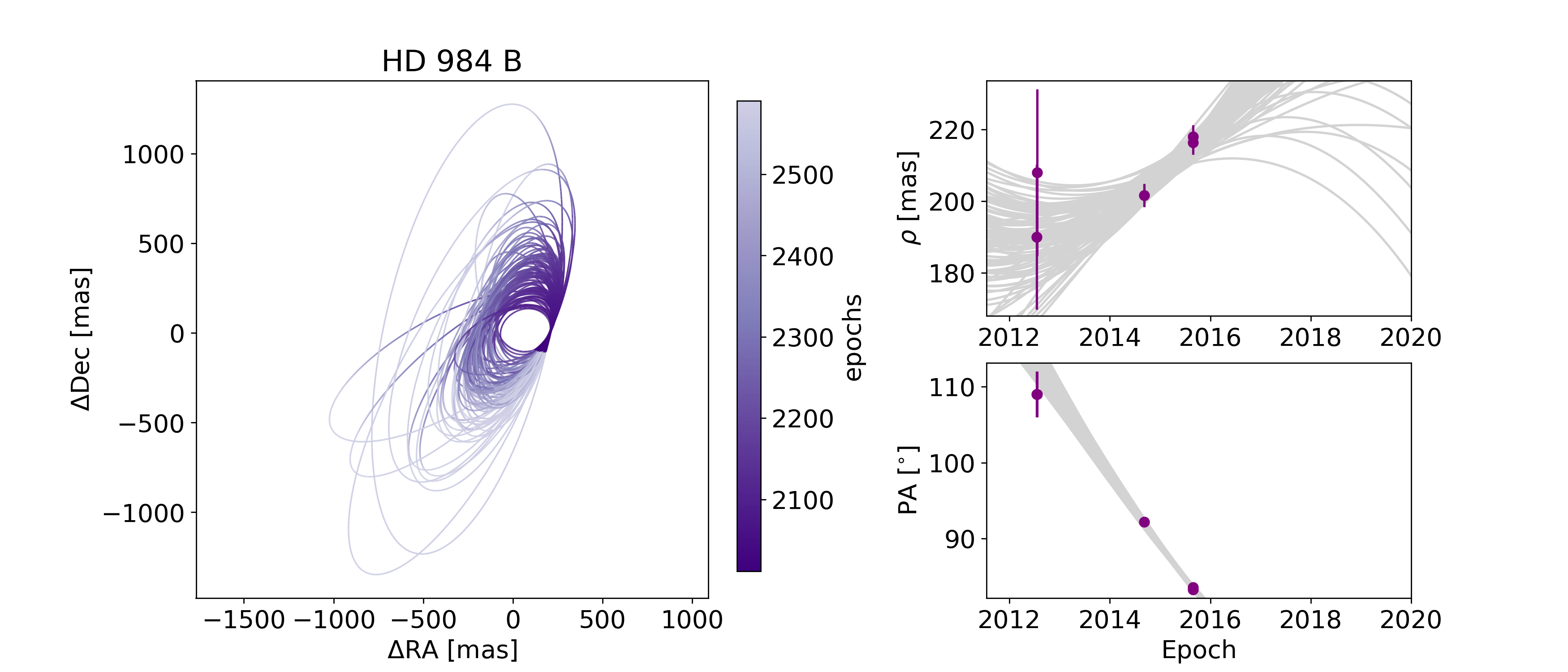}{0.5\textwidth}{(a) HD 984 B}
                \fig{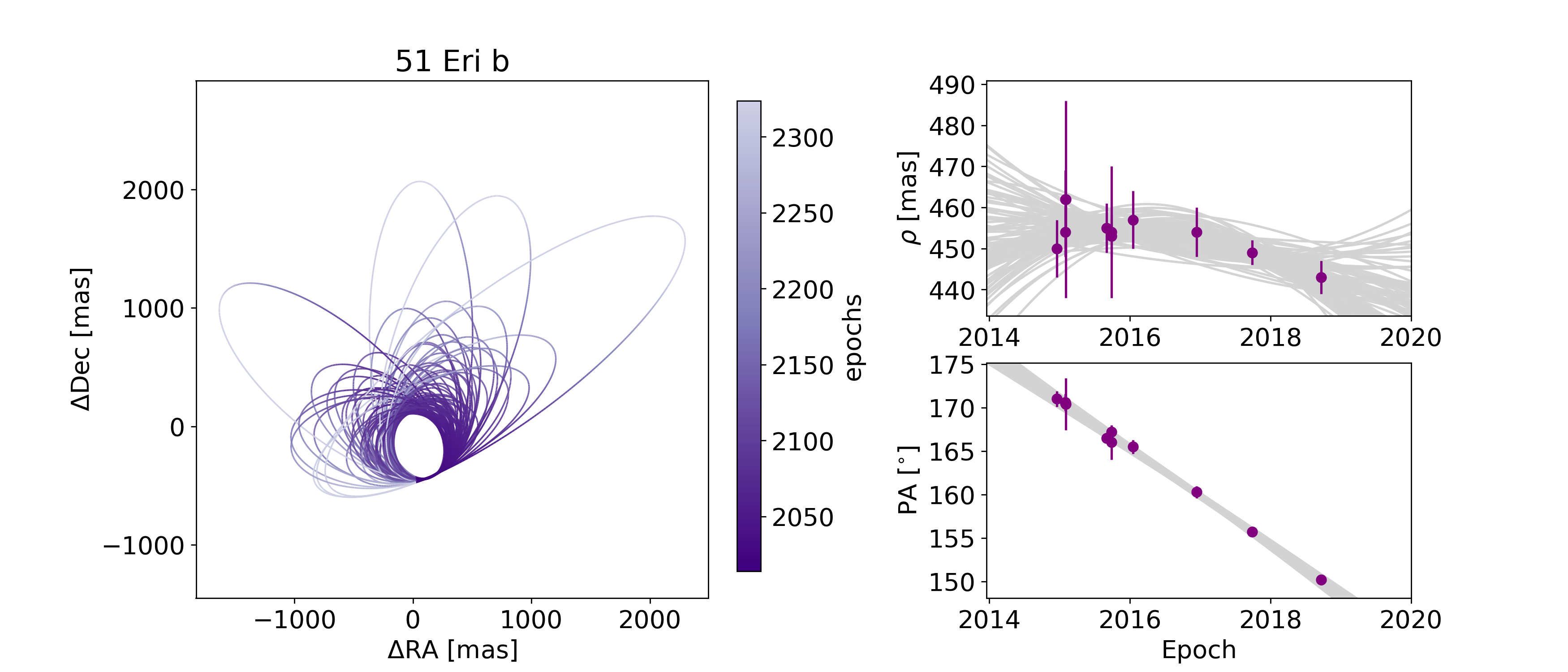}{0.5\textwidth}{(b) 51 Eri b}}                
  \gridline{\fig{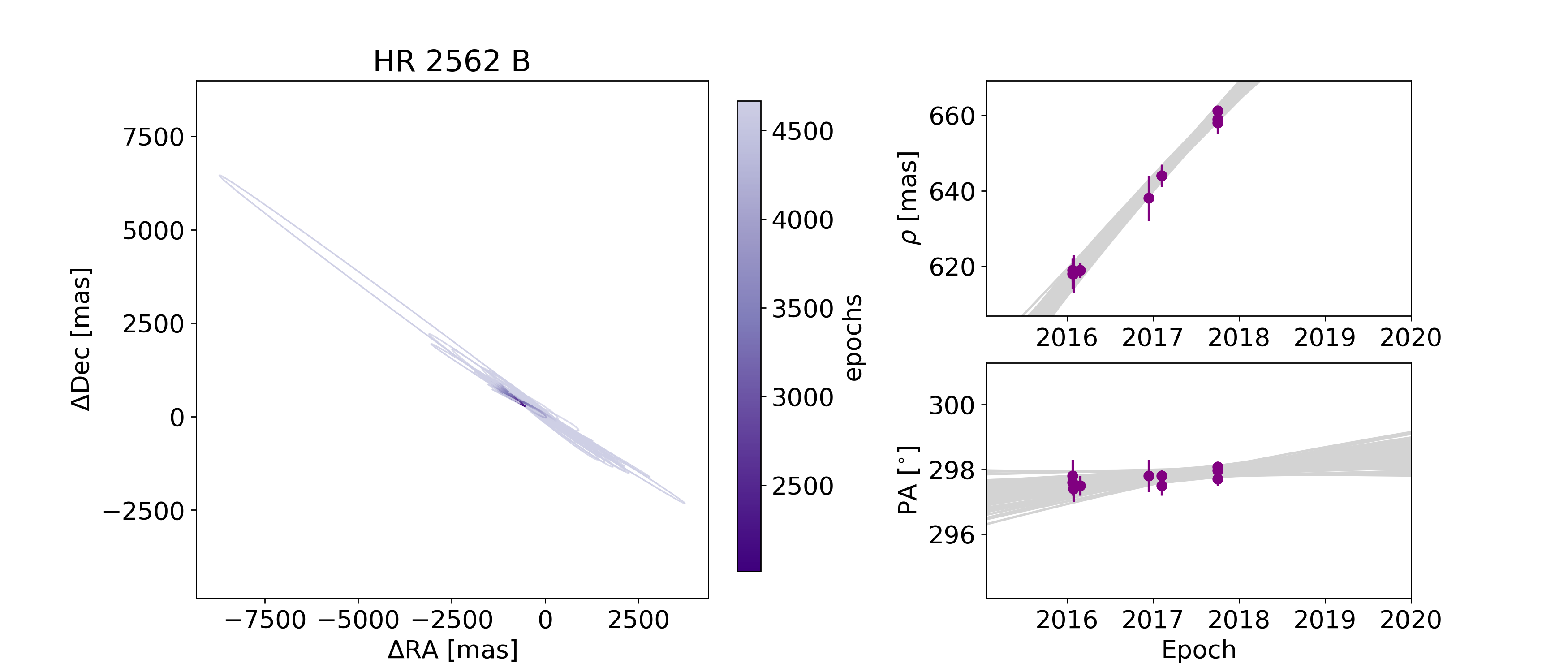}{0.5\textwidth}{(c) HR 2562 B}
                \fig{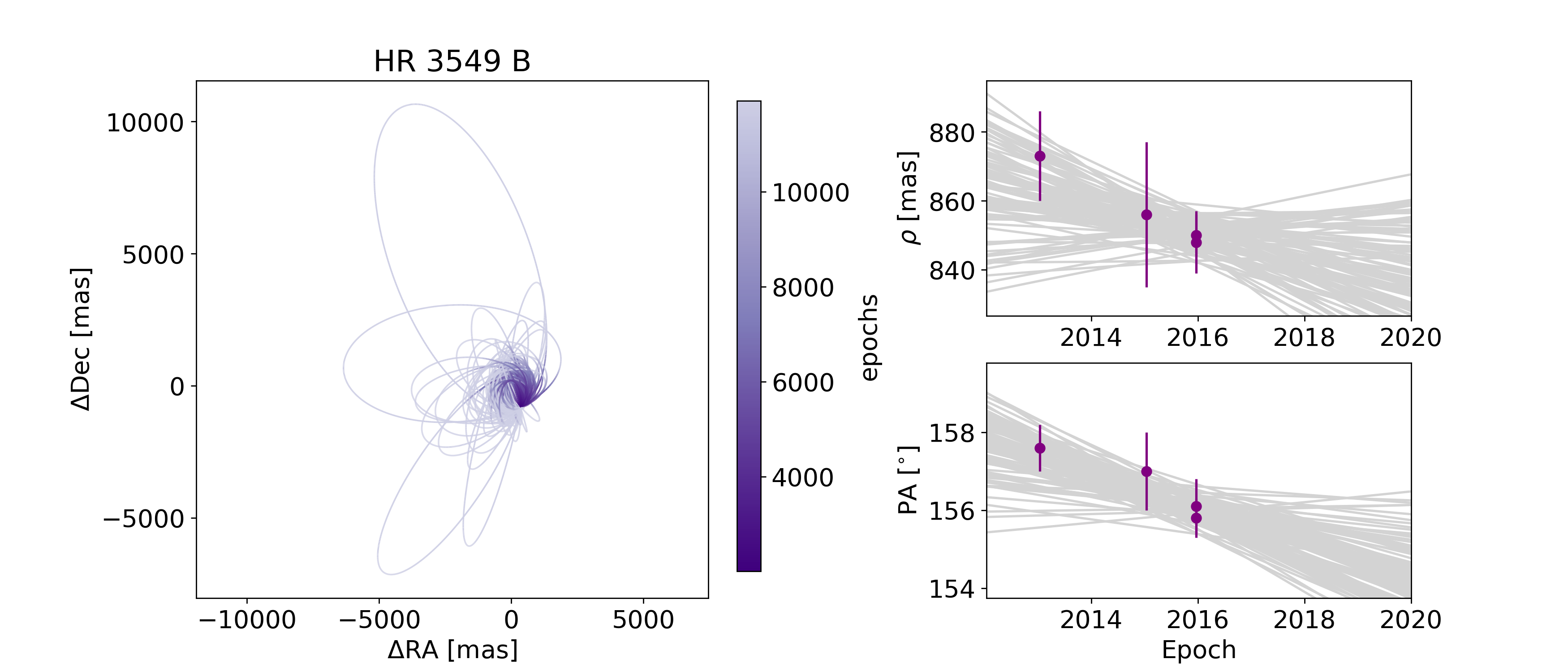}{0.5\textwidth}{(d) HR 3549 B}}
   \gridline{\fig{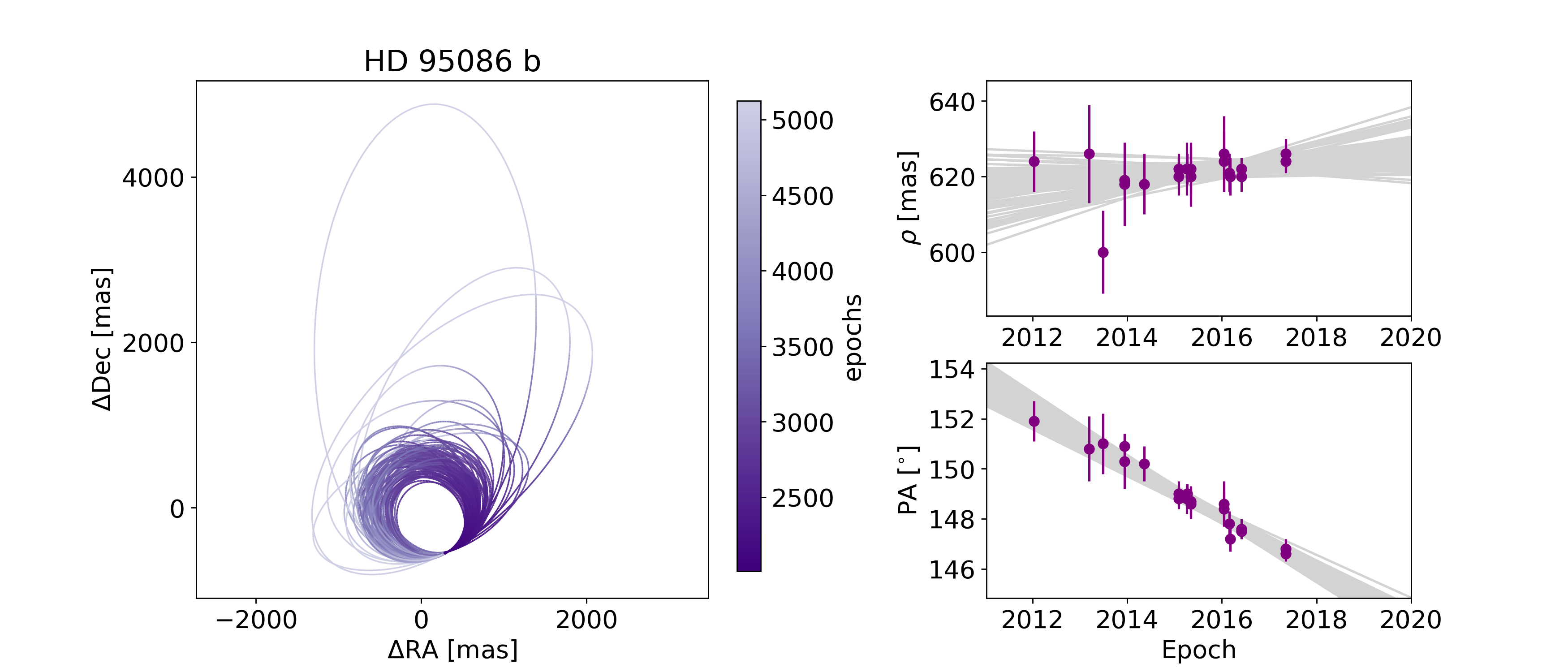}{0.5\textwidth}{(e) HD 95086 b}}
  \vskip 0 in
  \caption{Orbit fits for HD 984 B (a), 51 Eri b (b), HR 2562 B (c), HR 3549 B (d), and HD 95086 b (e),
  using \texttt{orbitize!}.
  For each object the left panel shows 100 randomly drawn orbits  
  from the posterior distributions.
  These are color coded to show the expected orbital location over time.  The right panels show measured separation (top) 
  and P.A. (bottom) of the companion compared to randomly drawn orbits from the posterior distributions.
    \label{fig:orbitfits3} } 
\end{figure*}

For the remaining nine systems not already discussed in Section \ref{sec:orbits}, we apply the same Bayesian 
priors and fitting approach as previously discussed.
Astrometric jitter for each system is assessed and implemented by identifying the amplitude of excess noise which,
when added in quadrature to quoted errors in separation and separately for P.A., results in a linear fit with a reduced $\chi^2$ value of 1.0 (Table~\ref{tab:linearfit}).
No additional uncertainty is added if 
$\chi^2_{\nu}$ is less than 1.0 using the raw unadjusted errors.  

\begin{figure*}[htb!]
   \gridline{\fig{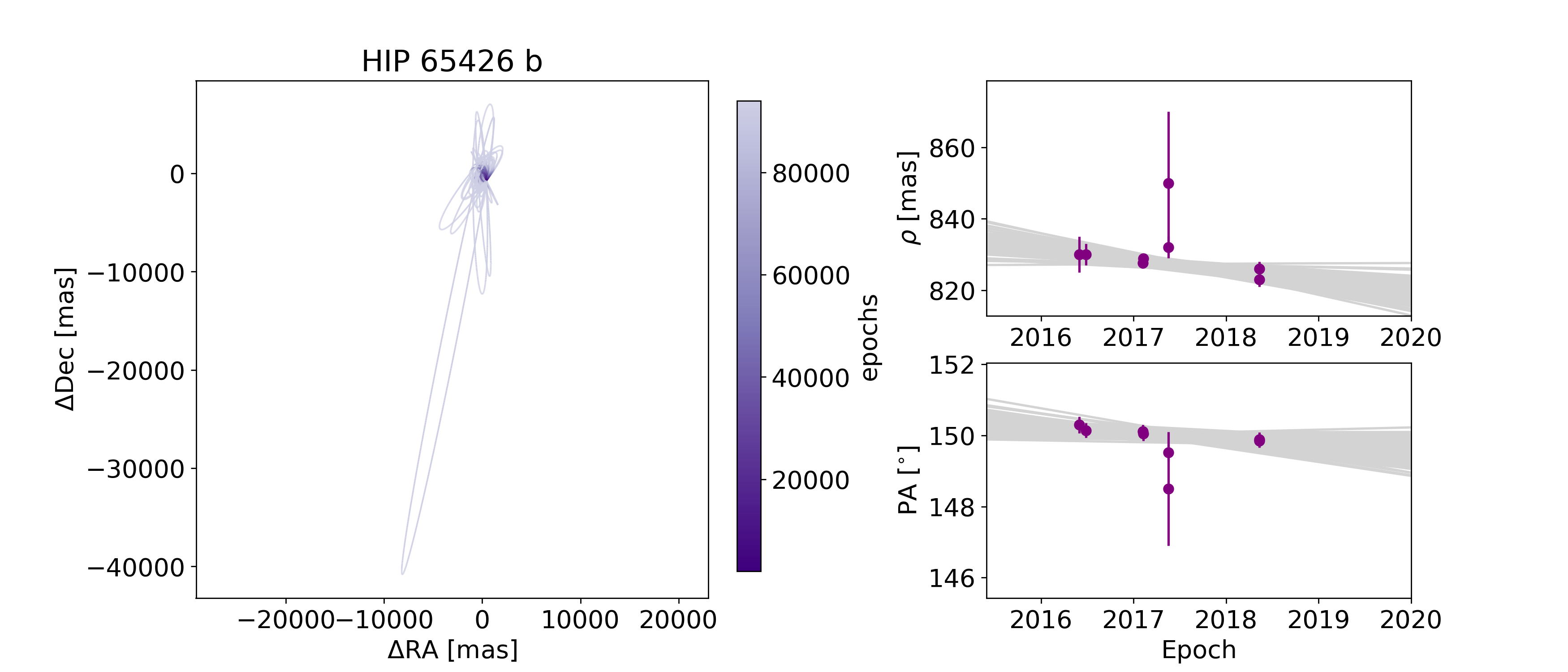}{0.5\textwidth}{(a) HIP 65426 b}
                 \fig{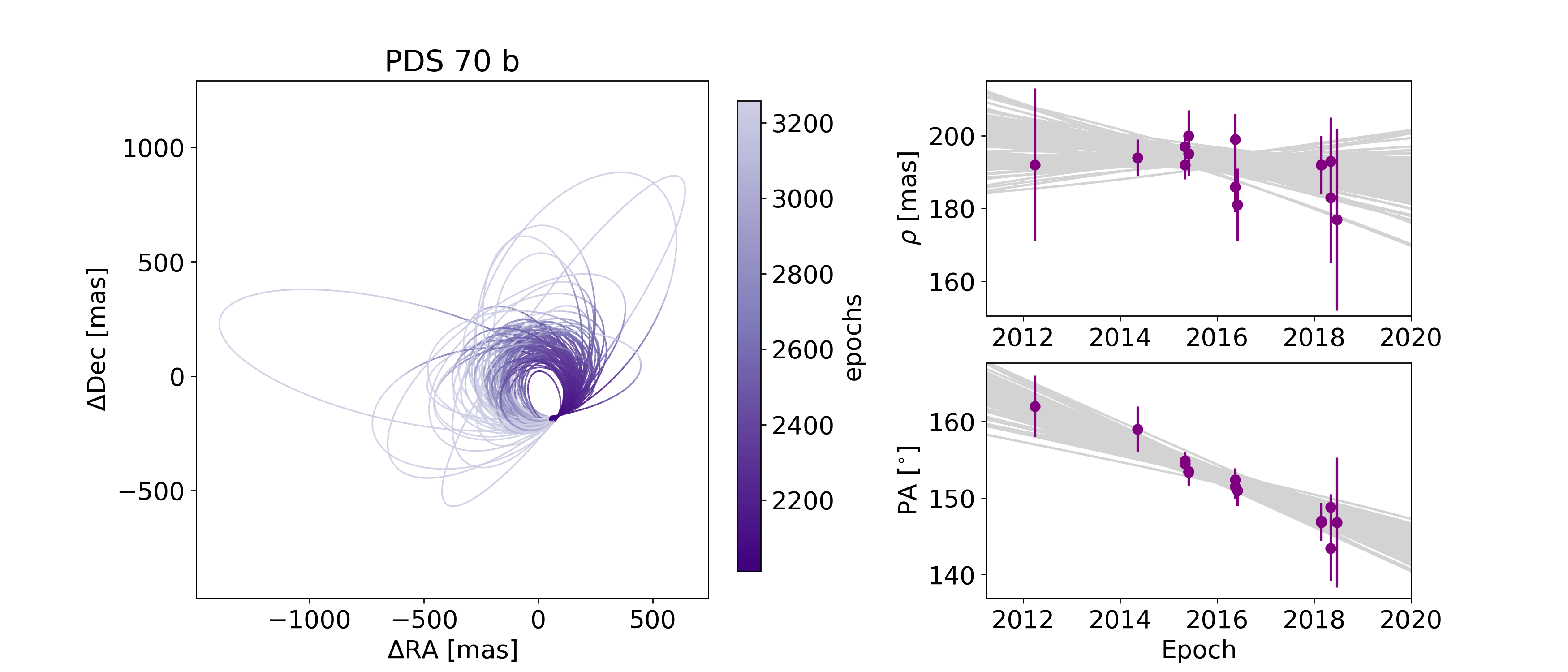}{0.5\textwidth}{(c) PDS 70 b}}
  \gridline{\fig{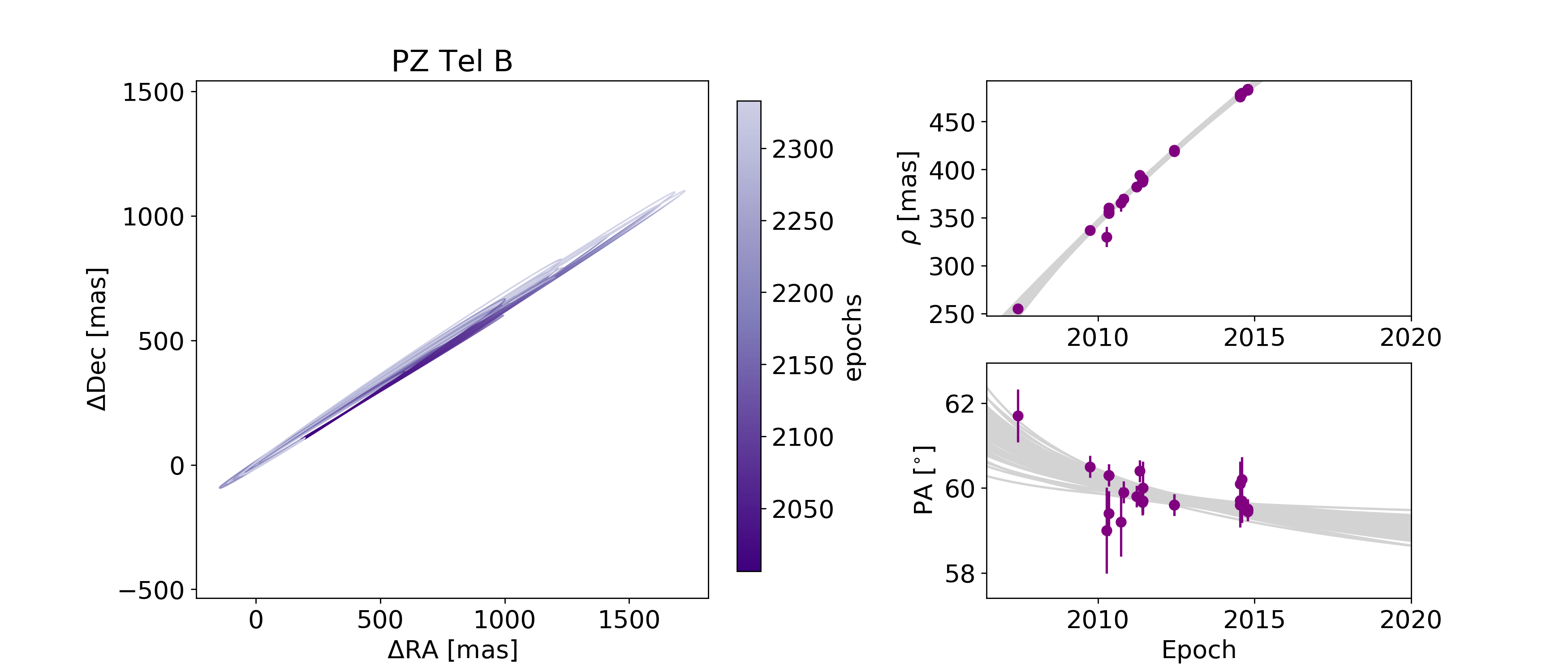}{0.5\textwidth}{(d) PZ Tel B}
                \fig{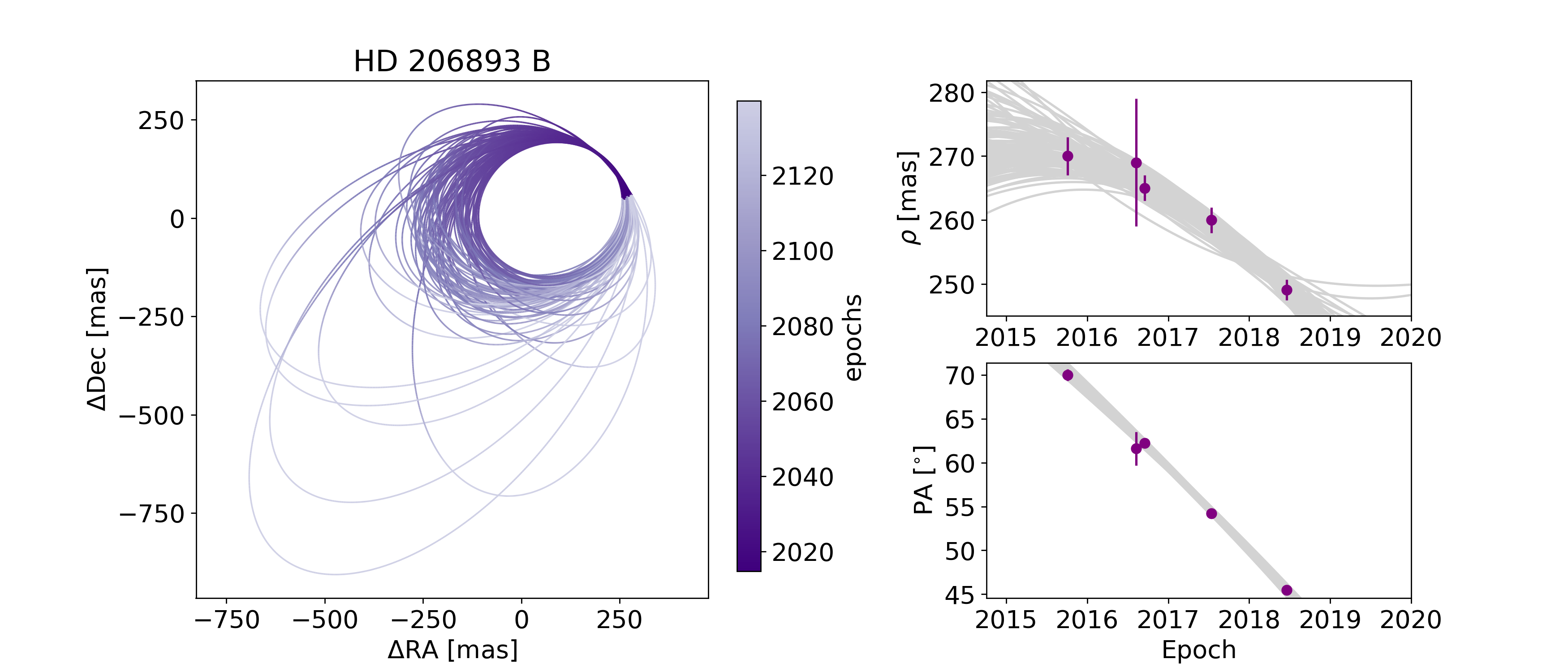}{0.5\textwidth}{(e) HD 206893 B}}
 \vskip 0 in
  \caption{Orbit fits for HIP 65426 b (a), PDS 70 b (b), PZ Tel B (c), and HD 206893 B (d)
  using \texttt{orbitize!}. 
  See Figure~\ref{fig:orbitfits3} for details.    \label{fig:orbitfits4} } 
\end{figure*}

Results from the orbit fits are shown in Figures \ref{fig:orbitfits3} and \ref{fig:orbitfits4} and summarized in Table \ref{tab:orbitize}.
Our constraints are generally consistent with published orbits that make use of the same astrometry.
Below we compare our fits with those in the literature with a particular focus on the eccentricity posteriors.

\emph{HD 984 B} --- 
This substellar companion was found by \citet{Meshkat:2015hd} and further characterized by \citet{JohnsonGroh:2017kh} with Gemini/GPI as part
of the GPIES survey.  We adopt a stellar mass of 1.15 $\pm$ 0.06 \Msun \ from \citet{Mints:2017di},
a total system mass of 1.21 $\pm$ 0.06 \Msun,
and a parallax of 21.781 $\pm$ 0.056 mas from $Gaia$ DR2 for our orbit fit.
Our orbital constraints are similar to those of Johnson-Groh et al. using 
the same five epochs from 2012 to 2016
(Figures~\ref{fig:orbitfits3} and \ref{fig:hd984corner}). 
We include a jitter term of 3.2 mas in separation but this does not significantly alter the results.
Our eccentricity posterior peaks near 0.0 with a long tail out to $e$ $\approx$ 0.8.
Although this constraint is broad, circular orbits are clearly favored for this system.

\emph{51 Eri b} --- 
This $\approx$2~\Mjup \ planet was discovered by \citet{Macintosh:2015fw}
and has been re-imaged several times with GPI and SPHERE since then.  
The M+M binary GJ 3305 AB orbits 51 Eri Ab at
$\approx$2000 AU (\citealt{Feigelson:2006tz}; \citealt{Montet:2015ky}).
A total system mass of 1.75 $\pm$ 0.05 \Msun \ (\citealt{Macintosh:2015fw})
and parallax of 33.577 $\pm$ 0.135 mas from $Gaia$ DR2 are fixed for our orbit fit.
We make use of 11 epochs for our analysis---five from \citet{DeRosa:2015jla}, four of which 
were originally presented in \citet{Macintosh:2015fw}, and six from \citet{Maire:2019kb}.
No additional jitter is included for this system.

Maire et al. find hints of curvature in the orbit and 
note a small ($\approx$1$\sigma$) offset in P.A. between GPI and 
SPHERE for observations  taken at similar epochs.  They proceeded to add a 1.0$\degr$ offset to the 
GPI data to reduce this difference and also removed the epoch from January 2015 from
consideration because the separation is somewhat larger than for other epochs.
For this study we take the astrometry at face value; no recalibrations are applied and we
consider all published astrometry.
Any slight differences in astrometry for individual 
epochs are at the $\approx$1$\sigma$ level and are well accounted for with 
larger uncertainties.  Indeed, linear fits to separation and P.A. over time yield 
$\chi^2_{\nu}$ values below unity (Table~\ref{tab:linearfit}), suggesting that reported uncertainties 
may even be slightly overestimated for this system. 
Nevertheless, our results are very similar to those of Maire et al. (Figures~\ref{fig:orbitfits3} and \ref{fig:51ericorner});  
we find a significant eccentricity
for 51 Eri b of $e$=0.50$^{+0.11}_{-0.08}$, in good agreement with their median value of $e$=0.45 and 
68\% credible interval of $e$=0.30--0.55.

\emph{HR 2562 B} ---
\citet{Konopacky:2016dk} presented four epochs of HR 2562 B with GPI in their discovery paper.  
They confirmed the companion shares common proper motion with the host star 
and found that its P.A. and the orientation of the disk are consistent with a nested orbit inside the disk.
\citet{Maire:2018ch} added six additional measurements over three epochs in 2016 and 2017 with SPHERE.
They found orbital motion and confirmed the orbital plane of HR 2562 B may be aligned with the debris disk.
They also identify  a wide range of eccentricities 
when using all the observations, but that values of $e$$\lesssim$0.3 were preferred when including a prior imposed by
the debris disk.

Our orbit fit using a stellar mass of 1.37 $\pm$ 0.02 \Msun \ (\citealt{Mesa:2018er}),
a total system mass of 1.40 $\pm$ 0.02 \Msun,
and $Gaia$ DR2 parallax of 29.377 $\pm$ 0.041 mas  
is broadly consistent with the results from Maire et al.
All eccentricities are allowed (the 2$\sigma$ credible interval spans $e$=0.037--1.0; 
see Figures~\ref{fig:orbitfits3} and \ref{fig:hr2562corner}).
We also find that HR 2562 B is on a nearly edge-on orbit, which is apparent from the 
large radial motion of HR 2562 B with respect to its host star,
whereas the P.A. is nearly unchanging.
Note that no additional jitter is included in our orbit fit for this system.
We do not incorporate additional constraints on orbital solutions from the debris disk
to avoid having to make assumptions about disk-planet coplanarity for HR 2562 and other systems
with disks in this study.


\emph{HR 3549 B} ---
Limited astrometry has been published for the substellar companion HR 3549 B.
\citet{Mawet:2015kk} obtained two epochs with VLT/NaCo from 2013 and 2015, and 
\citet{Mesa:2016kn} acquired two astrometric measurements taken at the same epoch in 2015 with SPHERE.
Mesa et al. examined orbital constraints for HR 3549 B with and without constraints from the disk.  
They found that high eccentricities ($e \gtrsim 0.3$) are preferred for uninformed priors, and orbits
informed by the presence of the inner disk suggest even larger values ($e \gtrsim 0.5$).

For this system we use a stellar mass of 2.3 $\pm$ 0.15 \Msun \ from \citet{Mawet:2015kk},
a total system mass of 2.34 $\pm$ 0.15 \Msun,
and a parallax of 10.486 $\pm$ 0.090 mas from $Gaia$ DR2. 
No additional jitter is added to the published astrometry.
Our orbit fit is shown in Figures~\ref{fig:orbitfits3} and \ref{fig:hr3549corner}.
Our results disagree with the eccentricity distribution from Mesa et al.; we find that 
all eccentricities are possible, and that the highest eccentricities above $e$$\approx$0.9 are least preferred (the 2$\sigma$
credible interval is 0.0--0.88).  The origin of this discrepancy is not immediately obvious, 
but may arise from the difference between the Bayesian orbit fitting of \texttt{orbitize!} 
compared to the Least-Squares Monte Carlo technique.

\emph{HD 95086 b} ---
HD 95086 b was identified by \citet{Rameau:2013vh}
and confirmed by \citet{Rameau:2013ds}.  This companion has been continuously monitored with 
GPI (\citealt{Rameau:2016dx}) and SPHERE (\citealt{Chauvin:2018ib}) since its discovery.
Both of these studies found
that the companion's modest orbital motion points to 
lower eccentricities ($e$$\lesssim$0.5), with the most likely values being near zero.

We adopt a stellar (and total system) mass of 1.7 $\pm$ 0.1 \Msun \ (see Table~\ref{tab:sample}) 
and a parallax of 11.568 $\pm$ 0.033 mas
from $Gaia$ DR2.  No additional jitter is needed for this companion.
Our results using all 20 epochs of published astrometry (Figures~\ref{fig:orbitfits3} and \ref{fig:hd95086corner})
are in good agreement with those of Rameau et al. and Chauvin et al.; our 2$\sigma$ credible interval for the eccentricity is 0.0--0.48 
with low values near zero being preferred.

\emph{HIP 65426 b} ---
\citealt{Chauvin:2017hl} discovered this long-period planet with SPHERE and demonstrated that this object 
shares a common proper motion with its host star; however no orbital motion was detected.
\citet{Cheetham:2019ju} presented follow-up multi-epoch astrometry with VLT/NaCo and SPHERE which enabled
them to carry out the first orbital analysis for this system.  They found that the eccentricity is essentially 
unconstrained and is dominated by the choice in prior rather than the likelihood (the data).

Our results are shown in Figures~\ref{fig:orbitfits4} and \ref{fig:hip65426corner}.
We adopt a stellar mass of 1.96 $\pm$ 0.04 \Msun \ (\citealt{Chauvin:2017hl}),
a total system mass of 1.97 $\pm$ 0.04 \Msun,
and parallax of 9.157 $\pm$ 0.063 mas from $Gaia$ DR2.
We find similar results as Cheetham et al. using a modest addition of 0.12$\degr$ of jitter in P.A. (but no adjustment
to the uncertainties in separation): the eccentricity distribution is essentially unconstrained and our 2$\sigma$ credible spans 0.03 to 0.97 as a 
result of the sparse orbital coverage for this system.

\emph{PDS 70 b} --- 
This young planet was found by \citet{Keppler:2018dd} nested in the transition disk of its host star
using NICI, NaCo, and SPHERE as part of the SHINE survey.
\citet{Muller:2018jr} presented additional observations with SPHERE and carried out preliminary orbit constraints,
finding an eccentricity posterior that peaks near $e$=0.0 with a tail to a value of about 0.6.
Additional astrometry were published by \citet{Wagner:2018hw} using Magellan/MagAO,
\citet{Christiaens:2019db} using VLT/SINFONI, and \citet{Haffert:2019ba} using VLT/MUSE.

We adopt a stellar host mass of 0.76 $\pm$ 0.02 \Msun \ from \citet{Muller:2018jr},
a total system mass of 0.77 $\pm$ 0.02 \Msun,
 and a parallax of 8.816 $\pm$ 0.041 mas from $Gaia$ DR2.
Our orbit fit using all 14 published epochs is shown in Figures~\ref{fig:orbitfits4} and \ref{fig:pds70corner}, and summarized in 
Table~\ref{tab:orbitize}.  No jitter is required.
Our 2$\sigma$ credible interval spans $e$=0.0--0.59, with lower values being preferred.

\emph{PZ Tel B} ---
This brown dwarf companion was independently found by \citet{Biller:2010ku} with NICI 
and \citet{Mugrauer:2010cp} with NaCo.  Despite having only two epochs available, Biller et al. 
were able to establish a high eccentricity of $>$0.6 for this system.  This has been bolstered
by additional astrometry and progressively more refined orbit constraints over the past decade
(\citealt{Mugrauer:2012ca}; \citealt{Ginski:2014ef}; \citealt{Beust:2016bf}; \citealt{Maire:2016go}).

We adopt a stellar host mass of 1.25 $\pm$ 0.10 \Msun \ from \citet{Biller:2010ku},
a total system mass of 1.3 $\pm$ 0.1 \Msun,
and a parallax of 21.219 $\pm$ 0.060 mas from $Gaia$ DR2 for our orbit analysis.
All 26 published epochs are used for our orbit fit.  There is some indication that on average the
reported uncertainties are underestimated; 
3.8 mas and 0.16$\degr$ of astrometric jitter are required to bring 
the separation and P.A. measurements in statistical agreement with a linear fit.
Note that we find significant curvature in both separation and P.A. (Figure~\ref{fig:orbitfits4}), 
suggesting we may be slightly overestimating the value of the jitter for this star.
This curvature was first noted by \citet{Mugrauer:2012ca}, \citet{Ginski:2014ef}, and \citet{Maire:2016go}.
With this additional adjustment we find that high values of eccentricity above 0.6 are strongly preferred,
with the 2$\sigma$ credible interval spanning 0.74---1.0.
Results for PZ Tel B can be found in Figures~\ref{fig:orbitfits4} and \ref{fig:pztelcorner}.

\emph{HD 206893 B} ---
Five astrometric epochs are available for this companion: two with SPHERE and NaCo as part of the discovery 
observations by \citet{Milli:2017fs} and three additional observations with SPHERE by \citet{Delorme:2017hl} and \citet{Grandjean:2019cv}.
Delorme et al. determined initial constraints on the orbital elements for this companion and found that all 
eccentricities are allowed with a preference against the highest values above about 0.9.  
When considering only orbits coplanar with the debris disk, this collapses to a range of about $e$=0.0--0.4.
More recently,  \citet{Grandjean:2019cv} combine relative astrometry, a radial acceleration,
and astrometric acceleration measured between \emph{Hipparcos} and \emph{Gaia} to constrain the 
orbit and dynamical mass of HD 206893 B.  They find a low eccentricity ($e$$\lesssim$0.4) and potential evidence
of a second companion based on the larger-than-expected radial acceleration.

For this study we fix the host mass at 1.32 $\pm$ 0.02 \Msun \ from \citet{Delorme:2017hl},
adopt a total system mass of 1.35 $\pm$ 0.02 \Msun,
and use a parallax of 24.506 $\pm$ 0.064 mas from $Gaia$ DR2.  No jitter is required in separation,
but modest jitter of 0.29$\degr$ is inferred for the P.A. measurements.
Results from our orbit fits are presented in Figures~\ref{fig:orbitfits4} and \ref{fig:hd206893corner}.
We determine an eccentricity of $e$=0.25$^{+0.17}_{-0.14}$ with a 2$\sigma$ credible
interval of 0.0--0.44.  
Our constraints are narrower than the general unrestricted fit from Delorme et al. and are similar to 
the recent results from \citet{Grandjean:2019cv}, who also included relative astromety, RVs, and absolute astrometry
of the host star from \emph{Hipparcos} and \emph{Gaia} in their orbit fit.


\begin{figure}
  \vskip -0.2 in
  \hskip -1.8 in
  \resizebox{6in}{!}{\includegraphics{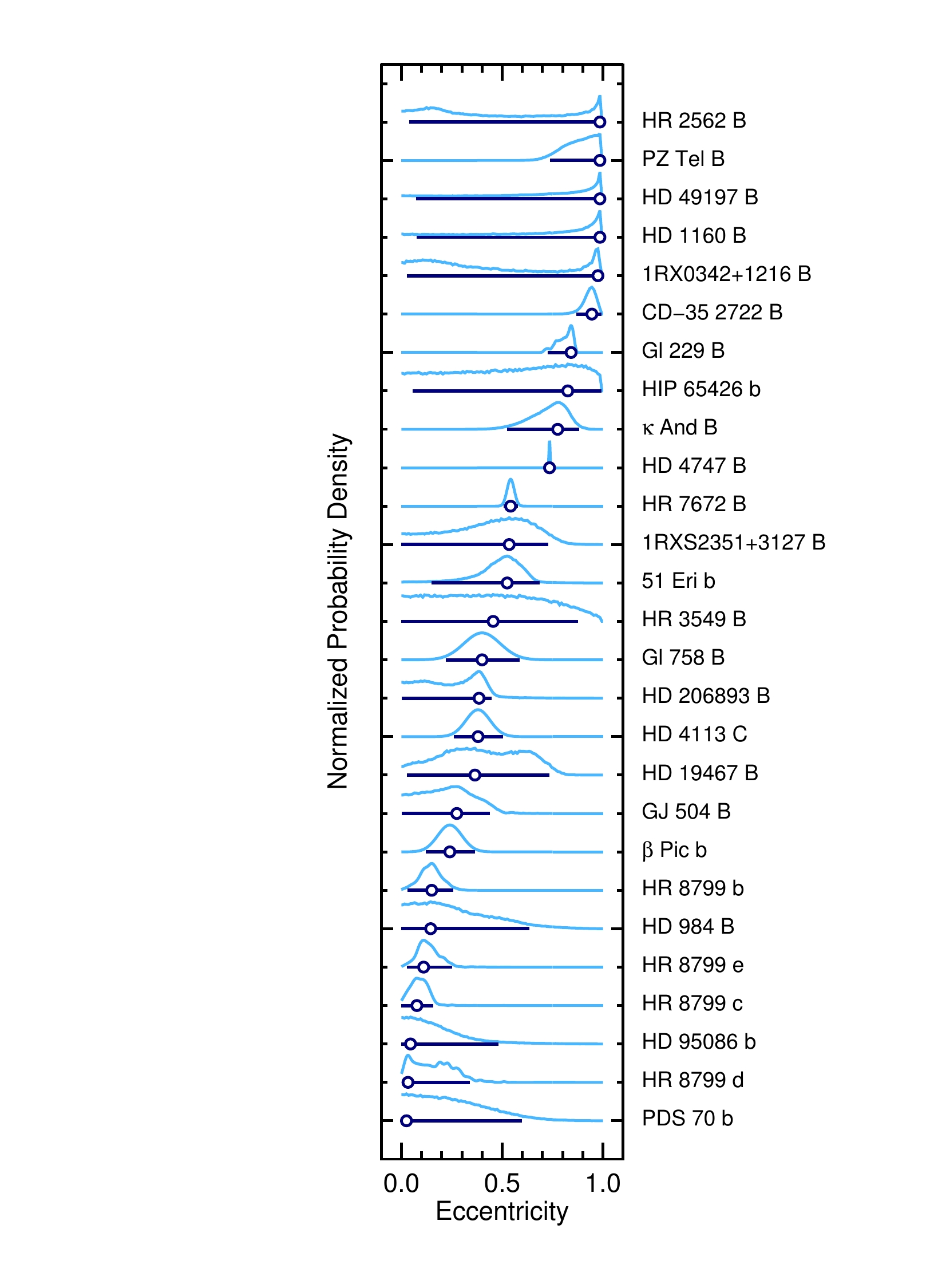}}
  \vskip -.3 in
  \caption{Posterior eccentricity distributions for the full sample of substellar companions.
  Objects are sorted from highest to lowest values of the peak of the distribution function, denoted
  with an open circle.  Dark blue lines indicate 95\% credible intervals.  The eccentricity constraints
  range from well-determined to completely unconstrained, which is especially true for four of the upper five
  systems in this figure.  The companions with the highest reliable eccentricities are PZ Tel B and CD--35~2722 B,
  which have $e$$>$0.7 and $e$=0.94$\pm$0.03, respectively.
    \label{fig:ecc} } 
\end{figure}

\begin{deluxetable*}{lcccccc}
\renewcommand\arraystretch{0.9}
\tabletypesize{\small}
\setlength{ \tabcolsep } {.1cm} 
\tablewidth{0pt}
\tablecolumns{7}
\tablecaption{Summary of Individual Eccentricity Distributions\label{tab:ecc}}
\tablehead{
 \colhead{Name} &  \multicolumn{4}{c}{$e$} &  \colhead{Data\tablenotemark{b}} &  \colhead{Orbit Fit}  \\
 \cline{2-5}
  & \colhead{Median}      & \colhead{MAP\tablenotemark{a}} &  \colhead{68\% C.I.}  & \colhead{95\% C.I.}    &  &     
        }   
\startdata
                  HD 984 B    &             0.23    &             0.15    &        0.0--0.33    &        0.0--0.63     &               DI     &                  This work    \\
                 HD 1160 B    &             0.74    &             0.98    &       0.51--0.99    &      0.067--0.99     &               DI     &                  This work    \\
                 HD 4113 C    &             0.38    &             0.38    &       0.32--0.44    &       0.26--0.50     &          DI + RV     &    \citet{Cheetham:2018ha}    \\
                 HD 4747 B    &             0.73    &             0.73    &       0.73--0.74    &       0.73--0.74     &    DI + RV + Ast     &      \citet{Brandt:2019ey}    \\
                HD 19467 B    &             0.39    &             0.36    &       0.22--0.65    &      0.032--0.73     &               DI     &                  This work    \\
            1RX0342+1216 B    &             0.34    &             0.97    &        0.0--0.59    &      0.022--0.98     &               DI     &                  This work    \\
                  51 Eri b    &             0.50    &             0.52    &       0.42--0.61    &       0.15--0.68     &               DI     &                  This work    \\
             $\beta$ Pic b    &             0.24    &             0.24    &       0.18--0.30    &       0.12--0.36     &    DI + RV + Ast     &       \citet{Dupuy:2019cy}    \\
              CD-35 2722 B    &             0.94    &             0.94    &       0.91--0.97    &       0.86--0.99     &               DI     &                  This work    \\
                  Gl 229 B    &             0.82    &             0.84    &       0.79--0.86    &       0.72--0.86     &    DI + RV + Ast     &      \citet{Brandt:2019kp}    \\
                HD 49197 B    &             0.70    &             0.98    &       0.48--0.99    &      0.067--0.99     &               DI     &                  This work    \\
                 HR 2562 B    &             0.42    &             0.98    &        0.0--0.69    &      0.032--0.99     &               DI     &                  This work    \\
                 HR 3549 B    &             0.43    &             0.46    &        0.0--0.58    &        0.0--0.87     &               DI     &                  This work    \\
                HD 95086 b    &             0.14    &            0.045    &        0.0--0.21    &        0.0--0.48     &               DI     &                  This work    \\
                  GJ 504 B    &             0.22    &             0.27    &      0.053--0.33    &        0.0--0.44     &               DI     &                  This work    \\
               HIP 65426 b    &             0.55    &             0.83    &       0.34--0.96    &      0.031--0.97     &               DI     &                  This work    \\
                  PDS 70 b    &             0.23    &            0.025    &        0.0--0.33    &        0.0--0.59     &               DI     &                  This work    \\
                  PZ Tel B    &             0.89    &             0.98    &       0.84--0.99    &       0.73--0.99     &               DI     &                  This work    \\
                  Gl 758 B    &             0.40    &             0.40    &       0.31--0.49    &       0.22--0.58     &    DI + RV + Ast     &      \citet{Brandt:2019ey}    \\
                 HR 7672 B    &             0.54    &             0.54    &       0.52--0.56    &       0.50--0.58     &    DI + RV + Ast     &      \citet{Brandt:2019ey}    \\
                 HR 8799 b    &             0.15    &             0.15    &      0.086--0.19    &      0.031--0.25     &               DI     &        \citet{Wang:2018fd}    \\
                 HR 8799 c    &            0.088    &            0.077    &      0.041--0.13    &     0.0040--0.16     &               DI     &        \citet{Wang:2018fd}    \\
                 HR 8799 d    &             0.15    &            0.033    &      0.011--0.22    &        0.0--0.34     &               DI     &        \citet{Wang:2018fd}    \\
                 HR 8799 e    &             0.13    &             0.11    &      0.070--0.18    &      0.024--0.25     &               DI     &        \citet{Wang:2018fd}    \\
               HD 206893 B    &             0.25    &             0.39    &       0.12--0.42    &        0.0--0.44     &               DI     &                  This work    \\
            $\kappa$ And B    &             0.74    &             0.78    &       0.67--0.84    &       0.53--0.88     &               DI     &                  This work    \\
           1RXS2351+3127 B    &             0.46    &             0.53    &       0.28--0.70    &     0.0010--0.73     &               DI     &                  This work    \\
\enddata
\tablenotetext{a}{Maximum \emph{a posteriori} probability.}
\tablenotetext{b}{``DI"=relative astrometry from direct imaging; ``Ast"=absolute astrometry from HGCA (\citealt{Brandt:2018dja}); ``RV"=relative radial velocities.}
\end{deluxetable*}

\subsection{Eccentricity Distributions with Hierarchical Bayesian Modeling}{\label{sec:hbi}}

Our goal is to determine the underlying behavior of the substellar eccentricity distribution at
the population level given a sample of measured eccentricity posteriors for individual systems---some 
precisely determined, others broadly constrained, and most of which are asymmetric and 
non-Gaussian in shape (Figure~\ref{fig:ecc} and Table~\ref{tab:ecc}).
Incorporating the structure of the posterior distribution in these constraints is especially important in this study  
because the number of systems under consideration is relatively modest: 9 long-period planets and 18 brown dwarfs, totaling 
27 substellar companions altogether.

Hierarchical Bayesian modeling offers a natural framework to incorporate this type of 
probabilistic information at multiple (individual and population) levels.
This tool is gaining popularity in astronomy (see, e.g., \citealt{Loredo:2013aa}) and particularly within the field of exoplanets; for example, 
it has been used to determine 
exoplanet eccentricity distributions (\citealt{Hogg:2010gh}; \citealt{Shabram:2016gb}; \citealt{VanEylen:2019cy}), 
the value of $\eta_{\earth}$ (\citealt{ForemanMackey:2014bj}), 
the planet mass-radius relationship (\citealt{Rogers:2015jn}; \citealt{Wolfgang:2016fi}),
and host star obliquities (\citealt{Morton:2014in}; \citealt{Campante:2016jq}).

Hierarchical Bayesian modeling enables simultaneous inference for parameters of individual systems ($\boldsymbol{\theta}$) and hyperparameters governing 
the underlying behavior of the population ($\boldsymbol{\Lambda}$), given the data $\boldsymbol{d}$.
Here variables in bold denote vectors of multiple values, parameters, or datasets.
Bayes' Theorem for this multi-level modeling becomes

\begin{equation}
p(\boldsymbol{\theta}, \boldsymbol{\Lambda}\mid\boldsymbol{d}) = \frac{p(\boldsymbol{d}\mid\boldsymbol{\theta}) p(\boldsymbol{\theta}\mid\boldsymbol{\Lambda}) p(\boldsymbol{\Lambda})}{p(\boldsymbol{d})},
\end{equation}

\noindent where $p(\boldsymbol{\theta}, \boldsymbol{\Lambda}\mid\boldsymbol{d})$ is the joint
posterior distribution for the individual and population parameters,
$p(\boldsymbol{d}\mid\boldsymbol{\theta})$ is the likelihood function of the data, 
$p(\boldsymbol{\theta}\mid\boldsymbol{\Lambda})$ is the 
prior on the individual systems conditioned on the set of population-level hyperparameters,
$p(\boldsymbol{\Lambda})$ is the hyper-prior---the prior distribution on the set of population-level parameters $\boldsymbol{\Lambda}$ --- and 
$p(\boldsymbol{d})$ is the marginalized likelihood.

For this problem the orbit fits are carried out separately and the individual posteriors for the eccentricity 
distributions (and all other orbital elements) are available.  
We therefore seek to constrain the hyperparameters of a parameterized model
for the underlying eccentricity distribution based on observations of $N$ systems.
The posterior probability distribution of $\boldsymbol{\Lambda}$ is simply

\begin{equation} \label{eqn:postprob}
p(\boldsymbol{\Lambda} \mid \boldsymbol{d}) \propto \mathcal{L(\boldsymbol{d}\mid\boldsymbol{\Lambda})} \pi(\boldsymbol{\Lambda}).
\end{equation}

\noindent where $\mathcal{L(\boldsymbol{d}\mid\boldsymbol{\Lambda})}$ is the likelihood function and $\pi(\boldsymbol{\Lambda})$ is
the set of priors on the hyperparameters.

\citet{Hogg:2010gh} describe an importance sampling approach to hierarchical Bayesian modeling with
a specific application to  exoplanet eccentricities.  
We follow their method by making use of the eccentricity posterior
distributions of individual systems from our orbit fits to inform the population-level likelihood function
for parameters in the underlying eccentricity distribution, $f(e\mid\boldsymbol{\Lambda})$.
In this case the sampling approximation to the likelihood function in Equation~\ref{eqn:postprob} is

\begin{equation}
\mathcal{L(\boldsymbol{d}\mid\boldsymbol{\Lambda})} \approx \prod_{n=1}^N \frac{1}{K} \sum_{k=1}^K \frac{f(e_{nk}\mid\boldsymbol{\Lambda})}{\pi (e_{nk})},
\end{equation}

\noindent where $N$ is the number of substellar companions being considered,
$K$ is the number of samples from the posterior eccentricity distribution,
$e_{nk}$ is the $k$th random draw for each eccentricity distribution $n$,
$f(e_{nk}\mid\boldsymbol{\Lambda})$ is the probability density of the population-level eccentricity distribution evaluated at $e_{nk}$
and conditioned on the hyperparameters $\boldsymbol{\Lambda}$,
and $\pi (e_{nk})$ is the probability density of the prior probability distribution evaluated at $e_{nk}$.

We follow the approach of \citet{Hogg:2010gh}, \citealt{Kipping:2013he}, and \citet{VanEylen:2019cy},
by adopting a 
standard Beta distribution for our population-level eccentricity distribution, $f(e\mid\boldsymbol{\Lambda})$.
The advantage of the standard Beta distribution's functional form is that it is flexible, spans a range of [0,1], and is described by 
only two shape parameters, $\boldsymbol{\Lambda}$$\equiv$($\alpha$, $\beta$), both of which take on 
values $>$0:

\begin{equation}
f(e\mid \alpha, \beta) = \frac{\Gamma (\alpha+\beta)}{\Gamma (\alpha) \Gamma (\beta)} e^{\alpha-1} (1-e)^{\beta-1}.
\end{equation}

\noindent Here $\Gamma$ represents the Gamma function.  
The Beta distribution can capture a wide range of shapes, including uniform ($\alpha$=1, $\beta$=1);
$\mathcal{U}$-shaped with a single anti-mode;  
unimodal with positive, negative, or no skew; 
$\mathcal{J}$-shaped or reverse $\mathcal{J}$-shaped; bell-shaped; and triangular.
$\alpha$ governs the function's behavior at small eccentricities and $\beta$ influences
its shape at high eccentricities.  Small values of $\alpha$ and $\beta$ correspond to high probability
densities near $e$=0 and $e$=1, while large values of those parameters correspond to low 
probability densities near zero.\footnote{For large values of $\alpha$ and $\beta$ ($>$100), we use
a normal approximation to the Beta distribution for computational efficiency: 
$B(\alpha,\beta)$$\approx$$\mathcal{N}(\mu=\alpha/(\alpha+\beta), \sigma=\sqrt(\alpha \beta/( (\alpha+\beta)^2 (\alpha+\beta+1) )$.}

This simple parameterization is especially convenient for this study because it can qualitatively reproduce a wide range
of potential physical outcomes of the planet formation and migration process: 
outward scattering, which is expected to excite eccentricities (\citealt{Veras:2009br}; \citealt{Scharf:2009eq});
cloud fragmentation, which should result in a broad range of eccentricities;
dynamically relaxed systems that follow the thermal eccentricity distribution ($f(e)$$\sim$2$e$; \citealt{ambartsumian:1937wd});
and formation in a disk, in which case companions should retain nearly circular orbits if they are 
 dynamically undisturbed.

Our primary goal is therefore to constrain the hyperparameters $\alpha$ and $\beta$ of the Beta distribution
for our sample of imaged substellar companions undergoing measured orbital motion.  
We also examine the eccentricity distribution
of other subsamples: a division of giant planets and brown dwarfs based on companion mass, 
a subdivision by mass ratio,
imaged planets including and excluding the HR 8799 system, 
and the full sample subdivided by 
orbital separation.  These are discussed separately in more detail below.

For each of these cases we use the Metropolis-Hastings Markov Chain Monte Carlo (MCMC) algorithm 
(\citealt{Metropolis:1953vj}; \citealt{Hastings:1970wm}) to sample the posterior
distributions of the model parameters $\alpha$ and $\beta$.  
Linearly uniform hyperpriors are chosen for $\alpha$ from 0 to 1000, $\beta$ from 0 to 1000, and $\pi (e)$ from 0 to 1.
With only two parameters for the parent model and uncomplicated covariance, we use a single chain 
typically comprising 10$^6$ links and $K$=1000 samples from each individual system-level posterior 
to explore the population-level eccentricity posterior\footnote{The two exceptions are 
the ``HR 8799 only'' and the ``Giant Planets Excluding HR 8799'' cases in Section~\ref{sec:beta_hr8799}.
For these we use 10$^7$ links to better sample posteriors.}.
Convergence is monitored using the Gelman-Rubin (GR) statistic (\citealt{Gelman:1992ts}),
which compares the variance within chains to the variance between chains (here 
our single chain divided into sub-chains).  In all cases the GR statistic is less
than 1.1 within 10$^5$ links and in most cases it is less than 1.01, indicating the chains are well mixed.
No burn-in is warranted because of the low dimensionality 
of the model and the starting points are all near the equilibrium point of the posterior distributions.  
We adopt normal proposal distributions with standard deviations set to avoid acceptance rates
that are too high (near 1), in which case step sizes become too small to efficiently map posterior space,
and too low (near 0) where large jumps mean few proposed values are accepted and convergence is slow.
Most of our final acceptance rates fall between 0.3 and 0.8.

\subsubsection{Example: Recovering the Short-Period Exoplanet Eccentricity Distribution}{\label{sec:beta_example}}

\begin{deluxetable}{lccc}
\renewcommand\arraystretch{0.9}
\setlength{ \tabcolsep } {.1cm} 
\tablewidth{0pt}
\tablecolumns{4}
\tablecaption{Experiments Recovering the RV Eccentricity Distribution\label{tab:exp}}
\tablehead{
       \colhead{$N$\tablenotemark{a}} & \colhead{$\sigma_e$\tablenotemark{b}}  &  \colhead{$\alpha$\tablenotemark{c}}  & \colhead{$\beta$\tablenotemark{c}}
        }   
\startdata
    5 &  0.20 & 15.94$^{+7.86}_{-12.46}$ & 31.75$^{+18.24}_{-8.08}$  \\
    5 &  0.05 &  2.03$^{+0.83}_{-1.35}$  &  3.65$^{+1.53}_{-2.33}$  \\
    5 &  0.01 &  2.23$^{+0.85}_{-1.27}$  & 14.01$^{+5.91}_{-8.35}$  \\
   10 &  0.20 &  8.13$^{+3.60}_{-5.39}$  & 35.91$^{+14.05}_{-6.59}$  \\
   10 &  0.05 &  4.72$^{+1.83}_{-2.97}$  & 21.24$^{+8.26}_{-12.87}$  \\
   10 &  0.01 &  1.97$^{+0.73}_{-0.86}$  &  9.43$^{+3.34}_{-4.73}$  \\
   20 &  0.20 &  8.69$^{+3.99}_{-5.55}$  & 31.35$^{+16.97}_{-9.05}$  \\
   20 &  0.05 &  1.44$^{+0.46}_{-0.54}$  &  4.73$^{+1.51}_{-1.86}$  \\
   20 &  0.01 &  0.80$^{+0.25}_{-0.36}$  &  4.89$^{+1.48}_{-2.32}$  \\
   50 &  0.20 &  8.60$^{+3.04}_{-3.16}$  & 37.70$^{+12.29}_{-6.23}$  \\
   50 &  0.05 &  1.25$^{+0.23}_{-0.32}$  &  3.84$^{+0.78}_{-0.97}$  \\
   50 &  0.01 &  0.66$^{+0.14}_{-0.15}$  &  2.22$^{+0.46}_{-0.59}$  \\
\enddata
\tablenotetext{a}{Number of draws from the underlying Beta distribution ($\alpha$=0.867, $\beta$=3.03) 
from \citet{Kipping:2013he}.}
\tablenotetext{b}{Measurement errors for each mock realization.  $\sigma_e$ denotes the standard deviation
of the Gaussian probability distribution for the orbital eccentricity of each system.}
\tablenotetext{c}{Recovered $\alpha$ and $\beta$ shape parameters of the standard Beta distribution
from hierarchical Bayesian modeling.}
\end{deluxetable}


\begin{figure}
  \vskip -0.1 in
  \hskip 0.2 in
  \resizebox{6.in}{!}{\includegraphics{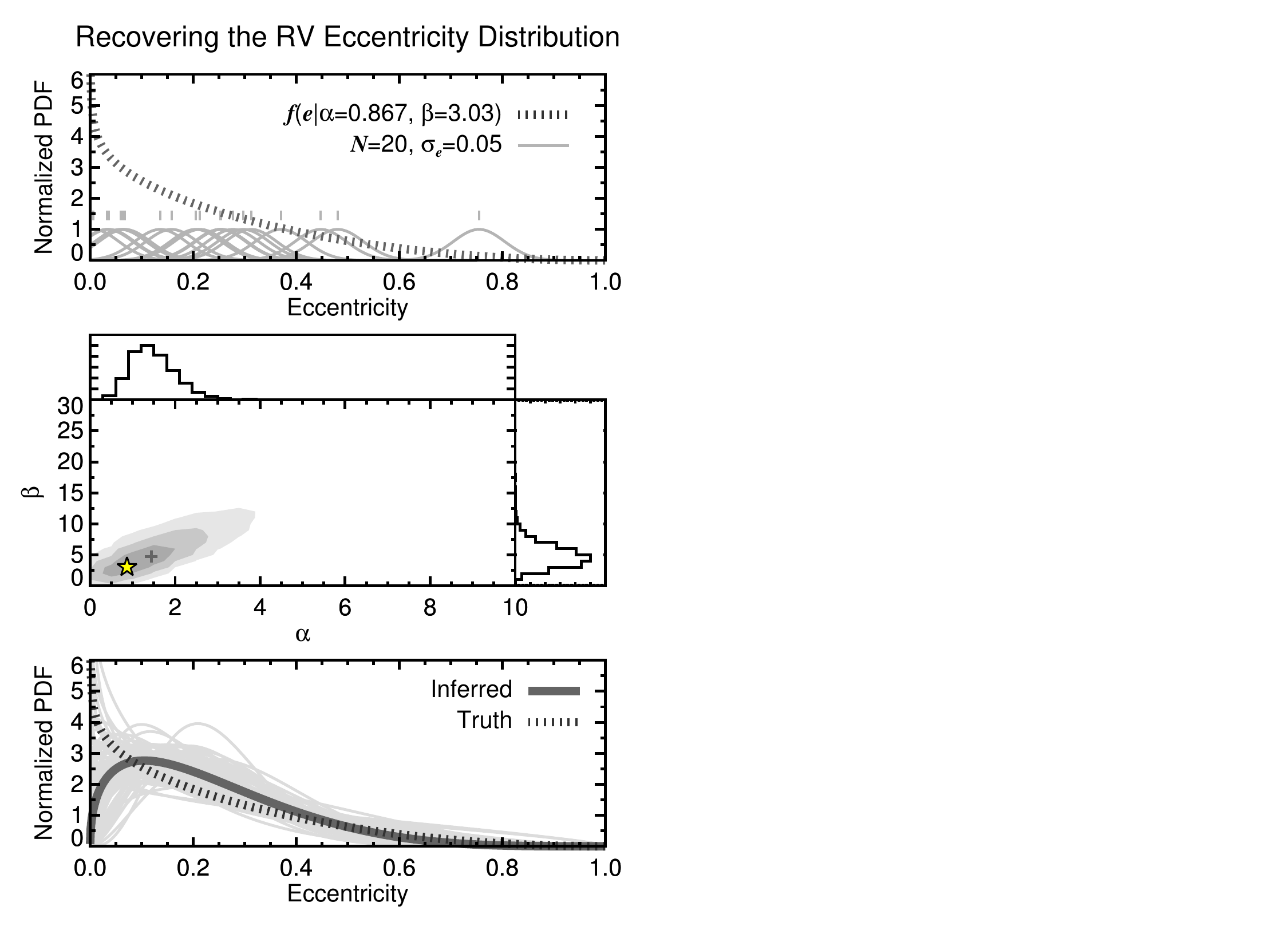}}
  \vskip -0.1 in
  \caption{Example of our hierarchical Bayesian modeling approach.  Here we apply this method to recover
  the known warm Jupiter eccentricity distribution given $N$ random mock ``measurements'' with various uncertainties ($\sigma_e$).
  \emph{Top panel:} The true underlying distribution (dotted curve), which is well described by a Beta distribution with
  $\alpha$=0.867 and $\beta$=3.03 (\citealt{Kipping:2013he}).  
  In this example 20 random values are drawn (thin gray curves) from the underlying parent distribution (thick dashed curve); 
  for each of these we assign a Gaussian uncertainty with a standard deviation of $\sigma_e$=0.05 (truncated at $e$$<$0.0 and
  $e$$>$1.0).   \emph{Middle panel:} Joint distribution from MCMC sampling of the posteriors of $\alpha$ and $\beta$.  
  Marginalized posteriors for each parameter are projected on the $x$ and $y$ axes.  Contours represent regions encompassing 
  1$\sigma$, 2$\sigma$, and 3$\sigma$ fractions of the joint posterior distributions.  The star denotes the value of the 
  true underlying distribution---within the 1$\sigma$ contour in this example.  
  \emph{Bottom panel:} The inferred underlying distribution (thick solid curve) compared to the true distribution.  Thin gray lines
  show 100 distributions randomly sampled from the posteriors of $\alpha$ and $\beta$.
 In this example the inferred distribution has captured the broad shape of the true distribution, with somewhat less fidelity (but more 
  uncertainty) at small eccentricities.  See Table~\ref{tab:exp} for results from experiments 
  varying $N$ and $\sigma_e$.  \label{fig:beta_example} } 
\end{figure}


\begin{figure*}
  \vskip -0.5 in
  \hskip 0.2 in
  \resizebox{7.in}{!}{\includegraphics{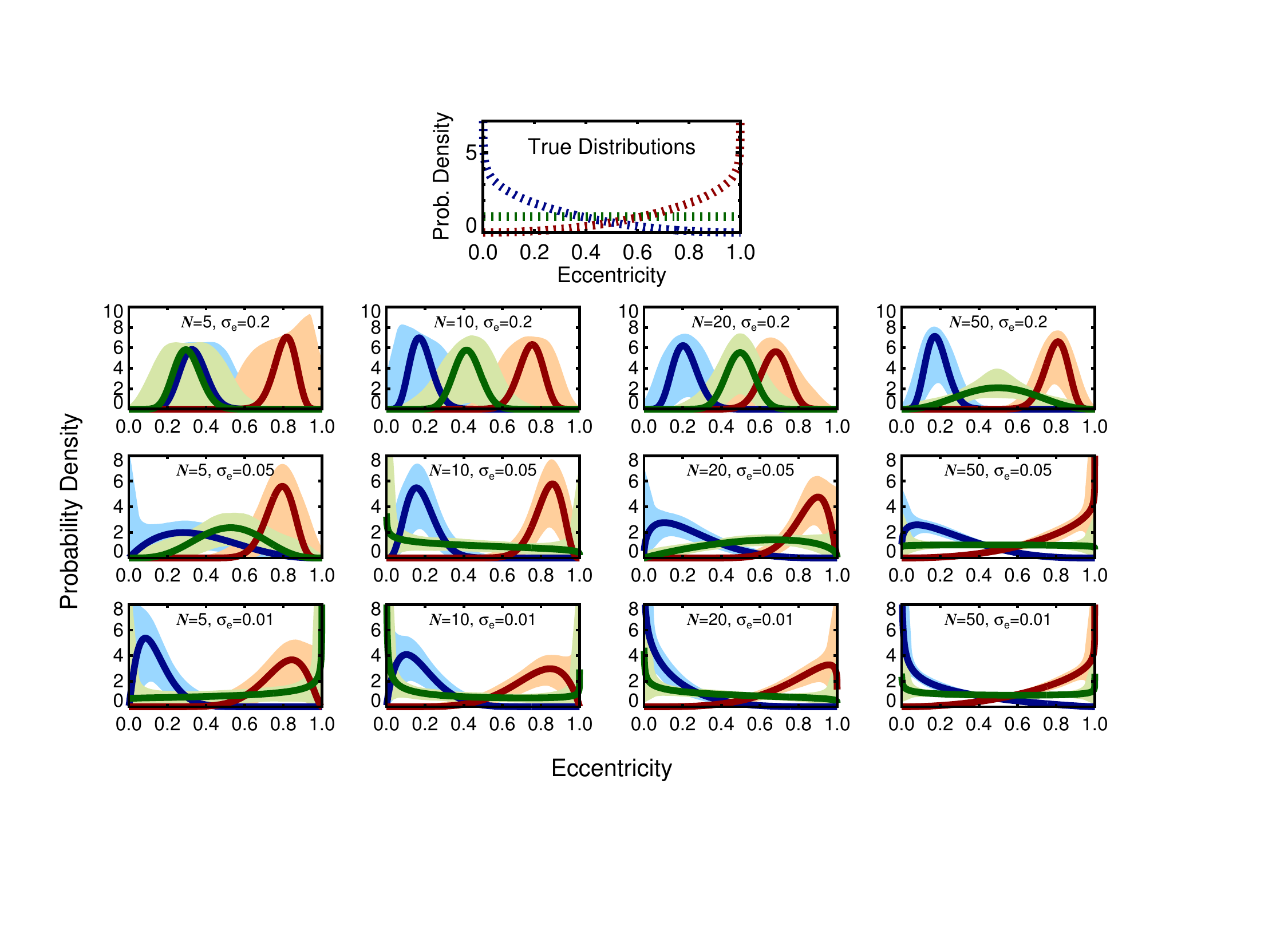}}
  \vskip -0.8 in
  \caption{Experiments recovering three underlying eccentricity distributions (top panel) as a 
  function of the number of randomly drawn measurements ($N$; increasing from left to right) 
  and Gaussian measurement uncertainty ($\sigma_e$; decreasing from top to bottom).  Here we test
  the warm Jupiter eccentricity distribution (Beta parameters $\alpha$=0.867, $\beta$=3.03; blue),
  a uniform distribution ($\alpha$=1.0, $\beta$=1.0; green), and the warm Jupiter distribution mirrored 
  at high eccentricities ($\alpha$=3.03, $\beta$=0.867, red).  The shape of the true distributions are 
  progressively better inferred with larger samples and smaller errors.   However, with precise enough measurements
  these distributions can be shown to be qualitatively distinct even with small samples of $N$$\approx$5.
  Shaded regions represent 2$\sigma$ credible intervals from posterior draws.  \label{fig:beta_example_summary} } 
\end{figure*}

We carried out a series of experiments 
using mock datasets drawn from the actual eccentricity distribution of 
warm Jupiters to establish how the individual-level eccentricity measurement precision ($\sigma_e$) and number 
of systems under consideration ($N$)
influence the ability to constrain the shape parameters $\alpha$ and $\beta$.
We adopt Beta parameters $\alpha$=0.867 and $\beta$=3.03 from \citet{Kipping:2013he}
as underlying ``truth'' for this exercise and test four samples comprising 5, 10, 20, and 50 systems.
For each sample we draw random eccentricity mock measurements from the population-level distribution and assign 
three sets of Gaussian uncertainties to these datasets, $\sigma_e$ = 0.2, 0.1, and 0.05. 
These individual synthetic measurements 
are meant to mimic random sampling from the parent distribution and are truncated below $e$=0 and 
above $e$=1.

Results of these 12 experiments are summarized in Table~\ref{tab:exp}, and one example with $N$=20 and 
$\sigma_e$=0.05 is shown in Figure~\ref{fig:beta_example}.
As expected, the fidelity with which the parameters of the true distribution are 
recovered strongly depends on the number of 
measurements as well as the precision of each measurement.  As more planets are ``observed'' and the 
constraints on the orbital eccentricity improve, the median value of the posteriors for $\alpha$ and
$\beta$ approach the true values and the uncertainties tend to decrease.  

However, it is also clear from this
exercise that randomness in the draws from the underlying eccentricity distribution can occasionally produce 
results that formally agree with the true parameters at the 2-3$\sigma$ level but which have 
the possibility of being mis- or over-interpreted.  This is especially
important for small samples, where stochasticity in the draws have a higher chance of 
producing results that qualitatively differ from the true underlying distribution.
For example, in the first experiment with $N$=5 and $\sigma_e$=0.2, we find values of 
$\alpha$=15.9$^{+7.9}_{-12.5}$ and $\beta$=31.8$^{+18.2}_{-8.1}$. 
The resulting Beta distribution peaks at higher eccentricities than the actual distribution, which
would overestimate how dynamically hot the RV exoplanet population really is.

Although the actual values of $\alpha$ and $\beta$ are within 3$\sigma$ of the joint constraints 
for all 12 experiments,
we conclude that the results for small sample sizes ($N$$\lesssim$10) should be interpreted as an indication of
the \emph{qualitative} behavior of the population.  Larger sample sizes are needed to more precisely uncover the 
detailed shape and quantitative constraints of the underlying distribution.  But even these rely on the assumption that
our input model $f(e)$ is correct.  It is certainly possible that another functional form better emulates the true
population-level eccentricity distribution, but we avoid this type of model comparison in our orbit study because of the limited
sample and the broad constraints on each individual system's eccentricity.  More targets 
and longer-term orbit monitoring will enable this type of model selection in the future.

To assess how readily different distributions can be distinguished from each other, we performed the same
experiment for a uniform distribution as well as the warm Jupiter distribution mirrored at high eccentricities 
(Beta shape parameters $\alpha$=3.03 and $\beta$=0.867). Because we are searching for population-level differences 
between giant planets and brown dwarfs, these results better reflect the goals of this study.
Results are summarized in Figure~\ref{fig:beta_example_summary}.  It is clear from these tests that larger
samples and better measurement precision more reliably reproduces the underlying distribution,
but differences \emph{between} distributions can readily be inferred from small samples.  For example,
even in the most pessimistic case ($N$=5, $\sigma_e$=0.2), the warm Jupiter and high-$e$ samples are
noticeably distinct, although in this case the uniform distribution happens to resemble the warm Jupiter sample.
We conclude that it is easier to determine whether two very different parent distributions are distinct from each
other than it is to establish the exact shape of the underlying distribution.

\subsubsection{Eccentricity Distribution for the Full Sample}{\label{sec:beta_full}}


\begin{figure}
  \vskip -0. in
  \hskip 0.2 in
  \resizebox{6.in}{!}{\includegraphics{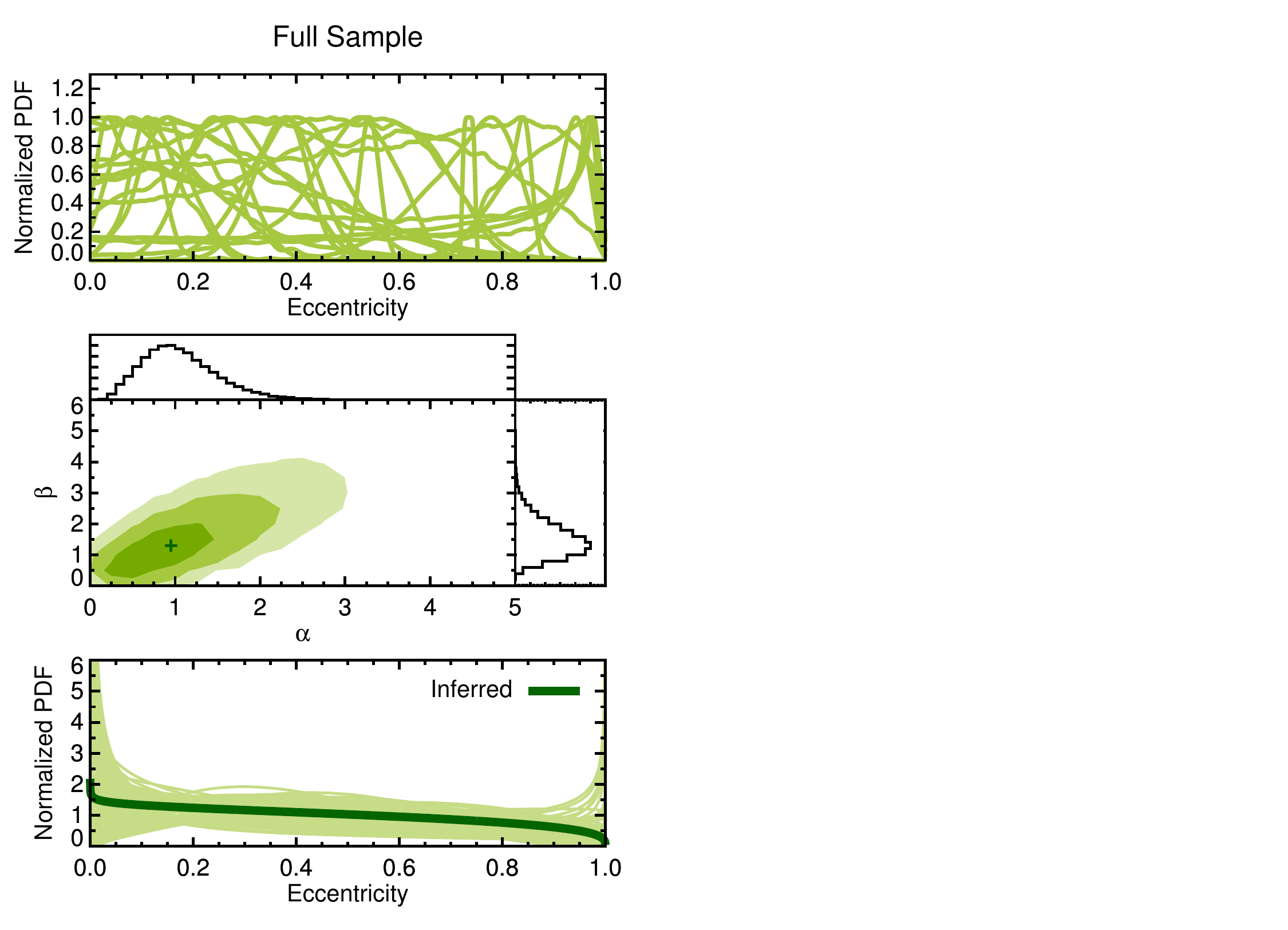}}
  \vskip 0 in
  \caption{Same as Figure~\ref{fig:beta_example} but for the full sample of 27 substellar companions
  (2--75~\Mjup, 5--100 AU).
  Individual eccentricity posteriors are plotted in the top panel.
  The middle and bottom panels show the best-fitting $\alpha$ and $\beta$ values of the 
  underlying Beta distribution,
  and the corresponding population-level posterior distribution for the full sample.
  The eccentricity distribution of substellar companions is approximately flat
  from $e$=0.0--1.0 with no significant evidence of a preference for high, intermediate, or low eccentricities.  \label{fig:beta_full} } 
\end{figure}

Our results for the underlying eccentricity distribution of the full sample 
of 27 substellar companions spanning 5--100 AU and 2--75~\Mjup \ are shown in Figure~\ref{fig:beta_full}
and summarized in Table~\ref{tab:popdist}.
The best-fitting values of $\alpha$ and $\beta$ are 0.95$^{+0.41}_{-0.43}$ and 1.30$^{+0.61}_{-0.46}$, 
respectively, with positive covariance between the two parameters.  
This corresponds to an approximately flat  distribution across
the entire range of eccentricities.  
Uncertainties in the posterior are larger at the lowest and highest eccentricities and allow for some
flexibility at both ends of the distribution, especially near $e$=0.

Four systems have
eccentricities that are peaked near 1.0 with substantial power spanning all values:
HD 49197 B, HD 1160 B,  HR 2562 B, and 1RXS0342+1216 B.  The origin of this shape is unclear, 
though it may be a result of an incorrect stellar mass or perhaps underestimated astrometric uncertainties.
To test whether these objects are biasing the results, we ran the same analysis except with uniform eccentricity
distributions for these systems.  The results for the full sample are nearly indistinguishable from the original case.
This also holds true for all of the additional experiments we carry out in the Sections below.

One of the practical implications of this result is that this flat distribution can reliably be used
as a Bayesian prior for  orbit fits of new substellar companions that are discovered in the future.  
A uniform eccentricity distribution has generally been adopted for orbit fits in the past as 
an uninformative prior, but this can now justifiably be used as an \emph{informed}
prior: in the absence of discovery bias (see Section~\ref{sec:discoverybias}), 
a newly identified substellar companion
is effectively equally likely to have a low, moderate, or high eccentricity.
A somewhat more precise prior would make use of the actual best-fitting Beta distribution we identify, 
which differs slightly from a uniform distribution.

The clearest implication of a flat posterior is that whatever formation or migration processes produce this population of low-mass companions
do not appear to imprint a strong preference for high, intermediate, or low eccentricities,
at least when marginalized over other parameters like stellar host mass, substellar mass, and system age.
This analysis for the full population also implicitly assumes that the underlying eccentricity distribution 
of this population does not vary as a function
of companion mass or separation.  That is, there is no strong gradient in the orbital  
properties of the sample within our sample spanning 5-100 AU and 2--75~\Mjup. 
Below we test these assumptions by subdividing this full sample based on companion mass,
mass ratio, orbital separation, and system age to determine whether there are signs of population-level changes
in the eccentricities of these companions.

\subsubsection{Giant Planets and Brown Dwarf Companions}{\label{sec:massdiv}}


\begin{figure}
  \vskip -0. in
  \hskip 0.2 in
  \resizebox{6.in}{!}{\includegraphics{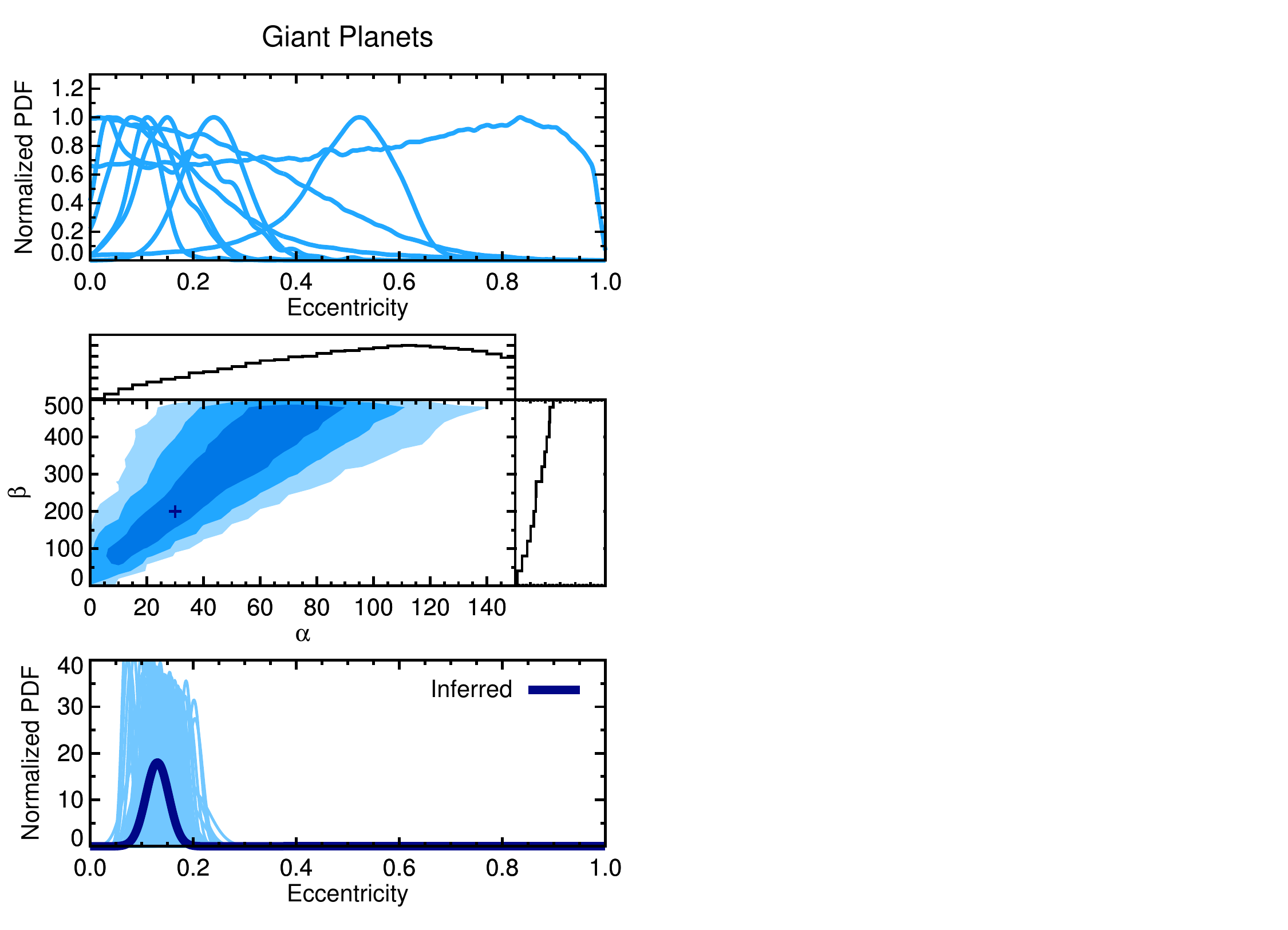}}
  \vskip 0 in
  \caption{Same as Figure~\ref{fig:beta_example} but for the sample of 9 imaged planets (2--15~\Mjup, 
  5--100 AU). The eccentricity distribution of giant planets indicates a preference 
  for low eccentricities ($e$$\approx$0.05--0.25). Following our cautionary results for 
  small samples in Section~\ref{sec:beta_example}, 
  we interpret this as a qualitative indication that long-period planets tend to have low eccentricities,
  not necessarily that circular orbits or moderate eccentricities are strongly disfavored.  \label{fig:beta_planet} } 
\end{figure}


\begin{figure}
  \vskip -0. in
  \hskip 0.2 in
  \resizebox{6.in}{!}{\includegraphics{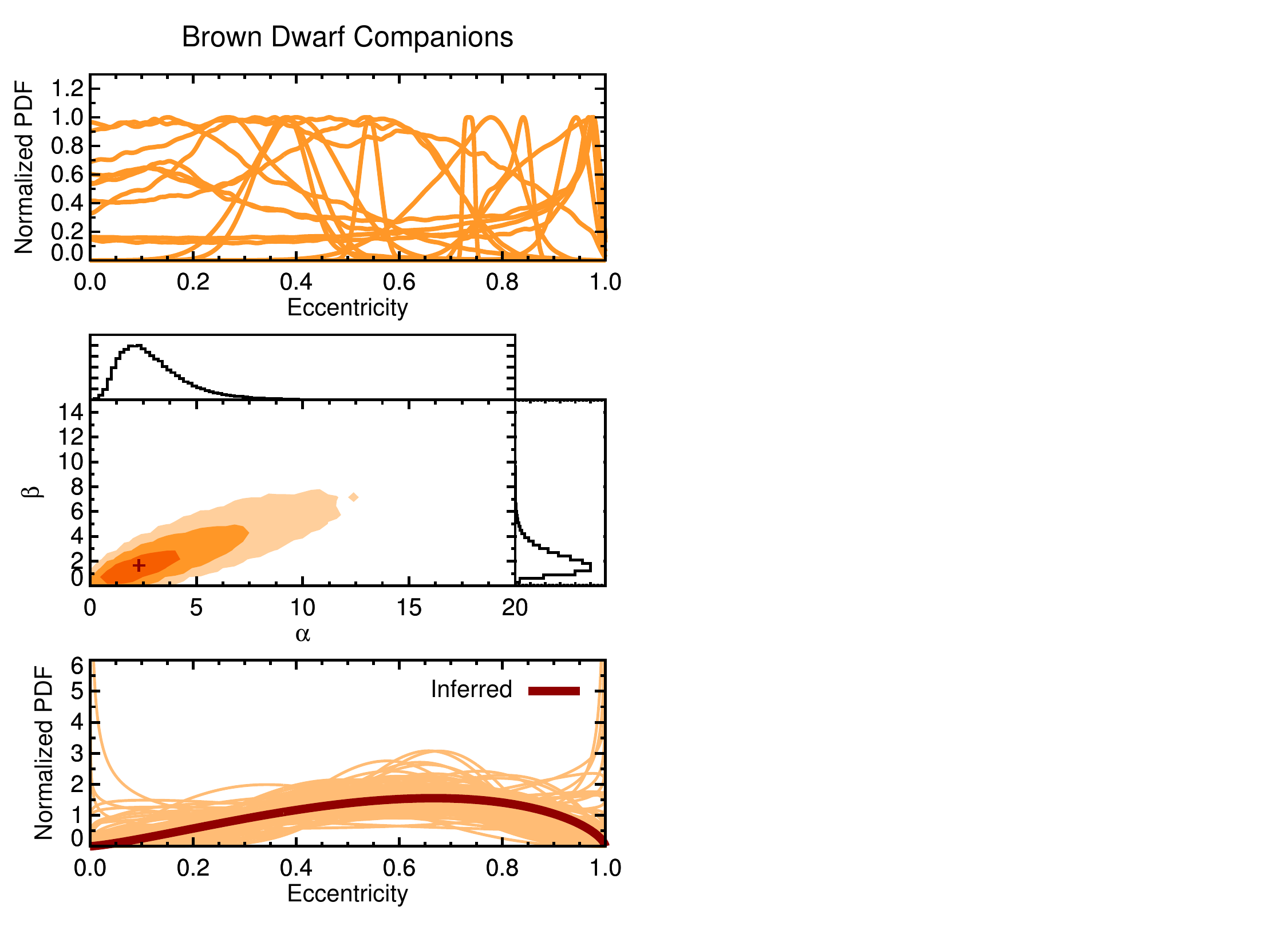}}
  \vskip 0 in
  \caption{Same as Figure~\ref{fig:beta_example} but for the sample of 18 brown dwarf companions (15--75~\Mjup, 
  5--100 AU). The eccentricity distribution of brown dwarfs indicates a broad peak towards high values, which differs  
  substantially from population-level distribution for giant planets in Figure~\ref{fig:beta_planet}. \label{fig:beta_bd} } 
\end{figure}


\begin{figure*}
  \vskip -0. in
  \hskip 0.4 in
  \resizebox{6.5in}{!}{\includegraphics{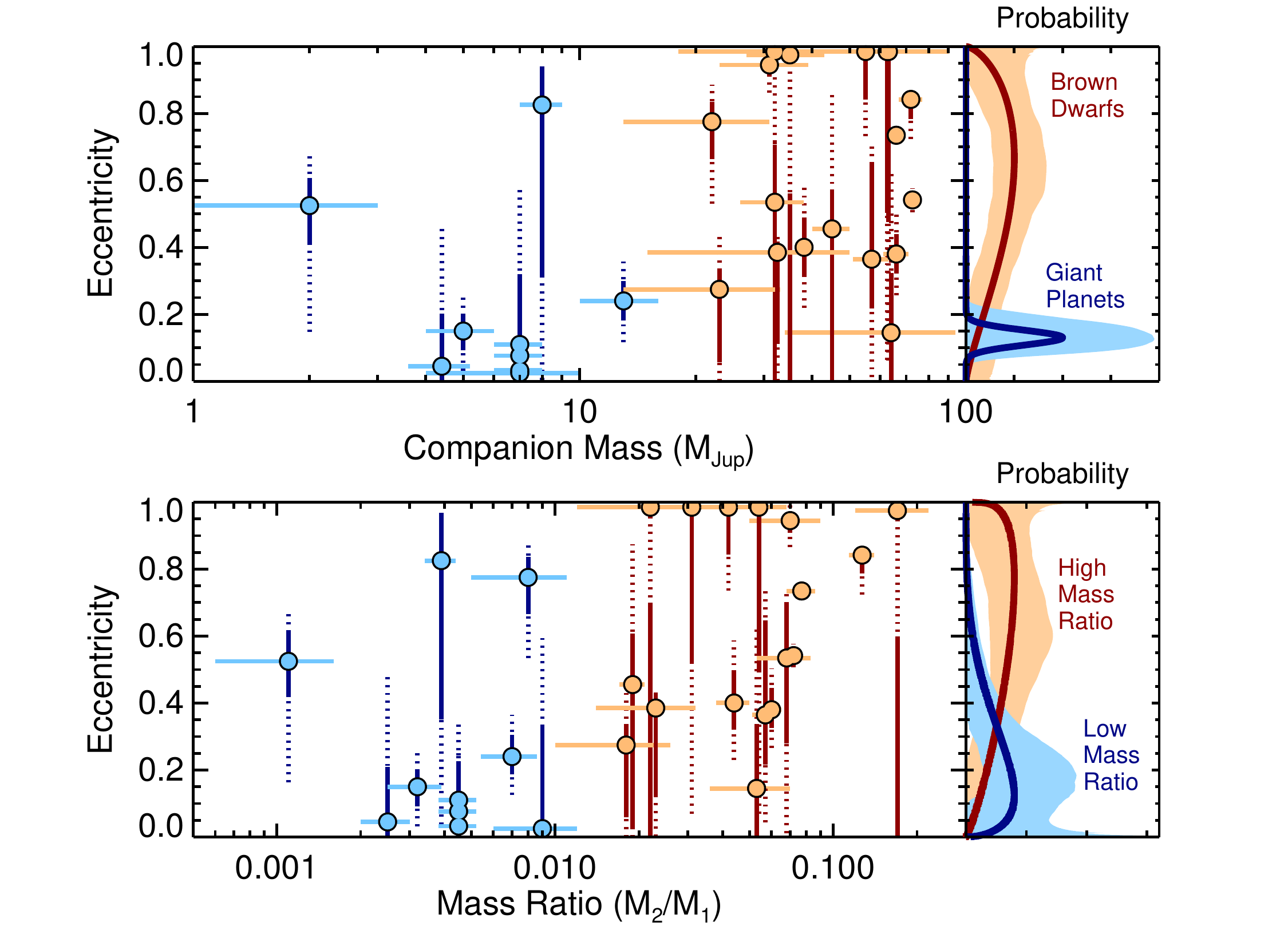}}
  \vskip 0 in
  \caption{Individual eccentricity distributions for substellar companions between 5--100 AU 
  as a function of mass (top) and mass ratio (bottom).
  The inferred population-level eccentricity distributions for subsamples of directly imaged planets and 
  brown dwarf companions are displayed in the panels on the right.  Dividing these by a mass threshold of 15~\Mjup \ 
  in the top panel or a mass ratio of 0.01 in the bottom panel does not influence the main conclusion that
  the eccentricity distribution of directly imaged planets is skewed to low values compared to brown dwarfs.
  Here blue and orange colors indicate the subsample divisions we have adopted.  
  The solid and dotted eccentricity uncertainties 
  represent 68\% and 95\% credible intervals, respectively.  Shaded regions in the right panels
  illustrate 2$\sigma$ credible intervals for the eccentricity posteriors. \label{fig:ecc_mass} } 
\end{figure*}

\begin{deluxetable*}{lcccccc}
\renewcommand\arraystretch{0.9}
\tabletypesize{\footnotesize}
\setlength{ \tabcolsep } {.1cm} 
\tablewidth{0pt}
\tablecolumns{7}
\tablecaption{Population-Level Eccentricity Distributions\label{tab:popdist}}
\tablehead{
       \colhead{}         & \colhead{Separation}  &  \colhead{Mass}     & \colhead{Other}       & \colhead{Sample} & \colhead{$\alpha$}          & \colhead{$\beta$} \\ 
       \colhead{Sample}   & \colhead{Range}       &  \colhead{Range}    & \colhead{Constraints} & \colhead{Size}   & \colhead{(1$\sigma$ C.I.)}  & \colhead{(1$\sigma$ C.I.)}    
        }   
\startdata
               Full Sample    &   5--100 AU    &                  2 $M_\mathrm{Jup}$ $\le$ $M_2$ $<$ 75 $M_\mathrm{Jup}$    &              $\cdots$    &          27     &        0.950 (0.520--1.36)     &         1.30 (0.840--1.91)    \\
             Giant Planets    &   5--100 AU    &                  2 $M_\mathrm{Jup}$ $\le$ $M_2$ $<$ 15 $M_\mathrm{Jup}$    &              $\cdots$    &           9     &           30.0 (58.2--156)     &            200 (542--1000)    \\
              Brown Dwarfs    &   5--100 AU    &                 15 $M_\mathrm{Jup}$ $\le$ $M_2$ $<$ 75 $M_\mathrm{Jup}$    &              $\cdots$    &          18     &          2.30 (1.02--3.68)     &         1.65 (0.860--2.55)    \\
           High Mass Ratio    &   5--100 AU    &                                            0.01 $<$ $M_2$/$M_1$ $<$ 0.2    &              $\cdots$    &          17     &         1.85 (0.930--3.34)     &         1.25 (0.800--2.44)    \\
            Low Mass Ratio    &   5--100 AU    &                                          0.001 $<$ $M_2$/$M_1$ $<$ 0.01    &              $\cdots$    &          10     &         1.50 (0.500--3.68)     &          4.50 (1.52--12.8)    \\
         Close Separations    &    5--30 AU    &                  2 $M_\mathrm{Jup}$ $\le$ $M_2$ $<$ 75 $M_\mathrm{Jup}$    &              $\cdots$    &          14     &         1.70 (0.830--2.87)     &          2.75 (1.52--4.45)    \\
          Wide Separations    &  30--100 AU    &                  2 $M_\mathrm{Jup}$ $\le$ $M_2$ $<$ 75 $M_\mathrm{Jup}$    &              $\cdots$    &          13     &        0.650 (0.260--1.28)     &        0.850 (0.510--1.56)    \\
 Planets Excluding HR 8799    &   5--100 AU    &                  2 $M_\mathrm{Jup}$ $\le$ $M_2$ $<$ 15 $M_\mathrm{Jup}$    &       No HR 8799 bcde    &           5     &           90.0 (86.9--311)     &            300 (555--1000)    \\
              Only HR 8799    &   5--100 AU    &                  2 $M_\mathrm{Jup}$ $\le$ $M_2$ $<$ 15 $M_\mathrm{Jup}$    &      Only HR 8799bcde    &           4     &            120 (43.8--127)     &            960 (570--1000)    \\
 Short-Period Brown Dwarfs    &    5--30 AU    &                 15 $M_\mathrm{Jup}$ $\le$ $M_2$ $<$ 75 $M_\mathrm{Jup}$    &              $\cdots$    &           9     &          3.25 (1.18--8.05)     &          3.25 (1.37--7.22)    \\
  Long-Period Brown Dwarfs    &  30--100 AU    &                 15 $M_\mathrm{Jup}$ $\le$ $M_2$ $<$ 75 $M_\mathrm{Jup}$    &              $\cdots$    &           9     &          3.12 (1.11--7.69)     &         1.75 (0.680--3.39)    \\
        Young Brown Dwarfs    &   5--100 AU    &                 15 $M_\mathrm{Jup}$ $\le$ $M_2$ $<$ 75 $M_\mathrm{Jup}$    &              $<$1 Gyr    &          11     &         1.50 (0.210--5.61)     &         1.05 (0.410--2.29)    \\
          Old Brown Dwarfs    &   5--100 AU    &                 15 $M_\mathrm{Jup}$ $\le$ $M_2$ $<$ 75 $M_\mathrm{Jup}$    &              $>$1 Gyr    &           7     &          4.70 (2.37--9.60)     &          4.35 (2.33--7.87)    \\
       Nearby Brown Dwarfs    &   5--100 AU    &                 15 $M_\mathrm{Jup}$ $\le$ $M_2$ $<$ 75 $M_\mathrm{Jup}$    &              $<$40 pc    &           8     &          4.50 (1.69--9.33)     &         2.25 (0.910--4.22)    \\
      Distant Brown Dwarfs    &   5--100 AU    &                 15 $M_\mathrm{Jup}$ $\le$ $M_2$ $<$ 75 $M_\mathrm{Jup}$    &              $>$40 pc    &          10     &         1.20 (0.500--7.06)     &         1.20 (0.680--6.91)    \\
                 Small IWA    &   5--100 AU    &                  2 $M_\mathrm{Jup}$ $\le$ $M_2$ $<$ 75 $M_\mathrm{Jup}$    &         IWA/$a$$<$0.5    &          12     &         1.00 (0.650--2.37)     &         1.20 (0.880--2.97)    \\
                 Large IWA    &   5--100 AU    &                  2 $M_\mathrm{Jup}$ $\le$ $M_2$ $<$ 75 $M_\mathrm{Jup}$    &         IWA/$a$$>$0.5    &          15     &        0.600 (0.430--1.46)     &        0.800 (0.770--2.25)    \\
\enddata
\end{deluxetable*}

We further dissect our full sample of 27 substellar companions by mass
to assess whether there is evidence for a difference between 
the population-level eccentricities of giant planets and brown dwarf companions.  
We adopt a mass dividing line of 15~\Mjup \ in this study,
which is motivated by the approximate transition at smaller separations 
between the planet mass function and that of stellar/substellar companions from RV samples
(e.g., \citealt{Udry:2007dq}; \citealt{Schneider:2011gf}).
These subsamples consist of 9 
directly imaged long-period giant planets (2--15~\Mjup) and 18 brown dwarf companions (15--75~\Mjup).

Fits to the giant planet subsample are presented in Figure~\ref{fig:beta_planet} and Table~\ref{tab:popdist}.  
The best-fitting values are $\alpha$=30 and $\beta$=200, with 1$\sigma$ credible
intervals spanning 58--156 and 540--1000, respectively. 
Note that the peak of this joint distribution does not fall within the marginalized 1$\sigma$
intervals because of the strong positive covariance between these parameters.
Nevertheless, despite this broad range of values, each \{$\alpha$, $\beta$\} pair
tends to balance each other  to produce a  fairly narrow range for the  
resulting eccentricity distribution function between $e$=0.05--0.25, and with a peak value at $\bar{e}$=0.13.

However, as discussed in Section \ref{sec:beta_example}, 
we caution that interpretations of the resulting eccentricity posterior for small samples should be made with care.
Here our sample size comprises nine systems, but several of the individual eccentricities are so broad that they
provide little meaningful constraint at the population level.
Moreover, the additional discovery of a single moderate- to high-eccentricity planet 
has the potential to substantially inflate this distribution.
So while the resulting posterior is expected to capture the overall qualitative trend of the underlying
eccentricities, our experiments in Section \ref{sec:beta_example} indicate that 
small sample sizes can influence the detailed behavior of the reconstructed distribution, especially 
near the endpoints at $e$=0 or $e$=1.  
It is therefore unclear whether the lack of power at the lowest eccentricities reflects something genuine about
this population or is perhaps just a reflection of small number statistics.
New discoveries and continued orbit monitoring of these systems are needed to address this question in the future.

Results for the sample of brown dwarf companions are summarized in Figure~\ref{fig:beta_bd}.
The best-fitting hyperparameters are $\alpha$=2.3$^{+1.4}_{-1.3}$ and $\beta$=1.7$^{+0.9}_{-0.8}$,
with significant positive covariance between the two parameters.
The eccentricity distribution for brown dwarfs peaks between $e$$\approx$0.6--0.9 with a broad range between
$e$$\approx$0.3--1.0.  If a single Beta distribution adequately describes this population, 
it appears that near-circular orbits are rare, and there is an indication that the highest
values above 0.9 are disfavored.

The most striking aspect of this underlying distribution is its dissimilarity with results for the giant planets.
As a population, widely-separated brown dwarf companions are significantly more eccentric and span
a wider range of eccentricities than their lower-mass counterparts. 
This tendency is evident in Figure~\ref{fig:ecc_mass}, which shows 
eccentricities for individual systems as a function of companion mass.
There is a noticeable dearth of planets at high eccentricities; most of 
the cumulative power is focused at low values.  On the other hand, brown dwarfs
span a wide range of eccentricities with fewer companions 
on near-circular orbits.
We revisit the implications of these distinct distributions in Section~\ref{sec:discussion}.

To quantify the difference between these two inferred eccentricity distributions,
we calculate the probability that random variables $p$ drawn from the brown dwarf distribution (BD)
are greater than the giant planet (GP) distribution (e.g., \citealt{Cook:2003vv}; \citealt{Raineri:2014cy}):
\begin{equation}\label{eqn:prob}
P(p_\mathrm{BD}>p_\mathrm{GP}) = \int_0^1 f_\mathrm{BD}(e\mid \alpha_\mathrm{BD}, \beta_\mathrm{BD}) F_\mathrm{GP}(e\mid \alpha_\mathrm{GP}, \beta_\mathrm{GP}) de,
\end{equation}
\noindent where 
$f_\mathrm{BD}(e\mid \alpha_\mathrm{BD}, \beta_\mathrm{BD})$
is the probability distribution function for brown dwarfs and 
$F_\mathrm{GP}(e\mid \alpha_\mathrm{GP}, \beta_\mathrm{GP})$
is the cumulative distribution function (CDF) for giant planets.
The parameterized model we adopt for the underlying population-level eccentricities is the Beta distribution.
The CDF of the Beta distribution is the regularized incomplete beta function:
\begin{equation}
F_\mathrm{GP}(e\mid \alpha_\mathrm{GP}, \beta_\mathrm{GP}) =  
\frac{\int_0^e t^{\alpha_\mathrm{GP} - 1} (1-t)^{\beta_\mathrm{GP} - 1} dt}
{\int_0^1 t^{\alpha_\mathrm{GP} - 1} (1-t)^{\beta_\mathrm{GP} - 1} dt}.
\end{equation}

Pairs of $\alpha$ and $\beta$ are randomly drawn from the joint posteriors for both the giant planet and brown dwarf 
MCMC results and Equation \ref{eqn:prob} is numerically integrated for each trial.  This procedure is repeated 10$^5$ times
to produce a distribution of probabilities.  We find that $P(p_\mathrm{BD}>p_\mathrm{GP})$ = 0.979
with a 2$\sigma$ credible interval of 0.85--1.0.
Two equivalent distributions will produce probabilities of 50\%, so this value of $\approx$98\% points to a significant difference
between these two populations.


\begin{figure*}
  \vskip -0.1 in
  \hskip .7 in
  \resizebox{6.in}{!}{\includegraphics{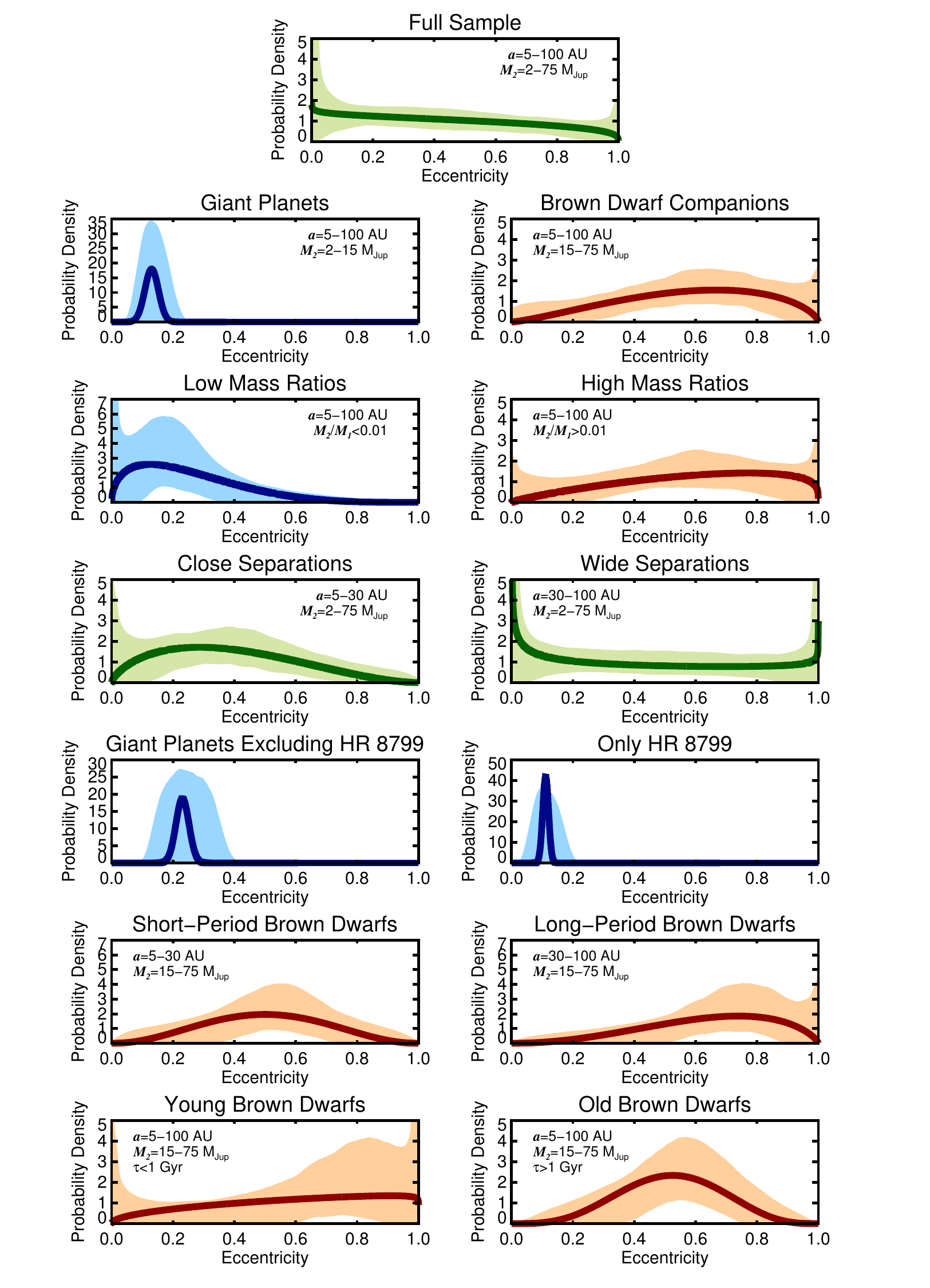}}
  \vskip -0.15 in
  \caption{Population-level eccentricity distributions for the full sample of substellar companions between 5--100 AU (top panel)
  and various subsamples divided by companion mass, system mass ratio, and orbital separation (second, third,
  and fourth rows). 
  Results for the planet population excluding HR 8799, and results only for HR 8799, are displayed in the fifth row.
  The sixth row shows brown dwarfs at small (5--30~AU) and large (30--100 AU) orbital separations, and the
  bottom row displays the recovered eccentricity distributions for young ($<$1 Gyr) and old ($>$1 Gyr) brown dwarfs.   
  The thick curve shows the best-fitting Beta distribution for each sample.  Shaded regions illustrate the 
  2$\sigma$ credible intervals for posteriors at each eccentricity.
  \label{fig:beta_summary} } 
\end{figure*}

\subsubsection{A Mass Ratio Threshold}

A single companion mass may not be a reliable threshold for distinguishing planets
from brown dwarfs.  Protoplanetary disk masses correlate with host star masses (\citealt{Andrews:2013ku}),
so the maximum mass of a planet can be expected to also scale with stellar mass.
Similarly, low-mass companions in high-mass ratio systems like 2M1207--3932 b indicate that 
companion mass is not likely to be the best criterion to distinguish giant planets from brown dwarfs.
Using companion mass ratio rather than companion mass may be a more appropriate 
way to divide the full sample.

We carried out the same procedure as described above except now using a mass ratio of 0.01 to divide
our sample into high-mass ratio and low mass ratio bins. This value corresponds to a 10~\Mjup \ companion
orbiting a Sun-like star, or a 1~\Mjup \ companion to a 0.1~\Msun \ host star.  
The difference between this experiment and the prior analysis with a mass cutoff is that 
a single high-$e$ system---$\kappa$ And B---has joined the previous giant planet sample.
Results are summarized in Figure~\ref{fig:ecc_mass} and Table~\ref{tab:popdist}. 
We measure hyperparameter values of
$\alpha$=1.5$^{+2.2}_{-1.0}$ and $\beta$=4.5$^{+8.3}_{-3.0}$ for the low mass ratio subsample and
$\alpha$=1.9$^{+1.5}_{-0.9}$ and $\beta$=1.3$^{+1.2}_{-0.5}$ for the high mass ratio systems.
The high mass ratio distribution is statistically indistinguishable from the brown dwarf subsample
in Section~\ref{sec:massdiv}.  The low mass ratio results are qualitatively similar to the giant planet 
subsample in that both generally point to low eccentricities.
However, the low mass ratio distribution
exhibits a broader peak and a 
longer tail to modest eccentricities ($e$$\approx$0.6), indicating that a substantial
portion of this population is located on distinctly non-circular orbits.
Using the same approach as in Section~\ref{sec:massdiv}, we find that
the high-mass ratio distribution results in higher eccentricities 87.0\% of the time (2$\sigma$ credible
interval: 67.8--100\%).  This difference is substantial, suggesting dynamically distinct populations, but 
it is not as strong as  the giant planet
and brown dwarf subsamples.

This underlying distribution for low mass ratio systems bears a close resemblance to the 
warm Jupiter eccentricity distribution---both qualitatively, with most of the posterior power
located at low eccentricities, and quantitatively with comparable values of $\alpha$ and
$\beta$ (which for warm Jupiters is 0.867 and 3.03, respectively; \citealt{Kipping:2013he}).
Altogether this indicates that the choice of mass or mass ratio to divide the parent sample does not significantly impact the
interpretation of the reconstructed population-level eccentricity distribution:
at wide separations, the tail end of the companion mass function appears to have dynamically distinctive properties.

\subsubsection{A Separation Threshold}

We also examine whether there are differences in the eccentricity distribution 
as a function of separation, which might be expected if, for example, the inner population of substellar
companions predominantly originates within a disk while the outer population largely represents 
the product of cloud fragmentation.
For this experiment we adopt a threshold of 30~AU, which is chosen so as to divide the full
sample of substellar companions into two approximately equal bins.

Results for the subsample of 14 companions between 5--30 AU and 13 companions between 30--100 AU are 
shown in Figure~\ref{fig:beta_summary} and summarized in Table~\ref{tab:popdist}.  
We find values of $\alpha$=1.70$^{+1.2}_{-0.9}$ and $\beta$=2.8$^{+1.7}_{-1.2}$  for the sample at 
close separations, which corresponds to a broad peak between $e$=0.1--0.4 with significant power from $e$=0.0--0.8.
At wide separations we find $\alpha$=0.7$^{+0.6}_{-0.4}$ and $\beta$=0.9$^{+0.7}_{-0.3}$.  
This corresponds to a roughly flat distribution with somewhat higher power at the bounded endpoints.
There is some evidence that the more closely separated population lacks companions at the highest eccentricities,
but these two distributions are otherwise broadly similar and are not nearly as distinct as the giant planet and brown dwarf
subsamples.  
The probability that wide companions have higher eccentricities than close companions
is 52.3\% (2$\sigma$ credible interval: 24.4--81.7\%).

\subsubsection{Exploring the Influence of HR 8799}{\label{sec:beta_hr8799}}

As a system of four giant planets with masses $\gtrsim$5~\Mjup \ and
separations between 15--70 AU, HR 8799 is atypical among directly imaged planetary systems.\footnote{HR 8799 
has long been the only multi-planet system to be imaged, but it may now
be joined with PDS 70 (\citealt{Haffert:2019ba}), 
$\beta$~Pic (\citealt{Lagrange:2019bi}), and LkCa15 (\citealt{Kraus:2012gk}; \citealt{Sallum:2015ej}).
Note that an additional unknown close-in substellar companion has the potential to bias the astrometry and inferred
eccentricities of wider imaged companions (\citealt{Pearce:2014ht}).  Astrometry is calculated with respect to the
primary, so a relatively massive inner object can perturb the host star and alter the apparent orbital elements of long-period companions.}
Based on the $\approx$1\% occurrence rate of planets at these separations (\citealt{Bowler:2016jk}; \citealt{Galicher:2016hg}),
the probability of randomly finding four such planets around one star 
is $\sim$0.01$^4$, or 10$^{-8}$, assuming each planet represents an independent probabilistic event.
Only about 10$^3$ stars have been observed in high-contrast imaging surveys to date,
 which makes it exceedingly unlikely that these planets are independent of each other.  HR 8799 is therefore a 
 special case in which the probability of an additional planet in this system is conditioned 
 on the presence of another one being there.
Other planets may of course reside in the apparently single systems at closer separations and 
lower masses, but HR 8799 appears to be unique in that it has four massive planets detected on wide orbits.
 The physical underpinning of this is, of course, likely to have been an unusually massive and physically large
 protoplanetary disk. 

The fact that the HR 8799 planets are in an apparently stable, near-circular, approximately coplanar 
orbital arrangement could bias the results of our eccentricity analysis for the giant planet subsample.
Larger eccentricities for any of these planets have the potential to destabilize the system, 
so we may be imposing an unintentional anthropic bias by including this system in our analysis.
For example, if the eccentricities had been high, the planets may not have formed, migrated,
or persisted at their present locations and therefore may not have been discovered.
It is not clear if the large masses and separations of the HR 8799 planets are what make 
this system unusual, or perhaps the close dynamical packing of the planets
relative to other systems.
Nevertheless, because of this unusual status we carry out two additional tests: one for the giant planet subsample excluding 
HR 8799 bcde, and one \emph{only} considering the HR 8799 planets to assess constraints on the
underlying eccentricity distribution for that planetary system alone.

Results of these experiments are shown in Figure~\ref{fig:beta_summary} and 
hyperparameter values are listed in Table~\ref{tab:popdist}. 
When we exclude HR 8799 and only consider five planets, we find hyperparameters of $\alpha$=90
(1$\sigma$ credible interval: 87--311) and $\beta$=300 (1$\sigma$ credible interval: 555--1000).
Compared to the full sample of giant planets, the corresponding underlying eccentricity distribution 
broadens and shifts to slightly higher values between $e$=0.1--0.4 with a peak at $\bar{e}$=0.23.
The hyperparameters for HR 8799 bcde alone are $\alpha$=120
(1$\sigma$ credible interval: 44--127) and $\beta$=960 (1$\sigma$ credible interval: 570--1000).
This yields a tighter distribution between $e$=0.0--0.2 with a peak at $\bar{e}$=0.11.
The full sample of giant planets in Section~\ref{sec:massdiv} reflects intermediate eccentricities between 
HR 8799 and the rest of the population.
Based on these reconstructions, the probability that the HR 8799 planets have lower eccentricities than the other systems
is 99.9\% (2$\sigma$ credible interval: 78.5--100\%).

The HR 8799 planets are therefore on somewhat more circular orbits than the other apparently single 
imaged planets, which is similar to the eccentricity dichotomy between multi-planet and single systems
at small separations (\citealt{Xie:2016dp}).  However, as with ``single'' planets at small separations, we note that
single directly imaged systems may harbor additional planets below the detection threshold.

\subsubsection{Short- and Long-Period Brown Dwarfs}

\citet{Tokovinin:2015dk} presented the first observational results demonstrating that the eccentricities 
of wide stellar binaries well outside of the eccentricity-period tidal circularization envelope 
increase with larger orbital period, from a mean value of $e$$\approx$0.4
for 10$^{2-3}$ d to $e$$\approx$0.6 for 10$^{5-6}$ d.  This may be a reflection of 
the dissipative interaction of closer binaries with circumstellar disk or envelope material 
during the formation of the pair (e.g., \citealt{Bate:2009br}).  
If brown dwarf companions have a shared origin with stellar companions then
they may have a similar period dependence on the mean eccentricity.

To test this we divide our brown dwarf sample into a subsample of 9 short-period 
companions between 5--30 AU and 9 long-period companions spanning 30--100 AU.
Results are presented in Figure~\ref{fig:beta_summary} and Table~\ref{tab:popdist}.
We find values of \{$\alpha$=3.25$^{+4.8}_{-2.1}$, $\beta$=3.25$^{+4.0}_{-1.9}$\} for the
short-period subsample, which corresponds to an approximately normally distributed 
density function with a peak eccentricity at $\bar{e}$=0.50 and a 
broad width from $e$$\approx$0.1--0.9.
The long-period subsample yields \{$\alpha$=3.1$^{+4.6}_{-2.0}$, $\beta$=1.75$^{+1.6}_{-1.1}$\},
which gives a strongly left-skewed distribution with a peak at $\bar{e}$=0.74 and a broad range
from $e$$\approx$0.2--1.0.  
The probability that long-period brown dwarfs have higher eccentricities than their short-period counterparts
is 70.0\% (2$\sigma$ credible interval: 43.5--98.4\%).
This increasing mean eccentricity with orbital period 
appears to follow the same trend observed with stellar binaries and supports a common
formation channel\footnote{Note that brown dwarfs at wider separations have longer orbital periods 
and are thus more susceptible to the orbit fits being influenced by systematic errors in the astrometry.
However, they are also easier to discover earlier than short-period brown dwarfs so more time has generally elapsed to 
monitor their orbits, which partly makes up for this potential bias (albeit only slightly).  
Altogether we expect significantly better constraints and more reliable fits for the shorter-period companions.  
Ultimately larger samples and continued 
astrometric monitoring of known systems are needed to confirm this trend of increasing eccentricity with separation.}

\subsubsection{Young and Old Brown Dwarfs}
The architectures of planetary systems can evolve through three-body Kozai-Lidov oscillations or 
dynamical scattering events, both of which can influence the observed eccentricities of giant planets.
Indeed, several of the systems in our sample have wide stellar companions (HD 1160, 51 Eri, and HD 4113)
which could perturb the eccentricities of inner objects over long timescales.
Our planet sample comprises
too few objects to explore age effects for that population, but our brown dwarf sample includes systems
spanning a wide age range from $\approx$23 Myr (for PZ Tel) to 6--10 Gyr (for Gl 758).

Here we explore whether there is evidence for a difference in the eccentricity distributions of 
young and old brown dwarfs.  To produce comparably sized bins we adopt an age of 1 Gyr as a threshold for 
each subsample.  
Altogether there are 11 young brown dwarf companions ($<$1 Gyr) and 7 old systems ($>$1 Gyr).
Results from our hierarchical Bayesian modeling are shown in Figure~\ref{fig:beta_summary} and Table~\ref{tab:popdist}.
For our young subsample we find hyperparameter values of $\alpha$=1.50$^{+4.11}_{-1.29}$ and $\beta$=1.05$^{+1.24}_{-0.64}$,
which correspond to a broad eccentricity distribution with a general preference for high values.  The older sample
yields $\alpha$=4.70$^{+4.90}_{-2.33}$ and $\beta$=4.35$^{+3.52}_{-2.02}$, which is approximately 
Gaussian-shaped centered at $\bar{e}$=0.52 with the highest power between $e$$\approx$0.2--0.8.
The uncertainties are relatively large on these inferred shapes, which is reflected in the 
probability that young brown dwarfs have higher eccentricities than old brown dwarfs: 59.7\% with a
2$\sigma$ credible interval of 35.2--99.7\%.  
We therefore do not find significant evidence for distinct distributions among brown dwarf companions when subdivided by system age.

\subsection{Discovery Bias}{\label{sec:discoverybias}}

A direct imaging survey 
will preferentially find close-in companions with higher eccentricities compared to companions that have the same semi-major
axes but are on circular orbits (e.g., \citealt{Dupuy:2011ip}; \citealt{Kane:2013coa}).  
This is because more eccentric orbits will reach wider apastron distances and therefore companions 
will spend more time at large separations compared to those with lower eccentricities.  
This preference produces a bias that can skew the apparent eccentricity distribution of discoveries towards high eccentricities.  
Discovery bias is strongest when the 
semimajor axis is much smaller than the IWA (at the same contrast as the companion) 
and it asymptotically disappears for companions with semi-major axes 
well beyond the IWA.
This metric---IWA/$a$---is therefore a useful tool to assess whether a given sample is skewed to
higher eccentricities from discovery bias.  

Our sample of substellar companions draws from an assortment of adaptive optics imaging 
surveys carried out over the past two decades 
which had a wide range of sensitivities, 
inner working angles, and resulting contrast curves.  
Moreover, targets in our sample span a broad range of distances from 5.8 pc (Gl 229) to 126 pc (HD 1160);
for a given contrast curve, more distant systems may be more heavily biased in favor of higher eccentricities
for the same reasons shorter-period companions
are preferentially selected against.

If our sample is strongly influenced by discovery bias then we would expect the eccentricities of our targets to 
exhibit several trends characteristic of this preferential selection.  Below we discuss five such correlations
in detail to assess the potential impact of this bias on our results.
\begin{enumerate}
\item The IWA at the time of discovery 
should be comparable to or larger than the 
semi-major axes of companions in our sample.  Large values of IWA/$a$ would suggest 
that a strong bias is likely present, while small
values below unity would indicate there is minimal bias.

The IWA here should be at the same contrast as the companion and must be determined at the time the 
companion was first imaged.
We therefore revisited the original discovery papers for each system in our sample and 
visually estimated the IWA from the discovery images.  These are then converted to physical units using the 
distance to the system and divided by the maximum \emph{a posteriori} of the semimajor axis posterior
from Table~\ref{tab:orbitize} (or values from the literature for systems that we did not refit in this study).

The distribution of IWA/$a$ for our sample is shown in Figure~\ref{fig:iwa_dist} and individual values are listed in Table~\ref{tab:sample}.  
13 systems from our full sample
have IWA/$a$ $<$ 0.5, 12 systems have  0.5 $\le$ IWA/$a$ $<$ 1.0, 
and two systems (HD 1160 B and HD 49197 B) have IWA/$a$ $\ge$ 1.0.
\citet{Dupuy:2011ip} simulated the impact of discovery bias as a function of both IWA/$a$ and eccentricity
(see their Figure~1) and found that there is minimal impact on the relative fraction of detected systems 
across all eccentricities for values of IWA/$a$ $\lesssim$ 0.5.  For IWA/$a$ values between $\approx$0.5--1,
there is a modest preferential suppression of low-eccentricity orbits.  For values above IWA/$a$ $\approx$1.0,
this suppression becomes severe and only the highest eccentricities are detectable.
We expect that about half of our sample suffers from minimal discovery bias and about half is moderately 
influenced by this effect. 


\begin{figure}
  \vskip -0. in
  \hskip -0.5 in
  \resizebox{4.5in}{!}{\includegraphics{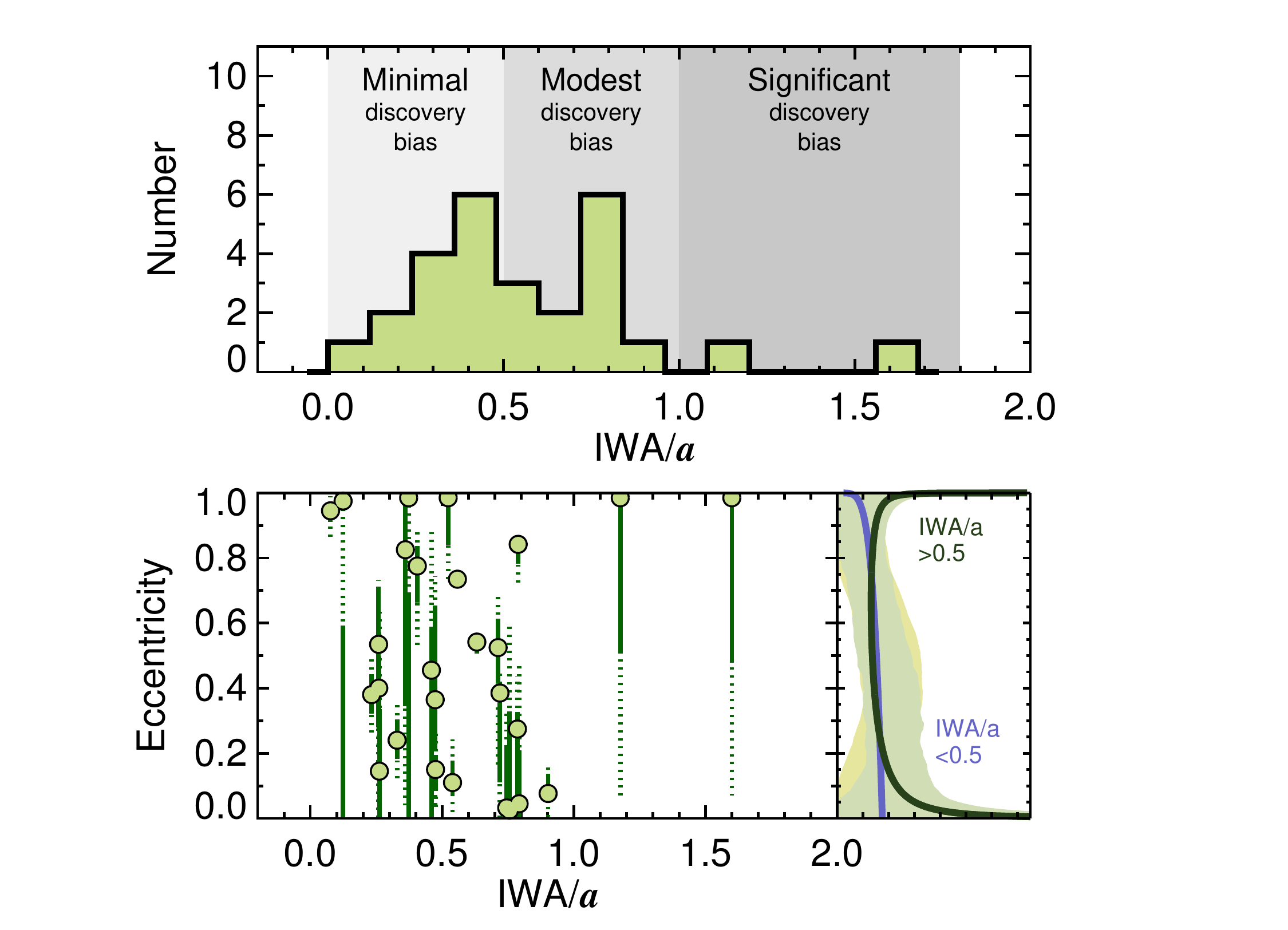}}
  \vskip 0 in
  \caption{\emph{Top:} Distribution of IWA/$a$ values for brown dwarf companions and giant planets  
  in our sample at the time of their discovery.  Discovery bias is expected to preferentially
  suppress low eccentricities for large values of IWA/$a$ ($\gtrsim$1).  Values of IWA/$a$
  between 0.5 and 1.0 are modestly affected by discovery bias, and values below 0.5 are minimally
  influenced (\citealt{Dupuy:2011ip}).   \emph{Bottom:} Eccentricity as a function of IWA/$a$ for targets
  in our sample.  Discovery bias is expected to imprint higher eccentricities for higher values of IWA/$a$.
  There is some evidence for this for the two systems with IWA/$a$ values above 1.0, whose
  eccentricities are poorly constrained but favor high values, but the inferred eccentricity distribution
  for IWA/$a$$<$0.5 and IWA/$a$$>$0.5 does not reveal any noticeable trend (bottom right).
  This suggests that the observed population-level eccentricity
  distribution is not substantially shaped by discovery bias.
  \label{fig:iwa_dist} } 
\end{figure}


\begin{figure*}
  \vskip -0. in
  \hskip 0. in
  \resizebox{7in}{!}{\includegraphics{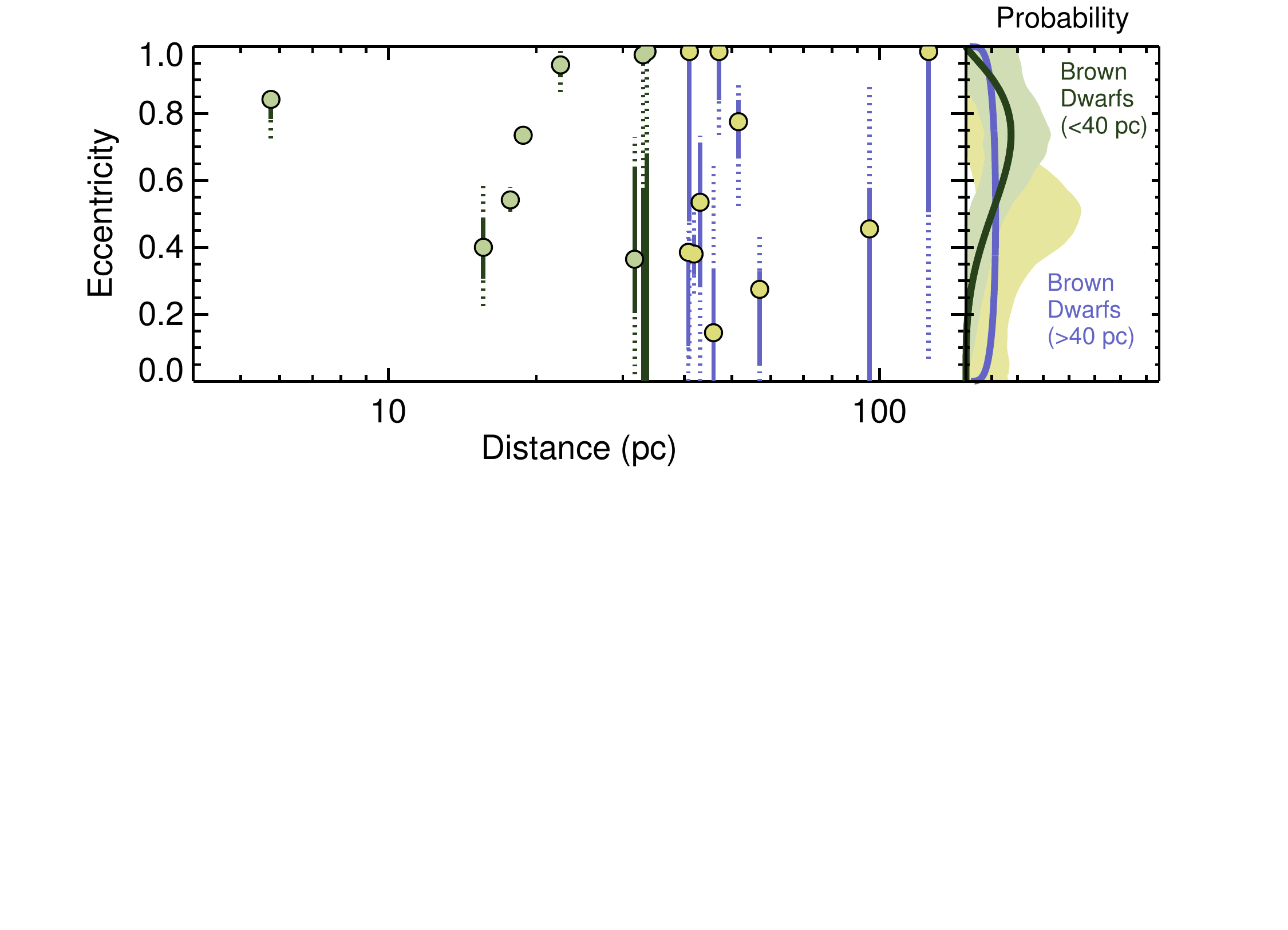}}
  \vskip -2.5 in
  \caption{Eccentricities of brown dwarf companions as a function of distance.  For a given contrast curve, discovery bias is 
  expected to more strongly select for higher eccentricities at larger distances.  We find the opposite trend
  for the recovered population-level eccentricity distributions for the nearby ($<$40 pc) and distant ($>$40 pc)
  subsamples (right panel): the nearby sample is skewed to higher eccentricities, while the best-fitting
  Beta distribution is flat for the more distant systems. Shaded regions represent 2$\sigma$ credible intervals
  for each eccentricity.  Note that the peak in the uncertainty profile for distant brown dwarfs at  
  $e$$\approx$0.5 results from a pile up of tighter posteriors in this region, even though the best-fit is approximately
  uniform.  \label{fig:near_far_bds} } 
\end{figure*}

\item There should be a correlation between IWA/$a$ and eccentricity.  Objects with semi-major axes well outside the IWA
should be minimally biased, whereas those at or inside the IWA should be preferentially more eccentric.

The lower panel of Figure~\ref{fig:iwa_dist} shows the eccentricity of each system as a function of IWA/$a$.
There is a broad range of eccentricities for IWA/$a$ values between 0.0--1.0.  The two systems with
the highest values of IWA/$a$ (HD 1160 B and HD 49197 B) have poorly constrained eccentricities, 
but these distributions favor higher values of $e$.  This suggests that these two companions are probably
influenced by this bias, but there is no obvious trend for the rest of the sample.  
This is reflected in the reconstructed eccentricity distributions for systems with IWA/$a$$<$0.5 and
IWA/$a$$>$0.5, which both have approximately flat shapes spanning $e$=0--1.

\item More distant systems should have higher eccentricities, on average, as 
IWA/$a$ becomes larger and this bias becomes stronger.
A positive correlation between the population-level eccentricity distribution and the distance to the system would be 
another indicator that discovery bias may be important.

To test whether our results for brown dwarf companions in Section~\ref{sec:eccentricities} may be skewed to high values
because of discovery bias, we reassess the population-level eccentricities for 
nearby ($<$40 pc) and distant ($>$40 pc) subsamples.
Results from this exercise are shown in Figure~\ref{fig:near_far_bds} and Table~\ref{tab:popdist}.
The recovered eccentricity distribution for nearby brown dwarfs peaks at high values ($\bar{e}$$\approx$0.7).
The best-fitting distribution for distant systems is flat across all eccentricities with an overdensity of posterior values
at modest eccentricities near $e$=0.5.
This apparent tendency for closer systems to have \emph{higher} eccentricities 
is opposite to the trend we would expect if discovery bias played a strong role in shaping the population-level 
eccentricity distribution.
With only eight objects in the nearby sample and nine in the distant sample, 
we expect that the slight differences we observe are caused by small number statistics.

\item Lower-mass companions with higher contrasts relative to their host stars should be more strongly biased 
towards high eccentricities compared to higher mass (lower contrast) companions.  This is because contrast curves
are not constant but typically curve to larger angular separations at higher contrasts.  That is, the effective IWA increases 
at lower companion masses.

In Section~\ref{sec:eccentricities} we showed that the inferred eccentricity distribution of 
giant planets in our sample is more circular than
the distribution for brown dwarf companions.
This is opposite of what we would expect if discovery bias played a significant role in shaping these observed
distributions.

\item Companions with shorter orbital periods should be more eccentric than their counterparts at wider separations.

In Section~\ref{sec:eccentricities} we derived the underlying eccentricity distributions for 
substellar companions at separations between 5--30 AU and 30--100 AU.
The close-separation subsample exhibited a broad peak at $\bar{e}$$\approx$0.3 with a paucity of power
at the highest eccentricities.  The wide-separation subsample was essentially flat across all eccentricities.
Once again, these trends are opposite of what we would expect if discovery bias played a dominant role
in shaping these distributions.

\end{enumerate}

Altogether these series of tests argue against discovery
bias playing a major role in shaping the population-level eccentricity distribution of 
substellar companions in our sample.  It is possible (and likely) that some bias is present in this sample, but 
the eccentricity distributions we inferred in Section \ref{sec:eccentricities} appear to predominantly 
reflect the intrinsic properties of substellar companions.
We conclude that our primary finding that brown dwarf companions are more eccentric than giant planets
is robust against discovery bias.

\section{Discussion}{\label{sec:discussion}}

One of the overarching motivations for large high-contrast imaging surveys is to determine the 
dominant pathway(s) through which giant planets and brown dwarfs form and subsequently evolve.
This is an especially challenging task at wide separations where occurrence rates are low and there are orders of
magnitude fewer discoveries compared to short-period planetary systems.
There have been many proposed mechanisms to form and 
preserve substellar companions
at separations of tens to hundreds of AU: pebble accretion (\citealt{Johansen:2017im}); 
disk fragmentation (e.g., \citealt{DodsonRobinson:2009bz}; \citealt{Kratter:2010gf}); 
outward scattering processes (e.g., \citealt{Boss:2006ge}; \citealt{Veras:2009br}; \citealt{Bromley:2014hg}); 
disrupted inward migration (e.g., \citealt{Nayakshin:2017aa}); 
direct cloud fragmentation (\citealt{Bate:2002iq}); 
outward scattering plus stellar flybys to generate ``Oort planets'' (\citealt{Bailey:2019js});
and dynamical recapture of free-floating planets (\citealt{Perets:2012cv}).

Several observational signatures of these formation pathways are expected to be imprinted
in the orbital properties and atmospheric compositions of planets and brown dwarfs at wide separations, including
their orbital architectures (e.g., \citealt{Boley:2009dk});
abundance ratios and metallicities (e.g., \citealt{Fortney:2008ez}; \citealt{Oberg:2011je}; \citealt{Spiegel:2012ea});
luminosities and entropy (e.g., \citealt{Marley:2007bf}; \citealt{Marleau:2013bh});
and companion mass function (\citealt{Reggiani:2016dn}).  
However, because the occurrence rate of giant planets and brown dwarf companions is so low---about 1\%
for the former and about 2--4\% for the latter (\citealt{Bowler:2018dq})---most individual surveys
typically find only a small handful of substellar companions  
(e.g., \citealt{Metchev:2009ky}; \citealt{Bowler:2015ja}; \citealt{Chauvin:2015jy}; 
\citealt{Galicher:2016hg}; \citealt{Stone:2018da}; \citealt{Nielsen:2019cb}).
Alone these are not sufficient for robust assessments of formation scenarios, 
but as an ensemble they provide clues about the physical processes, timescales, efficiency, and evolution
of planet formation at wide separations.

In this work we have aimed to test whether brown dwarfs and giant planets form in the same fashion
based on orbital expectations from two scenarios that most closely resemble the planet
and star formation processes: formation within a disk and fragmenting collapse of a molecular cloud core.
Our analysis of the population-level eccentricity distributions for substellar companions between 5--100 AU shows a clear 
difference between the orbital properties of planetary-mass companions  
and those in the brown dwarf mass regime. 
The low mass companions ($<$15~\Mjup) and low mass ratio systems ($M_2$/$M_1$$<$0.01) 
preferentially have lower eccentricities similar to the population of warm Jupiters at small separations.
The brown dwarf companions (15--75~\Mjup) and higher mass ratio systems ($M_2$/$M_1$=0.01--0.2)
exhibit higher eccentricities and there is evidence for a period dependence on the eccentricity distribution
analogous to results from \citet{Tokovinin:2015dk} for stellar binaries.
This is especially pronounced when compared with the roughly flat eccentricity distribution for brown dwarfs 
at small separations found by \citet{Ma:2014cc}.
Moreover, \emph{this difference in underlying eccentricity distributions based on mass and mass ratio 
are the clearest among any other subdivision we tested.}
The simplest explanation is that these populations predominantly form in distinct manners: 
the planetary-mass companions originate in disks, while brown dwarf companions represent the
low mass ratio end of binary star formation.
This complements the conclusions reached by \citet{Chabrier:2014up} that 
\emph{free-floating} brown dwarfs most likely form as the low-mass end of the star formation process based
on a range of kinematic, environmental, and statistical characteristics in common with stars.

There are more subtle details about the giant planet eccentricity distribution that may provide 
further insight into the formation and dynamical histories of these objects.
It is interesting to note that, like warm Jupiters, imaged planets and low-mass ratio binaries have an
extended, dynamically hot tail to modest eccentricities.
51 Eri b is a notable example: it is the lowest-mass directly imaged planet and has an unusually high
eccentricity ($e$=0.50$^{+0.11}_{-0.08}$) 
compared to the rest of the imaged planets in our sample with decent orbital constraints.
At small separations the shape of the warm Jupiter eccentricity distribution is 
generally interpreted as evidence for gravitational planet  
scattering and secular three-body interactions 
(e.g., \citealt{Juric:2008db}; \citealt{Ford:2008jo}; \citealt{Petrovich:2016ee}).
The same may be true for imaged planets: they may comprise a mix of low-eccentricity systems 
and companions that have been dynamically perturbed to modest eccentricities.  

Another feature of close-in planets is that they exhibit an eccentricity dichotomy
in which multi-planet systems have lower eccentricities, on average, compared to systems with
a single (known) planet (\citealt{Wright:2009jsa}; \citealt{Limbach:2015ho}; \citealt{Xie:2016dp}; \citealt{VanEylen:2019cy}).\footnote{Note
\citet{Bryan:2016cl} found that that this trend tends to reverse when considering long-term accelerations,
which are sensitive to planetary companions out to about 10 AU.  When this is taken into account, two-planet
systems tend to have \emph{higher} eccentricities than single-planet systems, perhaps a result of dynamical 
interactions.}
We find the first evidence of a similar phenomenon at wide separations:
the eccentricity distribution of the four HR 8799 planets peaks at $\bar{e}$=0.11,
whereas the distribution for single imaged planets shifts to $\bar{e}$=0.23 
and broadens when HR 8799 is excluded.
If confirmed with additional systems like PDS 70 bc and $\beta$~Pic bc in the future 
(\citealt{Haffert:2019ba}; \citealt{Lagrange:2019bi}),
this would suggest that long-period multi-planet systems probably have not
experienced strong scattering in their past,
whereas single imaged systems were probably excited at some point, perhaps by another planet.
Together this could indicate that most systems with Jovian planets have, or once harbored, multiple long-period planets.


\begin{figure}
  \vskip -0.2 in
  \hskip -0.4 in
  \resizebox{4.5in}{!}{\includegraphics{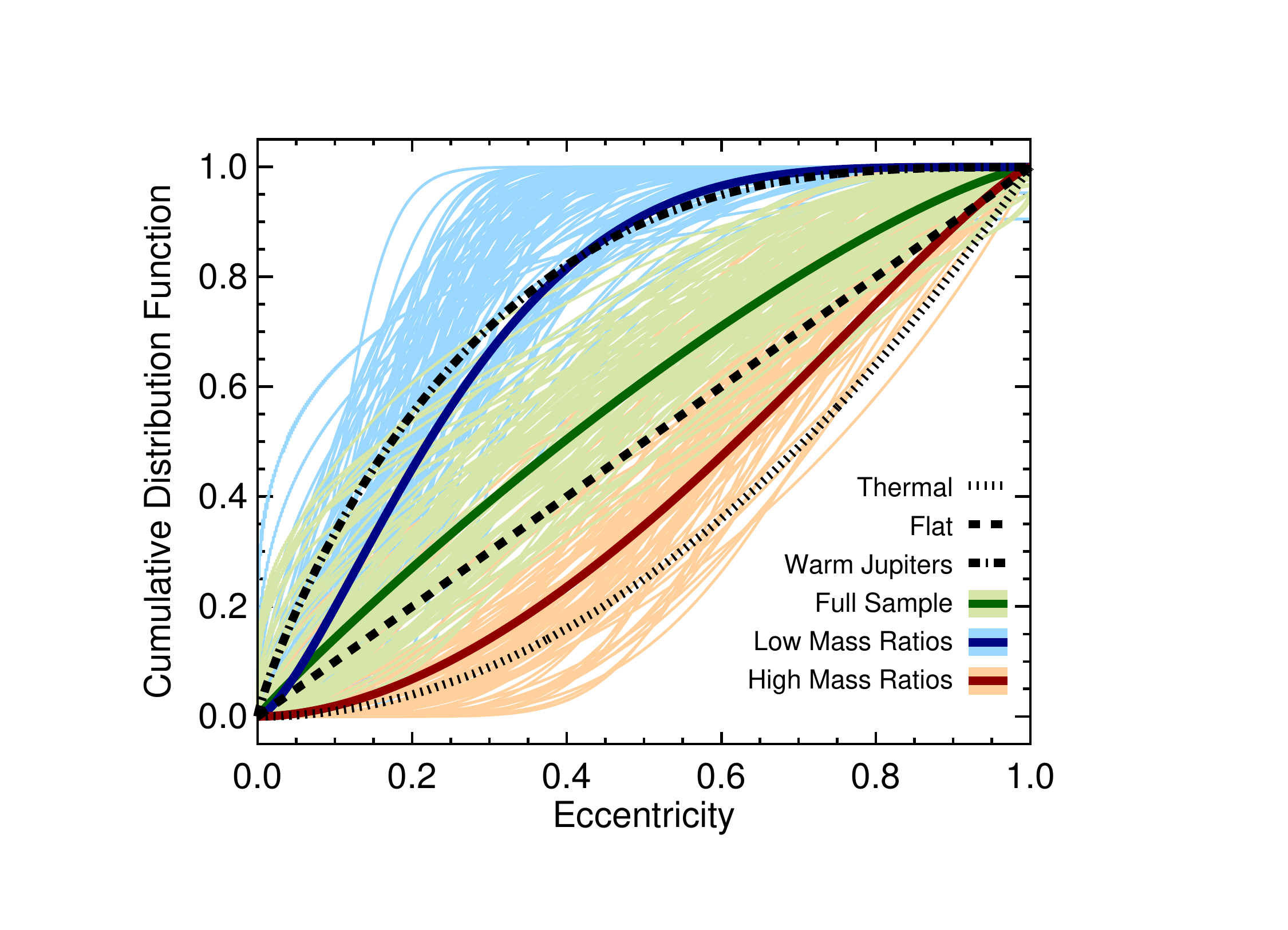}}
  \vskip -.3 in
  \caption{Cumulative distribution functions for our full sample of substellar companions (green),
  the subset of low mass ratio systems (blue), and the subset of high mass ratio systems (red).  
  The low mass ratio subsample resembles the warm Jupiter eccentricity distribution (dot-dash curve)
  from \citet{Kipping:2013he}, while the high mass ratio systems lie between a flat distribution 
  ($f$($e$) = const.; dashed curve) and thermal distribution ($f$($e$) = 2$e$; dotted curve). 
  Non-tidally circularized stellar binaries (not plotted) reside between the exoplanet and flat CDFs (\citealt{Duchene:2013il}).
  100 randomly drawn CDFs are displayed for each subsample.  \label{fig:cdf} } 
\end{figure}

In Figure~\ref{fig:cdf} we compare eccentricity cumulative distribution functions (CDFs) for our low mass ratio
subsample, high mass ratio subsample, and full sample of substellar companions to 
the eccentricity distribution of warm Jupiters from \citet{Kipping:2013he},
a flat distribution, and a thermal distribution (\citealt{ambartsumian:1937wd}).
The best-fitting low-mass ratio companion distribution between 5--100 AU is remarkably  
similar to that of close-in giant planets, which suggests that this functional form does not
appear to strongly vary from $\approx$0.1 AU out to 100 AU.  Note however that these populations
are quite different: the warm Jupiters have lower masses ($\approx$0.1--10~\Mjup \ with a bottom-heavy functional form) 
and older ages (typically several Gyr),
whereas the planets in our sample have high masses (2--15~\Mjup) and young ages ($\lesssim$50~Myr).
Nevertheless, the resemblance is noteworthy and may reflect something fundamental about 
the formation and early dynamical interactions of giant planets.

The eccentricity distribution of solar-type visual and spectroscopic binary stars outside of the tidal circularization radius 
(12 d $\lesssim$ $P$ $\lesssim$ 10$^6$ d; $a$$\approx$0.1--200 AU) is approximately flat, 
perhaps with a slight peak at modest eccentricities,
and falls between the CDFs for RV exoplanet and flat distributions (\citealt{Raghavan:2010gd}; \citealt{Duchene:2013il}).
\citet{Tokovinin:2015dk} showed that the eccentricities of solar-type stellar binaries progressively increase 
with increasing orbital period from 10$^{2}$ d to 10$^{6}$ d ($a$$\approx$0.4 AU to 200 AU)
toward the thermal distribution curve ($f(e)\sim2e$).
We found that the eccentricity distribution for all substellar companions is approximately uniform, which 
is consistent with (but slightly flatter than) close-in stellar binaries.
The brown dwarf and high mass ratio eccentricity CDFs fall below that of the uniform distribution and 
more closely follow a thermal distribution.
Most of the targets in our analysis are young ($\lesssim$200 Myr), so these results are largely probing the 
dynamical conditions of these systems soon after formation and generally before several Gyr of potential evolution.
As suggested by \citet{Geller:2019in}, these distributions are therefore likely to reflect the
intrinsic properties of brown dwarf companions
rather than long-term dynamical processing.

This work joins other recent studies that have found evidence for population-level differences
between brown dwarfs and imaged planets on wide orbits.
\citet{Nielsen:2019cb} presented statistical results from the first 300 stars in the GPIES
survey.  They found a positive correlation between the frequency of giant planets and stellar host mass,
a bottom-heavy planet mass distribution, and generally shorter orbital periods for the planetary companions.
The brown dwarfs in their sample exhibited the opposite trends.
Nielsen et al. interpret these observations as evidence in favor of core/pebble accretion
for giant planets and gravitational fragmentation of disks for brown dwarfs.
\citet{Wagner:2019iy} analyzed the underlying relative mass distribution of substellar companions 
using survival analysis and found a bottom heavy mass function which resembles that of close-in planets.
This similarity suggests formation from core accretion for planets below $\approx$10--20~\Mjup.
Our results that lower-mass companions have more circular orbits bolsters these conclusions.
A positive correlation with stellar mass, a bottom-heavy mass function, and  low eccentricity
orbits all point to bottom-up formation in disks.

Our relatively modest sample of 27 giant planets and brown dwarf companions 
has limited our assessment to broad 
trends in the shape of the eccentricity distribution function.  
This has left open a number of questions related to the formation and evolution of long-period substellar companions:
\begin{itemize}
\item \emph{Where is the dynamical distinction between giant planets and brown dwarfs?}
We have adopted thresholds of 15~\Mjup \ and $M_2$/$M_1$ = 0.01 for this study, but we do not have the statistical 
leverage to test the mass or mass ratio at which this difference arises, or 
whether it is a smooth transition towards more circular orbits at lower masses.

\item \emph{Do the population-level eccentricities of substellar companions evolve over time, or are they 
established at young ages?}
Most direct imaging surveys have preferentially focused on young stars.  This bias is reflected in the 
young ages of most targets in our sample.  Substellar companions spanning a wide range of ages (1 Myr to 10 Gyr)
will provide a probe of the dynamical evolution of these systems over time.

\item \emph{Does the observed period-eccentricity trend 
continue at wider separations?}
The current astrometric precision from AO observations at large telescopes is $\approx$1~mas.
Orbital motion of wider companions out to several hundred AO can readily be detected with this precision after a few years
of monitoring.  An expanded analysis to longer periods 
will establish whether the eccentricity distribution of brown dwarfs continues to peak at higher values at wider separations.

\item \emph{Is the eccentricity dichotomy at wide separations real?}
Our analysis of the HR 8799 multi-planet system and the rest of the imaged planet population within 100 AU
suggests that single systems appear to be dynamically hot.  More discoveries are needed 
to establish whether this is a universal property of single versus multiple planets on wide orbits, 
or whether it reflects peculiarities in the orbits of HR 8799 bcde.
In addition, higher-contrast observations at smaller angular separations will establish whether
these apparently single systems have closer-in companions.

\item \emph{How do these eccentricity distributions depend on stellar multiplicity, mass, and metallicity?}
We restricted our study to mostly include single stars to avoid the dynamical influence of stellar binaries at close and wide separations.
We also largely ignored the effect of host mass or composition in this work.  Yet all of these influence 
the properties of planets at close separations.  Broader samples that include a diversity
of host star properties are needed to explore how these characteristics influence long-period
substellar companions.

\end{itemize}

These questions can be addressed through continued orbit monitoring of known systems and 
larger samples of imaged planets and brown dwarf companions.
With few exceptions, these targets have largely been found in blind AO imaging surveys.
Transitioning towards ``informed targets'' with dynamical evidence of a long-period, low-mass companion
is a promising approach to improve the efficiency of these discoveries.  
In particular, the final $Gaia$ data release is expected to deliver tens of thousands of giant planets
within 10 AU (e.g., \citealt{Perryman:2014jra}).  
Astrometric accelerations from $Gaia$ will eventually point the way to wider substellar companions.

In the near term, the $Hipparcos$-$Gaia$ Catalog of Accelerations---a cross-calibrated catalog
of astrometry from \emph{Hipparcos} and \emph{Gaia} developed by \citet{Brandt:2018dja}---offers 
an especially promising pathway to identify long-period substellar companions over the next few years
(see also \citealt{Kervella:2019bw}).
This can be facilitated with the complementary nature of extreme AO systems on large telescopes  
(\citealt{Jovanovic:2013vw}; \citealt{Macintosh:2014js}; \citealt{Beuzit:2019wt})
and autonomous instruments like Robo-AO (\citealt{Baranec:2014jc}) on smaller telescopes 
to systematically and efficiently survey these new candidates.
Furthermore, in the future the next generation of 30-m class telescopes will substantially increase this landscape by
probing lower planet masses, closer separations, and older ages.

\section{Summary and Conclusions}{\label{sec:conclusions}}

In this study we have carried out the first population-level analysis of the eccentricities of directly 
imaged planets and brown dwarf companions.  We first presented new AO 
observations of 13 substellar companions from Keck/NIRC2 and Subaru/HiCIAO 
along with updated orbit fits to these systems using \texttt{orbitize!}, which is 
optimized for systems with small fractional orbit coverage (Sections~\ref{sec:obs} and \ref{sec:orbits}).
We identified a sample of 27 companions between 5--100 AU 
with masses under 75~\Mjup \ which are undergoing orbital motion;
for nine systems 
we assembled astrometry from the literature and 
uniformly refit their orbits (Section~\ref{sec:neworbits} and Table~\ref{tab:orbitize}).

Assuming a Beta distribution as a flexible model for the underlying population-level eccentricity distribution, 
we determined the overall behavior of substellar 
orbital eccentricities within the framework of hierarchical Bayesian inference.  
Following the importance sampling approach from \citet{Hogg:2010gh}, individual posterior eccentricities 
from our orbit fits are used to approximate the likelihood function, then hyperparameter posteriors
are sampled with MCMC.
This procedure was carried out for the full sample of substellar companions as well as various subdivisions by companion
mass, system mass ratio, separation, and system architecture.
Finally, we assessed the potential role of discovery bias in shaping our results (Section~\ref{sec:discoverybias}).
Below is a summary of our major conclusions.

\begin{enumerate}
\item The primary result from this study is that the underlying eccentricity distributions for 
directly imaged planets (2--15~\Mjup) and brown dwarf companions (15--75~\Mjup) between 5--100 AU 
are significantly different. 
Giant planets have low orbital eccentricities with a peak at $\bar{e}$=0.13, while brown dwarfs exhibit a broad distribution
with a preference for higher eccentricities and a peak in the distribution function between $\bar{e}$=0.6--0.9.
The corresponding Beta distributions have shape parameters of \{$\alpha$=30, $\beta$=200\}
for imaged planets and \{$\alpha$=2.30, $\beta$=1.65\} for brown dwarf companions.
The eccentricity trends hold whether subdivided by mass (15~\Mjup) or mass ratio ($M_2$/$M_1$=0.01).
We interpret this as evidence for formation within a disk for the imaged planets and from cloud
fragmentation for brown dwarfs.
These differences in dynamical properties based on mass or mass ratio are the 
clearest among any other subdivision we tested.

\item The underlying eccentricity distribution function for the low-mass ratio subsample bears a close
resemblance to the distribution of warm Jupiters outside of the tidal circularization radius
from radial velocity surveys.  
Low-mass ratio companions do not reside on circular orbits but have a dynamically hot tail out to modest eccentricities 
of $\approx$0.6.  This may indicate that dynamical heating from scattering events plays a role in shaping the orbital
properties of directly imaged planets.

\item We find evidence that the eccentricity dichotomy between single and multiplanet systems 
also exists at wide separations.  The HR 8799 planets have more circular orbits  ($\bar{e}$=0.11) with a narrower
range of eccentricities compared to systems with single long-period giant planets ($\bar{e}$=0.23). 

\item There is evidence that brown dwarfs have higher eccentricities at longer orbital periods.
Brown dwarf companions on closer orbits (5--30 AU) have a ``softer'' eccentricity distribution with a peak
at $\bar{e}$=0.50 compared to those at wider separations (30--100 AU), which peak at $\bar{e}$=0.74.
This is similar to results for wide ($>$50 AU)
stellar binaries from \citet{Tokovinin:2015dk}, suggesting a shared formation channel
for these populations.

\item The full substellar population from 2--75~\Mjup \ and 5--100 AU 
is approximately flat across all eccentricities.  This represents a 
balance between the more circular giant planets and more eccentric brown dwarfs.  
The best-fitting Beta shape parameters are \{$\alpha$=0.95; $\beta$=1.30\}.  

\item Our sample appears to be robust against 
discovery bias, which tends to 
preferentially select high eccentricities for systems with semi-major axes comparable to or below 
the IWA.  The distribution of IWA/$a$ for our sample at the time of discovery is largely below unity, and we find
no correlations with eccentricity, distance to the system, 
orbital distance, or companion mass in the way that would be expected if discovery bias 
played a major role in shaping the inferred population-level eccentricity distributions.

\item The linear evolution of separation and position angle over time has been uniformly measured  
for 21 systems with substellar companions
using all available astrometry to date.  These fits may be helpful for future orbit monitoring purposes and spectroscopic 
characterization, for example with
$JWST$ or new fiber injection units.  
Linear relations are presented in Table~\ref{tab:linearfit} for convenience.

\end{enumerate}

\appendix

\section{Literature Astrometry}{\label{sec:litastrometry}}

Published astrometry used in this analysis.  When previously published observations were re-reduced 
and presented in subsequent studies, we generally adopt 
the more recent measurements for this work. 

\startlongtable
\begin{deluxetable*}{lcccc}
\renewcommand\arraystretch{0.9}
\tabletypesize{\small}
\setlength{ \tabcolsep } {.1cm} 
\tablewidth{0pt}
\tablecolumns{5}
\tablecaption{Literature Astrometry\label{tab:astrometry}}
\tablehead{
       \colhead{Name} & \colhead{Epoch}  &  \colhead{Separation}  & \colhead{P.A.} & \colhead{Reference} \\
       \colhead{}    & \colhead{(UT)} &  \colhead{(mas)}  & \colhead{(deg)} & \colhead{}
        }   
\startdata
HD 984 B      & 2012.545 &    190 $\pm$ 20    &    109 $\pm$ 3    & \citet{Meshkat:2015hd}  \\
HD 984 B      & 2012.550 &    208 $\pm$ 23    &    109 $\pm$ 3    & \citet{Meshkat:2015hd}  \\
HD 984 B      & 2014.687 &  201.6 $\pm$ 0.4   &   92.2 $\pm$ 0.5  & \citet{Meshkat:2015hd}  \\
HD 984 B      & 2015.657 &  216.3 $\pm$ 1.0   &   83.3 $\pm$ 0.3  & \citet{JohnsonGroh:2017kh}  \\
HD 984 B      & 2015.657 &  217.9 $\pm$ 0.7   &   83.6 $\pm$ 0.2  & \citet{JohnsonGroh:2017kh}  \\
HD 1160 B     & 2002.570 &    770 $\pm$ 30    &  246.2 $\pm$ 1.0  & \citet{Nielsen:2012jk}  \\
HD 1160 B     & 2003.838 &    770 $\pm$ 30    &  245.6 $\pm$ 1.0  & \citet{Nielsen:2012jk}  \\
HD 1160 B     & 2005.975 &    760 $\pm$ 30    &  244.7 $\pm$ 1.0  & \citet{Nielsen:2012jk}  \\
HD 1160 B     & 2008.503 &    800 $\pm$ 60    &    245 $\pm$ 2    & \citet{Nielsen:2012jk}  \\
HD 1160 B     & 2010.710 &    770 $\pm$ 60    &    243 $\pm$ 2    & \citet{Nielsen:2012jk}  \\
HD 1160 B     & 2010.830 &    780 $\pm$ 30    &  244.3 $\pm$ 0.2  & \citet{Nielsen:2012jk}  \\
HD 1160 B     & 2010.890 &    760 $\pm$ 30    &  244.5 $\pm$ 0.2  & \citet{Nielsen:2012jk}  \\
HD 1160 B     & 2010.904 &    770 $\pm$ 20    &  244.9 $\pm$ 0.5  & \citet{Nielsen:2012jk}  \\
HD 1160 B     & 2011.523 &    780 $\pm$ 30    &  244.0 $\pm$ 1.0  & \citet{Nielsen:2012jk}  \\
HD 1160 B     & 2011.669 &    780 $\pm$ 30    &  244.9 $\pm$ 1.0  & \citet{Nielsen:2012jk}  \\
HD 1160 B     & 2011.803 &    770 $\pm$ 30    &  244.5 $\pm$ 0.2  & \citet{Nielsen:2012jk}  \\
HD 1160 B     & 2011.852 &    780 $\pm$ 30    &  244.4 $\pm$ 1.0  & \citet{Nielsen:2012jk}  \\
HD 1160 B     & 2014.613 &  780.9 $\pm$ 1.1   & 244.25 $\pm$ 0.13 & \citet{Maire:2016go}    \\
HD 1160 B     & 2014.613 &  781.0 $\pm$ 0.5   &  243.9 $\pm$ 0.2  & \citet{Maire:2016go}    \\
HD 1160 B     & 2017.936 &    784 $\pm$ 6     &  244.9 $\pm$ 0.3  & \citet{Currie:2018hj}   \\
HD 19467 B    & 2011.660 &   1663 $\pm$ 5     & 243.14 $\pm$ 0.19 & \citet{Crepp:2014ce} \\
HD 19467 B    & 2012.016 &   1666 $\pm$ 7     &  242.3 $\pm$ 0.3  & \citet{Crepp:2014ce} \\
HD 19467 B    & 2012.016 &   1657 $\pm$ 7     &  242.4 $\pm$ 0.4  & \citet{Crepp:2014ce} \\
HD 19467 B    & 2012.652 &   1662 $\pm$ 4     & 242.19 $\pm$ 0.15 & \citet{Crepp:2014ce} \\
HD 19467 B    & 2012.758 &   1653 $\pm$ 4     & 242.13 $\pm$ 0.14 & \citet{Crepp:2014ce} \\
HD 19467 B    & 2013.791 &   1640 $\pm$ 7     &  241.7 $\pm$ 0.3  & \citet{Crepp:2015gt} \\
1RXS0342+1216 B & 2007.951 &  883.0 $\pm$ 0.2   &  17.58 $\pm$ 0.09 & \citet{Bowler:2015ch}   \\
1RXS0342+1216 B & 2008.63  &    860 $\pm$ 8     &   17.3 $\pm$ 0.4  & \citet{Janson:2012dc}   \\
1RXS0342+1216 B & 2008.87  &    866 $\pm$ 8     &   17.8 $\pm$ 0.4  & \citet{Janson:2012dc}   \\     
1RXS0342+1216 B & 2010.659 &    851 $\pm$ 3     &   18.7 $\pm$ 0.1  & \citet{Bowler:2015ch}   \\
1RXS0342+1216 B & 2012.02  &    834 $\pm$ 57    &   17.6 $\pm$ 1.7  & \citet{Janson:2014gz}   \\
1RXS0342+1216 B & 2012.645 &    831 $\pm$ 2     &  18.71 $\pm$ 0.07 & \citet{Bowler:2015ja}   \\
1RXS0342+1216 B & 2013.044 &    822 $\pm$ 8     &   19.1 $\pm$ 0.7  & \citet{Bowler:2015ja}   \\
51 Eri b      & 2014.961   &   450 $\pm$ 7      &  171.0 $\pm$ 0.9 & \citet{DeRosa:2015jla} \\
51 Eri b      & 2015.079   &   454 $\pm$ 6      &  170.6 $\pm$ 1.0 & \citet{DeRosa:2015jla} \\
51 Eri b      & 2015.082   &   462 $\pm$ 7      &  170.5 $\pm$ 0.9 & \citet{DeRosa:2015jla} \\
51 Eri b      & 2015.085   &   462 $\pm$ 24     &  170.4 $\pm$ 3   & \citet{DeRosa:2015jla} \\
51 Eri b      & 2015.665   &   455 $\pm$ 6      &  166.5 $\pm$ 0.6 & \citet{DeRosa:2015jla} \\
51 Eri b      & 2015.74    &    453 $\pm$ 5     &  167.2 $\pm$ 0.6 & \citet{Maire:2019kb} \\
51 Eri b      & 2015.74    &    454 $\pm$ 16    &    166 $\pm$ 2   & \citet{Maire:2019kb} \\
51 Eri b      & 2016.04    &    457 $\pm$ 7     &  165.5 $\pm$ 0.8 & \citet{Maire:2019kb} \\
51 Eri b      & 2016.95    &    454 $\pm$ 6     &  160.3 $\pm$ 0.7 & \citet{Maire:2019kb} \\
51 Eri b      & 2017.74    &    449 $\pm$ 3     &  155.7 $\pm$ 0.4 & \citet{Maire:2019kb} \\
51 Eri b      & 2018.72    &    443 $\pm$ 4     &  150.2 $\pm$ 0.6 & \citet{Maire:2019kb} \\
CD--35 2722 B & 2009.041 &   3172 $\pm$ 5     &  244.1 $\pm$ 0.3  & \citet{Wahhaj:2011by}   \\
CD--35 2722 B & 2010.025 &   3137 $\pm$ 5     &  243.1 $\pm$ 0.3  & \citet{Wahhaj:2011by}   \\
HD 49197 B    & 2002.158 &    950 $\pm$ 5     &   78.3 $\pm$ 0.4  & \citet{Metchev:2004kl}  \\
HD 49197 B    & 2003.854 &    948 $\pm$ 2     &   77.6 $\pm$ 0.3  & \citet{Metchev:2004kl}  \\
HD 49197 B    & 2006.690 &    960 $\pm$ 100   &     77 $\pm$ 2    & \citet{Serabyn:2009ca}  \\
HD 49197 B    & 2015.890 &    862 $\pm$ 25    &   76.6 $\pm$ 1.8  & \citet{Bottom:2017bv}  \\
HR 2562 B     & 2016.066 &    619 $\pm$ 3     &  297.6 $\pm$ 0.4  & \citet{Konopacky:2016dk} \\
HR 2562 B     & 2016.066 &    618 $\pm$ 4     &  297.8 $\pm$ 0.5  & \citet{Konopacky:2016dk} \\
HR 2562 B     & 2016.074 &    618 $\pm$ 5     &  297.4 $\pm$ 0.4  & \citet{Konopacky:2016dk} \\
HR 2562 B     & 2016.151 &    619 $\pm$ 2     &  297.5 $\pm$ 0.3  & \citet{Konopacky:2016dk} \\
HR 2562 B     & 2016.95  &    638 $\pm$ 6     &  297.8 $\pm$ 0.5  & \citet{Maire:2018ch} \\
HR 2562 B     & 2017.10  &    644 $\pm$ 2     &  297.8 $\pm$ 0.2  & \citet{Maire:2018ch} \\
HR 2562 B     & 2017.10  &    644 $\pm$ 3     &  297.5 $\pm$ 0.3  & \citet{Maire:2018ch} \\
HR 2562 B     & 2017.75  &  661.2 $\pm$ 1.3   & 297.97 $\pm$ 0.16 & \citet{Maire:2018ch} \\
HR 2562 B     & 2017.75  &  658.9 $\pm$ 1.6   & 298.08 $\pm$ 0.17 & \citet{Maire:2018ch} \\
HR 2562 B     & 2017.75  &    658 $\pm$ 3     &  297.7 $\pm$ 0.2  & \citet{Maire:2018ch} \\
HR 3549 B     & 2013.033 &    873 $\pm$ 13    &  157.6 $\pm$ 0.6  & \citet{Mawet:2015kk} \\
HR 3549 B     & 2015.034 &    856 $\pm$ 21    &  157.0 $\pm$ 1.0  & \citet{Mawet:2015kk} \\
HR 3549 B     & 2015.964 &    850 $\pm$ 6     &  155.8 $\pm$ 0.5  & \citet{Mesa:2016kn} \\
HR 3549 B     & 2015.964 &    848 $\pm$ 9     &  156.1 $\pm$ 0.7  & \citet{Mesa:2016kn} \\
HD 95086 b    & 2012.030 &    624 $\pm$ 8     &  151.9 $\pm$ 0.8  & \citet{Chauvin:2018ib} \\
HD 95086 b    & 2013.197 &    626 $\pm$ 13    &  150.8 $\pm$ 1.3  & \citet{Chauvin:2018ib} \\
HD 95086 b    & 2013.485 &    600 $\pm$ 11    &  151.0 $\pm$ 1.2  & \citet{Chauvin:2018ib} \\
HD 95086 b    & 2013.939 &    619 $\pm$ 5     &  150.9 $\pm$ 0.5  & \citet{Rameau:2016dx} \\
HD 95086 b    & 2013.942 &    618 $\pm$ 11    &  150.3 $\pm$ 1.1  & \citet{Rameau:2016dx} \\
HD 95086 b    & 2014.361 &    618 $\pm$ 8     &  150.2 $\pm$ 0.7  & \citet{Rameau:2016dx} \\
HD 95086 b    & 2015.090 &    622 $\pm$ 4     &  148.8 $\pm$ 0.4  & \citet{Chauvin:2018ib} \\
HD 95086 b    & 2015.090 &    620 $\pm$ 5     &  149.0 $\pm$ 0.5  & \citet{Chauvin:2018ib} \\
HD 95086 b    & 2015.260 &    622 $\pm$ 7     &  148.8 $\pm$ 0.6  & \citet{Rameau:2016dx} \\
HD 95086 b    & 2015.266 &    622 $\pm$ 4     &  149.0 $\pm$ 0.4  & \citet{Rameau:2016dx} \\
HD 95086 b    & 2015.340 &    622 $\pm$ 7     &  148.6 $\pm$ 0.6  & \citet{Chauvin:2018ib} \\
HD 95086 b    & 2015.340 &    620 $\pm$ 8     &  148.7 $\pm$ 0.6  & \citet{Chauvin:2018ib} \\
HD 95086 b    & 2016.047 &    624 $\pm$ 8     &  148.4 $\pm$ 0.7  & \citet{Chauvin:2018ib} \\
HD 95086 b    & 2016.047 &    626 $\pm$ 10    &  148.6 $\pm$ 0.9  & \citet{Chauvin:2018ib} \\
HD 95086 b    & 2016.162 &    621 $\pm$ 5     &  147.8 $\pm$ 0.5  & \citet{Rameau:2016dx} \\
HD 95086 b    & 2016.178 &    620 $\pm$ 5     &  147.2 $\pm$ 0.5  & \citet{Rameau:2016dx} \\
HD 95086 b    & 2016.413 &    622 $\pm$ 3     &  147.5 $\pm$ 0.3  & \citet{Chauvin:2018ib} \\
HD 95086 b    & 2016.413 &    620 $\pm$ 4     &  147.6 $\pm$ 0.4  & \citet{Chauvin:2018ib} \\
HD 95086 b    & 2017.353 &    624 $\pm$ 3     &  146.6 $\pm$ 0.3  & \citet{Chauvin:2018ib} \\
HD 95086 b    & 2017.353 &    626 $\pm$ 4     &  146.8 $\pm$ 0.4  & \citet{Chauvin:2018ib} \\
GJ 504 B      & 2011.230 &   2479 $\pm$ 16    &  327.9 $\pm$ 0.4  & \citet{Kuzuhara:2013jz} \\
GJ 504 B      & 2011.386 &   2483 $\pm$ 8     &  327.5 $\pm$ 0.2  & \citet{Kuzuhara:2013jz} \\
GJ 504 B      & 2011.611 &   2481 $\pm$ 33    &  326.8 $\pm$ 0.9  & \citet{Kuzuhara:2013jz} \\
GJ 504 B      & 2011.619 &   2448 $\pm$ 24    &  325.8 $\pm$ 0.7  & \citet{Kuzuhara:2013jz} \\
GJ 504 B      & 2012.159 &   2483 $\pm$ 15    &  326.5 $\pm$ 0.4  & \citet{Kuzuhara:2013jz} \\
GJ 504 B      & 2012.279 &   2487 $\pm$ 8     &  326.5 $\pm$ 0.2  & \citet{Kuzuhara:2013jz} \\
GJ 504 B      & 2012.397 &   2499 $\pm$ 26    &  326.1 $\pm$ 0.6  & \citet{Kuzuhara:2013jz} \\
GJ 504 B      & 2015.340 &   2491 $\pm$ 3     & 323.46 $\pm$ 0.07 & \citet{Bonnefoy:2018ch} \\
GJ 504 B      & 2015.419 &   2496 $\pm$ 3     & 323.50 $\pm$ 0.07 & \citet{Bonnefoy:2018ch} \\
GJ 504 B      & 2015.424 &   2497 $\pm$ 4     & 323.60 $\pm$ 0.10 & \citet{Bonnefoy:2018ch} \\
GJ 504 B      & 2015.427 &   2495 $\pm$ 5     & 323.50 $\pm$ 0.14 & \citet{Bonnefoy:2018ch} \\
GJ 504 B      & 2015.427 &   2501 $\pm$ 3     & 323.49 $\pm$ 0.07 & \citet{Bonnefoy:2018ch} \\
GJ 504 B      & 2015.430 &   2499 $\pm$ 6     & 323.40 $\pm$ 0.14 & \citet{Bonnefoy:2018ch} \\
GJ 504 B      & 2016.241 &   2495 $\pm$ 2     & 322.48 $\pm$ 0.05 & \citet{Bonnefoy:2018ch} \\
GJ 504 B      & 2016.241 &   2493 $\pm$ 12    &  322.8 $\pm$ 0.3  & \citet{Bonnefoy:2018ch} \\
GJ 504 B      & 2017.110 &   2493 $\pm$ 3     & 321.74 $\pm$ 0.08 & \citet{Bonnefoy:2018ch} \\
HIP 65426 b   & 2016.411 &    830 $\pm$ 5     &  150.3 $\pm$ 0.2  & \citet{Chauvin:2017hl} \\
HIP 65426 b   & 2016.485 &    830 $\pm$ 3     & 150.14 $\pm$ 0.17 & \citet{Chauvin:2017hl} \\
HIP 65426 b   & 2017.101 &  827.6 $\pm$ 1.5   & 150.11 $\pm$ 0.15 & \citet{Chauvin:2017hl} \\
HIP 65426 b   & 2017.107 &  828.8 $\pm$ 1.5   & 150.05 $\pm$ 0.16 & \citet{Chauvin:2017hl} \\
HIP 65426 b   & 2017.375 &    832 $\pm$ 3     & 149.52 $\pm$ 0.19 & \citet{Cheetham:2019ju} \\
HIP 65426 b   & 2017.378 &    850 $\pm$ 20    &  148.5 $\pm$ 1.6  & \citet{Cheetham:2019ju} \\
HIP 65426 b   & 2018.359 &    823 $\pm$ 2     & 149.85 $\pm$ 0.15 & \citet{Cheetham:2019ju} \\
HIP 65426 b   & 2018.359 &    826 $\pm$ 2     & 149.89 $\pm$ 0.16 & \citet{Cheetham:2019ju} \\
PDS 70 b      & 2012.246 &   192 $\pm$ 21     &    162 $\pm$ 4    & \citet{Keppler:2018dd} \\
PDS 70 b      & 2014.353 &   194 $\pm$ 5      &    159 $\pm$ 3    & \citet{Christiaens:2019db} \\
PDS 70 b      & 2015.334 &   192 $\pm$ 4      &  154.5 $\pm$ 1.2  & \citet{Keppler:2018dd} \\
PDS 70 b      & 2015.334 &   197 $\pm$ 4      &  154.9 $\pm$ 1.1  & \citet{Keppler:2018dd} \\
PDS 70 b      & 2015.411 &   200 $\pm$ 7      &  153.4 $\pm$ 1.8  & \citet{Keppler:2018dd} \\
PDS 70 b      & 2015.411 &   195 $\pm$ 6      &  153.5 $\pm$ 1.8  & \citet{Keppler:2018dd} \\
PDS 70 b      & 2016.367 &   186 $\pm$ 7      &  152.4 $\pm$ 1.5  & \citet{Haffert:2019ba} \\
PDS 70 b      & 2016.367 &   199 $\pm$ 7      &  151.5 $\pm$ 1.6  & \citet{Keppler:2018dd} \\
PDS 70 b      & 2016.416 &   181 $\pm$ 10     &    151 $\pm$ 2    & \citet{Haffert:2019ba} \\
PDS 70 b      & 2018.148 &   192 $\pm$ 8      &  147.0 $\pm$ 2.4  & \citet{Muller:2018jr} \\
PDS 70 b      & 2018.148 &   192 $\pm$ 8      &  146.8 $\pm$ 2.4  & \citet{Muller:2018jr} \\
PDS 70 b      & 2018.334 &   183 $\pm$ 18     &  148.8 $\pm$ 1.7  & \citet{Wagner:2018hw} \\
PDS 70 b      & 2018.337 &   193 $\pm$ 12     &  143.4 $\pm$ 4.2  & \citet{Wagner:2018hw} \\
PDS 70 b      & 2018.465 &   177 $\pm$ 25     &  146.8 $\pm$ 8.5  & \citet{Haffert:2019ba} \\
PZ Tel B      & 2007.446 &   255 $\pm$ 3      &   61.7 $\pm$ 0.6  & \citet{Mugrauer:2012ca}  \\
PZ Tel B      & 2009.739 & 336.6 $\pm$ 1.2    &   60.5 $\pm$ 0.2  & \citet{Mugrauer:2012ca}  \\
PZ Tel B      & 2010.274 &   330 $\pm$ 10     &   59.0 $\pm$ 1.0  & \citet{Biller:2010ku}  \\
PZ Tel B      & 2010.340 & 356.4 $\pm$ 1.1    &   60.3 $\pm$ 0.2  & \citet{Mugrauer:2012ca}  \\
PZ Tel B      & 2010.345 & 354.7 $\pm$ 1.2    &   60.3 $\pm$ 0.2  & \citet{Mugrauer:2012ca}  \\
PZ Tel B      & 2010.350 &   360 $\pm$ 3      &   59.4 $\pm$ 0.5  & \citet{Biller:2010ku}  \\
PZ Tel B      & 2010.734 &   365 $\pm$ 8      &  59.2 $\pm$ 0.8  & \citet{Beust:2016bf}  \\
PZ Tel B      & 2010.821 & 369.3 $\pm$ 1.1    &   59.9 $\pm$ 0.2  & \citet{Mugrauer:2012ca}  \\
PZ Tel B      & 2011.227 & 382.2 $\pm$ 1.0    &   59.8 $\pm$ 0.2  & \citet{Mugrauer:2012ca}  \\
PZ Tel B      & 2011.334 &   394 $\pm$ 2      &  60.4 $\pm$ 0.2  & \citet{Beust:2016bf}  \\
PZ Tel B      & 2011.419 & 387.8 $\pm$ 1.2    &   59.7 $\pm$ 0.2  & \citet{Mugrauer:2012ca}  \\
PZ Tel B      & 2011.422 & 388.5 $\pm$ 0.8    &  59.66 $\pm$ 0.16 & \citet{Mugrauer:2012ca}  \\
PZ Tel B      & 2011.424 & 387.1 $\pm$ 1.4    &   59.7 $\pm$ 0.3  & \citet{Mugrauer:2012ca}  \\
PZ Tel B      & 2011.427 & 389.0 $\pm$ 1.0    &   59.7 $\pm$ 0.2  & \citet{Mugrauer:2012ca}  \\
PZ Tel B      & 2011.430 &   390 $\pm$ 5      &  60.0 $\pm$ 0.6  & \citet{Beust:2016bf}  \\
PZ Tel B      & 2012.435 & 420.1 $\pm$ 1.3    &  59.6 $\pm$ 0.2  & \citet{Ginski:2014ef}  \\
PZ Tel B      & 2012.435 & 418.8 $\pm$ 1.4    &  59.6 $\pm$ 0.2  & \citet{Ginski:2014ef}  \\
PZ Tel B      & 2014.53  & 478.2 $\pm$ 0.7    &  59.71 $\pm$ 0.19 & \citet{Maire:2016go}  \\
PZ Tel B      & 2014.53  &   478 $\pm$ 2      &   59.6 $\pm$ 0.5  & \citet{Maire:2016go}  \\
PZ Tel B      & 2014.53  &   476 $\pm$ 2      &   60.1 $\pm$ 0.5  & \citet{Maire:2016go}  \\
PZ Tel B      & 2014.60  & 479.5 $\pm$ 0.7    &  59.62 $\pm$ 0.14 & \citet{Maire:2016go}  \\
PZ Tel B      & 2014.60  & 479.7 $\pm$ 0.3    &   59.7 $\pm$ 0.5  & \citet{Maire:2016go}  \\
PZ Tel B      & 2014.60  & 479.6 $\pm$ 0.3    &   60.2 $\pm$ 0.5  & \citet{Maire:2016go}  \\
PZ Tel B      & 2014.78  & 482.6 $\pm$ 0.9    &  59.44 $\pm$ 0.15 & \citet{Maire:2016go}  \\
PZ Tel B      & 2014.78  & 483.9 $\pm$ 0.3    &  59.49 $\pm$ 0.16 & \citet{Maire:2016go}  \\
PZ Tel B      & 2014.78  & 483.9 $\pm$ 0.3    &  59.51 $\pm$ 0.16 & \citet{Maire:2016go}  \\
HD 206893 B   & 2015.758  &  270 $\pm$ 3      &   70.0 $\pm$ 0.6  & \citet{Milli:2017fs} \\
HD 206893 B   & 2016.602  &  269 $\pm$ 10     &   61.6 $\pm$ 1.9  & \citet{Milli:2017fs} \\
HD 206893 B   & 2016.709  &  265 $\pm$ 2      &  62.25 $\pm$ 0.11 & \citet{Delorme:2017hl} \\
HD 206893 B   & 2017.531  &  260 $\pm$ 2      &   54.2 $\pm$ 0.4  & \citet{Grandjean:2019cv} \\
HD 206893 B   & 2018.465  & 249.1 $\pm$ 1.6   &   45.5 $\pm$ 0.4  & \citet{Grandjean:2019cv} \\
$\kappa$ And B & 2012.000 & 1070 $\pm$ 10     &   55.7 $\pm$ 0.6  & \citet{Carson:2013fw} \\
$\kappa$ And B & 2012.518 & 1058 $\pm$ 7      &   56.0 $\pm$ 0.4  & \citet{Carson:2013fw} \\
$\kappa$ And B & 2012.841 & 1028 $\pm$ 14     &   55.4 $\pm$ 0.6  & \citet{Currie:2018hj} \\
$\kappa$ And B & 2013.627 & 1015 $\pm$ 14     &   54.8 $\pm$ 0.6  & \citet{Currie:2018hj} \\
$\kappa$ And B & 2017.676 &  914 $\pm$ 17     &   50.9 $\pm$ 0.7  & \citet{Currie:2018hj} \\
$\kappa$ And B & 2017.936 &  909 $\pm$ 14     &   50.3 $\pm$ 0.6  & \citet{Currie:2018hj} \\
HD 23514 B    & 2006.939  & 2640 $\pm$ 20     &  228.7 $\pm$ 1.0  & \citet{Rodriguez:2012ef} \\
HD 23514 B    & 2007.813  & 2640 $\pm$ 10     &  227.8 $\pm$ 0.3  & \citet{Rodriguez:2012ef} \\
HD 23514 B    & 2008.843  & 2620 $\pm$ 40     &  227.2 $\pm$ 0.5  & \citet{Rodriguez:2012ef} \\
HD 23514 B    & 2009.832  & 2642 $\pm$ 3      & 227.51 $\pm$ 0.04 & \citet{Rodriguez:2012ef} \\
HD 23514 B    & 2009.835  & 2642 $\pm$ 1      &  227.7 $\pm$ 0.03 & \citet{Rodriguez:2012ef} \\
HD 23514 B    & 2010.827  & 2644 $\pm$ 4      &  227.5 $\pm$ 0.1  & \citet{Rodriguez:2012ef} \\
HD 23514 B    & 2010.827  & 2644 $\pm$ 2      & 227.48 $\pm$ 0.05 & \citet{Rodriguez:2012ef} \\
HD 23514 B    & 2010.827  & 2642 $\pm$ 0.5    & 227.47 $\pm$ 0.09 & \citet{Rodriguez:2012ef} \\
HD 23514 B    & 2010.827  & 2645 $\pm$ 2      & 227.52 $\pm$ 0.02 & \citet{Rodriguez:2012ef} \\
HD 23514 B    & 2010.914  & 2646 $\pm$ 33     &  227.6 $\pm$ 0.7  & \citet{Yamamoto:2013gu} \\
DH Tau B      & 1999.044  & 2332 $\pm$ 10     & 138.68 $\pm$ 0.19 & \citet{Ginski:2014ef} \\
DH Tau B      & 2002.893  & 2340 $\pm$ 6      & 139.56 $\pm$ 0.17 & \citet{Itoh:2005ii} \\
DH Tau B      & 2004.019  & 2344 $\pm$ 3      & 139.83 $\pm$ 0.06 & \citet{Itoh:2005ii} \\
DH Tau B      & 2009.747  & 2339 $\pm$ 4      & 138.63 $\pm$ 0.14 & \citet{Ginski:2014ef} \\
DH Tau B      & 2012.058  & 2332 $\pm$ 6      & 138.76 $\pm$ 0.16 & \citet{Ginski:2014ef} \\
DH Tau B      & 2012.928  & 2343 $\pm$ 6      & 138.61 $\pm$ 0.15 & \citet{Ginski:2014ef} \\
DH Tau B      & 2014.934  & 2343 $\pm$ 1      & 140.25 $\pm$ 0.02 & \citet{Bryan:2016eo} \\
DH Tau B      & 2015.844  & 2339 $\pm$ 1      & 139.94 $\pm$ 0.02 & \citet{Bryan:2016eo} \\
Ross 458 B    & 2000.134  &  475.1 $\pm$ 7.1  &    81.4 $\pm$ 2.8   & \citet{Mann:2019ey} \\
Ross 458 B    & 2001.337  &  526.3 $\pm$ 8.2  &    66.9 $\pm$ 2.6   & \citet{Mann:2019ey} \\
Ross 458 B    & 2001.515  &  522.0 $\pm$ 6.2  &    65.3 $\pm$ 2.5   & \citet{Mann:2019ey} \\
Ross 458 B    & 2001.591  &  527.6 $\pm$ 5.0  &    64.1 $\pm$ 2.6   & \citet{Mann:2019ey} \\
Ross 458 B    & 2002.170  &    533 $\pm$ 12   &    56.2 $\pm$ 1.2   & \citet{Mann:2019ey} \\
Ross 458 B    & 2005.329  &    280 $\pm$ 50   &     357 $\pm$ 1     & \citet{WardDuong:2015ej} \\
Ross 458 B    & 2006.389  &  236.2 $\pm$ 3.9  &  304.62 $\pm$ 0.96  & \citet{Mann:2019ey} \\
Ross 458 B    & 2006.389  &  233.6 $\pm$ 5.7  &   304.3 $\pm$ 1.5   & \citet{Mann:2019ey} \\
Ross 458 B    & 2007.142  &  270.6 $\pm$ 8.2  &   269.3 $\pm$ 1.1   & \citet{Mann:2019ey} \\
Ross 458 B    & 2009.323  & 309.38 $\pm$ 0.67 & 203.182 $\pm$ 0.069 & \citet{Mann:2019ey} \\
Ross 458 B    & 2009.323  &  307.5 $\pm$ 1.1  & 202.950 $\pm$ 0.049 & \citet{Mann:2019ey} \\
Ross 458 B    & 2009.323  &  308.2 $\pm$ 1.8  &  203.22 $\pm$ 0.10  & \citet{Mann:2019ey} \\
Ross 458 B    & 2013.301  & 448.30 $\pm$ 0.66 &  86.777 $\pm$ 0.041 & \citet{Mann:2019ey} \\
Ross 458 B    & 2015.471  & 524.49 $\pm$ 0.27 &  60.087 $\pm$ 0.016 & \citet{Mann:2019ey} \\
2M1559+4403 B & 2008.24   & 5654  $\pm$ 4     & 284.2  $\pm$ 0.3   & \citet{Janson:2012dc} \\
2M1559+4403 B & 2009.13   & 5623  $\pm$ 4     & 284.9  $\pm$ 0.3   & \citet{Janson:2012dc} \\
2M1559+4403 B & 2009.42   & 5638  $\pm$ 4     & 284.8  $\pm$ 0.3   & \citet{Janson:2012dc} \\
2M1559+4403 B & 2012.02   & 5598  $\pm$ 56    & 284.7  $\pm$ 0.3   & \citet{Janson:2014gz} \\
2M1559+4403 B & 2012.357  & 5647  $\pm$ 15    & 284.46 $\pm$ 0.10 & \citet{Bowler:2015ja} \\
2M1559+4403 B & 2014.446  & 5670  $\pm$ 70    & 284.4  $\pm$ 0.6  & \citet{Bowler:2015ch} \\
TWA 5 B       & 1998.312  & 1960  $\pm$ 10    & 1.8    $\pm$ 0.4  & \citet{Lowrance:1999ck} \\
TWA 5 B       & 2000.15   & 1954  $\pm$ 8     & 359.16 $\pm$ 0.08 & \citet{Brandeker:2003bt} \\
TWA 5 B       & 2005.126  & 1940  $\pm$ 20    & 357.4  $\pm$ 0.6  & \citet{Galicher:2016hg} \\
TWA 5 B       & 2007.518  & 1902 $\pm$ 2      & 356.4  $\pm$ 0.2  & \citet{Kohler:2013im} \\
TWA 5 B       & 2010.11   & 1897 $\pm$ 11     & 354.6  $\pm$ 0.3  & \citet{Janson:2012dc} \\
TWA 5 B       & 2011.071  & 1888 $\pm$ 7      & 355.2  $\pm$ 0.2  & \citet{Kohler:2013im} \\
TWA 5 B       & 2012.003  & 1879 $\pm$ 2      & 355.0  $\pm$ 0.1  & \citet{Kohler:2013im} \\
TWA 5 B       & 2012.01   & 1869 $\pm$ 19     & 355.1  $\pm$ 0.3  & \citet{Janson:2014gz} \\
TWA 5 B       & 2012.049  & 1875 $\pm$ 3      & 354.8  $\pm$ 0.1  & \citet{Kohler:2013im} \\
TWA 5 B       & 2013.044  & 1873 $\pm$ 2      & 354.5  $\pm$ 0.1  & \citet{Kohler:2013im} \\
1RXS2351+3127 B & 2011.470 &   2392.2 $\pm$ 2.0 &  91.77  $\pm$  0.05  & \citet{Bowler:2012cs} \\
1RXS2351+3127 B & 2011.871 &   2386.3 $\pm$ 1.5 &  91.81  $\pm$  0.04  & \citet{Bowler:2012cs} \\
1RXS2351+3127 B & 2013.626 &     2391 $\pm$ 4   &  91.63  $\pm$  0.02  & \citet{Bowler:2015ja} \\
1RXS2351+3127 B & 2013.626 &     2391 $\pm$ 3   &  91.647 $\pm$  0.015 & \citet{Bowler:2015ja} \\
1RXS2351+3127 B & 2013.626 &   2390.7 $\pm$ 1.1 &  91.65  $\pm$  0.01  & \citet{Bowler:2015ja} \\
1RXS2351+3127 B & 2013.626 &   2391.2 $\pm$ 1.1 &  91.63  $\pm$  0.03  & \citet{Bowler:2015ja} \\
1RXS2351+3127 B & 2013.626 &   2391.6 $\pm$ 1.7 &  91.643 $\pm$  0.014 & \citet{Bowler:2015ja} \\
1RXS2351+3127 B & 2013.626 &     2390 $\pm$ 5   &  91.64  $\pm$  0.08  & \citet{Bowler:2015ja} \\
\enddata
\end{deluxetable*}


\section{Corner Plots}{\label{sec:corners}}

In this section we report corner plots displaying joint and marginalized distributions of orbital elements 
from our orbit fits using \texttt{orbitize!}.  These are also summarized in Table~\ref{tab:orbitize}.


\begin{figure*}
  \vskip -0.1 in
  \hskip 0.8 in
  \resizebox{5.5in}{!}{\includegraphics{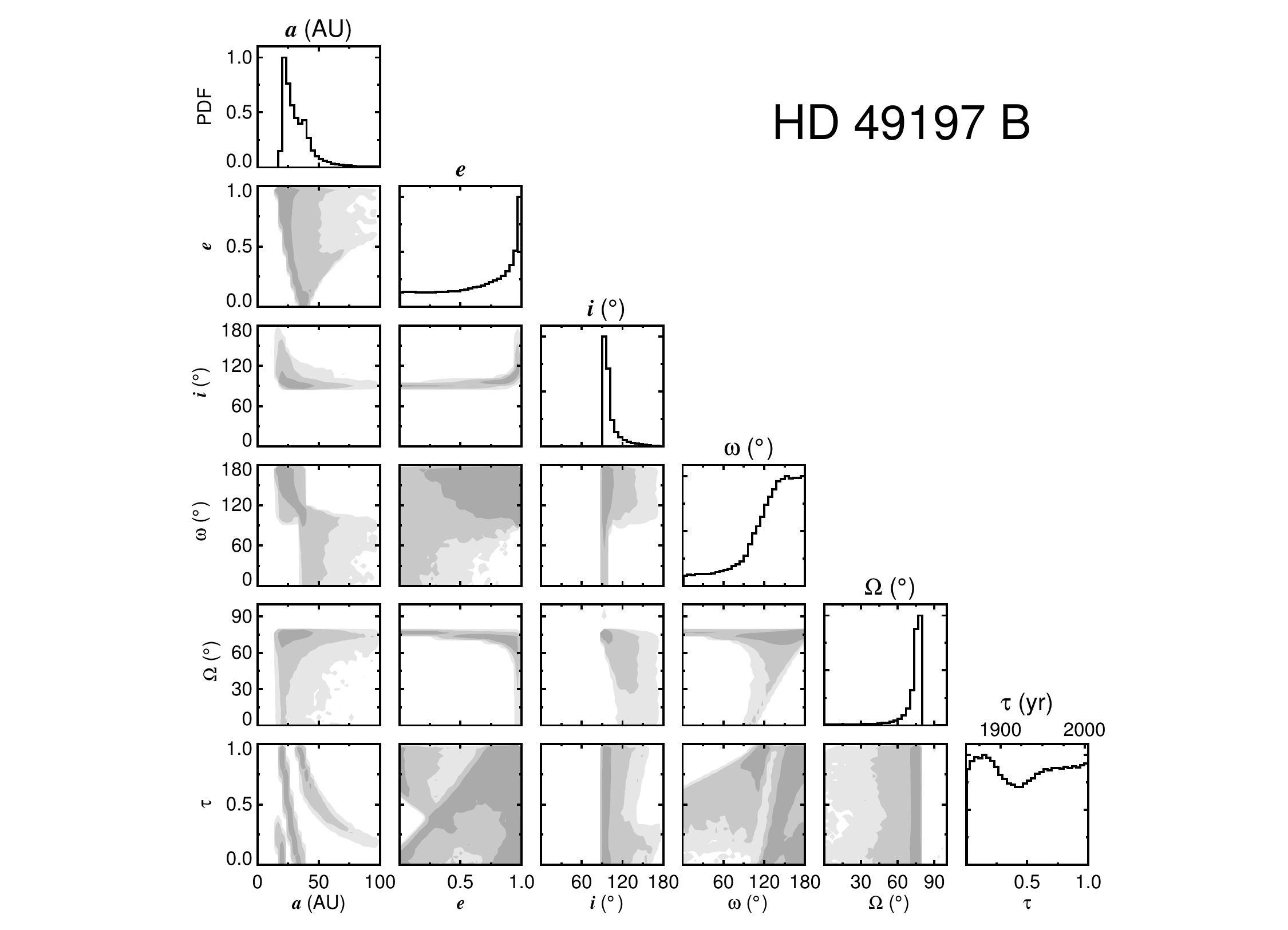}}
  \vskip 0 in
  \caption{Corner plot for HD 49197 B.  One-dimensional marginalized distributions are shown along the diagonal.
  Inclination ($i$), argument of periastron ($\omega$), and longitude of 
  ascending node ($\Omega$) are expressed in degrees.  Units for the time of periastron passage, $\tau$, are 
  fraction of the orbital period past MJD=0.  Gray contours show the 1, 2, and 3$\sigma$ regions encompassing the two-dimensional joint
  posterior distributions.  
    \label{fig:hd49197corner} } 
\end{figure*}

\begin{figure*}
  \vskip -0. in
  \hskip 0.8 in
  \resizebox{5.5in}{!}{\includegraphics{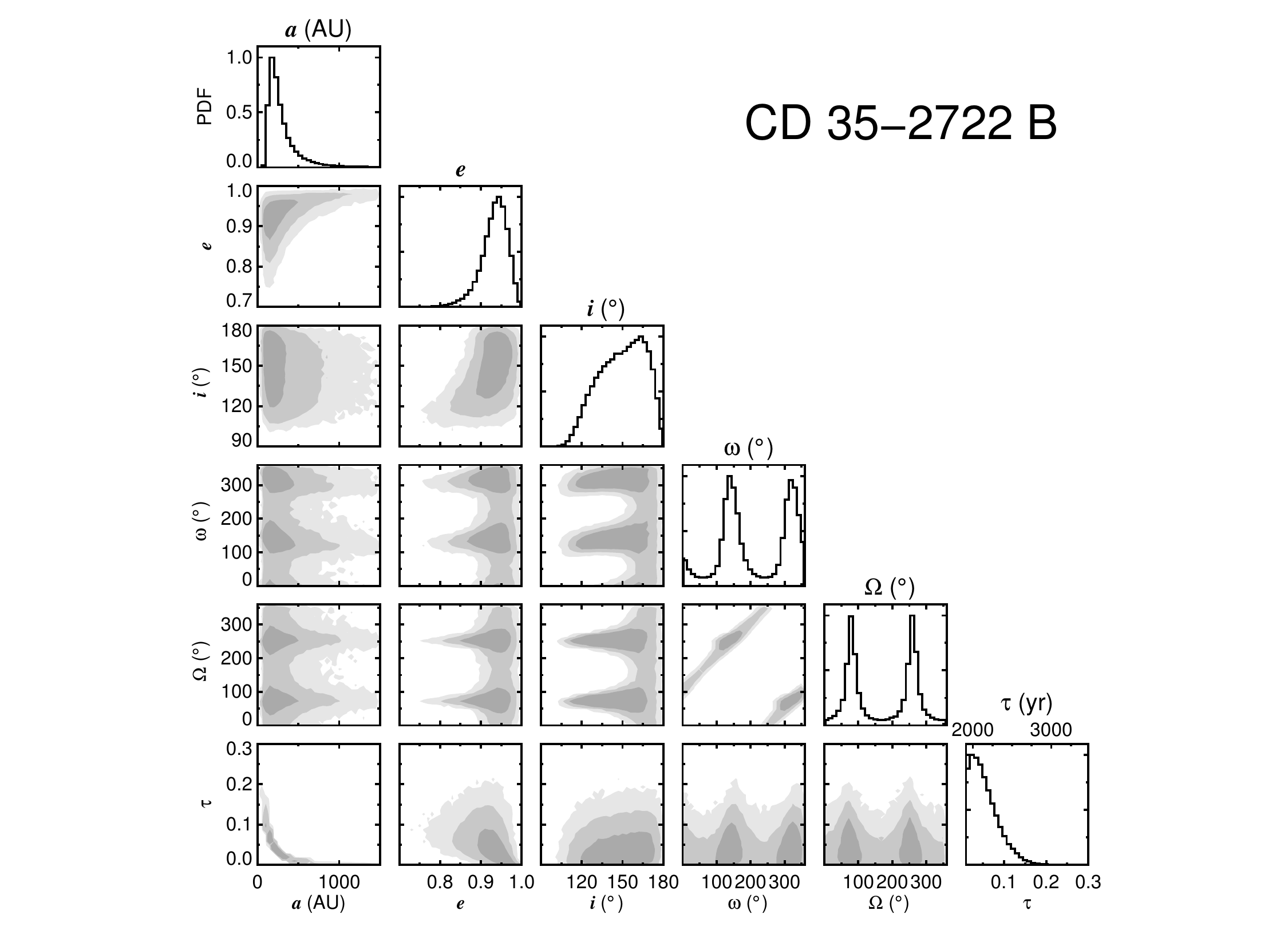}}
  \vskip 0 in
  \caption{Corner plot for CD--35 2722 B.  See Figure~\ref{fig:hd49197corner} for details.
    \label{fig:cd352722corner} } 
\end{figure*}

\begin{figure*}
  \vskip -0. in
  \hskip 0.8 in
  \resizebox{5.5in}{!}{\includegraphics{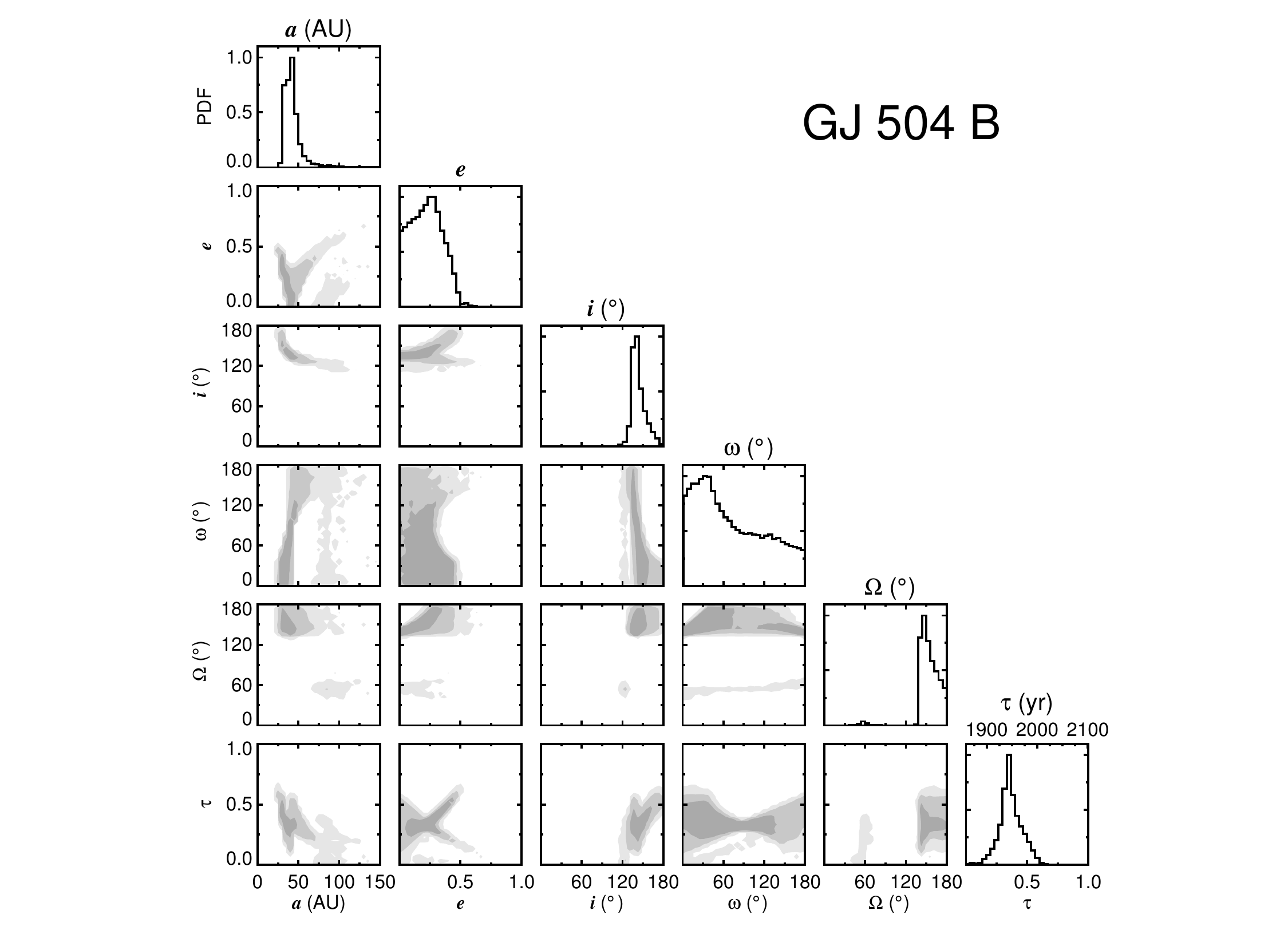}}
  \vskip 0 in
  \caption{Corner plot for GJ 504 B.   See Figure~\ref{fig:hd49197corner} for details.
    \label{fig:gj504corner} } 
\end{figure*}

\begin{figure*}
  \vskip -0. in
  \hskip 0.8 in
  \resizebox{5.5in}{!}{\includegraphics{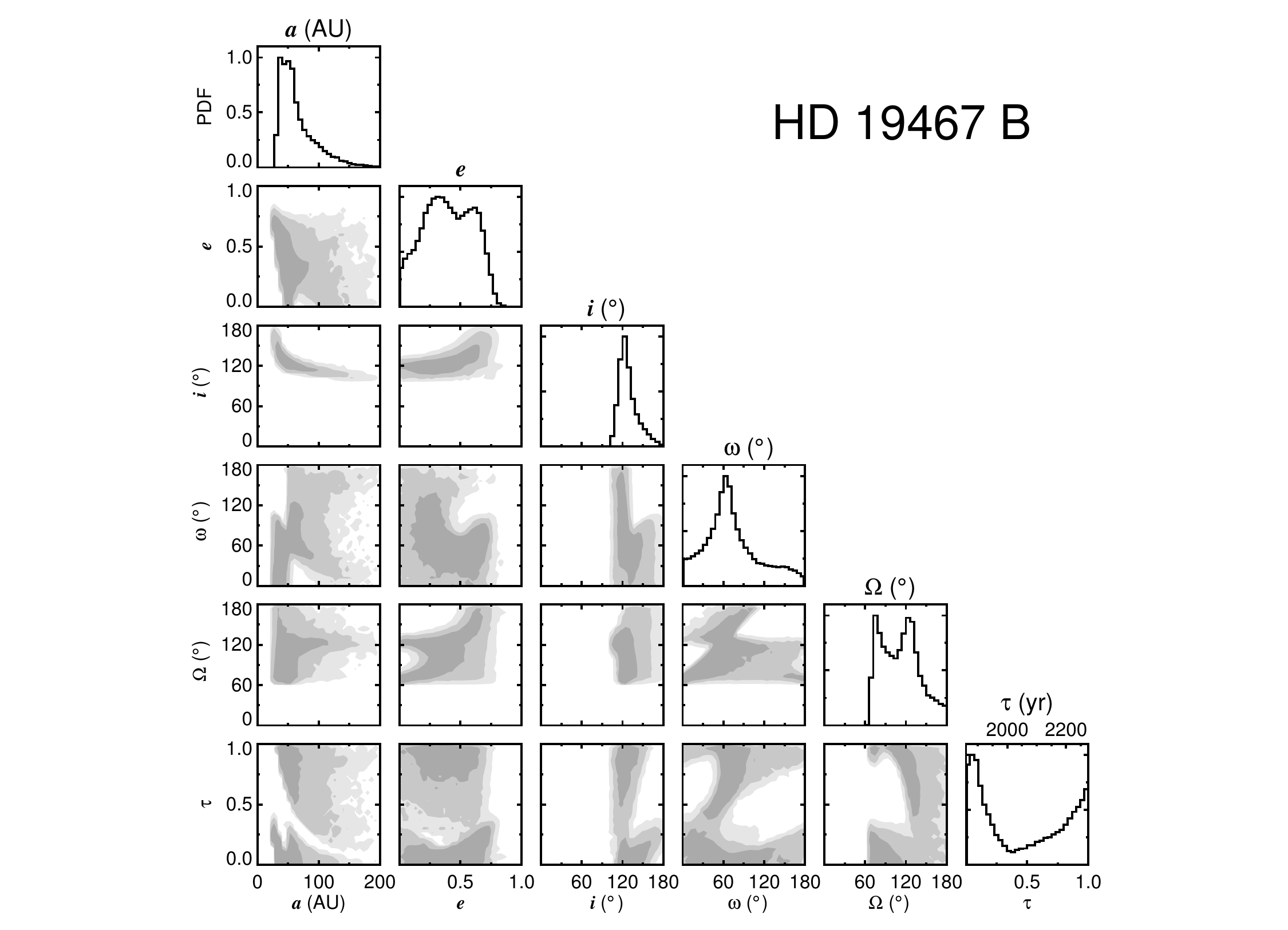}}
  \vskip 0 in
  \caption{Corner plot for HD 19467 B.  See Figure~\ref{fig:hd49197corner} for details.
    \label{fig:hd19467corner} } 
\end{figure*}

\begin{figure*}
  \vskip -0. in
  \hskip 0.8 in
  \resizebox{5.5in}{!}{\includegraphics{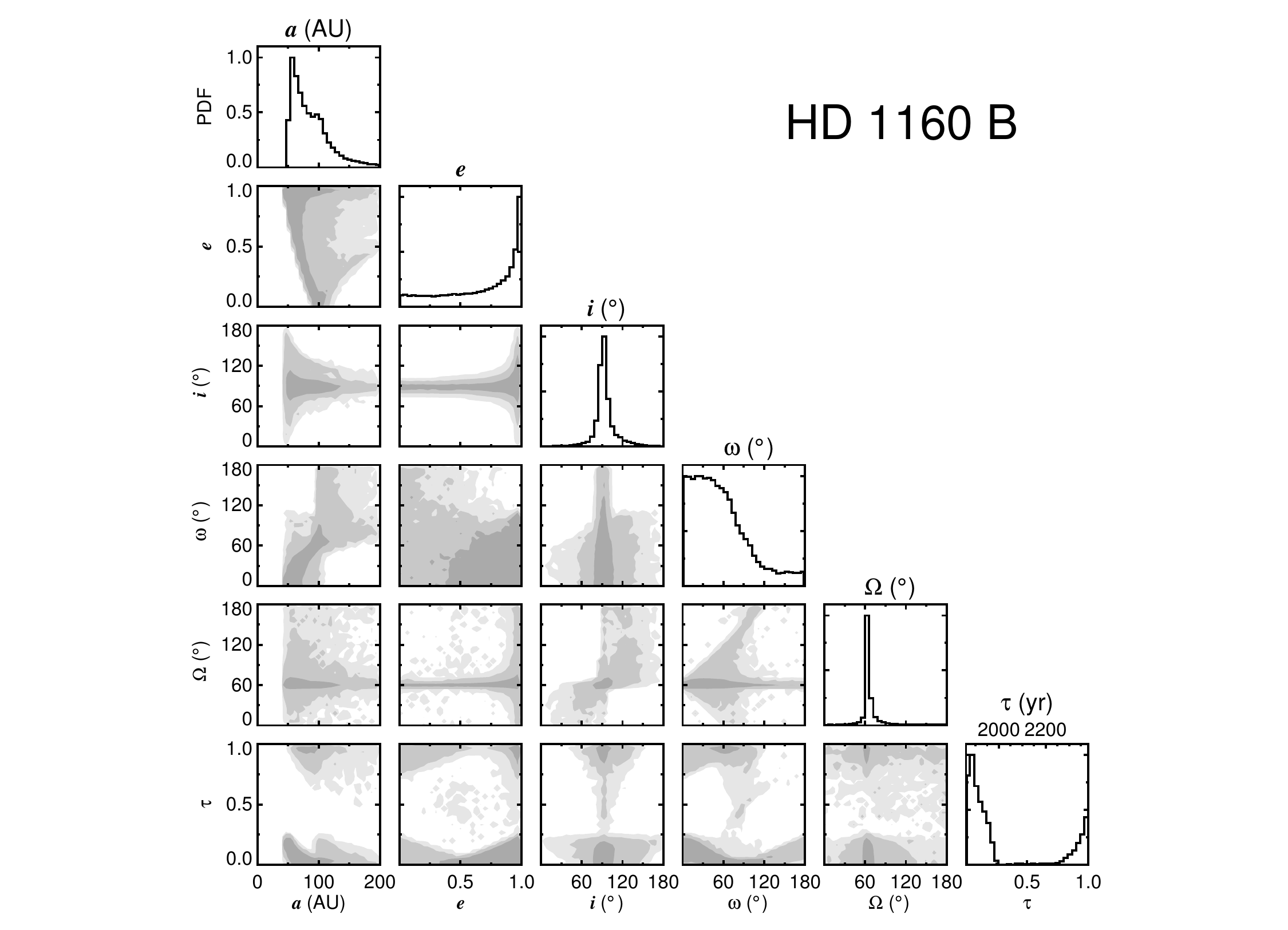}}
  \vskip 0 in
  \caption{Corner plot for HD 1160 B.   See Figure~\ref{fig:hd49197corner} for details.
    \label{fig:hd1160corner} } 
\end{figure*}

\begin{figure*}
  \vskip -0. in
  \hskip 0.8 in
  \resizebox{5.5in}{!}{\includegraphics{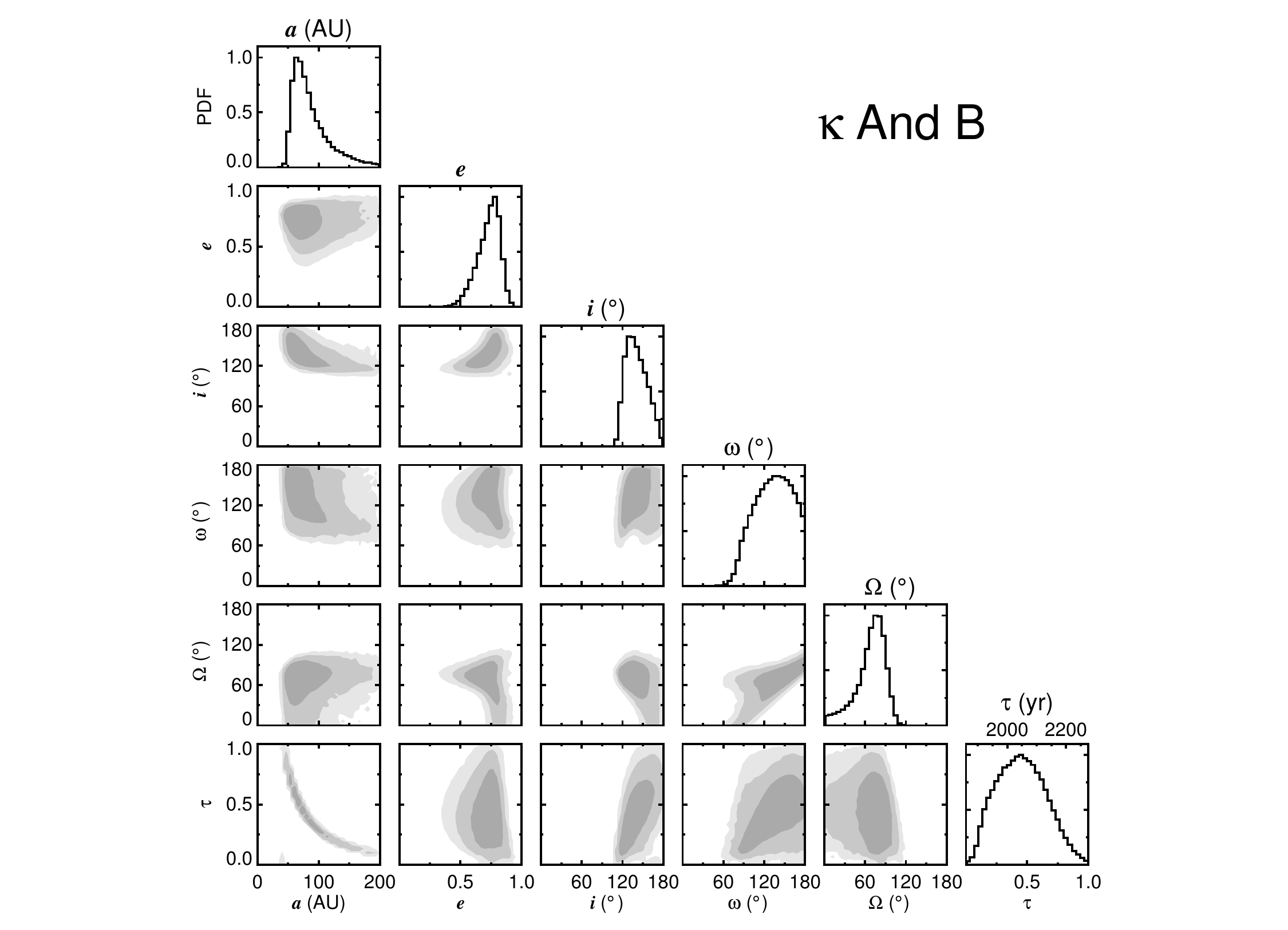}}
  \vskip 0 in
  \caption{Corner plot for $\kappa$ And B.   See Figure~\ref{fig:hd49197corner} for details.
    \label{fig:kappaandcorner} } 
\end{figure*}

\begin{figure*}
  \vskip -0. in
  \hskip 0.8 in
  \resizebox{5.5in}{!}{\includegraphics{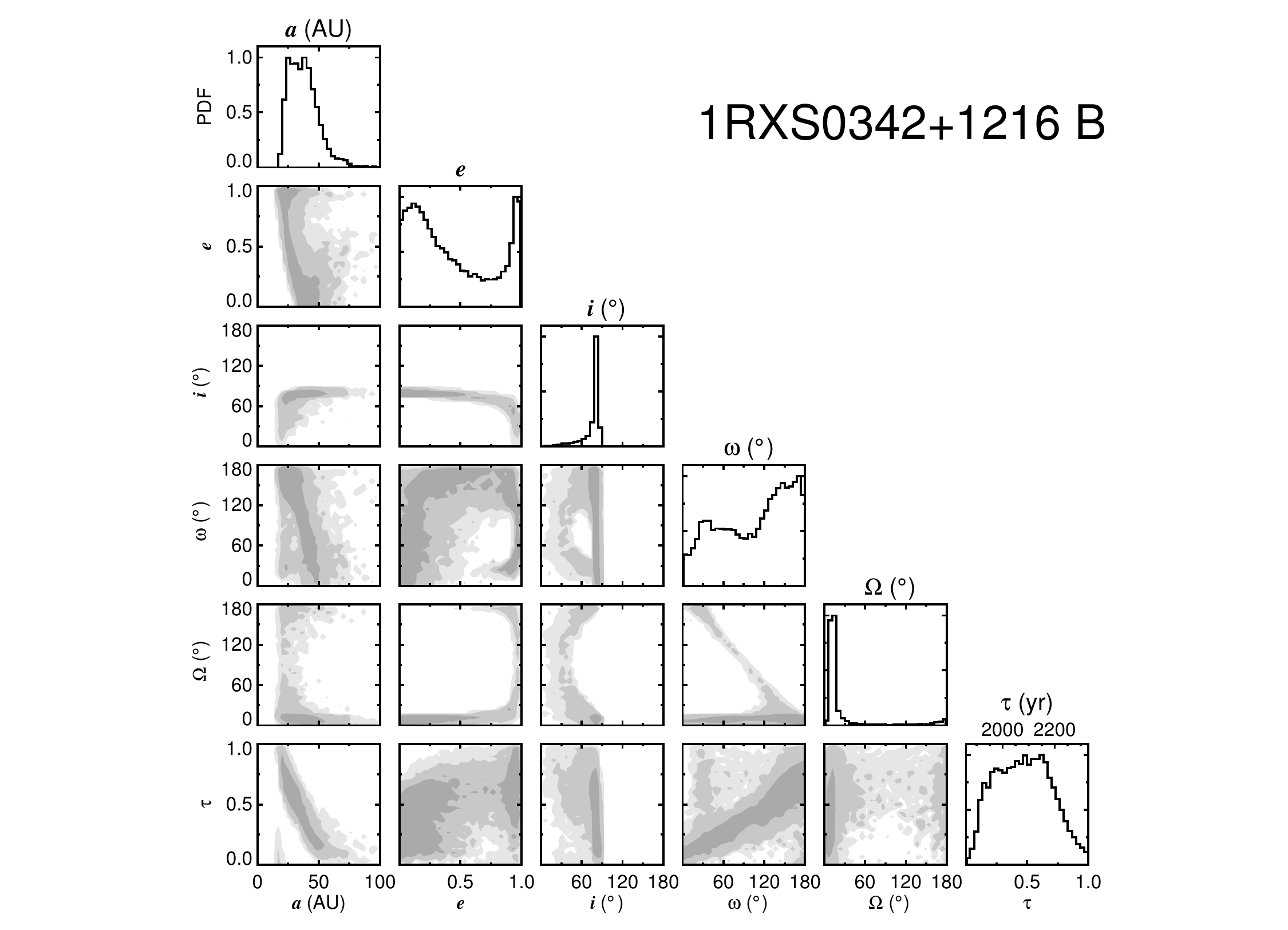}}
  \vskip 0 in
  \caption{Corner plot for 1RXS0342+1216 B.  See Figure~\ref{fig:hd49197corner} for details.
    \label{fig:rxs0342corner} } 
\end{figure*}

\begin{figure*}
  \vskip -0. in
  \hskip 0.8 in
  \resizebox{5.5in}{!}{\includegraphics{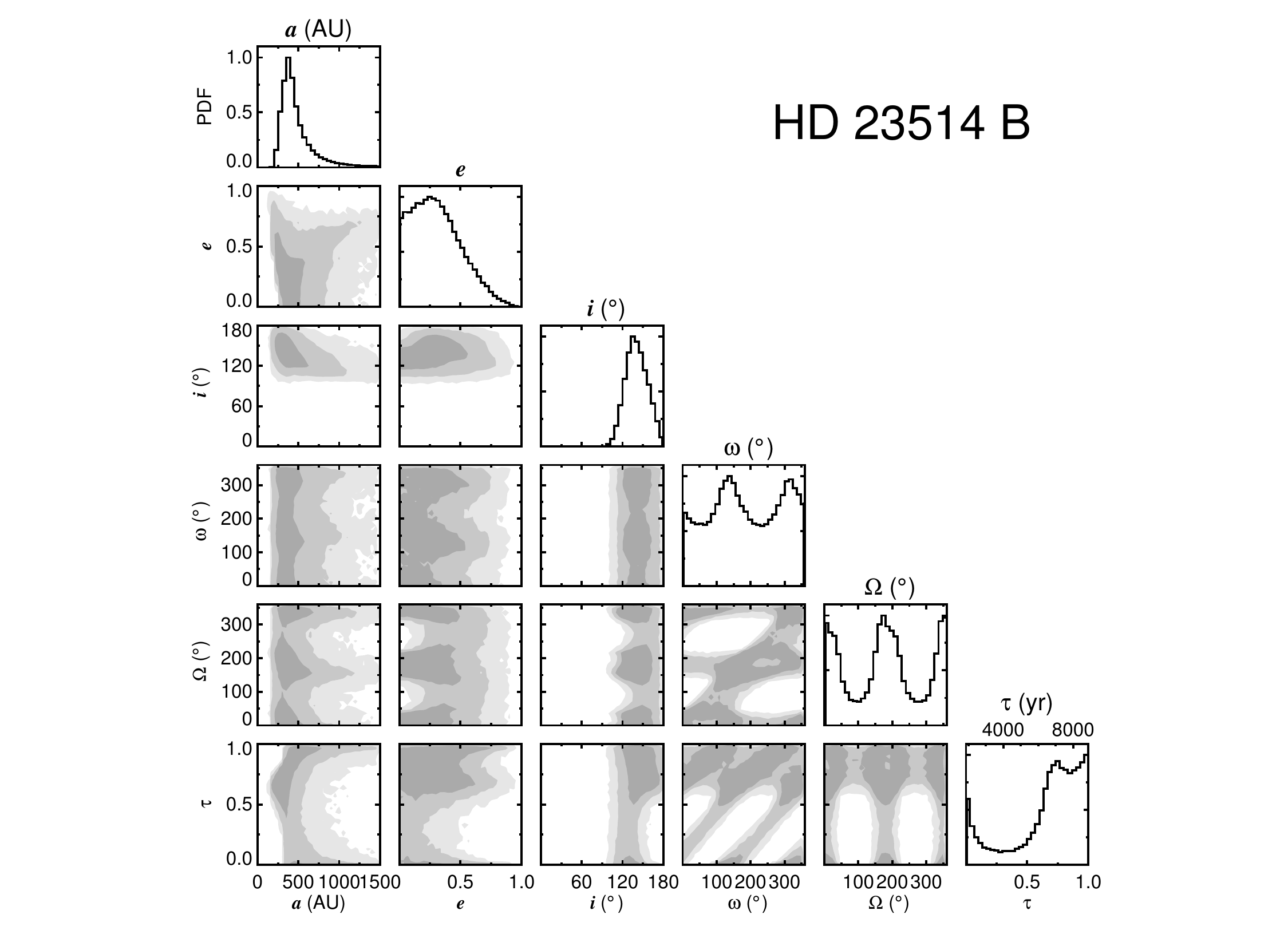}}
  \vskip 0 in
  \caption{Corner plot for HD 23514 B.   See Figure~\ref{fig:hd49197corner} for details.
    \label{fig:hd23514corner} } 
\end{figure*}

\begin{figure*}
  \vskip -0. in
  \hskip 0.8 in
  \resizebox{5.5in}{!}{\includegraphics{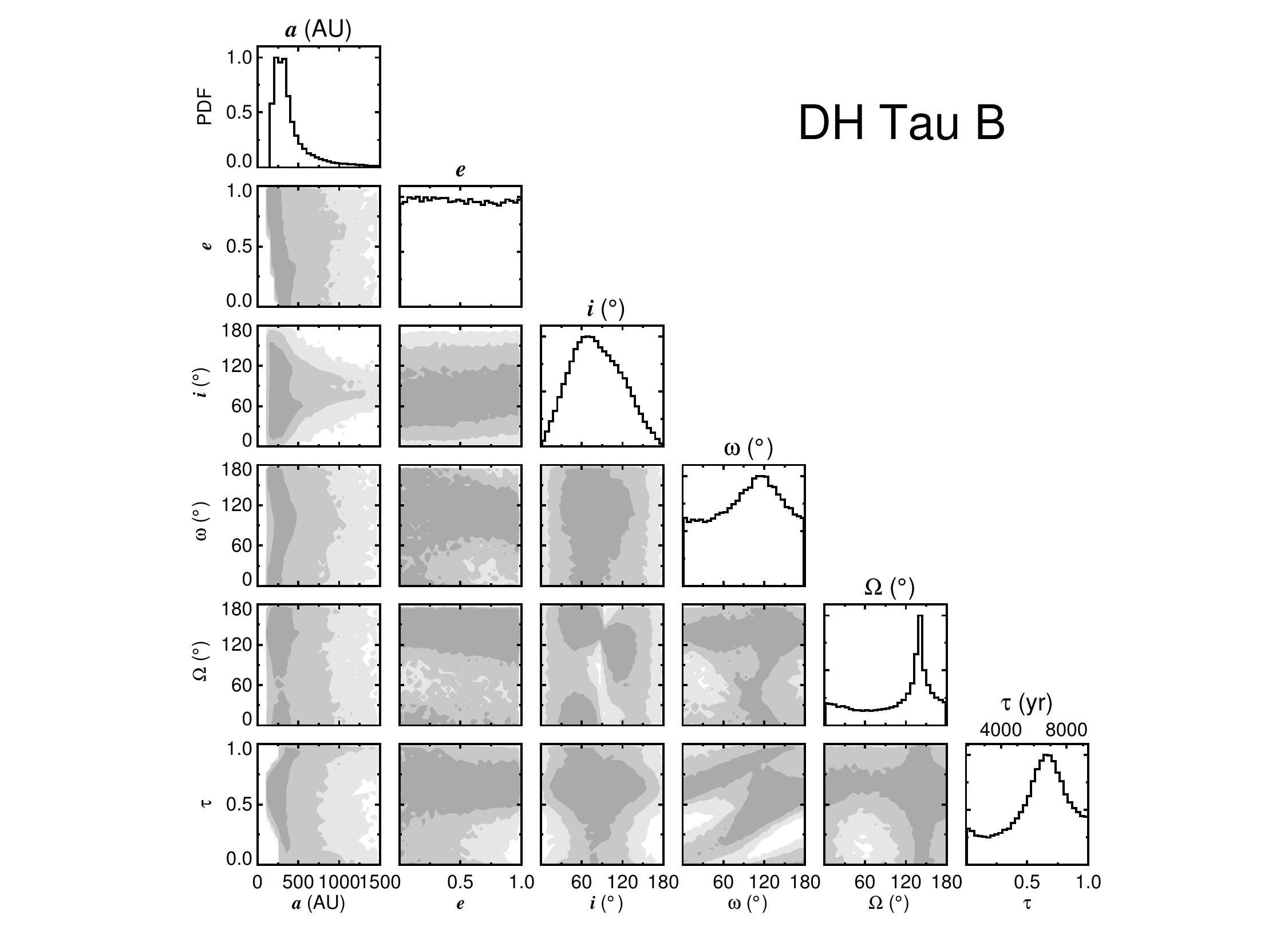}}
  \vskip 0 in
  \caption{Corner plot for DH Tau B.   See Figure~\ref{fig:hd49197corner} for details.
    \label{fig:dhtaucorner} } 
\end{figure*}

\begin{figure*}
  \vskip -0. in
  \hskip 0.8 in
  \resizebox{5.5in}{!}{\includegraphics{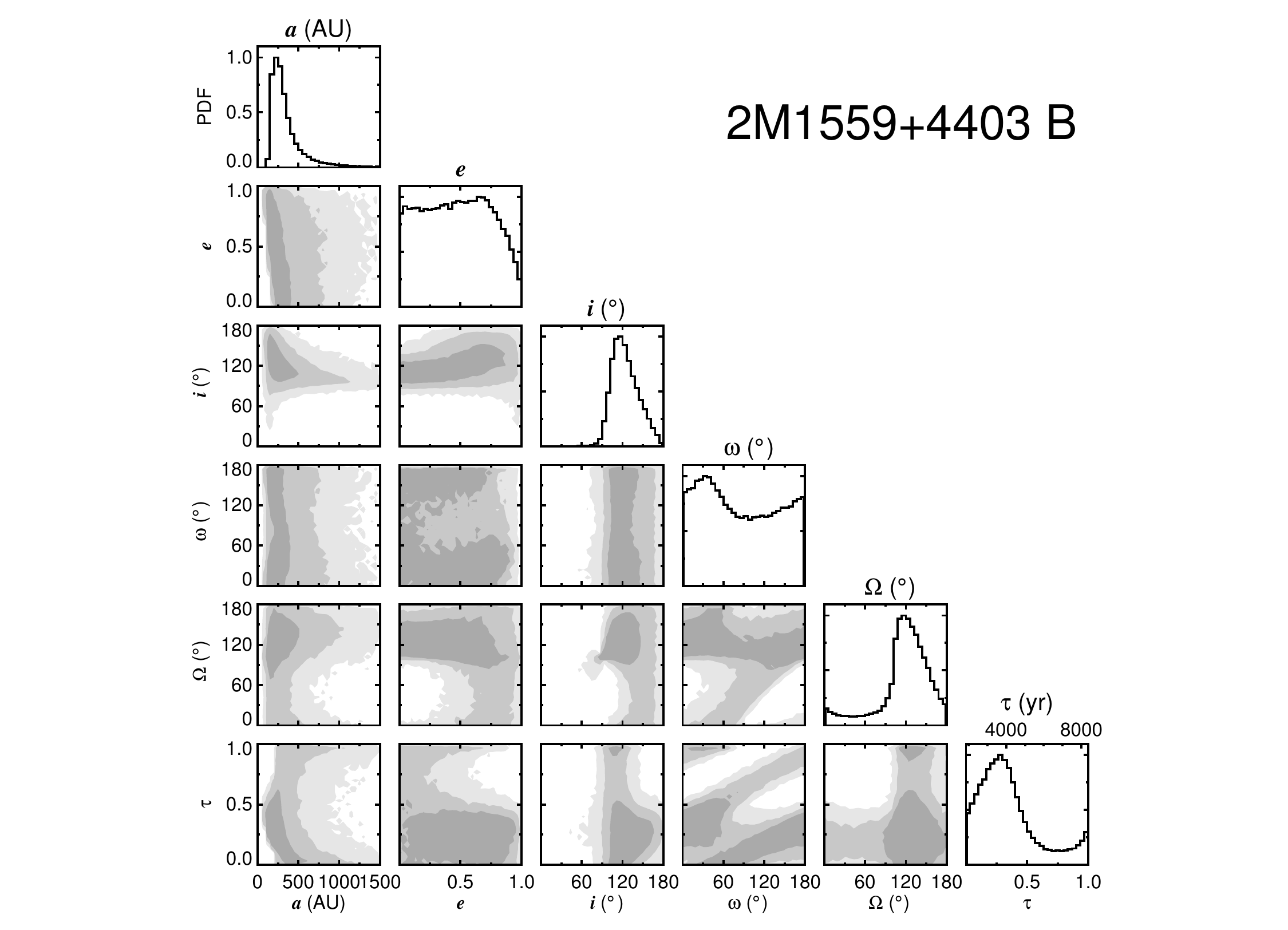}}
  \vskip 0 in
  \caption{Corner plot for 2M1559+4403 B.   See Figure~\ref{fig:hd49197corner} for details.
    \label{fig:2m1559corner} } 
\end{figure*}

\begin{figure*}
  \vskip -0. in
  \hskip 0.8 in
  \resizebox{5.5in}{!}{\includegraphics{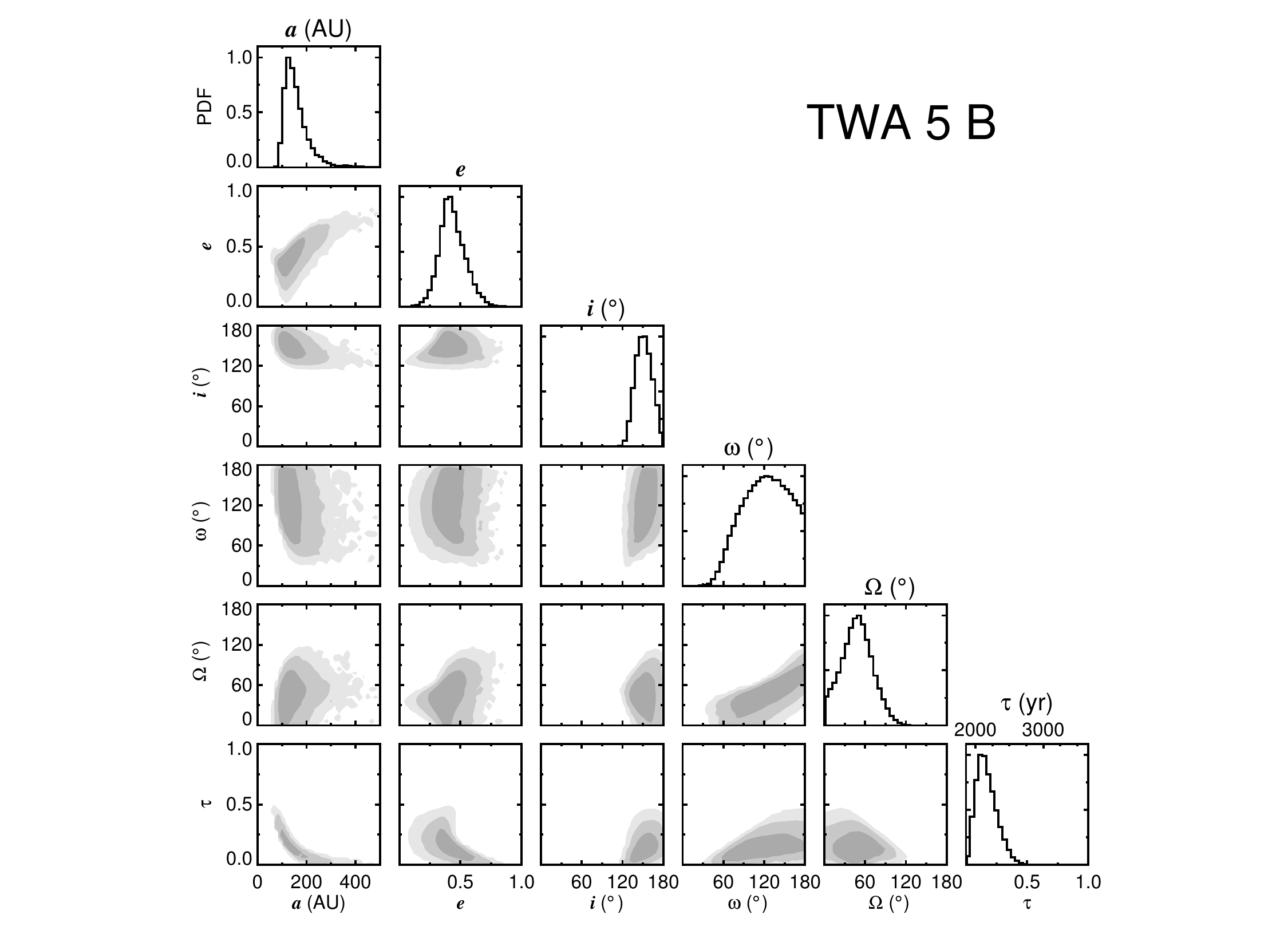}}
  \vskip 0 in
  \caption{Corner plot for TWA 5 B.   See Figure~\ref{fig:hd49197corner} for details.
    \label{fig:twa5corner} } 
\end{figure*}

\begin{figure*}
  \vskip -0. in
  \hskip 0.8 in
  \resizebox{5.5in}{!}{\includegraphics{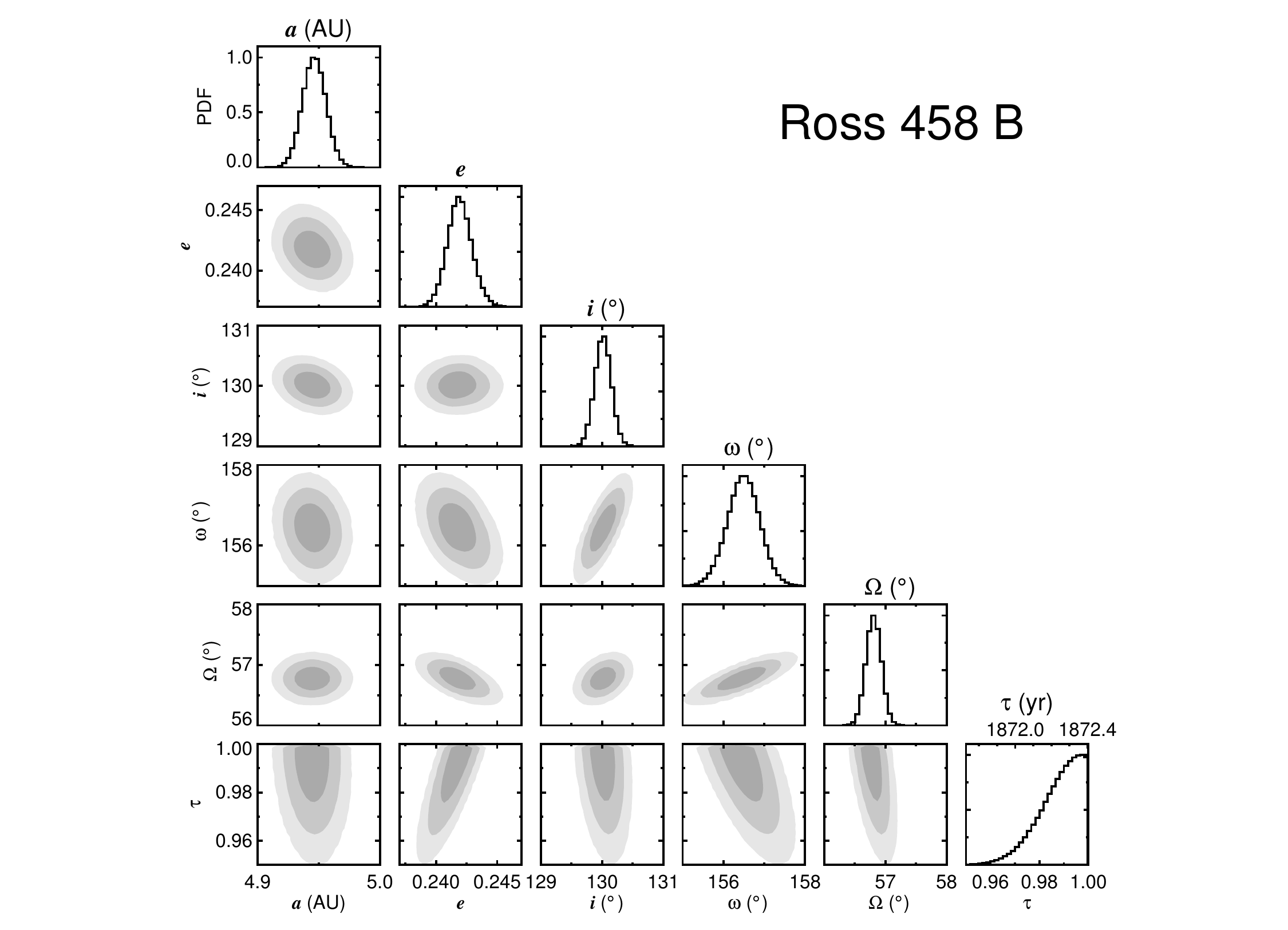}}
  \vskip 0 in
  \caption{Corner plot for Ross 458 B.   See Figure~\ref{fig:hd49197corner} for details.
    \label{fig:ross458corner} } 
\end{figure*}

\begin{figure*}
  \vskip -0. in
  \hskip 0.8 in
  \resizebox{5.5in}{!}{\includegraphics{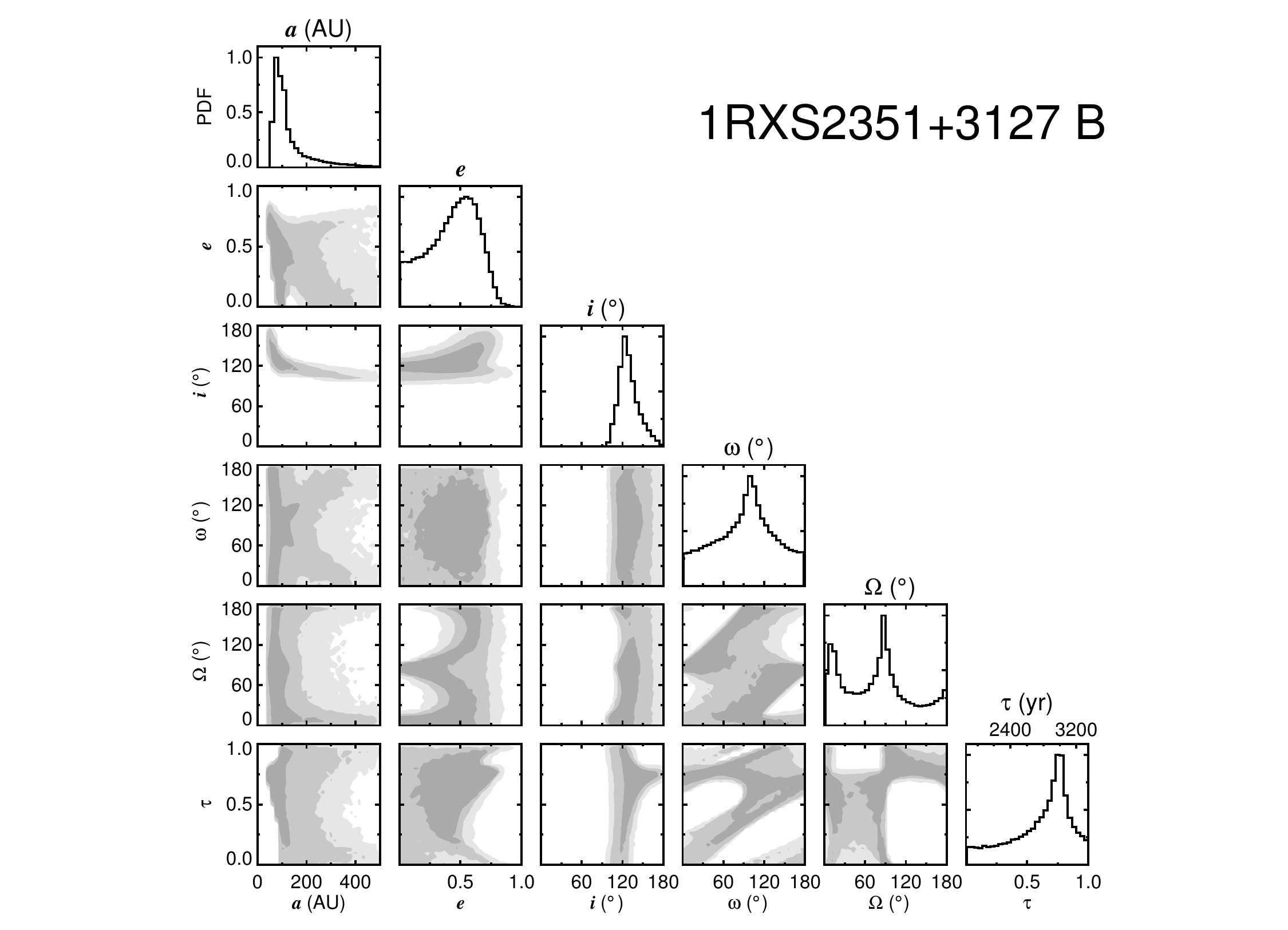}}
  \vskip 0 in
  \caption{Corner plot for 1RXS2351+3127 B.   See Figure~\ref{fig:hd49197corner} for details.
    \label{fig:ross458corner} } 
\end{figure*}

\begin{figure*}
  \vskip -0. in
  \hskip 0.8 in
  \resizebox{5.5in}{!}{\includegraphics{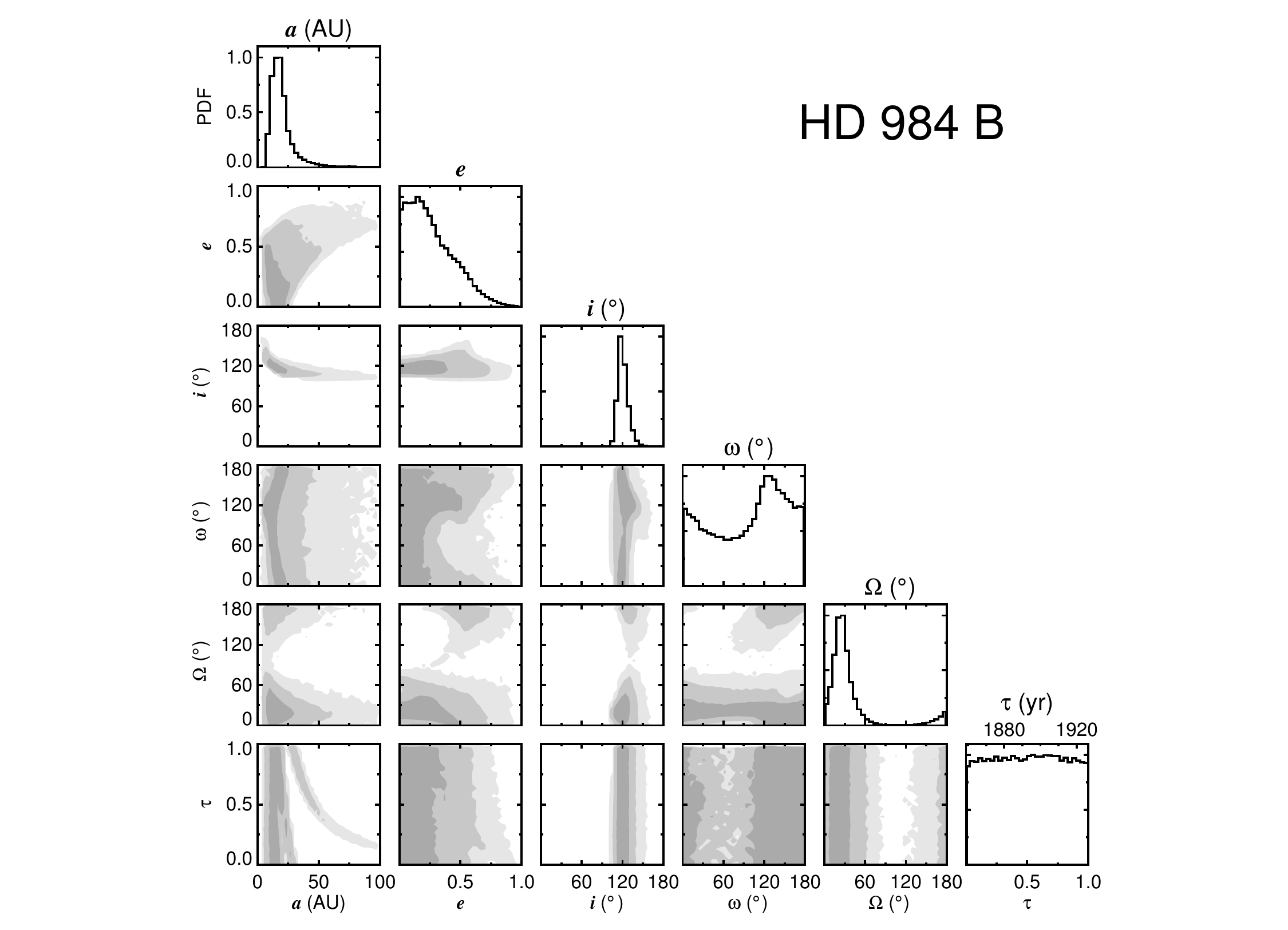}}
  \vskip 0 in
  \caption{Corner plot for HD 984 B.   See Figure~\ref{fig:hd49197corner} for details.
    \label{fig:hd984corner} } 
\end{figure*}

\begin{figure*}
  \vskip -0. in
  \hskip 0.8 in
  \resizebox{5.5in}{!}{\includegraphics{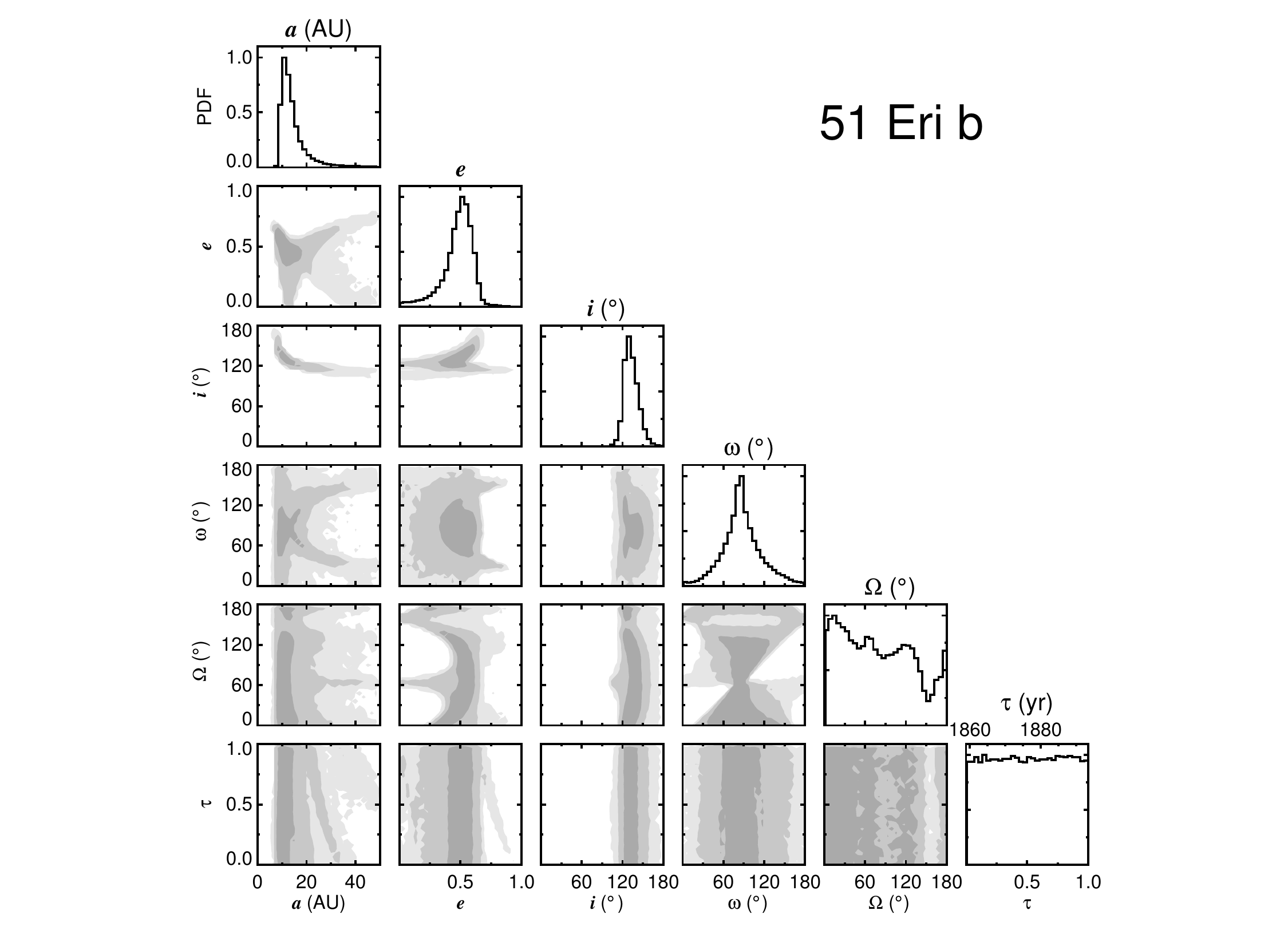}}
  \vskip 0 in
  \caption{Corner plot for 51 Eri b.   See Figure~\ref{fig:hd49197corner} for details.
    \label{fig:51ericorner} } 
\end{figure*}

\begin{figure*}
  \vskip -0. in
  \hskip 0.8 in
  \resizebox{5.5in}{!}{\includegraphics{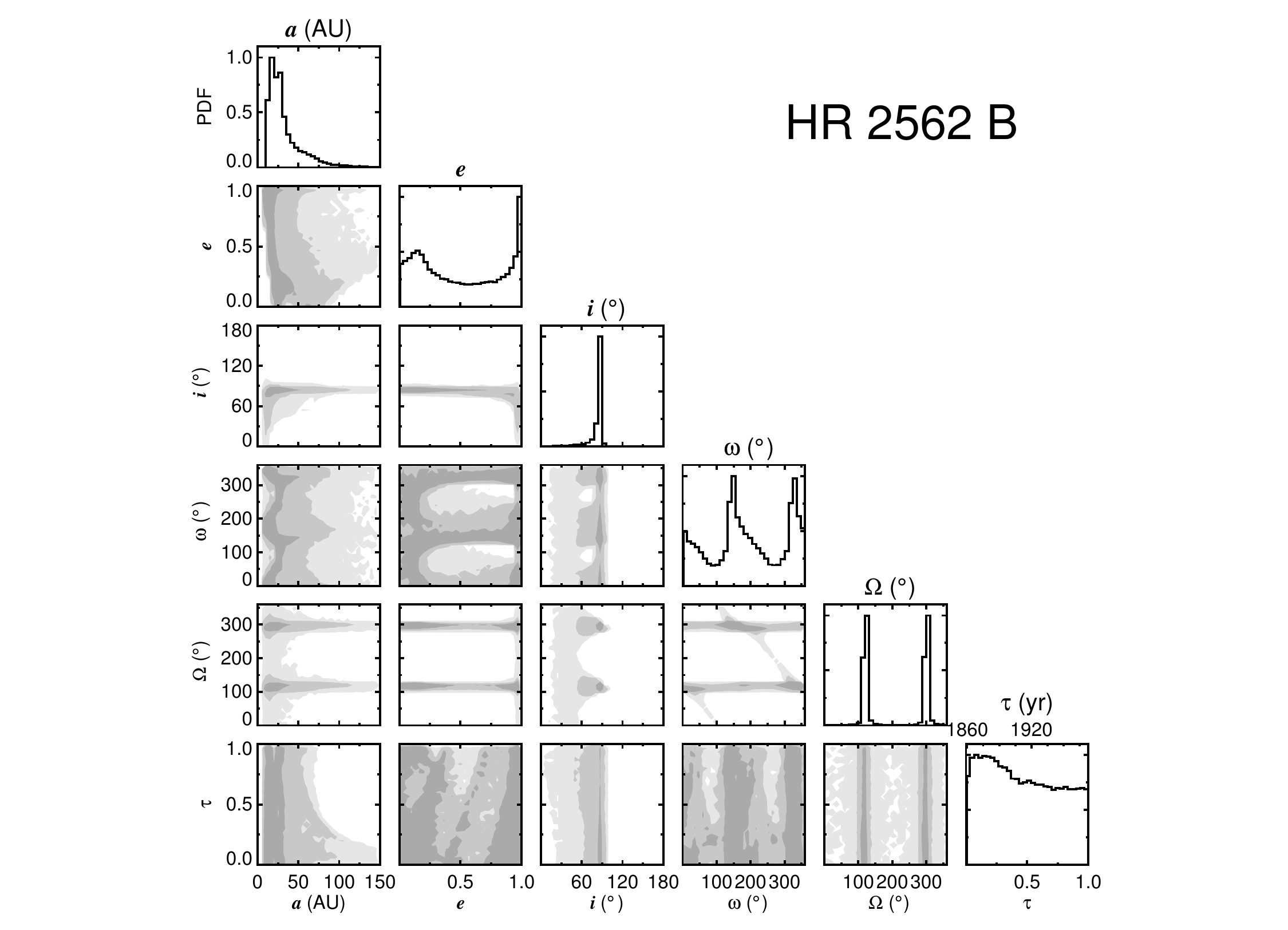}}
  \vskip 0 in
  \caption{Corner plot for HR 2562 B.   See Figure~\ref{fig:hd49197corner} for details.
    \label{fig:hr2562corner} } 
\end{figure*}

\begin{figure*}
  \vskip -0. in
  \hskip 0.8 in
  \resizebox{5.5in}{!}{\includegraphics{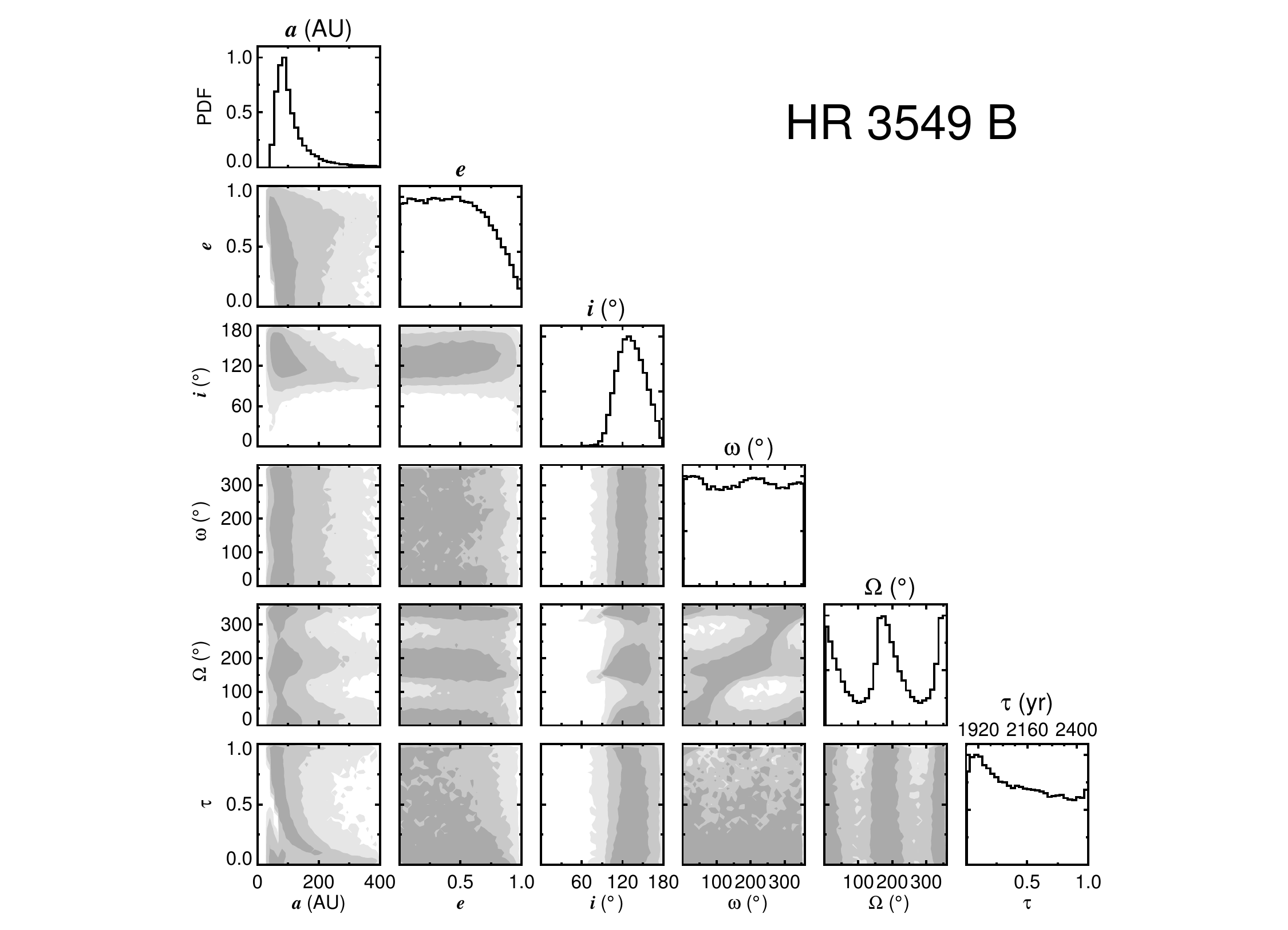}}
  \vskip 0 in
  \caption{Corner plot for HR 3549 B.   See Figure~\ref{fig:hd49197corner} for details.
    \label{fig:hr3549corner} } 
\end{figure*}

\begin{figure*}
  \vskip -0. in
  \hskip 0.8 in
  \resizebox{5.5in}{!}{\includegraphics{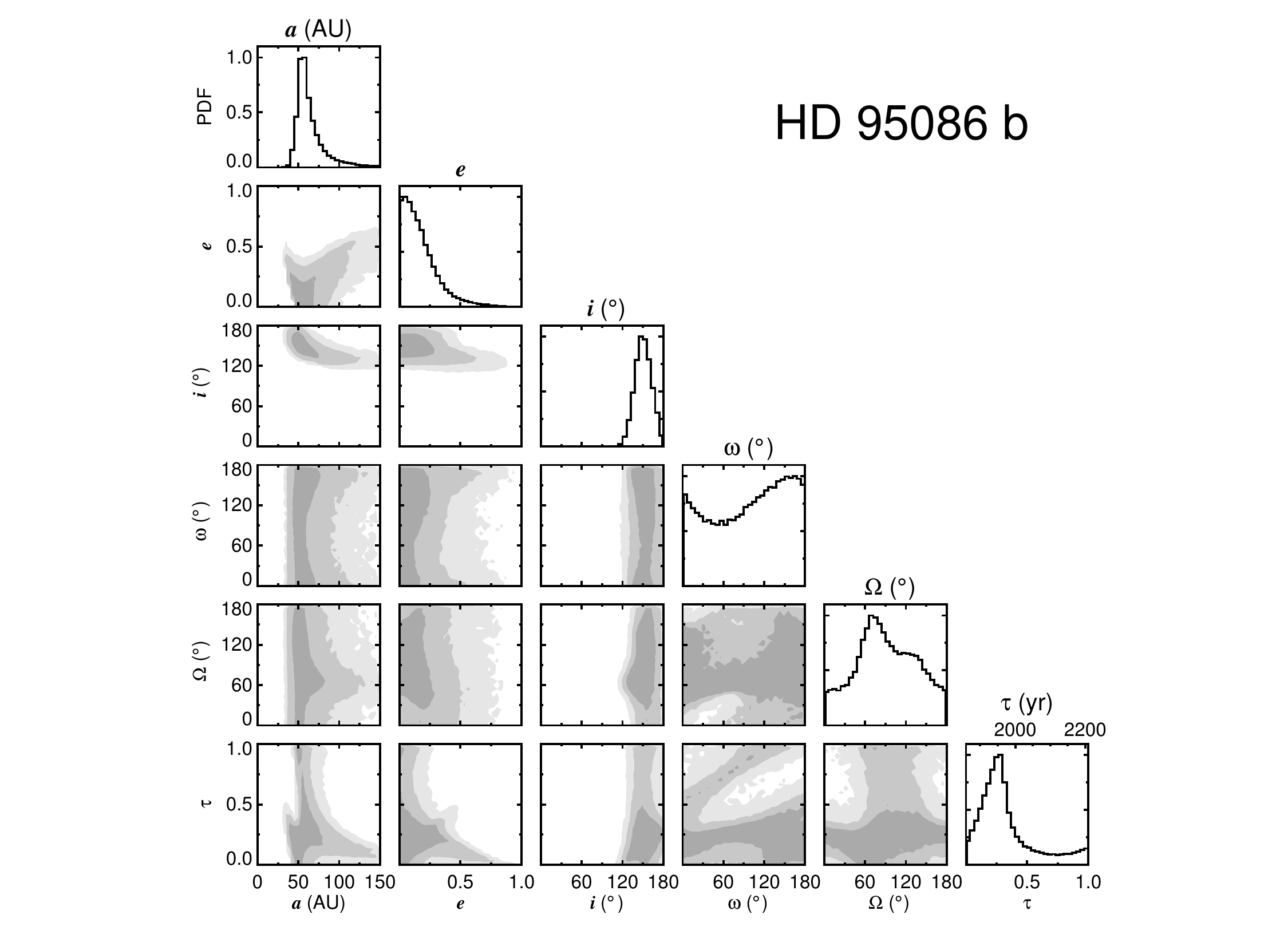}}
  \vskip 0 in
  \caption{Corner plot for HD 95086 b.   See Figure~\ref{fig:hd49197corner} for details.
    \label{fig:hd95086corner} } 
\end{figure*}

\begin{figure*}
  \vskip -0. in
  \hskip 0.8 in
  \resizebox{5.5in}{!}{\includegraphics{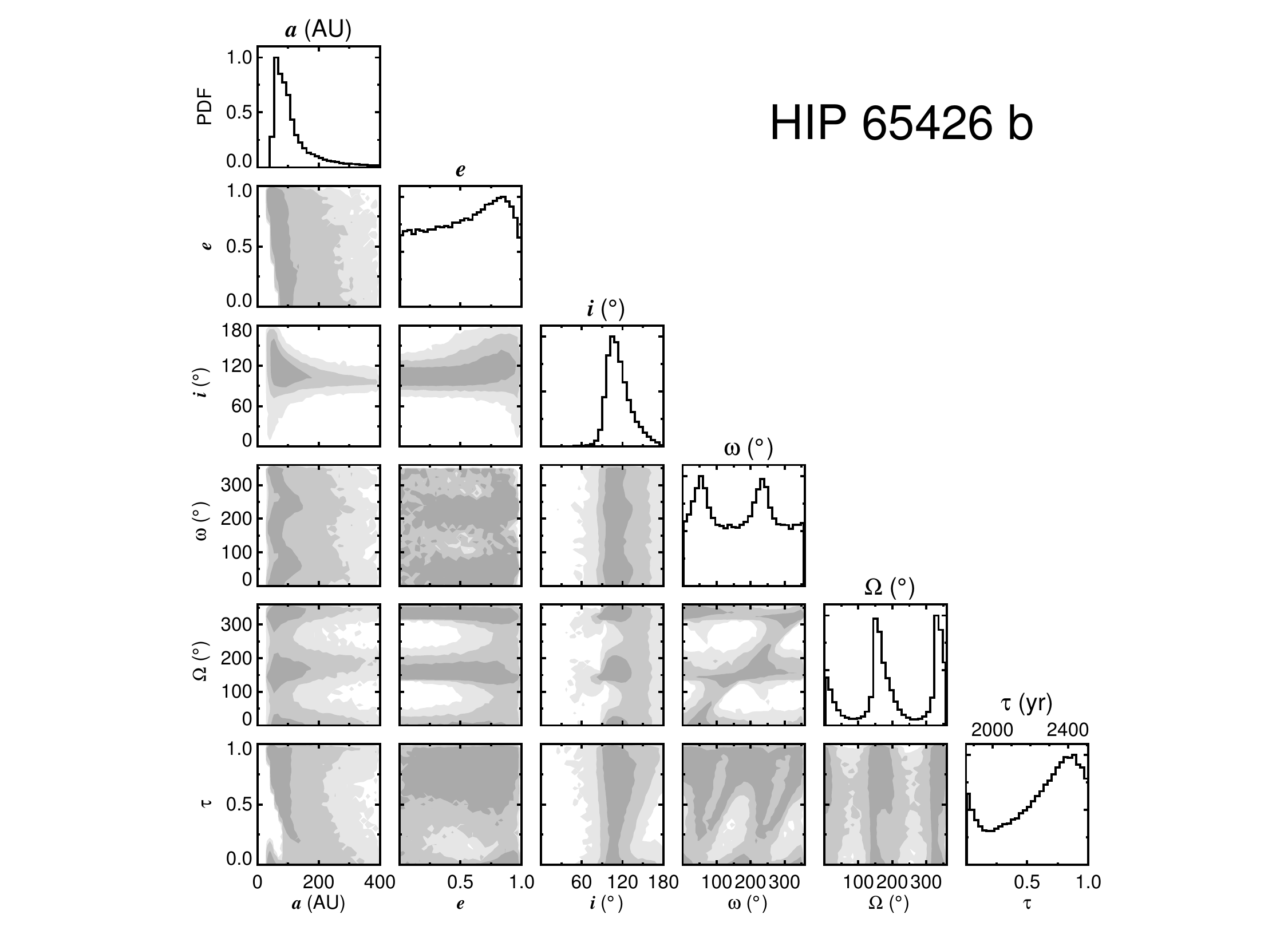}}
  \vskip 0 in
  \caption{Corner plot for HIP 65426 b.   See Figure~\ref{fig:hd49197corner} for details.
    \label{fig:hip65426corner} } 
\end{figure*}

\begin{figure*}
  \vskip -0. in
  \hskip 0.8 in
  \resizebox{5.5in}{!}{\includegraphics{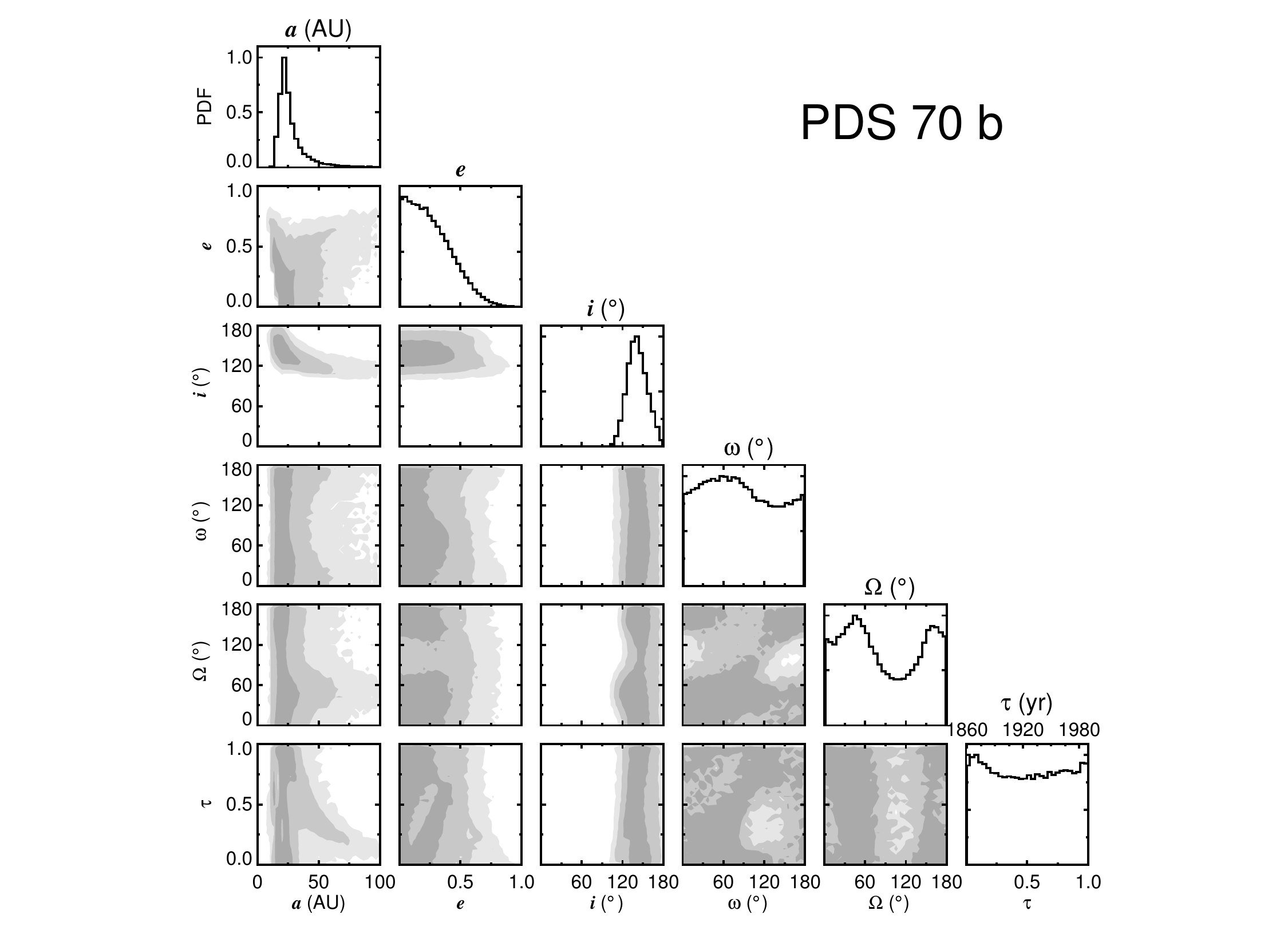}}
  \vskip 0 in
  \caption{Corner plot for PDS 70 b.   See Figure~\ref{fig:hd49197corner} for details.
    \label{fig:pds70corner} } 
\end{figure*}

\begin{figure*}
  \vskip -0. in
  \hskip 0.8 in
  \resizebox{5.5in}{!}{\includegraphics{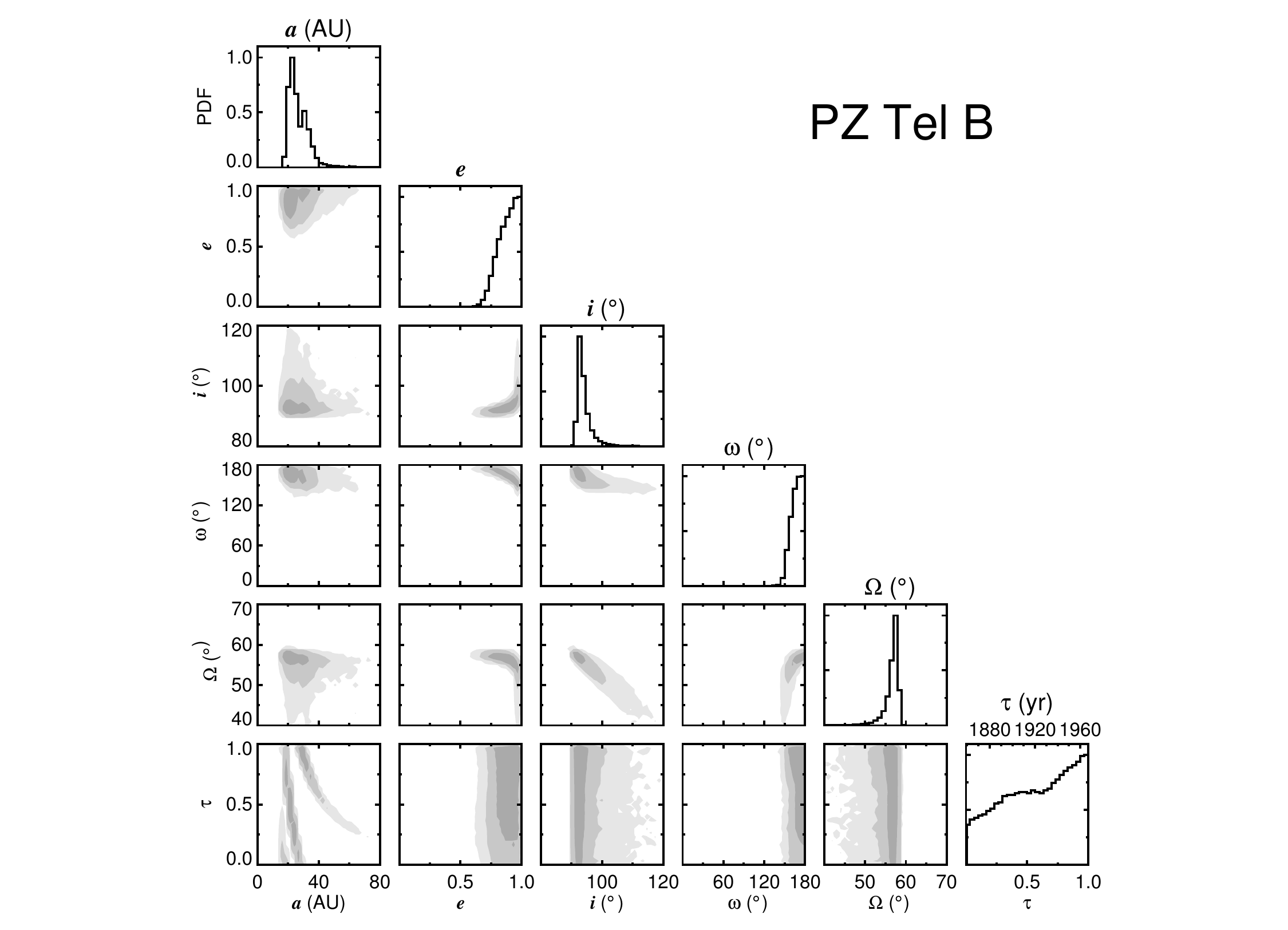}}
  \vskip 0 in
  \caption{Corner plot for PZ Tel B.   See Figure~\ref{fig:hd49197corner} for details.
    \label{fig:pztelcorner} } 
\end{figure*}

\begin{figure*}
  \vskip -0. in
  \hskip 0.8 in
  \resizebox{5.5in}{!}{\includegraphics{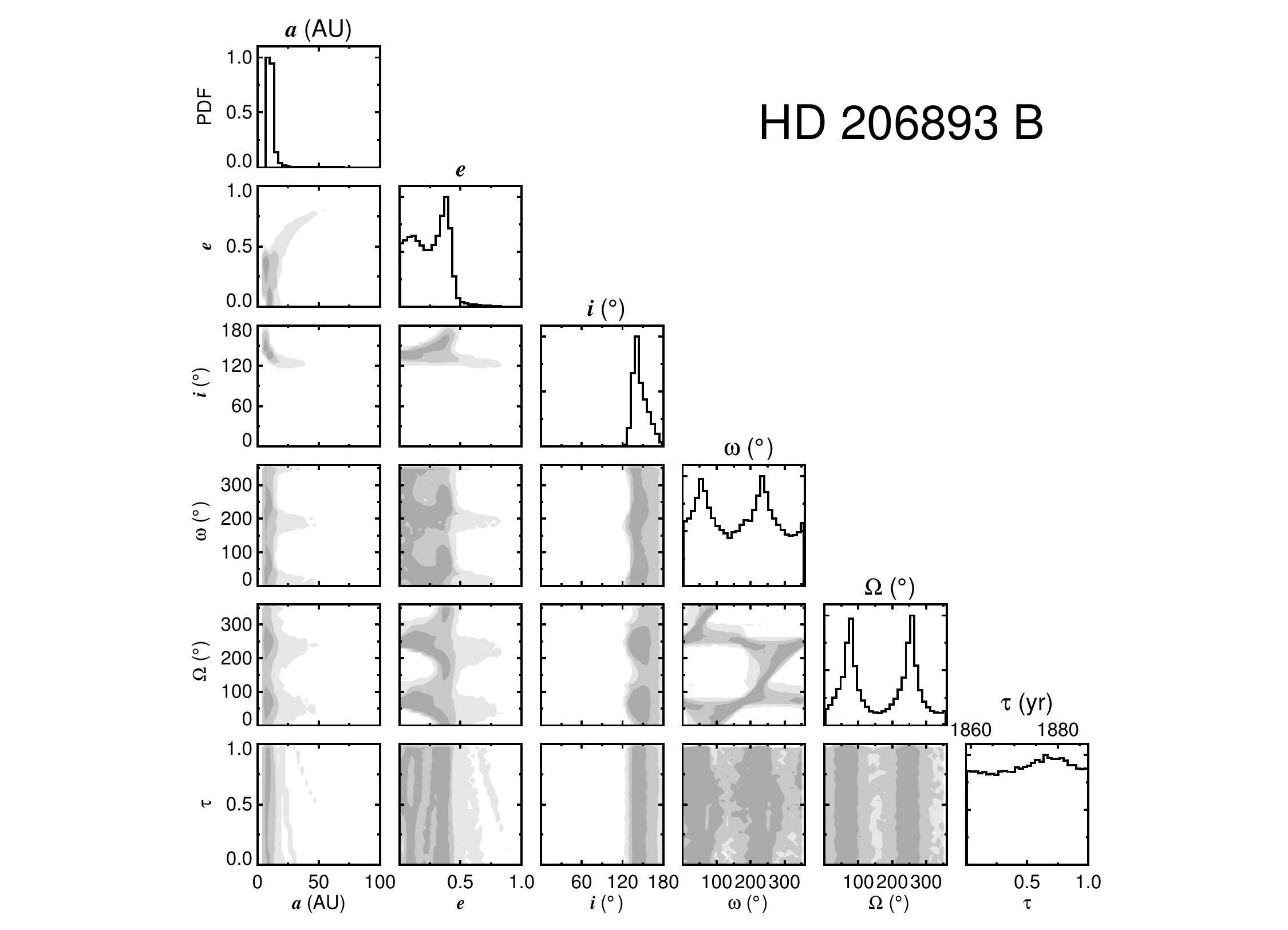}}
  \vskip 0 in
  \caption{Corner plot for HD 206893 B.   See Figure~\ref{fig:hd49197corner} for details.
    \label{fig:hd206893corner} } 
\end{figure*}

\clearpage
\newpage

\acknowledgments

We thank the anonymous referee for constructive suggestions that improved the quality of this manuscript;
Jason Wang, Erik Petigura, Michael Liu, and Quang Tran for helpful feedback on this study;
and 
Eugene Chiang for discussions about long-period planet formation.
The eccentricity posteriors for Gl 229 B and the HR 8799 planets were shared
by Tim Brandt and Jason Wang.
Brian Mason provided astrometry in the WDS catalog for Ross 458.
B.P.B. acknowledges support from the National Science Foundation grant AST-1909209.
S.B. is supported by the NSF Graduate Research Fellowship, grant No. DGE 1745303. 
This work was supported by a NASA Keck PI Data Award, administered by the NASA Exoplanet Science Institute. Data presented herein were obtained at the W. M. Keck Observatory from telescope time allocated to the National Aeronautics and Space Administration through the agency's scientific partnership with the California Institute of Technology and the University of California. The Observatory was made possible by the generous financial support of the W. M. Keck Foundation.

This publication makes use of data products from the Two Micron All Sky Survey, which is a joint project of the University of Massachusetts and the Infrared Processing and Analysis Center/California Institute of Technology, funded by the National Aeronautics and Space Administration and the National Science Foundation
This work has made use of data from the European Space Agency (ESA) mission
{\it Gaia} (\url{https://www.cosmos.esa.int/gaia}), processed by the {\it Gaia}
Data Processing and Analysis Consortium (DPAC,
\url{https://www.cosmos.esa.int/web/gaia/dpac/consortium}). Funding for the DPAC
has been provided by national institutions, in particular the institutions
participating in the {\it Gaia} Multilateral Agreement.
 NASA's Astrophysics Data System Bibliographic Services together with the VizieR catalogue access tool and SIMBAD database 
operated at CDS, Strasbourg, France, were invaluable resources for this work.
The authors wish to recognize and acknowledge the very significant cultural role and reverence that the summit of Maunakea has always had within the indigenous Hawaiian community.  We are most fortunate to have the opportunity to conduct observations from this mountain.

\facilities{Keck:II (NIRC2), Subaru (HiCIAO)}

\clearpage
\newpage


\end{document}